UNIVERSIDADE DE LISBOA
FACULDADE DE CIÊNCIAS

# The *BFCG* Theory and Canonical Quantization of Gravity

**Doutoramento em Física**

Miguel Ângelo Oliveira

Tese orientada por:
Professor Aleksandar Miković    Professor José Pedro Mimoso

Documento especialmente elaborado para a obtenção do grau de doutor

2017

# The *BFCG* Theory and Canonical Quantization of Gravity

## Resumo

Das quatro interacções fundamentais conhecidas na Natureza, a gravidade é a que melhor é acessível aos sentidos , bem como aquela que mais está presente na vida de todos os dias. A melhor descrição que a investigação científica foi capaz de produzir da gravitação é a Teoria da Relatividade Geral (RG) que foi criada por Albert Einstein no principio do século XX, mais precisamente em 1915.

O mesmo século vinte testemunhou também o nascimento daquela que seria outro dos pilares da Física contemporânea: a Teoria Quântica. Esta assenta sobre a compatibilidade das teorias físicas com o chamado Princípio de Incerteza formulado por Werner Heisenberg. Desta maneira foi criado um programa que tem como objetivo a quantização das interações físicas . Este programa foi bem sucedido em três das interações fundamentais, nomeadamente: a interação eletromagnética, a interação (nuclear) fraca e, a interação (nuclear) forte. No caso destas interações, este programa não só levou a cabo a dita quantização mas produziu também a unificação destas interações. A teoria que saiu deste processo dá pelo nome de Modelo Padrão da Física de Partículas (MP) e representa um dos grandes (se não o maior) triunfos da Física contemporânea. Só a interação gravítica permanece portanto fora deste programa.

A RG é, de facto, uma teoria clássica. Termo que em Física se toma por significar não-quântica. O método mais comummente utilizado para levar a cabo a quantização de uma teoria é a denominada Quantização Canónica (QC). Esta assenta na utilização do formalismo Hamiltoniano, baseado numa distinção entre o espaço e o tempo. Em algumas teorias aparecem relações entre as variáveis (e os momentos) chamadas constrangimentos. Estes constrangimentos implicam que nem todas as variáveis iniciais correspondem a graus de liberdade físicos da teoria. Assim sendo, torna-se necessário utilizar um método, o Procedimento de Dirac, para determinar quais são os graus de liberdade da teoria em questão, bem assim como para explicitar o numero e a natureza dos constrangimentos. No caso da RG, a separação entre espaço e tempo e a subsequente formulação Hamiltoniana dá pelo nome de formalismo Arnowitt Deser Misner (ADM). A QC aplicada à RG dá origem a equação de Wheeler–DeWitt (WdW) que para além de não estar rigorosamente definida é muito difícil de resolver. Esta dificuldade na resolução tem a ver com o facto de o Constrangimento Hamiltoniano (CH) não ser polinomial.

Numa tentativa de resolver este problema Abhay Ashtekar, encontrou um novo conjunto de variáveis para a RG. Esta teoria é comummente formulada em termos da métrica, ao passo que estas novas variáveis têm o carácter de conexões. O CH, em função destas novas variáveis torna-se de facto polinomial. No entanto, dado que as variáveis de Ashtekar são complexas, torna-se necessária a introdução de uma nova condição, a condição de realidade. Esta condição é por sua vez, também difícil de quantizar. Podem ser utilizadas variáveis análogas reais mas, neste caso, o CH torna-se de novo não polinomial.

As dificuldade encontradas na resolução do CH no contexto do formalismo canónico, levaram ao desenvolvimento de uma abordagem da quantização baseada em integrais de caminho. Mais precisamente, numa generalização destes. Esta abordagem é conhecida como modelos de spin foam. Estes modelos têm no entanto, o problema do limite clássico. Este limite deve ser a RG mas é difícil de calcular. Têm também o problema do acoplamento de fermiões à spin foam. Ambos estes problemas estão relacionados com o facto de os comprimentos das arestas, ou equivalentemente as tetradas, nem sempre estarem definidos na gravidade quântica de spin foams.

Para se introduzir os comprimentos das arestas no formalismo das spin foams, torna-se necessários introduzir as tetradas na formulação da Relatividade Geral baseada na teoria *BF* (as letras representam os campos que aparecem na ação). Isto é levado a cabo usando uma formulação da RG baseada no 2-grupo de Poincaé. A ideia geral é reformular a RG como uma teoria topológica do tipo *BFCG* (as letras representam os campos que aparecem na ação) constrangida. Este método é uma generalização categórica da formulação da RG baseada na teoria *BF* que por sua vez é a base dos modelos de spin foam.

A reformulação da RG, em termos da teoria *BFCG*, é especialmente apropriada para a quantização baseada em integrais de caminho. Assim como da teoria *BF* constrangida se derivam os modelos de spin foam, no caso da teoria *BFCG* constrangida obtêm-se os modelos spin-cube. Estes modelos representam uma generalização categórica dos modelos de Spin foam.

No que diz respeito à QC, o progresso está a ser obstaculizado pelo facto de a teoria *BFCG* ter uma estrutura canónica complicada. Uma estratégia razoável consiste em estudar em primeiro lugar uma teoria mais simples. Neste caso decidimos começar pela teoria *BFCG* topológica, ou seja sem graus de liberdade locais. A imposição posterior de um constrangimento conduz a uma teoria que é equivalente à Relatividade Geral. A analise canónica da teoria *BFCG*, primeiro passo para a sua quantização canónica, pode mostrar-se ser simples, este é um dos resultados deste trabalho. Outra característica digna de nota é o fato de que a teoria *BFCG* é equivalente á teoria *BF* para o grupo de Poincaré. Dado que a quantização canónica da teria *BFCG* se encontra muito dificultada pelo facto de se desconhecer, até ao momento, o equivalente para 2-grupos do teorema de Peter Weyl, a eqivalencia mencionada acima com o teoria *BF* torna-se particularmente útil. Sabendo que a teoria *BF* foi já quantizada usando quantização canónica. Torna-se possível, graças a esta equivalência fazer uma quantização canónica da teoria *BFCG* diretamente em termos das variáveis da teoria *BF*. Esta abordagem, alem de ser mais simples do que a quantização direta da teoria *BFCG*, pode também ajudar à compreensão da quantização na base spin foam que por sua vez é uma generalização categórica da quantização na base de spin network.

Em conclusão, nesta tese levamos a cabo análise canónica da teoria *BFCG* para o 2-grupo de Poincaré em termos das 2-conexões espaciais e os seus momentos canónicos conjugados. Encontramos uma equivalência dinâmica entre a ação de *BFCG* para o 2-grupo de Poincaré e a ação *BF* para o grupo de Poincaré e determinamos a transformação canónica que relaciona as duas teorias. Estudamos a quantização canónica da teoria *BFCG* passando à base da conexão de Poincaré. A quantização na base da 2-conexão pode então ser encontrada efetuando uma transformada de Fourier. Discutimos também muito brevemente o problema de como construir uma base de estados de spin foam. Estes estados são a generalização categórica dos estados de spin network.

Na primeira parte deste trabalho, utilizou-se um método abreviado para encontrar a estrutura canónica da teoria *BFCG*. Num artigo subsequente generalizamos o método utilizado. Para isto introduzimos uma ação que interpola entre a teoria *BFCG* para o 2-grupo de Poincaré e a teoria *BF* para o grupo de Poincaré, reduzindo-se a estas teorias em limites apropriados de um parâmetro escolhido. Fizemos em seguida a análise canónica desta teoria mais geral, utilizando o procedimento de Dirac. Desta maneira fomos capazes de reter toda a liberdade de gauge da teoria. Encontramos que, apesar de haver uma equivalência dinâmica entre o *BFCG* para o 2-grupo de Poincaré e o *BF* para o grupo de Poincaré, estas teorias têm estruturas canónicas muito diferentes.

# The *BFCG* Theory and Canonical Quantization of Gravity

Abstract

Of the four interactions known in Nature, gravity is the most accessible to the senses. The best description we have, so far, of the gravitational interaction is the theory of General Relativity (GR) developed by Albert Einstein in the beginning of the XX$^{th}$ century.

This same century also witnessed the birth of Quantum Theory. This means that the Principle of Uncertainty formulated by Werner Heisenberg must be taken into account in the construction of physical theories. This way a programme was set up to quantize interactions. This quantization was accomplished in all but the gravitational force.

GR in fact is a classical theory. The most commonly used method to carry forth the quantization of a theory is called Canonical Quantization (CQ). This method applied to GR gives rise to the Wheeler–DeWitt equation which is ill defined and notoriously hard to solve. This difficulty is for the most part connected to the non-polynomial character of the Hamiltonian Constraint (HC).

To alleviate this problem, Abhay Ashtekar found a new set of variables for GR. Written in these variables, the HC has a polynomial character. However, given that Ashtekar's variables are complex, the necessity arises for a new condition, the reality condition which is very hard to quantize. A real version of Ashtekar's variables may be used but the Hamiltonian Constraint is again non-polynomial.

The difficulties of solving the HC in the canonical formalism have led to the development of a path-integral quantization approach known as spin-foam models (SF). SF models have the problem of the classical limit and the problem of the coupling of fermionic matter. These problems are related to the fact that the edge-lengths, or the tetrads, are not always defined in a SF model of quantum gravity.

In order to introduce the edge lengths in the SF formalism, one has to introduce the tetrads in the *BF* (the letters represent the fields featuring in the action) theory formulation of GR. This can be done by using a formulation of GR based on the Poincaré 2-group. The idea is to reformulate GR as a constrained topological theory of the *BFCG* (the letters represent the fields featuring in the action) type. This approach is a categorical generalization of the constrained *BF* theory formulation of GR which is used for the SF models.

The *BFCG* reformulation of GR is useful for the path-integral quantization. In this case one obtains the spin-cube models, which represent a categorical generalization of the SF models. As far as the CQ is concerned, the progress has been hindered because the constrained *BFCG* theory has a complicated canonical structure. A reasonable strategy is to study first a simpler theory, which is the unconstrained *BFCG* theory. This is a topological gravity theory, and we will show that its canonical formulation is simple to understand. Another feature of this theory is that it is equivalent to the Poincaré group *BF* theory, so that one can perform a CQ in terms of the *BF* theory variables. This is mathematically simpler than performing a CQ in terms of the BFCG theory variables and it can also help to understand the quantization based on a spin-foam basis, which is a categorical generalization of the spin-network basis.

Dedicated to my parents and my wife.

# Acknowledgments

A very special thanks goes to my supervisor, Professor Aleksandar Miković. All this work was only possible because of his help, patience and trust.

A very special word of gratitude goes to my co-supervisor, Professor José Pedro Mimoso. For his invaluable help with bureaucratic matters, also for providing me with the opportunity to go to conferences, and for insightful discussions about cosmology.

I am also deeply indebted to Professor José Cidade Mourão for his precious help in the last and final version of this thesis.

I also would like to thank Doctor Marko Vojinović for help and useful discussions in the context of this work, and especially regarding the canonical formalism.

I am profoundly grateful to my parents, Fernando and Leonor, and to my wife Sónia to whom this thesis is dedicated, for all their support and patience.

This work was funded by the FCT PhD grant SFRH/BD/79285/2011.

I also acknowledge the use of the XeLaTeX typesetting system and of BibTeX for the bibliography as well as Ti*k*Z for the figures.

# Contents

# Part I

# General formalism

# Introduction

At the end of the nineteenth century the Newtonian view prevalent in Physics was challenged by two sets of problems. The first set contained among other issues the *black-body radiation*, the *photoelectric effect,* and, the stability of matter at the atomic level, whereas a second group incorporated the *electrodynamics of moving bodies,* or the possible existence of a preferred frame for the propagation of light. These two classes of scientific questions would develop into the two major theories of the last century. These theories Quantum Mechanics and General Relativity seem to be incompatible. Their harmonization is the subject of this thesis.

Also in mathematics, the last century witnessed some major developments. Among these we can count category theory. A category is the simplest framework where we can talk about objects and morphisms. These categories proved to be important in areas like physics, topology, logic and computation (see [1]). In physics, categories may be thought of as generalizations of Feynman diagrams.

This work revolves around the use of category theory as a way to generalize gauge theories, and the application of one such generalization — the *BFCG* theory to the problem of Quantum Gravity. It was carried out in the context of a Doctoral Programme at Faculdade de Ciências da Universidade de Lisboa (FCUL), and in the Grupo de Física Matemática da Universidade de Lisboa (GFM).

This work produced the following papers:

- A. Miković and M. A. Oliveira,
  "Canonical formulation of Poincaré *BFCG* theory and its quantization",
  Gen. Rel. Grav. **47**, no. 5, 58 (2015) [arXiv:1409.3751 [gr-qc]].

- A. Miković, M. A. Oliveira and M. Vojinović,
  "Hamiltonian analysis of the *BFCG* theory for the Poincaré 2-group",
  Class. Quant. Grav. **33**, no. 6, 065007 (2016) [arXiv:1508.05635 [gr-qc]].

- A. Mikovic, M. A. Oliveira and M. Vojinovic,
  "Hamiltonian analysis of the *BFCG* theory for a generic Lie 2-group",
  arXiv:1610.09621 [math-ph].

the third awaits publication.

This thesis is composed of two parts: Part I in which we deal with the general formalism and, Part II where we present the *BFCG* theory.

Specifically, concerning the general formalism, in chapter 1 we summarize the problem of Quantum Gravity and explain how it is rooted in the development of physics in the last century. In chapter 2 we review some mathematical tools and methods useful for the discussion of gauge theories. In this chapter we also present some relevant (both because of their similarity to the General Theory of Relativity and, because they are quantized theories) examples of gauge theories. In chapter 3 we give a brief account of the Dirac procedure for the Hamiltonian analysis of constrained theories. In this chapter we also apply this procedure to some of the examples given in chapter 2. In chapter 4 we discuss the Hamiltonian formulation of General Relativity. This is a crucial step for the canonical quantization of Einstein's theory of gravitation. And, in chapter 5 we introduce the Ashtekar variables, as well as holonomies, fluxes and some of the Loop Quantum Gravity (canonical) formalism.

Regarding the *BFCG* theory, we discuss in chapter 6 some aspects of category theory. We also generalize categories to higher categories. We give the notion of 2-group wich is of paramount importance for the *BFCG* theory. Also in this chapter we explain the meaning of the term *categorical generalization* and that of *higher gauge theory*. In chapter 7 we do the canonical analysis of the *BFCG* theory for a generic Lie 2-group. In chapter 8 we discuss the Hamiltonian structure of the *BFCG* theory for the Poincaré 2-group. We also study the relation between the *BFCG* theory and the topological Poincaré Gauge Theory. Finally in chapter 9 we present our conclusions.

Some of the chapters of this thesis represent original results. This is the case of chapters 7 and 8. Additionally appendix D also contains original results from an ongoing work related to the canonical analysis of the constrained *BFCG* theory. They are presented as an appendix because they are to this date unfinished.

# 1

# Quantum Gravity: a *bird's-eye view*

## 1.1 The twofold way of physics in the last century

The first group of questions gave rise to Quantum Mechanics (QM). In 1900 Max Planck put forth a hypothesis to account for the radiation function of a *black body* (see [2, 3]). This consisted in assuming that the blackbody emits energy in a discrete way, with each energy packet obeying a relation between energy $E$ and frequency $v$ given by,

$$E = hv \tag{1.1}$$

where $h = 6.626 \times 10^{-34} m^2\, kg\, s^{-1}$ $(Js)$ is Planck's constant. Later in 1905, Einstein in an effort to explain the photoelectric effect, hypothesised that light (and electromagnetic radiation in general) exists as discrete *quanta*, called *photons*, each of which obeys the same relation, now called the Planck-Einstein relation.

Quantum Mechanics grew out of this hypothesis. It is based on *Heisenberg's uncertainty principle* first introduced in 1927. This principle states that the uncertainty in the position of a particle $\Delta x$ and in the momentum $\Delta p_x$ are related by,

$$\Delta x \Delta p_x \geq \frac{\hbar}{2}, \tag{1.2}$$

where the reduced Planck constant is $\hbar = h/(2\pi)$.

Also pivotal in the development of QM was the de Broglie hypothesis, according to which, a particle (an electron for example) with momentum $p$ caries a wavelength $\lambda$ given by,

$$\lambda = \frac{h}{p}. \tag{1.3}$$

A similar concept of wavelength associated to a material particle was discovered by Arthur Compton when analysing the scattering of photons by charged particles. He found that a transfer of energy occurs between the photon and the charged particle (decrease in the energy of the photon and increase in that of the charged particle in the Compton effect) and that this particle (the photon) behaves in respect to this transfer as having a wavelength,

$$\lambda_c = \frac{h}{mc}\,, \tag{1.4}$$

called the Compton wavelength.

In the the first decades of the twentieth century, QM was developed by Schrödinger, Dirac, Bohr, Born and others. In QM states are vectors of a Hilbert space, the dynamical variables are hermitian operators and (in the Schrödinger picture) the evolution of the states is given by the Schrödinger equation. For one particle, in the Schrödinger representation, the Hilbert space is $\mathrm{L}^2(\mathbb{R}, dxdydz)$ and the Schrödinger equation reads:

$$i\hbar\frac{\partial}{\partial t}\Psi(\vec{r},t) = H\Psi(\vec{r},t)\,, \tag{1.5}$$

where $i = \sqrt{-1}$ and, $H$ is the Hamiltonian of the system.

The classical variables $x$, $p$ and $H_{cla}$ correspond to real values observables $X$, $P$, and $H$, according to the rule,

$$x \to X; \quad X\Psi = x\Psi \tag{1.6}$$

$$p \to P; \quad P_x\Psi = \frac{\hbar}{i}\frac{\partial}{\partial x}\Psi \tag{1.7}$$

$$H_{cla} \to H; \quad H = H_{cla}(X,P)\,. \tag{1.8}$$

The operators in QM are hermitian and obey commutation relations of the form,

$$[X, P_x] \equiv XP_x - P_xX = i\hbar\,. \tag{1.9}$$

This is an example of the so called canonical commutation relations which will play an important role in subsequent chapters. For a thorough formulation of QM see [4].

The problems in the second group mentioned above, were related to an incompatibility between the Maxwell theory of electromagnetism and classical mechanics. Maxwell's equations predict a constant value for the speed of light $c$ related to the (vacuum) magnetic permeability $\mu_0$, and electric permittivity $\varepsilon_0$ by $c = (\varepsilon_0\mu_0)^{-\frac{1}{2}}$.

Since in Newtonian mechanics a composition law for velocities exists, related to Galileo's principle of relativity, one may naturally ask: in which reference frame is the speed of light $c$ and, why is this frame singled out in Maxwell's equations? At the time, a purported frame for the propagation of luminous signals was postulated, and dubbed *aether*. Around 1887, the Michelson–Morley experiment was performed in an attempt to measure the speed of light in two perpendicular directions and therefore identify the *aether*. No velocity difference was ever found and, this experiment became one of the most surprising (null) results of the time. See [5] for details about this subject.

Einstein solved this issue in 1905, by postulating that: (1) Physical laws are the same in all inertial reference frames (*principle of relativity*); and (2) the speed of light (in the vacuum) $c = 2.997 \times 10^8 m\, s^{-1}$ is the same for all observers (*constancy of the speed of light*). These are the postulates of the Special Theory of Relativity (SR) (see [6] and references therein).

In SR, time intervals and lengths are not invariant by themselves, rather for any two events separated in time by $dt$ and in space by $dx\,, dy$, and $dz$, the invariant quantity is termed the space-time interval:

$$ds^2 = -c^2 dt^2 + dx^2 + dy^2 + dz^2\,. \tag{1.10}$$

This quadratic form is preserved by the set of Lorentz transformations. These transformations and the translation form the Poincaré group.

Using the Minkowski metric

$$\eta_{ab} = \begin{pmatrix} -1 & 0 & 0 & 0 \\ 0 & 1 & 0 & 0 \\ 0 & 0 & 1 & 0 \\ 0 & 0 & 0 & 1 \end{pmatrix} \tag{1.11}$$

the space-time interval may be written,

$$ds^2 = \eta_{ab} dx^a dx^b\,. \tag{1.12}$$

We are using the *Einstein summation convention* i.e. whenever an index is repeated up and down, a summation is implied. Also, in this thesis, lower-case latin letters from the beginning of the alphabet $a\,, b\,, c\ldots$ will be used when the metric involved is $\eta$. Therefore these indices will always be raised or lowered with the (flat) Minkowski metric. The points (events) in Minkowski space $M_4$ may be split into three sets, according to the sign of the interval between the point and the origin. If a point has coordinates $x^a = \left(x^0, x^i\right) \quad i = 1\,, 2\,, 3\,.$ (Latin letters from the middle of the alphabet $i, j, k \dots$ will be used for three-dimensional vectors) we have,

| | | |
|---|---|---|
| $\eta_{ab}x^a x^b > 0$ | spacelike | |
| $\eta_{ab}x^a x^b = 0$ | null or lightlike | (1.13) |
| $\eta_{ab}x^a x^b < 0$ | timelike | |

Points separated by spacelike vectors are causally disconnected that is, the event at the origin of the vector cannot influence the event at the tip since that would require a signal to travel at a velocity greater than $c$.

The relation, for a free particle, between energy and momentum in Newton's mechanics $E = p^2/(2m)$ becomes in SR

$$E^2 - (pc)^2 = (mc^2)^2 \,, \tag{1.14}$$

and the total energy is,

$$E = \frac{mc^2}{\sqrt{1 - \frac{v^2}{c^2}}} \,. \tag{1.15}$$

Whenever a body is at rest relative to an inertial reference frame, we find the well known relation $E_0 = mc^2$, between the rest energy and the mass.

The special theory of relativity, became the setting for non-gravitational physics. Physical laws are required to be invariant under Poincaré transformations, *Poincaré invariance.* QM was later harmonized with SR. First relativistic quantum mechanical equations were found. The Klein–Gordon equation reads,

$$\left(\Box + \left(\frac{mc}{\hbar}\right)^2\right) \varphi(t, \vec{r}) = 0 \,, \tag{1.16}$$

where the d'Alembert operator is define as,

$$\Box = -\eta^{ab}\partial_a\partial_b = \frac{1}{c^2}\frac{\partial^2}{\partial t^2} - \nabla^2 \,. \tag{1.17}$$

In the case of the Schrödinger equation a probability density can be built from a complex $\Psi$ it is,

$$\rho_{\text{Schrödinger}} = \Psi^*\Psi \,, \tag{1.18}$$

this quantity is obviously positive definite.

For the Klein–Gordon equation (1.16) the corresponding quantity is,

$$\rho_{_{\text{KG}}} = \frac{i\hbar}{2m}\left(\varphi^*\dot{\varphi} - \varphi\dot{\varphi}^*\right) \,, \tag{1.19}$$

and is a *not a positive definete* quantity, which forces us to abandon the interpretation of the Klein-Gordon equation as a single particle equation. Details may be found in any elementary relativistic quantum mechanics textbook e.g. [7].

The Dirac equation developed in 1928 is,

$$(i\hbar\gamma^a\partial_a - mc)\,\psi = 0\,, \tag{1.20}$$

where the $\gamma^a$ , $a = 0, \ldots, 3$, are a set of $4 \times 4$ matrices called the Dirac matrices (see [8]).

A solution of this equation, $\psi(t,\vec{x})$ is called a *Dirac spinor*. It describes a particle (e.g. an electron) with its associated *antiparticle*, both of which may have spin ”up” or ”down”. Antiparticles (the *positron* if the particle is an electron) were experimentally observed by Carl David Anderson in 1932, and spin (intrinsic angular momentum) a purely quantum property of matter had been proposed in 1925 by George Uhlenbeck and Samuel Goudsmit.

The pursuit of quantization led to a quantum theory of the electromagnetic field — Quantum Electrodynamics (QED). Moreover current experimental research has so far uncovered four fundamental interactions: strong, electromagnetic, weak, and gravitational. Of these, gravity is not only the most common-day interaction, in the sense that we experience it constantly and in an obvious way, for example in the way we are “puled to the ground” but (maybe because of its commonness) was the first to be investigated. The gravitational interaction is also the weakest of the four.

It is significant that the strong, electromagnetic and weak interactions, have been put into a quantum framework called (there are numerous textbooks on this subject, one of the most notable is [8] ) Quantum Field Theory (QFT). Also the known elementary particles in nature are described in the formalism of QFT in the so called Standard Model of particle physics (SM). QFT has had an experimental success that can hardly be overstated and, QM has led to the development of nuclear physics, atomic and molecular physics, solid state physics, with all the technological triumphs achieved by these fields.

Gravity however has remained outside of this equation!

After developing SR in the beginning of the twentieth century, Einstein turned to the problem of finding a relativistic theory of gravity that would replace the newtonian description of this interaction. After arduous labour, in November 1915 he presented to the Prussian Academy of Science, a theory — the General theory of Relativity (GR). In this theory, the “force of gravity” is identified with the curva-ture of spacetime. The gravitational field is described by a rank two tensor, the metric.

In the words of John Wheeler:"spacetime tells matter how to move; matter tells spacetime how to curve"([9] page 235). What is described in loose terms here (see the next section for a more precise account of GR) is the backreaction of matter on geometry.

In the course of his intellectual endeavour Einstein was guided by ideas such as the *equivalence principle*, which asserts that a gravitational field is locally equivalent to a field of accelerations. He eventually came (after having considered it for a first time and abandoning it see [10]) to the notion of *general covariance* i.e. the field equations for gravity must be invariant under general coordinate transformations. This is a crucial feature of GR also related to *background independence*. This means that unlike previous theories — which depended on the *a priori* existence of some fixed spacetime structures (e.g. the Minkowski metric $\eta$) both for the development of the formalism and for the interpretation of the results — GR does not rely on any fixed, non-dynamical space-time structures. Einstein's field equations are invariant under all diffeomorphisms of the underlying manifold, which has no space-time structure until a solution of the field equations is specified (see [11, 12]).

Background independence is a crucial characteristic of GR and it will be a constant source of issues in the subsequent presentation.

Despite being the weakest of all four interactions, gravity can in some situations become the dominant force. In the final stages of the evolution of some stars a *gravitational collapse* occurs terminating in a state of infinite density and curvature. Thus we have a *singularity* of spacetime. This physical situation is particularly evident in the Schwarzschild solution of the Einstein field equations. In this metric, a *singular* point exists that may be regarded as the final state of a collapsing star. There is also a *horizon*, which is a surface around the singularity, that can only be entered and never exited. It acts as a point of no return ([6]). This horizon is at the *Schwarzschild radius*[1] ,

$$r_s = \frac{2GM}{c^2} \, , \tag{1.21}$$

where $G = 6.674 \times 10^{-11} Nm^2 kg^{-2}$ is Newton's gravitational constant.

Surprisingly this view of a black hole is not the whole story. Stephen Hawking working in a fixed curved background geometry, an approach called *quantum field theory in curved spacetime,* (CSQFT) (refer to [13, 14] for textbooks on this subject) found the startling result that a black hole emits (the radiation is called the *Hawking radiation*) like a blackbody at a well defined temperature called the *Hawking temperature $T_H$*,

$$T_H = \frac{\hbar \kappa}{2\pi k_B c} \tag{1.22}$$

where $k_B = 1.380 \times 10^{-23}\, J/K$ is the Boltzmann constant and $\kappa$ is the surface gravity of the black hole, which for a spherically symmetric Schwarzschild black hole is given by,

$$\kappa = \frac{c^4}{4GM} = \frac{4GM}{r_s^2} \, . \tag{1.23}$$

The entropy of the black hole can be calculated and is found to be proportional to $A_{hor}$ the area of the event horizon,

$$S_{BH} = \frac{c^3}{4\hbar G} A_{hor} \, . \tag{1.24}$$

This is the Bekenstein-Hawking entropy. *Black hole thermodynamics* was developed with four laws in full analogy to the very well known theory of heat developed in the nineteenth century (refer to [15]). As an example, the area of the horizon never decreases therefore the second law states that the (black hole) entropy also shares in this property. Finally we mention that since the black hole is emitting it must loose mass and a phenomenon known as *black hole evaporation* occurs. Eventually the black hole may evaporate completely.

General Relativity brought with it another innovation: *relativistic cosmology*. Although the construction of physical theories suited to the description of the Universe was possible in pre-relativistic physics, GR introduced a novelty into the

[1]We mention that the Schwarzschild radius has a simple (heuristic) non-relativistic interpretation.
It may be thought of as the radius corresponding to an escape velocity equal to the speed of light

$$v_e \equiv \sqrt{\frac{2GM}{r_s}} = c$$

.

picture, a nonstatic universe! In fact, the Newtonian universal law of gravitation seemed to imply ([16] contains some History and the description of the contemporary discussions) an infinite, static, perfectly uniform Universe. GR on the contrary, has an exact solution, the *Friedmann–Lemaître–Robertson–Walker metric* (FLRW), for which the Einstein field equations (called Friedmann equations in the context of cosmology) predict an expanding or contracting Universe. Einstein tried to escape this prediction by adding the *cosmological constant* term to have a static Universe.The explanation of the origin of this constant became a problem in itself, the *cosmological constant problem*, one of the most important in contemporary physics.

However Hubble's observations of the relation between the distance to galaxies $d$ and their velocities $v$ (redshifts to be precise) led him to propose a linear law relating the two $v = Hd$, where $H$ is the Hubble *constant*[2]. These observational data awoke the necessity for a dynamical model of the Universe. Subsequently the work by Friedmann, Lemaître, Robertson and Walker led to the development of *Big Bang cosmology* (refer to [17] for a recent review), in which the Universe is not only expanding but exhibits a *singularity* at a finite time in the past.

The observational evidence for the Big Bang cosmological model comes not only from the expansion of the Universe (which is presently in an accelerated expansion) but also from the Microwave Background Radiation (CMBR), the *relic radiation* left over from the cooling of the Universe to temperatures below those corresponding to the recombination of electrons and protons into bound atoms, and the *Big Bang nucleosynthesis* (BBN), which accounts for the relative abundances hydrogen, deuterium (an isotope of hydrogen) and helium. Recently in 1998 two independent research groups, the Supernova Cosmology Project and the High-Z Supernova Search Team, found that the expansion of the Universe is accelerating. This was a major scientific discovery that not only contradicted the decelerating expansion of the models of that time, but that itself requires a physical explanation. The first candidate for this explanation is the cosmological constant $\Lambda$, originally introduced by Einstein to create a static universe, and now brought back to make the universe accelerate! This constant, albeit a problem in itself, is central to the standard model of cosmology the $\Lambda_{\mathrm{CDM}}$ model. Also important in cosmology is the question of *inflation*. This is a proposed phase of accelerated expansion happening in the *early universe* (for details about this subject see [18, 19, 20]).

The cosmological solution, like the black hole solution, has singularities. These represent a loss of predictability of GR, the curvature (and other invariants) becomes infinite, and so they mark the failure of this theory at the points in question, or even in a sufficiently small neighbourhood of these points. Moreover, these sin-gularities are not an accident, there are theorems — *the singularity theorems* — that enforce the existence of these points given some reasonable set of hypotheses, plus the existence of what is called a *trapped surface* (for details about these theorems, and global methods we refer the reader to [6, 21, 22]).

[2]It is a time dependent function $H(t)$ to be precise.

## 1.2 What is Quantum Gravity?

We are therefore presently in a situation in which the best description of the gravitational interaction — namely GR — is incomplete. There are physical systems (black holes, the Universe) for which the theory fails at (and around) some points. Someone could argue that black holes are not real physical systems, or better stated that through the use some mechanism we could remove the singularity, or that the FLRW solution does not adequately account for the Universe. Even if these arguments are proposed, the fact remains that in some solutions of GR singular points exist where the theory ceases to be applicable.

Furthermore in GR, the geometry interacts with matter trough the Einstein equations, and one can show that treating matter quantum mechanically and the geometry classically leads to inconsistencies. Einstein made a similar remark (although not directly linked to Quantum Gravity (QG)) when he said that the left hand side of his equations (the geometry) is like marble whereas the right hand side (the matter part) is akin to wood. The use of expectation values in the matter part only highlights the problem.

Also worthy of mention is the view held by most researchers, that some insight into QG is gained from the CSQFT treatment of black holes and their singularities. Namely, the Hawking temperature (1.22) and the Bekenstein-Hawking entropy (1.24) may be considered quantum gravitational in nature in the sense that they involve both the Planck $\hbar$ and Newton's constant $G$, which characterize QM and gravity respectively. These phenomena must therefore be a part of any theory of QG . Furthermore, the development of black hole thermodynamics motivates the search for a "microscopic statistical mechanics" that would explain it, in parallel to the situation in the usual thermodynamics.

The existence of singularities together with the quantum nature of all the other interactions strongly motivates the research into the quantization of gravity. A related problem is that of the unification of the fundamental interactions. QFT and the SM present a unified view of particle physics and their interactions. A first kind of unification comes from the fact that we are using the same formalism to describe different interactions. However a second type of unification comes from the fact that there are unified theories for which (above some energy scale) two interactions become different aspects of the same force. This is the case of the Glashow Weinberg Salam theory (GSW) that unifies the electromagnetic and weak interactions. In this model a single interaction called electroweak gives rise, through symmetry breaking to the two interactions we know to exist at lower energy scales. More ambitious is the case of Grand Unified Theories (GUT), in the context of which the strong, weak and electromagnetic interactions become one unified interaction. These however lack experimental confirmation. The unification of the gravitational force with (one or all of) the other interactions may also be a motivation to do research in quantum gravity.

Additionally questions like the cosmological constant problem, with its associated *dark energy problem* (the problem of the nature of a possible dynamical field driving the accelerating expansion of the Universe) are sometimes invoked as quantum phenomena that would be explained by a theory of quantum gravity.

In this thesis, by *Quantum Gravity* we mean a Quantum Field Theory of geometry and matter which is background independent[3], takes fully into account the backreaction of quantum matter on quantum geometry and reduces to GR and QM in appropriate limits ([23] gives a good overview of this area of research). An indication that there is something special about the physical domain where QM and GR are both relevant can be found if we take the uncertainty principle (1.2) written in the form $mvr \sim \hbar$, and take the velocity to be of the order of the speed of light $v \sim c$ and the radius to be of the order of the Schwarzschild radius (1.21) $r \sim \frac{GM}{c^2}$ then we find that the distance will be of the order of the Planck length $l \sim l_p$ where

$$l_p = \sqrt{\frac{\hbar G}{c^3}} = 1.616229(38) \times 10^{-35} m \,, \tag{1.25}$$

this astonishingly small distance is the relevant length scale for QG.

A related order of magnitude calculation comes from equating the Compton wavelength (1.4) to the Schwarzschild radius (1.21) $\lambda_c \sim r_s$, we find a mass[4],

$$m_p = \sqrt{\frac{\hbar c}{G}} = 2.176 \times 10^{-8} kg \,, \tag{1.26}$$

this is called the Planck mass.

Although the above arguments do not prove anything, they seem to indicate that a scale exists (and is related to relevant quantities in QM and GR) in which quantum gravity is non-negligible. This scale is characterized by the plank units (refer to

[3]This is not a requirement in all approaches to QM.

[4]Sometimes written $\sqrt{\frac{\hbar c}{8\pi G}}$.

[24] for an interesting account). In this thesis we will use $c = \hbar = G = 1$, but we will reinstate these fundamental constants when they are necessary to clarify the physics involved.

It must be said at the outset that the disagreement about QG seems to be larger than the points of consensus. This may be a result of the lack of direct experimental probing of the physics at the Planck length. There is a question however, that stands sharply athwart agreement about what a QG theory should be: *is the central problem in QG one of physics, mathematics or philosophy?* Can we build a theory that addresses these questions without first dealing with the nature of space, time and matter? How fundamental are these concepts? For reviews about this refer to [25, 26]. There are open issues related to the interpretation both of QM and of GR. On the QM side a complete interpretation was never fully developed. The *Copenhagen interpretation* for example relies on a fixed background structure, and gives only a partial answer.

Among the intuitions researchers have had, a persistent one claims that we must abandon *continuum* concepts, and that they must be replaced with some discrete notions. Also the linearity of QM, is sometimes believed to fail at Planck scales. In general the possibility of *new physics*, meaning perhaps, presently unknown fundamental interactions, principles or phenomena, at the Planck length is occasionally proposed as an explanation for the disparity between the principles of QM and those of GR .

As was mentioned above, physics in the last century evolved along a dual path: on one hand we have GR with its insistence on general covariance and, on the other QFT based on a fixed flat background. This being the case it is only natural that a similar polarization appears in the field of QG, with field theorists and particle physicists, on one side and researchers with a background on GR on the other. In fact, researchers coming from a QFT background are more likely to emphasize Poincaré group based techniques where the founding concepts come from SR and a flat fixed Minkowski spacetime where quanta of the gravitational field propagate. The expectation of a particle physicist is that the theory should produce scattering amplitudes and should be renormalizable or, better still, finite. In the best case scenario the theory should be part of a theory that unifies gravity with the other interactions.

For a general relativist everything revolves around the geometrical features of GR, with a great weight placed on background independence and a consequent reluctance in employing any methods that rely *a priori* on the use of a background structure, like perturbative methods. In general researchers in this category expect that a theory of QG may solve the problem of space-time singularities, both in the case of black holes — where the theory is expected to incorporate Hawking's prediction of the production of particles — and in cosmology where a resolution of the Big Bang singularity is awaited.

The construction of a theory of QG faces some key questions that arise even before the onset of its construction. One of these is the extent to which this theory can maintain the picture of spacetime given by GR and the framework and interpretation of QM. Another question, more concrete than the last is related to the role of the diffeomorphism group *Diff*($M$) (see the definition in the text following (2.3) ) of the space-time manifold $M$. General relativity, as was already mentioned is invariant (the equations are covariant under these transformations) under this group. Elements of *Diff*($M$) are active transformations, they move points not coordinates. This implies two things: (1) that a theory compatible with this group should be built using tensorial objects in spacetime and, (2) that points in spacetime have no direct physical significance.

The diffeomorphism group plays a role similar to the Yang-Mills gauge group. The main difference lies in the fact that whereas Yang-Mills transformations occur at a fixed spacetime point, the diffeomorphism group actively changes the points. This is related to the Einstein *hole argument,* which he devised as a line of reasoning against general covariance. Using a region of space devoid of matter (a hole) and a diffeomorphism reducing to the identity outside the hole Einstein purported to show that the specification of the matter sources in the exterior of the hole (and on the boundary of it) together with the field equations were not sufficient to determine the gravitational field in the interior. Einstein later realized that "in the absence of a metric tensor field, a coordinate system on a differentiable manifold has no intrinsic significance"[10]. This is of course a restatement of point (2) above.

A related question is that of the construction of physical observables, an issue on debate for decades. If we define an observable as a quantity (an operator) that commutes with the gauge group, then given the action of a representation of an element $U(f)$ of $f \in$ *Diff*($M$), i.e. $U(f)\phi(x)U^{-1}(f) = \phi(f^{-1}(x))$ we see that $\phi$ is not an observable in this sense. A particularly obvious way to construct a diffeomorphism invariant quantity out of a scalar function of $g_{\mu\nu}(x)$ is to integrate it over all spacetime (see [25]), in the following way:

$$\int_M R(x)(\det|g|)^{\frac{1}{2}}d^4x\,, \quad \int_M R_{\mu\nu}(x)R^{\mu\nu}(x)(\det|g|)^{\frac{1}{2}}d^4x \tag{1.27}$$

This would be a non-local quantity and the issue of non-locality finds its way into QG. Another way of achieving the same goal is to note that although the quantity $\phi(x)$ does not transform as an observable $\phi(X)$ does if $X$ is the location on the manifold of a concrete material object. That is we must use "material reference frames" to ascertain our location on the manifold. This highlights the importance of matter in QG. Vacuum theories more then just a simplification may turn out to be a oversimplistic approach to the problem of QG.

Related to the question of diffeomorphisms is the well known *problem of time*. This problem arises in any approach to QG that considers GR as a starting point. The problem is related to the roles time plays in QM and in GR. In quantum theory, time is not a physical observable in the sense that it is not represented by an operator. It is a background parameter that labels the evolution of a system. This idea of time can be carried over to SR and consequently to relativistic quantum mechanics. Here, the set of relativistic inertial reference frames replaces the Newtonian absolute concept of time. A theory in these conditions must thus be endowed with a unitary representation of the Poincaré group of isometries of the Minkowski spacetime. We note that this parameter cannot be measured by any clock, there is always a probability that a real clock will run backwards with respect to it[5] (see [28] and references therein).

Very different is the situation in classical general relativity, where time (instead of af a background parameter) is a coordinate that (like the spatial metric) is influence by the matter present. If the space-time manifold can be foliated as a one-parameter family of spacelike hypersurfaces there is in general no way to single out as natural any subset of such foliations. Each of the time directions orthogonal to these families of spacelike hypersurfaces could be considered a notion of time. These definitions of time are in general unphysical, in that they provide no hint as to how this time might be measured by physical clocks.

## 1.3 Do all roads lead to Quantum Gravity?

We explicitly assume here that a theory of Quantum Gravity exists. That is, whoever discovers it is not simply constructing a scientifically coherent set of statements with predictions that coincide with observations and experiments, but is rather factually uncovering an unknown part of the natural world. What we do not assume, either overt or covertly is that our approach is the only one that leads to *the* theory of QG.

Einstein himself was the first to recognize in his first paper on gravitational waves in 1916 [29] that,

[5]This means that a dynamical variable that correlates monotonically with time cannot be found, even in the Schrödinger theory. For details we refer the reader to [27].

> the atoms would have to emit, because of the inner atomic electronic motion, not only electromagnetic, but also gravitational energy, although in tiny amounts. Since this hardly holds true in nature, it seems that quantum theory will have to modify not only Maxwell's electrodynamics, but also the new theory of gravitation[6].

He would later change his position and work on the unification of gravity and electromagnetic theory.

The three positions about the quantization of gravity and the methodology to be used were already expressed by the mid-1930's [31]. They are:

1. In analogy with QED, quantum gravity should be formulated by quantizing the linearised field equations using the same methods. Investigators that held this viewpoint (Rosenfeld, Pauli, Fierz) thought that the problem which would arise would be substantially similar to those of the quantization of the electromagnetic field and they would be presumably solved *pari passu*.

2. The gravitational field has such distinctive features that the full non-linear set of field equations must be quantized. Although, according to this view, the techniques to be utilized come from QFT they must be properly generalized in order to suit the needs of this novel problem (Bronstein, Solomon).

3. GR is only a macroscopic theory, "e.g. a sort of thermodynamics limit of a deeper, underlying theory of interactions between particles"[32]. This was a position held by Frenkel, van Dantzig.

These are essentially the same positions that we find in today's approaches to QG. Perhaps the fact that more then three quarters of century later no fundamentally new ideas have appeared marks the need for some truly creative approaches as well as more experimental data.

A classification of the types of avenues to QG was given in [25]. The four types proposed are the following

**I** *The quantisation of general relativity.* We start with classical GR and apply some quantization scheme to this theory. This further subdivides into (i) *canonical quantization,* in which splitting of spacetime into three-dimensional space and time is made and, (ii) *covariant quantization* where a quantization method is applied to the four-dimensional space-time manifold . Note that both positions 1. and 2. mentioned above are of this type. Moreover the Rosenfeld-Pauli-Fierz stance is a covariant quantization scheme which we will explore succinctly in the next section, the Bronstein-Solomon position is also of this type and may be (but not necessarily) a canonical form of quantization. This subtype is the subject of this thesis and chapters will be devoted to the space-time splitting and to the new variables discovered by Ashtekar which prompted a renaissance in the research into canonical quantum gravity.

[6]Translation from [30] page 24.

**II** *Quantise Some Other Theory That Can Give General Relativity as the Low-Energy Limit.* We start, with some other classical theory, quantize it and given an appropriate notion of *classical limit*, we find GR. There are (at least) two difficulties here. The notion of classical limit can be: a special state whose evolution in time follows classical laws; or some quantum quantities taking values in a range where classical theory is successful. And, it is not clear from what theory we should begin.

The main example of an approach of this type is *superstring theory,* where a classical string theory given by the Polyakov action is second quantized, and GR or a generalization thereof is obtained in the low energy limit.

About the notion of GR as the classical limit of a quantum theory we remark following [25, 33] that "within the context of the particle-physics approach to quantum gravity (see the next section), there are a series of, more or less, rigorous theorems which show that any Lorentz-invariant theory of a spin-2 graviton coupled to a conserved energy-momentum tensor will necessarily yield the same low-energy scattering results as those obtained from the tree graphs of a weak-field perturbative expansion of the Einstein Lagrangian."

**III** *"General-Relativise"*[7] *Quantum Theory* This is in a way the inverse of type **I** Here we start with quantum theory and try to harmonize it with GR. This is exemplified in the already mentioned quantum field theory in curved space-time. This consists in building QFT on a fixed curved background, and although this is not a QG theory proper (backreaction is difficult to account for in this setting) it may help clear up some issues.

**IV** *General Relativity and Quantum Gravity Both Emerge From Something Quite Different.* This type involves starting *ab initio* from radically new concepts. The relation of QG and GR and/or QM is in this category analogous to the relation between quantum theory and classical mechanics. That is the leap from GR and QG is of the same iconoclastic kind as the one from classical to quantum physics. This is the most exciting avenue although arguably it may not be the most promising.

[7]The verb *General-Relativise* does not exist it is a creation of C. Isham.

The most natural starting point in the search for QG is the type **I** idea of quantizing GR. Since this option, in its covariant form using tools analogous to those used in QED, was the first to be tried, we next describe GR in more detail, and introduce the notion of graviton.

## 1.4 The General Theory of Relativity

Our best theory of the gravitational interaction is GR, so it is logical that we should describe this theory before attempting even to speak of quantization.

We define space-time, (see [6, 21, 34]) that is the set of all events, to be a connected four-dimensional Hausdorff $C^{\infty}$[8] manifold $M$ in which a connection, $\Gamma^{\lambda}{}_{\mu\nu}$, and a symmetric non-degenerate Lorentzian metric $g_{\mu\nu}$ are defined.

The connection is related to parallel transport and, can be used to define a covariant derivative:

$$\nabla_{\mu}A^{\nu}{}_{\sigma} = \partial_{\mu}A^{\nu}{}_{\sigma} + \Gamma^{\nu}{}_{\mu\alpha}A^{\alpha}{}_{\sigma} - \Gamma^{\alpha}{}_{\mu\sigma}A^{\nu}{}_{\alpha}. \tag{1.28}$$

The Riemann tensor is:

$$R^{\mu}{}_{\nu\sigma\lambda} = -\partial_{\lambda}\Gamma^{\mu}{}_{\nu\sigma} + \partial_{\sigma}\Gamma^{\mu}{}_{\nu\lambda} + \Gamma^{\mu}{}_{\alpha\sigma}\Gamma^{\alpha}{}_{\nu\lambda} - \Gamma^{\mu}{}_{\alpha\lambda}\Gamma^{\alpha}{}_{\nu\sigma}\,. \tag{1.29}$$

The Ricci tensor is:

$$R_{\mu\nu} \equiv R^{\sigma}{}_{\mu\sigma\nu} = -R^{\sigma}{}_{\mu\nu\sigma}, \tag{1.30}$$

or in terms of the connection,

$$R_{\mu\nu} = \partial_{\lambda}\Gamma^{\lambda}{}_{\mu\nu} - \partial_{\nu}\Gamma^{\lambda}{}_{\mu\lambda} + \Gamma^{\lambda}{}_{\sigma\lambda}\Gamma^{\sigma}{}_{\mu\nu} - \Gamma^{\lambda}{}_{\sigma\nu}\Gamma^{\sigma}{}_{\mu\lambda}. \tag{1.31}$$

In GR the torsion is identically zero:

$$T_{\mu\nu}{}^{\lambda} \equiv \Gamma^{\lambda}{}_{[\mu\nu]} = 0\,, \tag{1.32}$$

which is equivalent to the symmetry of the connection and, the metric is covariantly conserved,

$$\nabla_{\lambda}g_{\mu\nu} = 0\,, \tag{1.33}$$

[8]Infinite differentiability is in reality not a physical requisite, see [22] and reference therein. The requirement of $C^4$ differentiability or even $C^2$ with a $C^2$ piecewise $C^4$ differential structure is sufficient according to the author.

this quantity is sometimes called the non-metricity tensor $Q_{\mu\nu\lambda}$ which in GR is null.

Using these two conditions (1.32) and (1.33) a relation between the connection and the metric exists,

$$\Gamma^{\lambda}{}_{\mu\nu} = \frac{1}{2} g^{\lambda\rho} \left( \partial_\nu g_{\rho\mu} + \partial_\mu g_{\rho\nu} - \partial_\rho g_{\mu\nu} \right) , \tag{1.34}$$

this is the Levi-Civita connection.

The Einstein-Hilbert action

$$S_{EH} = \frac{c^4}{16\pi G} \int_M (R - 2\Lambda) \sqrt{-g} \mathrm{d}^4 x - \frac{c^4}{8\pi G} \int_{\partial M} \sqrt{h} K \mathrm{d}^3 x \, , \tag{1.35}$$

where $h$ is the determinant of the 3-metric on the boundary $\partial M$ (here assumed to be spacelike) $k$ is the trace of the second fundamental form or extrinsic curvature, and $\Lambda$ is the cosmological constant mentioned above.

The second term in this action, is a surface term needed to cancel the result of the variation with respect to the metric (about this surface term refer for example, to [35, 36]).

A matter action of the form,

$$S_M = \int_M \mathrm{d}^4 x \sqrt{-g} L_M(g_{\mu\nu}, \psi), \tag{1.36}$$

may be added to (1.35).

If we take the variation with respect to the metric we obtain,

$$G_{\mu\nu} \equiv R_{\mu\nu} - \frac{1}{2} R \, g_{\mu\nu} + \Lambda g_{\mu\nu} = \frac{8\pi G}{c^4} T_{\mu\nu} \, , \tag{1.37}$$

these are the Einstein equations. The stress-energy tensor $T_{\mu\nu}$ is given by

$$T_{\mu\nu} \equiv -\frac{2}{\sqrt{-g}} \frac{\delta S_M}{\delta g^{\mu\nu}} \, . \tag{1.38}$$

This is a second order formalism called the metric formalism. If we assume no *a priori* relation between $\Gamma$ and $g$, or in other words if we consider these to be two independent variables, we may build the *Palatini action*[9] [6, 34] (for $\Lambda = 0$)

$$S_P = \frac{c^4}{16\pi G} \int_M R(\Gamma, g) \; \sqrt{-g} \mathrm{d}^4 x \tag{1.39}$$

[9]Sometimes also called the Einstein-Palatini or Hilbert-Palatne action.

here $R$ is built using an independent connection and metric. Varying this action with respect to both the variables we recover the Einstein equations plus the definition of the Levi-civita connection. This is the Palatini variational formalism. We will rediscover this action in the following chapters written in terms of tetrads and spin connections.

In the Palatini approach the independent connection $\Gamma$ is not coupled to the matter part of the action. If however we decide to introduce such a coupling, we have what is termed *metric affine gravity*(see [37] and references therein). A good example is Einstein–Cartan Theory where the spin of matter is coupled to the torsion ([38]).

### 1.4.1 Plane gravitational waves and the graviton

The discovery that material particles have wavelike properties was the starting point for QM. It is natural therefore to try to take the same path in GR and look for wave solutions of Einstein's equations and try to quantize them. Gravitational waves were first detected in 2015 [39], however they had been predicted one hundred years before by Einstein himself. In the following we briefly review a set of solutions of Einstein's equation called plane gravitational waves and the quantization of GR based on the *graviton*.

A striking difference between GR and QM is that whereas the latter is linear, the former is highly non-linear. Thus we start (see [6, 30] for a details) by expanding the metric $g_{\mu\nu}$ around a fixed flat background $\eta_{\mu\nu}$[10] plus a perturbation $f_{\mu\nu}$ considered to be small (i.e. the components are small)

$$g_{\mu\nu} = \eta_{\mu\nu} + f_{\mu\nu} \,, \tag{1.40}$$

inserting this into (1.37) we have (for $c = 1$),

$$\Box f_{\mu\nu} = -16\pi G \left( T_{\mu\nu} - \frac{1}{2}\eta_{\mu\nu} T \right) \,, \tag{1.41}$$

where $T = \eta_{\mu\nu} T^{\mu\nu}$ the harmonic condition was used,

$$f_{\mu\nu,}{}^{\nu} = \frac{1}{2} f^{\nu}{}_{\nu,\mu} \,. \tag{1.42}$$

[10]We are making an exception in this subsection, to our Latin index convention for the sake of consistency.

If we make the coordinate transformation,

$$x^{\mu} = x'^{\mu} + \varepsilon^{\mu}(x) \, , \tag{1.43}$$

then $f_{\mu\nu}$ changes (to first order in $\varepsilon$) in the following way,

$$f_{\mu\nu} \rightarrow f_{\mu\nu} - \varepsilon_{\mu,\nu} - \varepsilon_{\nu,\mu} \tag{1.44}$$

If we take the combination,

$$\tilde{f}_{\mu\nu} = f_{\mu\nu} - \frac{1}{2} T_{\mu\nu} f^{\alpha}{}_{\alpha} \tag{1.45}$$

the field equations can be cast in the form,

$$\Box \tilde{f}_{\mu\nu} = -16\pi G T_{\mu\nu} \, , \tag{1.46}$$

this is the linear approximation for the Einstein field equations. The harmonic condition becomes $\partial_{\nu} \tilde{f}_{\mu}{}^{\nu} = 0$, in analogy with the Lorenz condition $\partial_{\mu} A^{\mu} = 0$. This condition is consistent with $\partial_{\nu} T^{\mu\nu} = 0$ but not with $\nabla_{\nu} T^{\mu\nu} = 0$. This implies that the energy momentum tensor is conserved but not in a covariant way, and that there is no energy transfer between matter and gravitational field in spite of the fact that $T_{\mu\nu}$ acts as a source for $f_{\mu\nu}$.

In this approximation, vacuum solutions (i.e. $T_{\mu\nu} = 0$) that are *planar*[11] *gravitational waves* can be found. They are given by,

$$f_{\mu\nu} = e_{\mu\nu} e^{ikx} + e^{*}_{\mu\nu} e^{-ikx} \tag{1.47}$$

where, $e_{\mu\nu}$ is the polarization tensor. We have the relations $k^{\mu} k_{\mu} = 0$ and $k^{\nu} e_{\mu\nu} = 1/2 k_{\mu} e^{\nu}{}_{\nu}$, see [30] for details. By making a new coordinate transformation of the form (1.43) we find the condition,

$$f'_{\mu\nu,}{}^{\nu} - \frac{1}{2} f'^{\nu}{}_{\nu,\mu} = -\Box \varepsilon_{\mu} \, . \tag{1.48}$$

This way we can consistently take the condition,

$$\Box \varepsilon_{\mu} = 0 \, , \tag{1.49}$$

and consequently reduce the number of components of $f_{\mu\nu}$ from 10 to $10-4-4 = 2$. We therefore see that in the linear approximation the gravitational field has two local degrees of freedom (two degrees of freedom per space point) corresponding to the two polarizations of the plane gravitational waves, (refer to [6, 30] for details).

[11]Sometimes also caled plane gravitational waves.

We furthermore mention that this linearised theory is the same that would be obtained by asking for the classical field corresponding to quantum-mechanical particles of (1) zero rest mass and (2) spin two in (3) flat spacetime, moreover from this "spin-2" approach one can recover GR albeit in a non-geometric way (refer to [6] for this point, see also [40] about the "spin-2 massless field in flat spacetime").

We see therefore that the gravitational field can be identified (in flat spacetime anyway) with a spin-2 massless field, the quantization of which gives rise to a (quantum) particle called the *graviton*. The arguments for the existence of a spin-2 particle come from the representations of the Poincaré group which are used in the SM to classify elementary particles (see [30]). The main argument for the masslessness is the long range of the gravitational interaction combined with the discontinuous effect of a non-vanishing mass on the deflection of light.

Using a method similar to the one used for the electromagnetic field we can quantize the perturbation $f_{\mu\nu}$. The field operator for the graviton is,

$$f_{\mu\nu}(x) = \sum_{\sigma=\pm 2} \int \frac{\mathrm{d}^3 k}{\sqrt{2|\mathbf{k}|}} \left[ a(\mathbf{k}\,,\sigma) e_{\mu\nu}(\mathbf{k}\,,\sigma) e^{ikx} + a^\dagger(\mathbf{k}\,,\sigma) e^*_{\mu\nu}(\mathbf{k}\,,\sigma) e^{-ikx} \right] \tag{1.50}$$

where $\sigma = \pm 2$ are the possible helicities for this particle. The $a(\mathbf{k}\,,\sigma)$ and $a^\dagger(\mathbf{k}\,,\sigma)$ operators are interpreted respectively as the annihilation and creation operators for a graviton having momentum $\hbar\mathrm{k}$ and helicity $\sigma$. These are bosonic operators and therefore obey the commutation rules

$$\left[ a(\mathbf{k}\,,\sigma)\,, a^\dagger(\mathbf{k}'\,,\sigma') \right] = \delta_{\sigma\sigma'} \delta(\mathrm{k} - \mathrm{k}')\,. \tag{1.51}$$

This quantization scheme ultimately hit the wall of the non-renormalizability of the gravitational field. Goroff and Sagnotti [41, 42] demonstrated that in order to obtain a finite S-matrix, the action should contain a counterterm,

$$\Gamma^{(2)}_{\mathrm{div}} = \frac{1}{\varepsilon} \frac{209}{2880} \frac{1}{(16\pi^2)^2} \int \mathrm{d}^4 x \sqrt{g}\, C_{\mu\nu\rho\sigma} C^{\rho\sigma\lambda\tau} C_{\lambda\tau}{}^{\mu\nu}\,, \tag{1.52}$$

cubic in the Weyl tensor (refer to [6] for the definition), the regularization parameter $\varepsilon$ (the deviation from four dimensions in dimensional regularization) must be taken as going to zero $\varepsilon \to 0$ at the end of the calculation. The difference between the Riemann tensor and the Weyl tensor is immaterial in this case since, all terms containing $R$ or $R_{\mu\nu}$ can be absorbed in redefinitions of the metric [43]. At higher loop order, counterterms involving the fourth and greater powers of the Riemann would appear. In general therefore, the theory would involve an infinite number of counterterms and associated coupling constants and thus becomes useful only as an effective theory of QG [44].

The existence of non-renormalizable infinities in GR may lead to very distinct conclusions. A considerable group of researchers think that a modification of Einstein's theory at short distances is needed to tame the infinities, this was the view in *Supergravity* and later in *String Theory*. Others consider that a proper non-perturbative quantization of Einstein's theory, will resolve the problem of the infinities since these were an effect of the methods just described. From this point of view the single most important feature of GR is its general covariance and background independence and relinquishing this central pillar of the theory will teach us nothing about quantum gravity.

f

# 2

# Mathematical toolbox for gauge theories

In this chapter we review some background from differential geometry. The following is not meant to be self contained, for a thorough treatment refer to a standard text like [45].

## 2.1 Manifolds vectors, tensors and differential forms

Manifolds are spaces that locally resemble $\mathbb{R}^n$, in a sense that will be made precise in the following subsection.

### 2.1.1 Manifolds and diffeomorphisms

Let $M$ be a topological space (see [34, 45, 46] for the definition, refer also to [47] upon which a part of this chapter is based). The space $M$ is an $m$-dimensional manifold if there exists a family of open sets $U_I$ for $I \in \mathcal{I}$ that cover $M$, that is $M = \bigcup_{I\in\mathcal{I}} U_I$, and homeomorphisms,

$$\begin{array}{rcl} x_I : U_I & \longrightarrow & x_I(U_I) \subset \mathbb{R}^m \\ p & \mapsto & x_I(p) \end{array} \tag{2.1}$$

with the following property: for every $I, J \in \mathcal{I}$ with $U_I \cap U_J \neq \emptyset$ the map,

$$\varphi_{IJ} = x_J \circ x_I^{-1} : x_I(U_I \cap U_J) \longrightarrow x_J(U_I \cap U_J)\,, \tag{2.2}$$

is a $C^\infty$ map between subsets of $\mathbb{R}^m$.

The pair $(U_I, x_I)$ is called a chart, while the set of charts $\{(U_I, x_I)_{I\in\mathcal{I}}\}$ is an atlas[1]. The set $U_I$ is called the coordinate neighbourhood while $x_I$ is a coordinate function.

Let the map,

$$\phi : M \longrightarrow N\,, \tag{2.3}$$

be a homeomorphism [2], and $x_I$ and $\tilde{x}_I$ coordinate functions. If $x_I\circ\phi\circ\tilde{x}_I^{-1}$ is invertible i.e. the map $\tilde{x}_I \circ \phi^{-1} \circ x_I^{-1}$ exists, and if both these functions are $C^\infty$ the map $\phi$ is called a $C^\infty$ diffeomorphism, and $M$ is said to be diffeomorphic to $N$. The dimensions of $M$ and $N$ are equal, in this case.

The set of diffeomorphisms of a manifold forms a group denoted $Diff(M)$.

### 2.1.2 Vector one-form and tensor fields

A smooth vector field is a linear map:

$$\begin{aligned} v : C^\infty(M) &\longrightarrow C^\infty(M) \\ f &\mapsto v[f]\,, \end{aligned} \tag{2.4}$$

that satisfies the Leibniz rule[3]:

$$v[fg] = v[f]\cdot g + f\cdot v[g]\,. \tag{2.5}$$

As a consequence we have the properties,

$$\text{for constant } c \qquad v[c] = 0\,, \tag{2.6}$$

$$\text{for } f \in C^\infty(M) \qquad (fv)[f'] = f\cdot v[f']\,. \tag{2.7}$$

The vector fields $\partial^I_\mu$ may be defined on $U_I$ through the condition

$$\partial^I_\mu[x^\nu_I](p) = \delta^\nu_\mu\,, \tag{2.8}$$

[1] The terminology is not uniform, we follow in this respect [46].

[2] A map between topological spaces, that is continuous and has a continuous inverse.

[3] A linear map $v : C^\infty(M) \longrightarrow C^\infty(M)$ that satisfies the Leibniz rule is a *derivation* in $C^\infty(M)$. Here we are defining vector fields as derivations (see [48] page 99), however the more intuitive path of defining tangent vectors and subsequently defining a vector field as a smooth assignment of a vector to each point of $M$ is usually taken.

we have from this definition,

$$\partial^I_\mu = \frac{\partial \varphi^\nu_{IJ}(x_I)}{\partial x^\mu_I}\partial^J_\nu \,, \tag{2.9}$$

for $IJ \in \mathcal{I}$. Here $\varphi_{IJ}$ is a diffeomorphism.

A vector field restricted to a point $p$ is a tangent vector. The set of tangent vectors at $p$ form the tangent space to $M$ at $p$ denoted $T_p(M)$. At each point $p$ the set vectors $\partial^I_\mu$ are a basis of $T_p(M)$. A tangent vector $v$ has components $v^\mu$ given by,

$$v = v^\mu \partial_\mu \,. \tag{2.10}$$

The basis $\{\partial_\mu\}$ is called the coordinate basis of $T_p(M)$. A non-coordinate basis $\mathrm{e_a}$ can be used in $T_p(M)$ and we have,

$$\mathrm{e_a} = e_a{}^\mu \partial_\mu \,, \tag{2.11}$$

where $e_a{}^\mu \in \mathrm{GL}(n, \mathbb{R})$.

The space of vector fields over $M$ is denoted $\chi(M)$ and forms a Lie algebra.

A smooth one form is a linear map,

$$\omega : \chi(M) \longrightarrow C^\infty(M) \tag{2.12}$$

such that for $f, f' \in C^\infty(M)$ and $v\,, v' \in \chi(M)$, we have

$$\omega\left[fv + f'v'\right] = f\omega\left[v\right] + f'\omega\left[v'\right] \,. \tag{2.13}$$

Given a function $f, \in C^\infty(M)$ we may define the one form $df$ by the rule $df[v] = v[f]$, we have as a consequence $dx^\mu_I[\partial^I_\nu](p) = \delta^\mu_\nu$.

The cotangent space $T^*_p(M)$ at point $p$ is the set of one-forms (sometimes also called dual vectors) at $p$ i.e. the fields defined in (2.12) restricted to $p$.

That is, the $dx^\mu$ form a basis of $T^*_p(M)$ dual to the coordinate basis of the tangent space at $p$. The components of a form $\omega$ in this basis are $\omega_\mu$ given by,

$$\omega = \omega_\mu dx^\mu \,. \tag{2.14}$$

A non-coordinate basis $\{\theta^a\}$ can also be use in $T^*_p(M)$

$$\theta^a = e^a{}_\mu dx^\mu \,, \tag{2.15}$$

where $e^a{}_\mu$ is the inverse of $e_a{}^\mu$ above.

The space of 1-form fields, (defined in (2.12)) is denoted $\mathcal{T}_1(M)$ or in the context of differential forms $\Omega^1(M)$.

A smooth $(a,b)$-tensor field ($a$ times contravariant and $b$ times covariant) is a multilinear (i.e. linear separately in each variable) functional of the form

$$t : [\times_{r=1}^{a}\mathcal{T}_1(M)] \times \left[\times_{s=1}^{b}\chi(M)\right] \longrightarrow C^\infty(M)\,, \tag{2.16}$$

with components (we drop the point and the chart indices)

$$t^{\mu_1\cdots\mu_a}_{\nu_1\ldots\nu_b} = t\left(dx^{\mu_1},\ldots,dx^{\mu_a};\partial_{\nu_1},\ldots,\partial_{\nu_b}\right)\,. \tag{2.17}$$

The vector space of $(a,b)$-tensors at point $p$ (i.e. the restriction of tensor fields to the point $p$) is denoted $\mathcal{T}^a_{b,p}(M)$. The space of tensor fields of type $(a,b)$ on $M$ is denoted by $\mathcal{T}^a_b(M)$.

The space of functions $C^\infty(M)$ is just $\mathcal{T}^0_0(M)$.

The tensor product of $s \in \mathcal{T}^a_b(M)$ and $t \in \mathcal{T}^c_d(M)$ is a tensor $s\otimes t \in T^{a+c}_{b+d}(M)$ given in terms of the components by

$$(s\otimes t)^{\mu_1\cdots\mu_{(a+c)}}_{\nu_1\ldots\nu_{(b+d)}} = s^{\mu_1\cdots\mu_a}_{\nu_1\ldots\nu_b}\times t^{\mu_{(a+1)}\cdots\mu_c}_{\nu_{(b+1)}\cdots\nu_d}\,. \tag{2.18}$$

### 2.1.3 Forms of degree $p$ exterior derivative and interior product

A differential $r$-form is a $(0,r)$-tensor antisymmetric in its indices that is,

$$\beta_{\mu_1\mu_2\ldots\mu_r} = \beta_{[\mu_1\mu_2\ldots\mu_r]}\,. \tag{2.19}$$

The anti-symmetrizer is given by,

$$A_{[\mu_1,\ldots,\mu_r]} = \frac{1}{r!}\sum_{\pi\in S_p} sgn(\pi)A_{\mu_{\pi(1)},\ldots,\mu_{\pi(r)}}\,, \tag{2.20}$$

and the symmetrizer by,

$$A_{(\mu_1,\ldots,\mu_r)} = \frac{1}{r!}\sum_{\pi\in S_p} A_{\mu_{\pi(1)},\ldots,\mu_{\pi(r)}}\,, \tag{2.21}$$

where $\pi\in S_r$ is a permutation of $\{1,\ldots,r\}$ and $sgn(\pi)$ is its sign.

The space of $r$-forms if denoted by $\Omega^r(M)$, functions are 0-forms i.e. $\Omega^0(M) = C^\infty(M)$.

The exterior (wedge) product is a map,

$$\wedge : \Omega^r(M) \times \Omega^s(M) \longrightarrow \Omega^{(r+s)}(M)\,. \tag{2.22}$$

That is from a $r$-form $\alpha$ and a $s$-form $\beta$ we form a $(r+s)$-form

$$(\alpha \wedge \beta)_{\mu_1 \dots \mu_{(r+s)}} = \frac{(r+s)!}{r!s!} \alpha_{[\mu_1 \dots \mu_r} \beta_{\mu_{(s+1)} \dots \mu_{(p+q)}]}\,, \tag{2.23}$$

that is, one takes the tensor product of the tensors corresponding to the forms and antisystematizes the result.

The wedge product has the following properties

$$\begin{aligned} \alpha \wedge \beta &= (-1)^{rs} \beta \wedge \alpha && (2.24)\\ (\alpha \wedge \beta) \wedge \gamma &= \alpha \wedge (\beta \wedge \gamma) && (2.25)\\ (c_1\alpha + c_2\beta) \wedge \gamma &= c_1\alpha \wedge \gamma + c_2\beta \wedge \gamma && (2.26)\\ \alpha \wedge (d_1\beta + d_2\gamma) &= d_1\alpha \wedge \gamma + d_2\beta \wedge \gamma\,. && (2.27) \end{aligned}$$

where $\alpha$ is a $r$-form, $\beta$ a $s$-form, $c_1$ , $c_2$, and $d_1$ , $d_2$ are functions.

The basis of the space of 1-forms at $p$ denoted $\Omega^1_P(M)$ is just $dx^\mu$ as stated above. For two forms, the basis of $\Omega^2_P(M)$ is $dx^\mu \wedge dx^\nu \equiv dx^\mu \otimes dx^\nu - dx^\nu \otimes dx^\mu$, and so on. Since for $dx^\mu \wedge dx^\mu = 0$ it follows that for an $n$-dimensional manifold $M$ there are no $r$-forms with $r > n$. Hence the space of forms is the direct sum of these spaces i.e. $\Omega^*_p(M) = \bigoplus_{r=1}^n \Omega^r_p(M)$. $\Omega^*_p(M)$ is the set of all differential forms at $p$ and is closed under the exterior product.

For differential forms, the appropriate concept of derivative is the exterior derivative,

$$d : \Omega^r(M) \longrightarrow \Omega^{(r+1)}(M)\,, \tag{2.28}$$

given by,

$$(d\alpha)_{\mu_1 \dots \mu_r} = (r+1)\partial_{[\mu_1} \alpha_{\mu_2 \dots \mu_r]} \tag{2.29}$$

this is just (2.23) for the partial derivative $d$ and the $p$-form $\alpha$ i.e. $\partial \wedge \alpha$. The exterior derivative is both independent of the choice of the usual derivative $\partial$ and, nilpotent $d^2 \equiv d \circ d = 0$.

The interior product of a $r$-form with a vector field $v$ is the map,

$$i_v : \Omega^r(M) \longrightarrow \Omega^{(r-1)}(M) \tag{2.30}$$

given in terms of components by,

$$(i_v\alpha)_{\mu_1 \dots \mu_r} = \frac{1}{(r-1)!} v^\mu \alpha_{\mu \mu_1 \dots \mu_{(r-1)}}\,. \tag{2.31}$$

We have $i_v f = 0$ for $f \in C^\infty(M)$, and $i_v(df) = v[f]$.

Given a manifold $M$ with dimension $m$, a metric $g \in T^0_2(M)$ is a symmetric non-degenerate two-times contravariant tensor field. The signature of $g$ i.e. the number of negative and positive eigenvalues of its component matrix, $(n, m-n)$ is constant on the manifold. If $n = 0$ then $g$ is said to be Riemannian and if $n = 1$ Lorentzian.

### 2.1.4 Hodge $*$ operator and adjoint exterior derivative

The dimensions of the $\Omega^r_p(M)$ spaces of differential $r$-forms, at point $p$, over an $n$-dimensional manifold $M$ are:

$$\dim(\Omega^r_p(M)) = \binom{n}{r} = \frac{n!}{r!(n-r)!}\,, \tag{2.32}$$

so given the equality of the binomial coefficients $\binom{n}{r} = \binom{n}{n-r}$, we have $\dim\left(\Omega^r_p(M)\right) = \dim\left(\Omega^{(n-r)}_p(M)\right)$ and the spaces $\Omega^r_p(M)$ and $\Omega^{(n-r)}_p(M)$ are isomorphic. If $M$ is endowed with a metric a natural isomorphism, given by the Hodge star $*$ can be defined between these spaces.

The Hodge $*$ is a linear map:

$$* \,:\, \Omega^r(M) \longrightarrow \Omega^{(m-r)}(M) \tag{2.33}$$

whose action on a basis vector of $\Omega^r(M)$ is defined by,

$$*(dx^{\mu_1} \wedge \ldots \wedge dx^{\mu_r}) = \frac{\sqrt{|g|}}{(n-r)!}\varepsilon^{\mu_1 \ldots \mu_r}{}_{\nu_{(r+1)} \ldots \nu_n} dx^{\nu_{(r+1)}} \wedge \ldots \wedge dx^{\nu_n}\,, \tag{2.34}$$

where,

$$\varepsilon_{\alpha_1, \ldots \alpha_m} \begin{cases} +1 & \text{if } (\alpha_1, \ldots \alpha_m) \text{ is an even permutation of } (1\,, \ldots\,, m) \\ -1 & \text{if } (\alpha_1, \ldots \alpha_m) \text{ is an odd permutation of } (1\,, \ldots\,, m) \\ 0 & \text{otherwise} \end{cases}\,. \tag{2.35}$$

In components the general $r$-form $\alpha$ is given by ,

$$\alpha = \frac{1}{r!}\alpha_{\mu_1 \ldots \mu_r} dx^{\mu_1} \wedge \ldots \wedge dx^{\mu_r}\,. \tag{2.36}$$

and using (2.34) $*\beta$ acts on an $r$-form like,

$$*\alpha = \frac{\sqrt{|g|}}{r!(n-r)!}\varepsilon^{\mu_1 \ldots \mu_r}{}_{\nu_{(r+1)} \ldots \nu_n}\alpha_{\mu_1 \ldots \mu_r} dx^{\nu_{(r+1)}} \wedge \ldots \wedge dx^{\nu_n}\,, \tag{2.37}$$

that is $(*\alpha)_{\nu_{(p+1)}\cdots\nu_n} = \frac{\sqrt{|g|}}{p!}\varepsilon^{\mu_1\cdots\mu_p}{}_{\nu_{(p+1)}\cdots\nu_n}\alpha_{\mu_1\cdots\mu_p}$.

The Hodge operator has the following property,

$$**\alpha = (-1)^{r(m-r)}\alpha \quad \text{if} \quad (M,g) \text{ is Riemannian}\,, \tag{2.38}$$
$$**\alpha = (-1)^{1+r(m-r)}\alpha \quad \text{if} \quad (M,g) \text{ is Lorentzian}\,, \tag{2.39}$$

for a $r$ form $\alpha$.

An important form is the volume form [46] given by

$$vol = *1 = \sqrt{|g|}dx^1 \wedge \ldots \wedge dx^n = \frac{\sqrt{|g|}}{n!}\varepsilon_{\mu_1\cdots\mu_n}dx^{\mu_1} \wedge \ldots \wedge dx^{\mu_n}\,, \tag{2.40}$$

the $\varepsilon_{\mu_1\cdots\mu_n}$ is the $n$-dimensional Levi-Civita object[4].

An inner product of forms can be defined for $\alpha$ and $\beta$ $r$-forms,

$$(\alpha,\beta) = \int_M \alpha \wedge *\beta\,, \tag{2.41}$$

(we will not give the definition of the integration of differential forms, however we refer the reader to [46]) if both forms are expanded in the basis we have,

$$(\alpha,\beta) \;=\; \frac{1}{r!}\int_M \alpha^{\mu_1\cdots\mu_r}\beta_{\mu_1\cdots\mu_r}vol\,, \tag{2.42}$$
$$\int_M \alpha \wedge *\beta \;=\; \int_M \langle\alpha,\beta\rangle\, vol\,, \tag{2.43}$$

where $\langle\alpha,\beta\rangle$ is the usual inner product. This last equality is sometimes used to define the $*$ operator.

For a given exterior derivative $d$ the adjoint exterior derivative is a map,

$$d^* : \Omega^r(M) \longrightarrow \Omega^{(r-1)}(M)\,, \tag{2.44}$$

constructed in the following way,

$$d^*\alpha = (-1)^{mr+m+1} * d * \alpha \quad \text{if} \quad (M,g) \text{ is Riemannian}\,, \tag{2.45}$$
$$d^*\alpha = (-1)^{mr+m} * d * \alpha \quad \text{if} \quad (M,g) \text{ is Lorentzian}\,, \tag{2.46}$$

for a $r$ form $\alpha$. The adjoint derivative has the property,

$$(d\alpha,\beta) = (\alpha,d^*\beta)\;, \tag{2.47}$$

for a $(r-1)$ form $\alpha$ and a $r$ form $\beta$.

[4]Sometimes also called Levi-Civita symbol.

### 2.1.5 Connection covariant derivative curvature and torsion

We mentioned in Section 1.4 that in a manifold we can define an affine connection and a metric. In general, given a manifold $M$ vectors are in the tangent space $T^1(M)(p)$ at point $p$. There is however no way to relate vectors on tangent spaces at different points. This is precisely the role of the affine connection and the covariant derivative.

A covariant derivative is an operator,

$$\nabla : \mathcal{T}^a_b(M) \longrightarrow \mathcal{T}^a_{b+1}(M) \tag{2.48}$$

with the following properties,

1. linearity
$$\nabla_X [aS + bT] = a\nabla_X S + b\nabla_X T\,, \quad a, b \in \mathbb{R} \tag{2.49}$$

2. the Leibniz rule
$$\nabla_X (TU) = \nabla_X (T)U + T\nabla_X (U) \tag{2.50}$$

3. commutes with contraction (in indices)
$$\nabla_\rho(T^{\mu_1\ldots\sigma\ldots\mu_{a-1}}_{\nu_1\ldots\sigma\ldots\nu_{b-1}}) = \nabla_\rho T^{\mu_1\ldots\sigma\ldots\mu_{a-1}}_{\nu_1\ldots\sigma\ldots\nu_{b-1}} \tag{2.51}$$

4. reduces to the usual derivative on functions,
$$\nabla_X f = X[f] = X^\mu \partial_\mu f\,, \quad f \in C^\infty(M)\,, \tag{2.52}$$

where $X \in \chi(M)$ and $S, T \in \mathcal{T}^a_b(M)$.Moreover, the vector structure of $T^1(M)$ implies

$$\nabla_{aX+bY} T = a\nabla_X T + b\nabla_Y T\,, \quad a, b \in \mathbb{R}\,, X, Y \in T^1(M)\,. \tag{2.53}$$

The connection components are defined as,

$$\nabla_\mu \partial_\nu = \Gamma^\rho{}_{\mu\nu}\partial_\rho\,, \tag{2.54}$$

where $\nabla_\mu = \nabla_{\partial_\mu}$. Using $\nabla_\rho(d^\mu[\partial_\nu]) = \nabla_\rho(\partial_\nu[d^\mu]) = 0$ we find,

$$\nabla_\mu dx^\nu = -\Gamma^\nu{}_{\mu\rho} d^\rho\,, \tag{2.55}$$

The coordinate expression for the covariant derivative of a general tensor field is the generalization of eq. (1.28),

$$\begin{aligned}\nabla_\mu T^{\nu_1\ldots\nu_a}{}_{\sigma_1\ldots\sigma_b} &= \partial_\mu T^{\nu_1\ldots\nu_a}{}_{\sigma_1\ldots\sigma_b} + \\ &+ \sum_{i=1}^{a} \Gamma^{\nu_i}{}_{\mu\alpha} T^{\nu_1\ldots\alpha\ldots\nu_a}{}_{\sigma_1\ldots\sigma_b} - \sum_{j=1}^{b} \Gamma^{\alpha}{}_{\mu\sigma_j} T^{\nu_1\ldots\nu_a}_{\sigma_1\ldots\alpha\ldots\sigma_b} \,.\end{aligned} \tag{2.56}$$

The curvature $R \in T^1_3(M)$ is defined as,

$$R\,[\cdot; w, u, v] = \left([\nabla_u\,, \nabla_v] - \nabla_{[u,v]}\right) w\,, \tag{2.57}$$

and the torsion $T \in T^1_2(M)$ is,

$$T\,[\cdot; u, v] = \nabla_u v - \nabla_v u - [u, v]\ , \tag{2.58}$$

where $[\cdot, \cdot]$ is defined as,

$$[u, v] f = u[v[f]] - v[u[f]]\,. \tag{2.59}$$

Evaluation of (2.57) in the bases $\{\partial_\mu\}$ and $\{dx^\nu\}$ results in (1.29) and (2.58) in the same basis gives (1.32).

### 2.1.6 Pull-back and push-forward

For a diffeomorphism $\psi : M \longrightarrow M$ and $f \in C^\infty(M)$ the function,

$$(\psi^* f)\,(p) = (f \circ \psi)\,(p) = (f(\psi(p)))\ , \tag{2.60}$$

is called the pull-back function.

Fora a vector field $v$ its push-forward is,

$$((\psi_* v)\,[f])\,(\psi(p)) = (v[\psi^* f])\,(p)\,, \tag{2.61}$$

for $f \in C^\infty(M)$.

For a one-form $\omega$ the pull-back is,

$$((\psi^* \omega)\,[v])\,(p) = (\omega[\psi_* v])\,(\psi\,(p))\ . \tag{2.62}$$

For tensors, we have:

$$\begin{aligned}&((\psi^* t)\,[\omega_1 \ldots \omega_a, v_1 \ldots v_b])\,(p) = \\ &= \left(t\left[(\psi^{-1})^* \omega_1 \ldots (\psi^{-1})^* \omega_a, \psi_* v_1 \ldots \psi_* v_b\right]\right)(\psi(p))\end{aligned} \tag{2.63}$$

$$\begin{aligned}&((\psi_* t)\,[\omega_1 \ldots \omega_a, v_1 \ldots v_b])\,(p) = \\ &= \left(t\left[\psi^* \omega_1 \ldots \psi^* \omega_a, (\psi^{-1})_* v_1 \ldots (\psi^{-1})_* v_b\right]\right)(\psi(p))\,.\end{aligned} \tag{2.64}$$

### 2.1.7 Lie derivative

Let $M$ be a manifold and $X$ a vector field, an integral curve $x(t)$ of $X$ is a solution of the equation,

$$\frac{dx^\mu}{dt} = X^\mu(x(t)) \,, \tag{2.65}$$

for coordinates $x^\mu$. This means that the tangent vector to the curve is $X$.

If $\sigma(t, x_0)$ is an integral curve of $X$ which passes a point $x_0$ at $t = 0$, the map $\sigma : \mathbb{R} \times M \longrightarrow M$ (provided it exists) is called the flow generated by $X$ [46]. Flows satisfy the equation

$$\sigma(t, \sigma(s, x_0)) = \sigma(t + s, x_0) \,, \tag{2.66}$$

for $s, t \in \mathbb{R}$ such that the previous equation makes sense.

For fixed $t$ a flow $\sigma(t, x)$ is a diffeomorphism from $M$ to $M$ denoted $\sigma_t : M \longrightarrow M$.

The set of flows satisfies:

1. $\sigma_t \circ \sigma_s = \sigma_{t+s}$;
2. $\sigma_0 = id$;
3. $\sigma_{-t} = (\sigma_t)^{-1}$;

where *id* is the identity map. This therefore definees a one-parameter subgroup of *Diff*$(M)$.

Under the action of $\sigma_\varepsilon$ with infinitesimal $\varepsilon$, a point with coordinates $x^\mu$ is mapped to,

$$\sigma^\mu_\varepsilon(x) = x^\mu + \varepsilon X^\mu \,. \tag{2.67}$$

The vector $X$ is termed the infinitesimal generator of the transformation $\sigma_t$.

Let $X, Y$ be vectors the Lie derivative $\mathcal{L}_Y X$ of $X$ along $Y$ is defined as,

$$\mathcal{L}_X Y = \lim_{\varepsilon \to 0} \frac{1}{\varepsilon} \left[ (\sigma_{-\varepsilon})_* Y|_{\sigma_\varepsilon(x)} - Y|_X \right] \,. \tag{2.68}$$

This definition can be extended to arbitrary tensors see [46].

## 2.2 Fibre bundles

A fibre bundle $(E \,, \pi \,, M \,, F \,, G)$ is composed of:

1. a manifold $E$ called the total space;
2. a manifold $M$ called the base space;
3. a manifold $F$ called the typical fibre;
4. a surjection[5]
$$\pi : E \longrightarrow M\,. \tag{2.69}$$
The inverse image $\pi^{-1}(p) = F_p$ is called the fibre at $p$;
5. a Lie group $G$ called the structure group which acts on $F$ on the left i.e.
$$\begin{aligned} \lambda : G \times F &\longrightarrow F \\ (h,f) &\mapsto \lambda(h,f) = \lambda_h(f) \end{aligned} \tag{2.70}$$
with $\lambda_h \circ \lambda_{h'} = \lambda_{hh'}$ and $\lambda_{h^{-1}} = (\lambda_h)^{-1}$.
6. an open covering $U_I\,, I \in \mathcal{I}$ and diffeomorphisms
$$\varphi_I : U_I \times F \longrightarrow \pi^{-1}(U_I)\,, \tag{2.71}$$
such that $\pi \circ \varphi_I(p,f) = p$, called a local trivialization.
7. The map
$$\begin{aligned} \varphi_{I,p} : F &\longrightarrow F_p \\ f &\mapsto \varphi_{I,p} = \varphi_i(p,f)\,, \end{aligned} \tag{2.72}$$
is a diffeomorphism.
8. maps called transition functions exist,
$$h_{IJ} = \varphi_{I,p}^{-1} \circ \varphi_{J,p} : U_I \cap U_J \neq \emptyset \quad \longrightarrow \quad G\,, \tag{2.73}$$
and we have
$$\varphi_J(p,f) = \varphi_I(p, h_{IJ}(p)f)\,. \tag{2.74}$$

[5]That is every element $p \in M$ has a corresponding element $F_p$ in $E$ such that $\pi(F_p) = p$.

Sometimes the shorthand notation

$$E \xrightarrow{\pi} M$$

is used for the bundle just defined.

The fibre bundle defined in this way depends on the open covering and the coordinates it is therefore a coordinated fibre bundle. Two coordinate bundles are equivalent if the bundle on the union of the atlases of the two is again a bundle. A fibre bundle is an equivalence class of coordinate bundles.

The tangent bundle $TM$ is the bundle formed from the base $M$ and with fibre $T^1_x(M)$ at $x \in M$.

The transition functions $h$ must satisfy the following conditions,

$$h_{II}(p) = Id_{U_I} \quad p \in U_I , \tag{2.75}$$
$$h_{IJ}(p) = h_{JI}^{-1}(p) \quad p \in U_I \cap U_J , \tag{2.76}$$
$$h_{IJ}(p)h_{JK}(p) = h_{IK}(p) \quad p \in U_I \cap U_J \cap U_K , \tag{2.77}$$

Where $Id_{U_I}$ is the identity map. These are consistency conditions that ensure that all local pieces of fibre bundle can be glued consistently.

A trivial bundle is one where there the transition function is independent of the label $I$ and therefore there is only one of them. In this case the bundle is diffeomorphic to the direct product $E = M \times F$.

A local section of $E$ is a map

$$s_I : U_I \subset M \longrightarrow P , \tag{2.78}$$

such that $\pi \circ s_I = Id_{U_I}$. We call a global section (i.e. one that is defined everywhere on $M$) a cross section.

Starting from $M$, $\{U_I\}$, $h_{IJ}(p)$, $F$ and $G$ it is possible to reconstruct the bundle $(E , \pi , M , F , G)$. This implies finding a unique $\pi$, $E$ and $\varphi_I$, see [46] for the process.

A principal $G$-bundle is a fibre bundle where the typical fibre and structure group coincide with $G$.

On a principal $G$-bundle $P$ (we will denote such bundles by their total space or by $P(M, G)$) a right action can be defined as a map

$$\rho : G \times P \longrightarrow P , \tag{2.79}$$

such that $\rho_h(u) = \varphi_I(\pi(u), h_I(u)h)$ for $u \in \pi^{-1}(U_I)$. With $h_I : P \to G$, $h_I(u) = \varphi^{-1}(u)$.

This right action is transitive in every fibre and fibre-preserving. In addition canonical local sections and trivializations may be constructed.

Unlike the case of general fibre bundles, for a principal $G$-bundle it can be shown, using transitivity of the right action that triviality is equivalent to the existence of a global section.

A vector bundle $E$ is a fibre bundle where the typical fibre is a vector space. Given a principal $G$-bundle and a left representations $\tau$ of the group $G$ on $F$, a vector bundle exists called the associated fibre bundle denoted by $E = P \times_\tau F$. It is given by the set of equivalence classes,

$$[(p,f)] = \left\{ \left( \rho_{h(p)}, \tau(h^{-1})f \right) : h \in G \right\} \tag{2.80}$$

for $(p,f) \in P \times F$. The projection is given by $\pi_E([p,f]) = \pi(p)$ while the local trivializations are $\psi_I(x,f) = [(s_I(x),f)]$.

## 2.3 Connections on principal fibre bundles

For a principal $G$-bundle $P(M,G)$. Let $u \in P$ and $G_p$ be the fibre at $p = \pi(u)$, the vertical subspace $V_u(p)$, is a subspace of the tangent space $T_u(P)$ which is tangent to $G_p$ at $u$.

It may be constructed using the right action,

$$\rho_{\exp(tA)} u = u \exp tA\,, \quad A \in \mathfrak{g}\,, \tag{2.81}$$

to define a curve passing through $u$ in $P$. Here $\exp : (g) \longrightarrow G$ is the exponential map.

Since $\pi(u) = \pi(u \exp tA) = p$, this curve is in $G_p$ and the vector,

$$A^{\#}f(u) = \frac{d}{dt} f(u \exp tA)\bigg|_{t=0}, \quad f \in C^\infty(P) \tag{2.82}$$

is tangent t $P$ at $u$ and therefore $A^{\#} \in V_u(P)$.

The map,

$$\begin{aligned} \# : \mathfrak{g} &\longrightarrow V_u(p) \\ A &\mapsto A^{\#} \end{aligned} \tag{2.83}$$

is a vector space isomorphism.

The horizontal subspace $H_u(P)$ is the complement of $V_u(P)$ in $T_u(P)$.

For a principal bundle $P(M, G)$,

$$P \xrightarrow{\pi} M$$

a connection on $P$ is a unique separation of the tangent space $T_u(P)$ into the vertical subspace $V_u(P)$ and the horizontal subspace $H_u(P)$, such that,

1. $T_u(P) = V_u(P) \oplus H_u(P)$;
2. a smooth vector field $X$ on $P$ is split $X = X^H + X^V$, into $X^H \in H_u(P)$ and $X^V \in V_u(P)$, and both $X^H$, and $X^V$ are smooth vector fields;
3. the space $H_{ug}(P) = \rho_{g^*} H_u(P)$, for $u \in P$ and $g \in G$.

This definition is equivalent to the introduction of a Lie algebra valued one form (see Appendix A) $\omega \in \mathfrak{g} \otimes T_1(P)$ satisfying the conditions below, which is called the connection one-form.

A connection one-form $\omega \in \mathfrak{g} \otimes \mathcal{T}_1^0(P)$ is a projection of $T_u(P)$ onto the vertical component $V_u(P) \simeq \mathfrak{g}$ with the following properties,

1. $\omega(A^\#) = A\,, \quad A \in \mathfrak{g}$;
2. $\rho_g^* \omega = Ad_{g^{-1}} \omega$;
3. $H_u(P) = \{X \in T_u(P) | \omega(X) = 0\}$.

For an open covering $\{U_I\}_{I \in \mathcal{I}}$ of $M$ and $s_I$ a local section defined on each $U_I$, the Lie algebra valued one-form $A_I$ defined on $U_I$ by,

$$A_I = s_I^* \omega \in \mathfrak{g} \otimes \Lambda^1(U_I)\,, \tag{2.84}$$

are called the connection potentials.

The $A_I$ obey the following identity,

$$\pi^* A_I = \pi^* \left[ Ad_{h_{IJ}} A_J - dh_{IJ} h_{IJ}^{-1} \right] \tag{2.85}$$

or pulling this expression back to $M$ we get,

$$A_I = Ad_{h_{IJ}} A_J - dh_{IJ} h_{IJ}^{-1}\,, \tag{2.86}$$

which is the transformation under a change of section (or trivialization or gauge).

Given a principal $G$-bundle $P(M, G)$, and a curve $c : [0\,, 1] \longrightarrow M$, the curve $\bar{c} : [0\,, 1] \longrightarrow P$ is the horizontal lift of $c$ if:

1. $\pi \circ \bar{c} = c$;
2. $\dfrac{d\bar{c}}{dt} \in H_{\bar{c}}(P)\,, \quad t \in [0\,, 1]\,.$

The horizontal lift can be proven to be unique.

For $g_c \in G$ the parallel transport equation is the following ordinary differential equation,

$$\dot{g}_c(t) = g_c(t) A_a(c(t)) \dot{c}^a(t)\,, \tag{2.87}$$

where $A_a$ is the connection potential. Given initial data $c(0)$ a solution $g_c$ exists, called the holonomy of $A$ along $c$. It is given by,

$$h_c(A) = \mathcal{P} \exp \left( \int_c A \right)\,, \tag{2.88}$$

where $\mathcal{P}$ is the path ordering operator it orders the smallest path parameter to the left. The connection $A$ is written in the form,

$$A \equiv A^i_a \dot{c}^a \tau_i\,, \tag{2.89}$$

with $\tau_i$ a Lie algebra basis.

Let $V$ be a vector space and $\alpha$ an vector valued $n$-form (i.e. $\alpha \in \Lambda^n(P) \otimes V$) the covariant derivative $\nabla\alpha$ is

$$\nabla\alpha[X_1 \ldots X_{n+1}] = d\alpha_p[X^H_1 \ldots X^H_{n+1}]\,, \tag{2.90}$$

where $X^H$ is the horizontal component of $X$ and $d$ is the exterior derivative.

The covariant derivative of the connection one form $\omega$ is the curvature two-form,

$$\Omega = \nabla\omega \in \Lambda^2(P) \otimes \mathfrak{g}\,. \tag{2.91}$$

The curvature two form $\omega$ has the property,

$$\rho^*_g \Omega = Ad_{g^{-1}} \Omega = g^{-1} \Omega g\,, \quad g \in G\,. \tag{2.92}$$

The connection one-form $\omega$ and the curvature two form $\Omega$ satisfy the Cartan structure equation,

$$\Omega = d_P \omega + \omega \wedge \omega\,, \tag{2.93}$$

see Appendix A for the definition of $\omega \wedge \omega$ and its relation to $[\omega, \omega]$.

The curvature and connection forms also satisfy the Bianchi identity,

$$\nabla\Omega = 0\,. \tag{2.94}$$

The local curvature form[6] $F_I$ of the curvature $\Omega$ is,

$$F_I = s_I^*\Omega\,, \tag{2.95}$$

and $F$ may be written in terms of the local connection as,

$$F = dA + A \wedge A\,. \tag{2.96}$$

Notions of connection and curvature may also be defined in the associated bundle $E = P \times_\rho V$ mentioned above. Roughly, a connection $A$ on a principal bundle $P$ completely determines the covariant derivative in the associated bundle $E$, modulo representations, refer to [46].

In the rest of this chapter, we present some examples of physical theories relevant to the quantization of gravity. We begin with electromagnetism (a $U(1)$ gauge theory) and continue to the Yang-Mills theory where we focus on the $SU(2)$ symmetry group. We then go on to describe a first order formulation of GR that uses the Cartan formalism. Finally we present a topological theory, the $BF$ model, that has relations with three-dimensional GR and with the full four-dimensional Einstein theory as well as with Yang-Mills theory.

## 2.4 Electrodynamics

The Maxwell theory (the standard reference is [49]) is the classical theory of the electromagnetic field. In the four-dimensional Minkowski space $M_4$, with coordinates $x^a$ $a = 0\,,1\,,2\,,3$, we define a (one form) potential $A = A_a dx^a$, this is a $U(1)$ gauge potential. The curvature is

$$F = dA\,, \tag{2.97}$$

this is also the field strength or the Faraday electromagnetic two form. The following Bianchi identity holds,

$$dF = ddA = 0\,. \tag{2.98}$$

[6] We will drop the index $I$ in some formulas.

The action is given by,

$$S = \int_{M_4} F \wedge \star F\,, \tag{2.99}$$

and its variation gives together with (2.98) the field equations ,

$$d \star F = 0\,, \qquad dF = 0\,. \tag{2.100}$$

It is noteworthy that $F$ is not uniquely determined by $A$ since for a 0-form $\chi$ we can make the change,

$$A \to A' = A + d\chi\,, \tag{2.101}$$

that leaves the $F$ unaltered, since $F' = d(A + d\chi) = dA$.

Expanding the forms in components we have,

$$F = \frac{1}{2} F_{ab} dx^a \wedge dx^b\,, \tag{2.102}$$

$$F_{ab} = \partial_a A_b - \partial_b A_a\,. \tag{2.103}$$

The Bianchi identity becomes,

$$\varepsilon^{abc} \partial_a F_{bc} = 0\,, \tag{2.104}$$

and the action is now,

$$S = \frac{1}{4} \int_{M_4} F_{ab} F^{ab}\,, \tag{2.105}$$

variation of this action gives,

$$\partial_a F^{ab} = 0\,. \tag{2.106}$$

We can write Maxwell's equations in their usual (vector notation) form by splitting $F_{ab}$ in its time and space components (using the $(- + ++)$ signature for the flat metric), we have for the electromagnetic potential,

$$A^a = \left(A^0, A^i\right) = (\varphi, \mathbf{A}) \tag{2.107}$$

where we are denoting the vector $A^i$ in boldface. The electromagnetic tensor splits in the following way,

$$F^{0i} = E^i\,, \tag{2.108}$$

$$F^{ij} = \frac{1}{2} \varepsilon^{ijk} B_k\,, \tag{2.109}$$

where $E^i \equiv \mathbf{E}$ is the electric field and, $B^i \equiv \mathbf{B}$ is the magnetic (induction) field.

With this splitting the field equations take the familiar form of the vacuum (i.e. in the absence of charges and currents) Maxwell equations,

$$\nabla \cdot \mathbf{E} = 0 \tag{2.110}$$
$$\nabla \times \mathbf{B} - \frac{\partial \mathbf{E}}{\partial t} = 0 \tag{2.111}$$
$$\nabla \times \mathbf{E} + \frac{\partial \mathbf{B}}{\partial t} = 0 \tag{2.112}$$
$$\nabla \cdot \mathbf{B} = 0\,. \tag{2.113}$$

## 2.5 The Yang-Mills theory

The Yang-Mills theory (YM) is a generalization of the Maxwell theory for non-abelian symmetry groups. It is a theory of paramount importance in particle physics — the standard model of particle physics is a YM theory for the group $SU(3) \times SU(2) \times U(1)$, for a survey of the enormous impact this theory had in the past decades refer to [50, 51]

Let $P \overset{\pi}{\longrightarrow} M$ be a principal $G$-bundle, $M$ be an $n$-dimensional manifold, $G$ a simple compact Lie group. Let $A$ be the local connection and $F_A = dA + A \wedge A$ the curvature.

The Yang-Mills action is given by,

$$S = tr \int_M F_A \wedge *F_A \tag{2.114}$$

where the trace indicates the symmetric bilinear non-degenerate form on the Lie algebra of $G$, denoted $\mathfrak{g}$, and $*$ the Hodge operator. We are interested here in the case of Minkowski flat metric $\eta$ for an application of general Lorentzian metric see [52]. The field equation is:

$$*\nabla_A * F_A = 0\,. \tag{2.115}$$

this equation together with the Bianchi identity

$$\nabla_A F_A = 0\,, \tag{2.116}$$

is a nonlinear version of Maxwell's equations.

The connection transforms in the usual way under gauge transformations,

$$A \to ad_{g^{-1}}A + g^{-1}dg\,, \tag{2.117}$$

where $g : M \longrightarrow G$ is the local form of a map in the gauge group see equation (2.85) and (2.86).

A class of important solutions is formed by the (anti)selfdual solutions. These are the connections satisfying,[7]

$$F_A^{\pm} = \pm * F_A^{\pm} \,. \tag{2.118}$$

These solutions are related to instantons, and are absolute minima of the Yang-Mills (Euclidean action) see [53] about this subject.

We further specialize to the case of a $SU(2)$ theory on flat Minkowski space-time. The relevant bundle is $P(M_4, SU(2))$ and the (local) connections is,

$$A = A^a{}_\mu T_a dx^\mu \,, \quad a = 1, 2, 3 \,, \tag{2.119}$$

where $T_a = \sigma_a/(2i)$ are the generators of the Lie algebra $\mathfrak{su}(2)$. Their commutator is,

$$[T_a, T_b] = \varepsilon_{abc} T_c \,, \tag{2.120}$$

where $\varepsilon_{abc}$ is the totally antisymmetric object in three dimensions.

We have also,

$$Tr(T_a T_b) = 2\delta_{ab} \,. \tag{2.121}$$

The action may thus be written,

$$S = -\frac{1}{4} \int_{M_4} d^4x F^a{}_{\mu\nu} F_a{}^{\mu\nu} \,, \tag{2.122}$$

where we have expressed the curvature in components and,

$$F^a{}_{\mu\nu} = \partial_\mu A^a{}_\nu - \partial_\nu A^a{}_\mu - g\varepsilon^{abc} \, A^b{}_\mu A^c{}_\nu \ . \tag{2.123}$$

The field equations (which we call the Yang-Mills equations) and the Bianchi identity become respectively,

$$\nabla_\mu F^{a\mu\nu} = 0 \,, \tag{2.124}$$

and,

$$\varepsilon^{\sigma\mu\nu\rho} \nabla_\mu F^a{}_{\nu\rho} = 0 \tag{2.125}$$

We use the following identifications,

$$F^{a\,0i} = E^{a\,i} \equiv \mathbf{E} \,, \tag{2.126}$$

$$\frac{1}{2}\varepsilon^{ijk} F^a{}_{ij} = B^{a\,k} \equiv \mathbf{B} \,, \tag{2.127}$$

[7] Where the $+$ sign is for the sefdual connection $A^+$, and the $-$ for the anti-seftdual one $A^-$.

note that the components of **E** and of **B** are $3 \times 3$-matrices.

Taking the time and space components of (2.124) and (2.125) we find,

$$\nabla \cdot \mathbf{E} = g\left(\mathbf{A} \cdot \mathbf{E} - \mathbf{E} \cdot \mathbf{A}\right) \tag{2.128}$$

$$\nabla \times \mathbf{B} = \frac{\partial \mathbf{E}}{\partial t} - g\left(A_0\mathbf{E} - \mathbf{E}A_0\right) + g\left(\mathbf{A} \times \mathbf{E} - \mathbf{E} \times \mathbf{A}\right) \tag{2.129}$$

$$\nabla \times \mathbf{E} + \frac{\partial \mathbf{B}}{\partial t} = g\left(A_0\mathbf{B} - \mathbf{B}A_0\right) + g\left(\mathbf{A} \times \mathbf{B} - \mathbf{B} \times \mathbf{A}\right) \tag{2.130}$$

$$\nabla \cdot \mathbf{B} = g\left(\mathbf{A} \cdot \mathbf{B} - \mathbf{B} \cdot \mathbf{A}\right) . \tag{2.131}$$

These are analogous to Maxwell's equations (see [54, 55]) but the non-linear character of the Yang-Mills theory introduces quantities that behave like sources of the fields even in the vacuum.

## 2.6 First order formalism for General Relativity

In Section 1.4 we considered the *Palatini formalism* defined in terms of two independent quantities the metric $g$, and the affine connection $\Gamma$. The action functional is to be varied with respect to both these fields.

We now consider this same variational method but defined in terms of two other quantities the coframe $e$ amd the connection $\omega$, which we will relate to $g$ and $\Gamma$.In terms of these variables, the gravitational field is represented by $e$. The reasons for this are to be found in the fact that in the standard model of particle physics the coupling to fermions is done using this field [12] and also the fact that the variables are better suited to describe gravity as a gauge theory (see [56] about this).

Let space-time be a 4-dimensional smooth manifold $M$. The co-frame field $e$ is a map of vector bundles (see [57, 58])

$$\begin{array}{ccc} TM & \xrightarrow{\;e\;} & \mathcal{T} \\ & {\scriptstyle P}\searrow \quad \swarrow{\scriptstyle \pi} & \\ & M & \end{array}$$

where $\mathcal{T}$, an internal bundle isomorphic to the tangent bundle $TM$. If $\mathcal{T}$ is trivializable and $e : TM \longrightarrow \mathcal{T} = M \times \mathbb{R}^{1,3}$ is a choice of trivialization then on each tangent space $e_x : T_x(M) \longrightarrow \mathbb{R}^{1,3}$ is a co-frame. The bundle $\mathcal{T}$ is equipped with a fixed metric $\eta$ and we can pull this back to $TM$ using $e$,

$$g(u, v) = \eta(eu, ev)\,, \tag{2.132}$$

for $u\,, v$ vector in $T_x(M)$. In this way, we have a relation between $e$ and $g$.

Using the non-coordinate basis (2.11) and (2.15) we have that (2.132) becomes in indices,

$$g_{\mu\nu} = e^a{}_\mu e^b{}_\nu \eta_{ab}\,. \tag{2.133}$$

If we have a connection $\omega$ defined in $\mathcal{T}$ we can pull it back to $TM$ using $e$ if it is an isomorphism. For a local section $s$ of $\mathcal{T}$ the covariant derivative is,

$$\nabla^\omega_\mu s^a = \partial_\mu s^a + \omega^a{}_{b\mu} s^b\,, \tag{2.134}$$

where $\nabla^\omega_\mu = \nabla^\omega_{\partial_\mu}$ the derivative in the direction of the basis vector.

In $TM$ we can differentiate a section in the usual way,

$$\nabla^\Gamma_\mu s^\nu = \partial_\mu s^\nu + \Gamma^\nu{}_{\mu\beta} s^\beta\,, \tag{2.135}$$

using the connection $\Gamma$.

If we use $e$ to pull back the vector $u$ in $TM$, differentiate it using $\nabla^\omega$ and use $e^{-1}$ to turn the result back into $TM$ we have,

$$\nabla^\Gamma_\nu u = e^{-1}\left(\nabla^\omega_\nu\,(eu)\right) \tag{2.136}$$

For $\nabla^\omega_\mu = \nabla^\omega_{\partial_\mu}$

$$\nabla^\Gamma_\mu u = e^a{}_\nu \left(\nabla^\omega_\mu\,(e_\alpha{}^\nu u^\alpha)\right) \tag{2.137}$$

whence

$$\partial_\mu e^a{}_\nu - \Gamma^\sigma{}_{\mu\nu} e^a{}_\sigma + \omega^a{}_{b\mu} e^b{}_\nu = 0\,. \tag{2.138}$$

This condition has generated a lot of misunderstandings, refer to [59] for the clarification of some of these.

In these variables the action fror GR can be written (see [12, 56, 60]) in the following way

$$S_{EC} \;=\; \frac{1}{4k}\int_M \varepsilon_{abcd}\left(e^a\wedge e^b\wedge R^{cd} - \frac{\Lambda}{6} e^a\wedge e^b\wedge e^c\wedge e^d\right)\,, \tag{2.139}$$

Where the curvature $R$ and torsion $T$ are given by eqs. (2.57) and (2.58) respectively, and $\Lambda$ is the cosmological constant. We call this action the Einstein-Cartan action although it is just the Palatini action (1.39) written in terms of $\omega$ and $e$ and in the case where the tetrads are invertible i.e. $\det(e^a{}_\mu) \neq 0$ (for the equivalence of the two formulations see Appendix C).

This action is invariant for diffeomorphisms and for (local) Lorentz transformations (refer to [61]).

Variation with respect to $\omega$ and $e$ yields respectively,

$$\varepsilon_{abcd}\nabla\left(e^a \wedge e^b\right) = 0 \tag{2.140}$$

$$\varepsilon_{abcd}e^a \wedge R^{bc} - \frac{\Lambda}{3}\varepsilon_{abcd}e^a \wedge e^b \wedge e^c = 0 \tag{2.141}$$

from the first equation above, if the tetrads are invertible i.e. $det(e^a{}_\mu) \neq 0$, we recover the no torsion condition $T^a = \nabla e^a = 0$, which constrains the connection to be the Levi-Civita connection. The second is equivalent to Einstein's equation with cosmological term

$$R_{\mu\nu} - \frac{1}{2}g_{\mu\nu}R + \Lambda g_{\mu\nu} = 0\,. \tag{2.142}$$

To couple GR to fermions we add the matter action[8],

$$S_D = i\kappa_1 \int \varepsilon_{abcd}\, e^a \wedge e^b \wedge e^c \wedge \bar{\psi}\left(\gamma^d \overset{\leftrightarrow}{d} + \{\omega, \gamma^d\} + \frac{im}{2}\, e^d\right)\psi\,, \tag{2.143}$$

where $\omega = \omega_{ab}[\gamma^a, \gamma^b]/8$ and $\kappa_1 = 8\pi l_p^2/3$.

In this case, variation of $S_{EC} + S_D$ with respect to $e$ gives,

$$T_a \equiv \nabla e_a = -\kappa_2 s_a\,, \tag{2.144}$$

where the spin 2-form is,

$$s_a = i\varepsilon_{abcd}\, e^b \wedge e^c\, \bar{\psi}\gamma_5\gamma^d\psi\,, \tag{2.145}$$

and $\kappa_2 = -3\kappa_1/4$.

## 2.7 The *BF* theory

A theory that does not have *local* degrees of freedom is called a *topological theory*, see [62] for an axiomatization. Topological theories are in a sense the simplest kind of gauge theories. These theories do not depend on any background geometry, the metric does not feature in the action or the field equations. They are therefore a realization of background independent theories. Since these theories do not have local degrees of freedom all interesting observables are global (see [63] for the construction of observables in the case of the *BF* theory).

[8]Note the importance of tetrads in this regard. We are not aware of any other way of coupling fermionic matter to gravity.

The *BF* theory is a topological theory, it can be defined in any dimension and for a general (under some restrictions) Lie group. This theory is related to other areas of Physics like electromagnetism [64], condensed matter [65] and hydrodynamics [66]. We are however, interested in *BF* theory for its relation to gravity (refer to [67] for a review). Since the theory has a lot of symmetry to extract metric degrees of freedom from the fields a constraint has to be added, this is the Plebanski approach [68] or a symmetry breaking term is added to the theory, this is the case of the MacDowell and Mansouri action [69].

Take a Lie group $G$ whose Lie algebra $\mathfrak{g}$ that has an invariant nondegenerate bilinear form (the Cartan-killing form $\langle x, y\rangle = tr(xy)$), $M$ an $n$-dimensional smooth manifold and a principle $G$-bundle $P$ over $M$.

A connection $A$ is defined on P, and we also need an $ad(p)$-valued ( $ad(p)$ is the vector bundle associated to P via the adjoint action of G in its Lie algebra) $(n-2)$-form $B$.

Given a local trivialization of P we may regard $A$ as a $\mathfrak{g}$-valued one-form on $M$, its curvature $F$ as a $\mathfrak{g}$-valued two-form and $B$ as a $\mathfrak{g}$-valued $(n-2)$-form.

The *BF* action is

$$S = \int_M \mathrm{tr}\, B \wedge F\,, \tag{2.146}$$

where $F = dA + A \wedge A$ Variation of the action with respect ot the connection $A$ and $B$ gives the field equations,

$$\begin{array}{rcl} \nabla_A B &=& 0 \\ F &=& 0 \end{array}\,. \tag{2.147}$$

Where $\nabla_A$ is the exterior covariant derivative.

The equations must be completed with the Bianchi identity,

$$\nabla_A F = 0\,. \tag{2.148}$$

The second equation of (2.147) says that the connection $A$ is flat. All flat connections are the same modulo gauge transformations. The infinitesimal gauge transformations are,

$$A \mapsto A + \nabla_A \alpha\,, \qquad B \mapsto [B, \alpha] \tag{2.149}$$

where $\alpha : M \longrightarrow \mathfrak{g}$.

The first of the field equations (2.147) has another symmetry,

$$A \mapsto A\,, \qquad B \mapsto B + \nabla_A \eta\,, \tag{2.150}$$

where $\eta$ is some $ad(P)$-valued $(n-3)$-form.

The relation with three-dimensional gravity is evident if we take $n = 3$, $G = SO(2,1)$ and choose minus the trace for the invariant form mentioned above. A Lorentzian metric on $M$ can be defined by[9],

$$g(u,v) = \langle Bu\,, Bv\rangle\ , \quad u\,,v \in T_p(M)\,, \tag{2.151}$$

provided that $B : T(M) \longrightarrow ad(P)$ is one-to-one. The map $B$ can also be used to pull back the connection $A$ so that we have a metric preserving connection $\Gamma$ is obtained on the tangent bundle $T(M)$. In this case the equation $\nabla_A B = 0$ turns into the no torsion condition for the metric $B$ which is effectively the Levi-Civita connection on $M$. The equation $F = 0$ enforces the flatness of $\Gamma$ and therefore of $g$. We therefore have a (vacuum) theory of flat metric in 3 dimensions. Since in $(2+1)$-dimensions GR is also a theory of flat metrics, and since modulo gauge transformations all flat metrics are equal, the *BF* theory is just an alternate formulation of Lorentzian General Relativity.

This is true, as stated above, if the map $B$ is one-to-one if it is not, the metric becomes degenerate and the *BF* theory with gauge group $SO(2,1)$ may be regarded as an extension of the Einstein theory to the case of degenerate metrics.

In the four dimensional case, GR can be recovered by constraining the form $B$ to be,

$$B = \star\,(e \wedge e)\ , \tag{2.152}$$

where $e \equiv e^a{}_\mu$ is the co-tetrad and $\star$ is the Hodge dual with respect to the internal metric $\eta_{ab}$. This is called the *simplicity constraint*.[10]

The Plebanski action enforces the simplicity constraint, it reads

$$S_{Pl} = \int_M B^{ab} \wedge F_{ab} + \varphi_{abcd} B^{ab} \wedge B^{cd}\,, \tag{2.153}$$

where the Lagrange multiplier $\varphi$ is by definition symmetric under the exchange of the first and second pair of indices, and antisymmetric within each pair.

The *BF* theory can allso be deformed in such a way that we recover Yang-Mills theory. The action is,

$$S = \int_M B \wedge F_A + qB \wedge *B\,, \tag{2.154}$$

where $q$ is a coupling constant.

---

[9]We use the notation $Bu = B[u]$, since $u$ is a vector and $B$ is a form.

[10]A simple form is one that may be written as the exterior product of form of degree one, hence the name.

Variation of the action with respect to the forms $B$ and $A$ gives

$$F_A + q * B = 0 \tag{2.155}$$
$$\nabla_A B = 0\,, \tag{2.156}$$

taking $q^{-1} * \nabla_A *$ of the first equation[11] and using the second we find $g^{-1} * \nabla_A * F_A = 0$ the Yang-Mills field equation (2.115).

[11]Note that, with a Lorentzian metric $** = -1$ when acting on 2-forms.

# 3
# The Dirac Procedure

In this chapter, after a brief review of Lagrangian and Hamiltonian mechanics, we present the Dirac-Bergman iterative algorithm for the calculation of all the constraints, and the Dirac classification of constraints. This method is considered as a first step for the canonical quantization of gauge theories. This subject is presented in many textbooks, we cite those we used in this Chapter [55, 70, 71, 72, 73]. We also apply this method to some of the examples of Chapter 2. In the case of topological theories, this procedure will enable us both to confirm the topological character of the theory (it will have zero local degrees of freedom) and to find the complete set of the constraints as well as their characterization.

*Gauge theories,* are those in which a physical system is described by more variables than the independent degrees of freedom of the system. The relevant degrees of freedom are those invariant under transformations connecting the variables. These are called *gauge transformations.* In other words, gauge transformations change the variables but leave the physical degrees of freedom unaltered.

As a consequence, gauge theories have the property that the general solution of the equations of motion contains arbitrary functions of time. Since we have 'too many variables', we will need to find relations among these variables. These relations are called *constraints*. Furthermore we have that a gauge system is always a constrained system, the converse is not always true: not all constraints arise from a gauge invariance. (Refer to [71] for details).

We start by reviewing the Lagrangian and Hamiltonian formulation of classical mechanics.

## 3.1 Review of Lagrangian and Hamiltonian mechanics

Let us assume that a system has a finite number $N$ of degrees of freedom. The equations of motion are obtained by varying the action functional:

$$S = \int_{t_1}^{t_2} L(q, \dot{q}) dt\,, \tag{3.1}$$

with respect to the generalized coordinates $q^n \quad n = 1, \ldots, N$. In this equation $L(q^n, \dot{q}^n)$ is called the Lagrangian function. That is, the equations of motion are those that make the above action equation (3.1) stationary under variations of the form $\delta q^n$ that vanish at the end points $t_1$ and $t_2$.

This variation gives the Euler-Lagrange equations:

$$\frac{\delta S}{\delta q^n} \equiv \frac{\partial L}{\partial q^n} - \frac{d}{dt}\left(\frac{\partial L}{\partial \dot{q}^n}\right) = 0\,, \quad \text{with } n = 1 \ldots N \tag{3.2}$$

here $\frac{\delta L}{\delta q^n}$ is a functional derivative, that is,

$$\delta S = \int_{t_1}^{t_2} \frac{\delta L(q, \dot{q})}{\delta q^n} \delta q^n dt\,, \tag{3.3}$$

for $n = 1 \ldots N$. We are assuming the usual summation convention, whenever an index is repeated up and down a summation is implied.

The Euler-Lagrange equations (3.2) can be written in the form,

$$\frac{\partial^2 L}{\partial \dot{q}^i \partial \dot{q}^j} \ddot{q}^j = \frac{\partial L}{\partial q^i} - \frac{\partial^2 L}{\partial \dot{q}^i \partial q^j} \dot{q}^j\,, \quad i, j = 1 \ldots N \tag{3.4}$$

or in matrix notation,

$$M_{ij} \ddot{q}^j = K_i\,, \tag{3.5}$$

with,

$$M_{ij} = \frac{\partial^2 L}{\partial \dot{q}^i \partial \dot{q}^j} \tag{3.6}$$

$$K_i = \frac{\partial L}{\partial q^i} - \frac{\partial^2 L}{\partial \dot{q}^i \partial q^j} \dot{q}^j\,. \tag{3.7}$$

Thus we see that if the matrix $M$ (the Hessian of $L$) is invertible that is,

$$\det(M_{ij}) = \det\left(\frac{\partial^2 L}{\partial \dot{q}_i \partial \dot{q}_j}\right) \neq 0\,, \tag{3.8}$$

we can write,

$$\ddot{q}^i = (M^{-1})^{ij} K_j = F^i(q,\dot{q})\,, \tag{3.9}$$

which means that, under the conditions of the existence and uniqueness theorems for the solutions of systems of differential equations, and given appropriate boundary (initial) conditions the equations are solvable with respect to the highest derivative, the accelerations $\ddot{q}$ (refer to [72, 73]).

A theory is called *singular* if

$$\det(M_{ij}) = \det\left(\frac{\partial^2 L}{\partial\dot{q}_i\partial\dot{q}_j}\right) = 0\,, \tag{3.10}$$

and non-singular if the condition (3.8) holds.

The canonical momenta are given by,

$$p_n = \frac{\partial L}{\partial\dot{q}^n}\,, \quad n = 1\dots N\,. \tag{3.11}$$

We see that the Hessian matrix mentioned above is equal to $\frac{\partial p_i}{\partial\dot{q}_j}$ so the above requirement (3.8) means that the $p_i$ are invertible as functions of the 'velocities' $\dot{q}^i$.

We call *configuration space,* the $N$-dimensional space whose coordinates are the $N$ generalized coordinates $\dot{q}^n$. The $2N$ dimensional space whose coordinates are the $\dot{q}^n$ and $p^n$ given by (3.11) is called *phase space*.

The Hamiltonian, for a non-singular theory is given by the Legendre transform of the Lagrangian with respect to the generalized velocities $\dot{q}^n$ that is:

$$H(q^n,p_n) = \sum_{n=0}^{N} p_n\dot{q}^n - L(q^n,\dot{q}^n)\,. \tag{3.12}$$

Note that $H_c$ is a function of $q^n$ ,$p_n$, so that is the reason we need the velocities expressed as functions of the coordinates and momenta.

The equations of motion in the Hamiltonian formalism are:

$$\begin{aligned}\dot{q}^n &= \frac{\partial H}{\partial p_n}\\ \dot{p}_n &= -\frac{\partial H}{\partial q^n}\end{aligned}\,, \tag{3.13}$$

for $n = 1, \ldots, N$.

The Poisson bracket is defined in the following way,

$$\{f, g\} \equiv \sum_{i=1}^{N} \left( \frac{\partial f}{\partial q^i} \frac{\partial g}{\partial p_i} - \frac{\partial f}{\partial p_i} \frac{\partial g}{\partial q^i} \right) , \tag{3.14}$$

here $f = f(p_n, q^n, t)$ and $g = g(p_n, q^n, t)$.

The Poisson bracket is antisymmetric, linear (in each variable), obeys the product law, and the Jacobi identity. And the time evolution of a phase-space function is given by

$$\frac{df}{dt} = \{f, H\} + \frac{\partial f}{\partial t} . \tag{3.15}$$

The equations of motion are therefore,

$$\begin{aligned} \dot{q}^n &= \{q^n, H\} \\ \dot{p}_n &= \{p_n, H\} \end{aligned} , \tag{3.16}$$

In the continuum case we have fields that have an infinite number of degrees of freedom. The action for a field $\varphi^A$ (where the index $A$ denotes the set of indices of the field) without higher order time derivatives is

$$S = \int L dt , \tag{3.17}$$

where the Lagrangian $L$ is given by,

$$L = \int \mathcal{L}(\varphi, \partial_\mu \varphi) d^3x , \tag{3.18}$$

where $\mathcal{L}(\varphi, \partial_\mu \varphi)$ is the Lagrangian density.

We may then write,

$$S = \int \mathcal{L}(\varphi, \partial_\mu \varphi) d^4x . \tag{3.19}$$

Variation with respect to $\varphi^A$ gives

$$\frac{\delta S}{\delta \varphi^A} \equiv \frac{\delta_t L}{\delta \varphi^A} - \frac{d}{dt} \left( \frac{\delta_t L}{\delta \dot{\varphi}^A} \right) \tag{3.20}$$

where $\frac{\delta_t L}{\delta \varphi^A}$ and $\frac{\delta_t L}{\delta \dot{\varphi}^A}$ are variational derivatives with fixed time (see [72]).

We may also write for a theory with no explicit time dependence,

$$\frac{\delta S}{\delta \varphi^A} \equiv \frac{\partial \mathcal{L}}{\partial \varphi^A} - \partial_\mu \left( \frac{\partial \mathcal{L}}{\partial(\partial_\mu \varphi^A)} \right) . \tag{3.21}$$

The criterion for the non-singularity of a theory is the invertibility of the Hessian,

$$M_{AB}(x, x') \equiv \frac{\delta_t^2 L}{\delta \dot{\varphi}^A(x) \delta \dot{\varphi}^B(x')} , \tag{3.22}$$

and takes the form,

$$M_{AB}(x, x') = \frac{\partial^2 \mathcal{L}}{\partial \dot{\varphi}^A(x) \partial \dot{\varphi}^B(x')} \delta^{(3)}(x - x') . \tag{3.23}$$

The momenta conjugate to the 'variables' $\varphi^A$ are in this case[1]:

$$\pi_A = \frac{\delta L}{\delta \dot{\varphi}^A} , \tag{3.24}$$

with our assumptions we have,

$$\pi_A = \frac{\partial \mathcal{L}}{\partial \dot{\varphi}^A} , \tag{3.25}$$

The Hamiltonian is,

$$H = \sum_A \int \pi_A(x) \dot{\varphi}^A(x) d^3x - L . \tag{3.26}$$

The Poisson bracket in this case is defined by,

$$\{f(x), g(x')\} \equiv \sum_A \int d^3y \left( \frac{\delta f(x)}{\delta \varphi^A(y)} \frac{\delta g(x')}{\delta \pi_A(y)} - \frac{\delta f(x)}{\delta \pi_A(y)} \frac{\delta g(x')}{\delta \varphi^A(y)} \right) . \tag{3.27}$$

And Hamilton's equations become,

$$\begin{aligned} \dot{\varphi}^A &= \{\varphi^A, H\} \\ \dot{\pi}_A &= \{\pi_A, H\} \end{aligned} , \tag{3.28}$$

In the next sections we describe the Dirac procedure using the case for a finite number of degrees of freedom for simplicity.

[1]We drop the index $t$ in the derivative.

## 3.2 Primary and Secondary constraints

If we are unable to obtain the velocities as functions of the momenta and coordinates, that is if we have (3.10) the theory is singular. We then have that the set of momenta (3.11) are not all independent and there are some relations,

$$\varphi_m(q^n, p_n) = 0\,, \quad m = 1, \dots M\,. \tag{3.29}$$

These relations are called *primary constraints* (PC).

We assume that rank $\left(\frac{\partial^2 L}{\partial \dot{q}_i \partial \dot{q}_j}\right) = N - M'$ is constant and so the Primary Constraints define a submanifold embedded in the phase space called *primary constraint surface*. The dimension of this submanifold is $2N - M'$ because there are $M'$ independent PCs. There exists the possibility that the constraints may be dependent in which case $M > M'$. We then have, $M$ constraints, $M'$ of which are independent and $M - M'$ dependent of the others.

The $2N - M'$-dimensional constraint surface $\varphi_m = 0$ should be coverable with open regions on each of which the constraint surface can be split into independent constraints $\varphi_{m'} = 0 \quad m' = 1, \dots, M'$ such that the Jacobian matrix $\frac{\partial(\varphi_{m'})}{\partial(q^n, p_n)}$ has rank $M'$ on the constraint surface, and $M - M'$ dependent constraints $\varphi_{m''} = 0 \quad m'' = M' + 1, \dots, M$ that hold as consequences of the others.

There are other ways to state the condition on the Jacobian:

1. The functions $\varphi_1 \; \dots \; \varphi_{m'}$ can be taken as the first $M'$ coordinates of a regular coordinate system in an open set in the constraint surface;
2. the gradients $d\varphi_1 \dots d\varphi_{M'}$ are linearly independent at every point of the constraint surface;
3. the variations $\delta\varphi_{m'}$ are of order $\varepsilon$ for variations $\delta q^i$ and $\delta p_i$ of order $\varepsilon$.

The following results may be proved:

- **Theorem:** If a smooth Phase space function $G$ vanishes *on the constraint surface* $\varphi_m = 0$ then
$$G = g^m \varphi_m = 0\,, \tag{3.30}$$
for some phase-space functions $g^m$.

- **Theorem:** If $\lambda_n \delta q^n + \mu^n \delta p_n = 0$, for arbitrary variations $\delta q^n$ , $\delta p_n$ tangential to the constraint surface then

$$\lambda_n = u^m \frac{\partial \varphi_m}{\partial q^n} \tag{3.31}$$

$$\mu^n = u^m \frac{\partial \varphi_m}{\partial p_n} \,. \tag{3.32}$$

where $u^m$ are some functions.

We refer the reader to [71] for the proofs.

## 3.3 Canonical Hamiltonian

The canonical Hamiltonian is defined as:

$$H_c = p_n \dot{q}^n - L \,, \quad n = 1 \,, \ldots N \,. \tag{3.33}$$

We have that $H_c$ is well defined on the primary constraint surface, it can be arbitrarily extended off this surface and, the formalism remains unchanged by the replacement:

$$H \to H + c^m(q,p)\varphi_m \,. \tag{3.34}$$

and the $c^m$ will be determined below.

The Legendre transform from $(q, \dot{q})$-space to the surface $\varphi_m(q,p) = 0$ of the $(q,p,u)$-space is given by

$$\begin{aligned} q^n &= q^n \\ p_n &= \frac{\partial L}{\partial \dot{q}^n}(q, \dot{q}) \\ u^m &= u^m(q, \dot{q}) \,, \end{aligned} \tag{3.35}$$

note these are now spacees of the same dimensionality $2N$.

The inverse transform is:

$$\begin{aligned} q^n &= q^n \\ \dot{q}^n &= \frac{\partial H}{\partial p_n} + u^m \frac{\partial \varphi_m}{\partial p_n} \\ u^m &= u^m(q, \dot{q}) \end{aligned} \tag{3.36}$$

we recover invertibility of the Legendre transform (when $\det(\frac{\partial^2 L}{\partial \dot{q}_i \partial \dot{q}_j}) = 0$) at the cost of the introduction of extra parameters.

We can write Hamilton's equations for constrained systems as:

$$\begin{aligned} \dot{q}^n &= \frac{\partial H}{\partial p_n} + u^m \frac{\partial \varphi_m}{\partial p_n} \\ \dot{p} &= -\frac{\partial H}{\partial q^n} - u^m \frac{\partial \varphi_m}{\partial q^n} \\ u^m &= u^m(q, \dot{q}) \,, \end{aligned} \tag{3.37}$$

they can also be written in the form,

$$\dot{F} = \{F, H\} + u^m \{F, \varphi_m\} \tag{3.38}$$

where $F(p, q)$ is a function in phase space, and $m = 1, \ldots, M$.

The Total Hamiltonian is defined as,

$$H_T = H_c + u^m \varphi_m(q^n, p_n) \,, \tag{3.39}$$

where the summation convention is used.

The equations (3.38) may be written using (3.39) as,

$$\dot{F} = \{F, H_T\} \,. \tag{3.40}$$

## 3.4 Secondary Constraints

A basic requirement, is that the time evolution of the PC must vanish, these are the *consistency conditions*

$$\dot{\varphi}_m = \{\varphi_m, H_c\} + u^{m'} \{\varphi_m, \varphi_{m'}\} = \{\varphi_m, H_T\} = 0 \,. \tag{3.41}$$

Equation (3.41) may:

1. be identically satisfied, in which case the process stops;
2. give rise to new constraints, called *Secondary Constraints*(SC);
3. determine some of the $u$'s.

If any SC's $\chi(q^n, p_n) = 0$ arise, new consistency condition similar to (3.41) must be imposed that is,

$$\dot{\chi} = \{\chi, H_c\} + u^{m'} \{\chi, \varphi_{m'}\} = \{\chi, H_T\} \approx 0 \,. \tag{3.42}$$

This process will continue until no new SC's arise. We are then left with $M+K$ constraints:

$$\varphi_j(q^n, p_n) = 0 \,, \quad j = (1 \ldots M, M+1 \ldots M+K = J) \tag{3.43}$$

where $M$ is the number of primary constraints, and $K$ is the number of secondary constraints. This is a complete set of constraints. The process is depicted schematically in figure (3.1).

## 3.5 Weak and Strong equalities

Two functions $F$ and $G$ that coincide on the submanifold defined by the constraints $\varphi_j \,, \quad j = (1 \ldots M, M+1 \ldots M+K)$ are said to be weakly equal an denoted $F \approx G$.

In particular the constraint equations (3.43) are written,

$$\varphi_j(q^n, p_n) \approx 0 \,, \quad j = (1 \ldots M, M+1 \ldots M+K) \,. \tag{3.44}$$

Using (3.30) with $\varphi_m$ replaced by $\varphi_j$ we have,

$$F \approx G \Longleftrightarrow F - G = C^j(p, q)\varphi_j \,. \tag{3.45}$$

If a quantity vanishes in all of phase space we say that it is strongly equal to zero and use $=$ in this case. We will also refer weak equality $\approx$ and *on-shell* and strong equality $=$ as *off-shell*.

## 3.6 Determination of the Lagrange Multipliers

The consistency conditions for this set (3.43) now results only in equations determining the $u$'s. That is,

$$\left\{\varphi_j, H_c\right\} + u^{m'} \left\{\varphi_j, \varphi_{m'}\right\} \approx 0 \,, \tag{3.46}$$

with the summation index $m = 1, \ldots, M$ and, $j = 1, \ldots, J$. The equations (3.46) form a system of $J$ linear equations on the $M$ unknowns $u^m$. Accordingly, the most general solution is given by,

$$u^m = U^m + V^m \,, \tag{3.47}$$

where, $U^m$ is a particular solution of the full equation and, $V^m$ is the general solution of the homogeneous equation i.e.

$$V^m \left\{\varphi_j, \varphi_m\right\} \approx 0 \,, \quad j = 1, \ldots, J \,, \tag{3.48}$$

and, the most general solution to this homogeneous system is a linear combination of independent solutions namely, $V_a^m$ with $a = 1, \ldots, A$.

Putting this together, most general solution of (3.46) can then be written:

$$u^m = U^m + v^a V_a^m \,, \tag{3.49}$$

the $v^a \equiv v^a(t)$ are arbitrary functions of time. The total Hamiltonian can now be written,

$$\begin{aligned} H_T &= H_c + u^m \varphi_m & (3.50)\\ &= H_c + \left(U^m + v^a V_a^m\right) \varphi_m \\ &= \left(H_c + U^m \varphi_m\right) + v^a V_a^m \varphi_m \\ &= H' + v^a \varphi_a \,. & (3.51) \end{aligned}$$

Where we have defined,

$$\varphi_a = V_a^m \varphi_m \,, \tag{3.52}$$
$$H' = H_c + U^m \varphi_m \,. \tag{3.53}$$

The index $m$ is summed in the range $1, \ldots, M$, and $a = 1, \ldots, A$.

Here we have a dependence of the Hamiltonian on arbitrary functions of time $v^a$, and we therefore conclude that initial conditions alone are not sufficient to determine state of the system (that is, the values of dynamical variables) for all time.

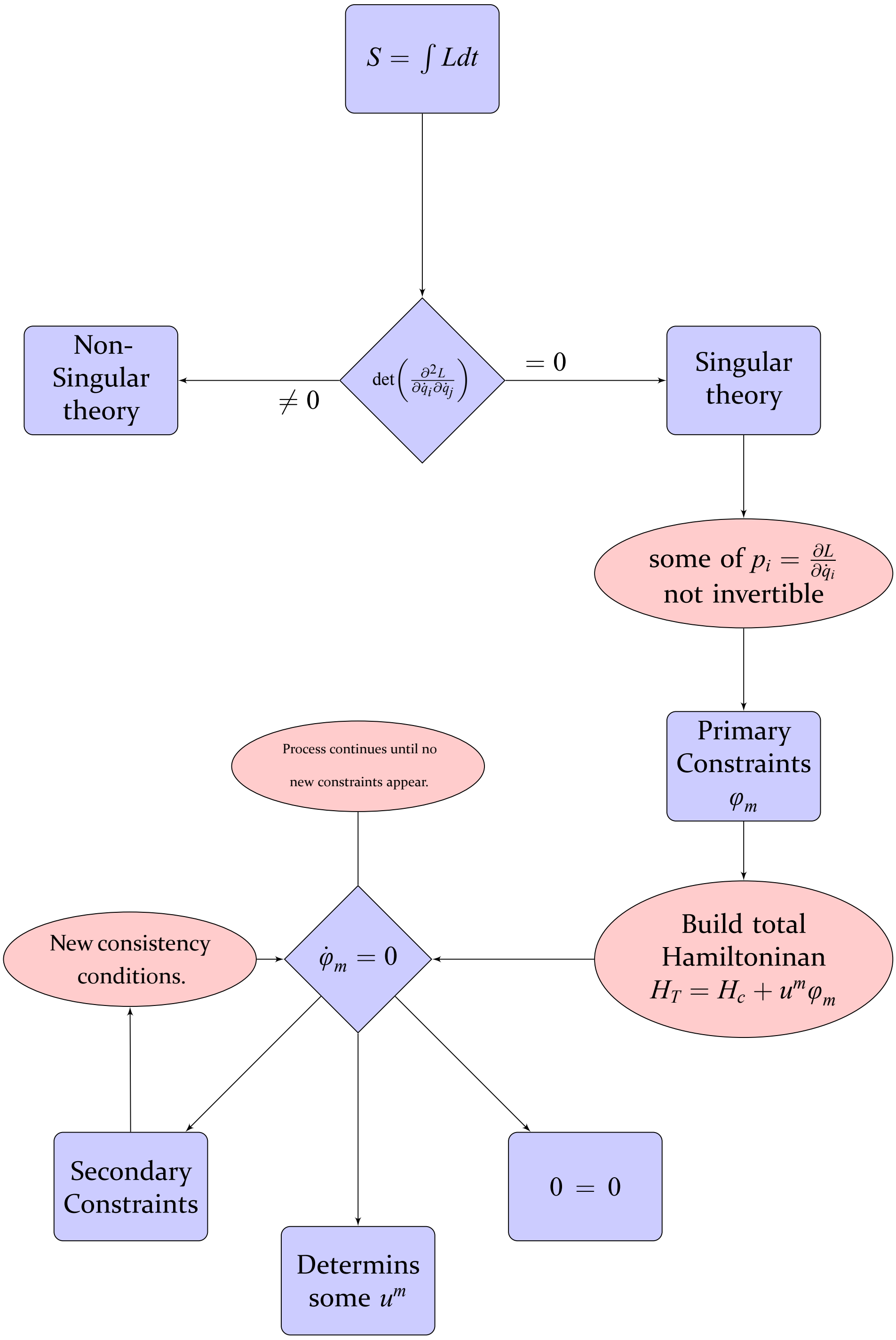

**Figure 3.1:** Schematic depiction of the Dirac procedure steps needed to obtain all the constraints in a singular theory. For detail and the meaning of the symbols see the main text.

## 3.7 First and Second Class constraints

The distinction between primary and secondary constraints is not fundamental in the construction of the Hamiltonian formalism.

A more useful classification of constraints (and phase space functions in general) distinguishes between *first and second class* constraints.

A function $F(q^n, p_n)$ is *First Class* (FC) if it has a weakly vanishing Poisson Bracket with every constraint:

$$\left\{F(q^n, p_n), \varphi_j\right\} \approx 0 \quad j = 1, \ldots, J. \tag{3.54}$$

A function that is not FC (that is, a function that has at least one constraint such that its Poisson bracket does not vanish weakly) is called *Second Class* (SC).

We note that the FC property is preserved by the Poisson bracket, this means the Poisson bracket of two FC functions if FC.

The quantities $H'$ and $\varphi_a$ in (3.51) are FC as a result of eqs. (3.46) and (3.48) In this way the total Hamiltonian can be (non uniquely) split into an FC Hamiltonian $H'$ and an FC (complete) set of Primary Constraints $\varphi_a$ multiplied by arbitrary functions.

As stated earlier, the presence of arbitrary functions of time means the canonical variables and their momenta cannot be uniquely determined from their initial values. These variables ($q$'s and $p$'s) therefore, do not have a direct physical significance.

Consider a dynamical variable $F(t)$ at time $t_0$ and its change after $\delta t$. The initial value $F(t_0)$ is determined by the canonical variables $(q^n(t_0), p_n(t_0))$ at time $t_0$. The value at time $\delta t$ can be calculated by,

$$\begin{aligned} F(\delta t) &= F(t_0) + \delta t \dot{F} \\ &= F(t_0) + \delta t \{F, H_T\} \\ &= F(t_0) + \delta t \left(\{F, H'\} + v^a \{F, \varphi_a\}\right), \end{aligned} \tag{3.55}$$

where in the second line (3.40) was used.

The functions $v^a$ are arbitrary, which means that if we choose different values for the $v^a$ we will find different $F(\delta t)$. We find that the difference in the values of a dynamical variable $F$ between $t$ and $t + \delta t$ and for two different choices of the arbitrary functions that is, $v^a$ and $\tilde{v}^a$ is,

$$\delta F = \delta v^a \{ F \, , \, \varphi_a \} \tag{3.56}$$

note that $\varphi_a$ with $a = 1, \ldots, A$ was defined in (3.52), and $\delta v^a = (v^a - \tilde{v}^a)\delta t$.

The transformations $\delta F$ does not correspond to a change in physical state. These transformations are called *gauge transformations* precisely because, although they change the canonical variables they leave the physical state unaltered. The $\varphi_a$ (primary first class constraints) generate these transformations.

If the equations $\varphi_j = 0 \, , j = 1 \, , \ldots \, , J$ are not independent the constraints are called reducible. When all the constraints are independent they are called irreducible.

The gauge transformations (3.56) are independent if and only if the constraints are irreducible. When the constraints are reducible some of the gauge transformations will lead to $\delta F = 0$. (See [71]).

It can be shown that:

1. the quantity $\{\varphi_a, \varphi_b\}$ where $\varphi_{a,b}$ are any FC primary constraints, generates a transformation thet does not change the physical state, a *gauge transformation*.
2. $\{\varphi_a, H'\}$ where $\varphi_a$ is FC primary constraints also generates a gauge transformation.

These results show that we can expect that at least some secondary first class constraints will act as gauge generators.

Dirac conjectured that *all* first class constraint in a theory generate gauge transformations. This is called the *Dirac conjecture.*

To include all the gage freedom the Hamiltonian is,

$$H_E = H' + u^a \gamma_a \, , \tag{3.57}$$

where $\gamma_a$ incorporates all (both primary and secondary) FC constraints. This is called the *extended Hamiltonian*.

## 3.8 Second Class constraints and Dirac Bracket

If a theory has Second Class (SC) constraints, they indicate that some of the degrees of freedom in the theory can be neglected. The presence of SC is known by the fact that the matrix $C_{jj'} = \{ \varphi_j \, , \, \varphi'_j \} \, , jj' = 1 \, \ldots \, , J$, does not vanish on the constraint surface.

After finding all the FC constraints which we will call $\gamma_a \, , a = 1 \, , \ldots \, N_1$, the remaining $\chi_\alpha \quad \alpha = 1, \ldots, N_2$ are SC, and the matrix $C_{\alpha\beta} = \left\{ \chi_\alpha, \chi_\beta \right\}$ is invertible (refer to [70, 71]). We then define the *Dirac bracket* as

$$\{f, g\}_D = \{f, g\} - \{f, \chi_\alpha\} \, C^{\alpha\beta} \left\{ \chi_\beta, g \right\} \, . \tag{3.58}$$

where $C^{\alpha\beta}$ is the inverse of $C_{\alpha\beta}$ that is, $C^{\alpha\beta} C_{\beta\gamma} = \delta^\alpha_\gamma$.

By construction the Dirac Bracket of any second class constraint $\chi_\alpha$ with any variable vanishes,

$$\{\chi_\alpha, F\}_D = 0 \, . \tag{3.59}$$

We have the following situation: the Poisson bracket served to distinguish between FC and SC constraints and once the Dirac bracket is introduced the PB may be abandoned. The SC constraints become strong equalities and all the equations of the theory may be formulated in terms of Dirac brackets. In principle, we may always use the SC constraints to eliminate some of the degrees of freedom of the theory, but in practice this may be extremely difficult.

So to sum up, FC constraint generate gauge transformations whereas the SC may be treated as strong equalities after one introduces Dirac brackets.

## 3.9 Gauge fixing and counting of degrees of freedom

Since the presence of first class constraint implies the existence of gauge transformations and that the the canonical variables are not in on-to-one correspondence with the physical states. Sometimes conditions are introduced to restore this correspondence. These are called *gauge (fixing) conditions*.

A gauge condition,

$$C_b(p, q) \approx 0 \, , \tag{3.60}$$

must satisfy two conditions:

1. Accessibility: given any set of canonical variables $(p^n,\, q_n)$ there must exist a gauge transformation that maps the given set of variables onto a set satisfying (3.60).

2. The gauge conditions (3.60) must fix the gauge completely. This implies that no gauge transformations except the identify will preserve the above gauge conditions. Or stated differently the equations,

$$\delta u^a \{ C_b \,, \gamma_a \} \approx 0 \,, \tag{3.61}$$

   must only be solved by $\delta u^a = 0$, where $\gamma_a$ are the first class constraints.

   In particular this means that we must have,

$$\det\{ C_b \,, \gamma_a \} \neq 0 \,. \tag{3.62}$$

These two conditions together imply that in order to completely fix the gauge, the number of independent gauge conditions must be equal to the number of independent first class constraints. We also see from (3.62) that the gauge conditions together with the FC constraints form a second class set $C_b,\ \gamma_a$.

After the gauge fixing only second class constraints remain, we will have no arbitrary function in the Hamiltonian, and a set of canonical variables that satisfies the constraint equations determines one and only one pysical state.

The following counting of independent degrees of freedom therefore holds. Given $N$ initial independent variables, the dimension of the phase space is $2N$. From this one subtracts the total number $F$ of first class constraints, the total number $S$ of second class constraints, and the total number $F$ of gauge fixing conditions, which we know to be equal to the number of first class constraints. The result is the dimension of the physical phase space, $2n$, where $n$ is the number of physical degrees of freedom.

Thus we have the general formula,

$$n = N - F - \frac{S}{2} \,. \tag{3.63}$$

In the continuum case, the phase space is infinite dimensional, the previous formula holds *only locally*. That is per point.

## 3.10 Generators of gauge transformations

Consider a system determined by the total Hamiltonian (3.39) and a complete set of constraints $\varphi_j \approx 0\,, j = 1\,, \ldots\,, J$.

Suppse we have a trajectory in phase space $(q^i(t), p_i(t))$ which starts from the point $(q^i(0), p_i(0))$ on the constraint surface.

The equations of motion for some choice of $v^a$ take the form,

$$\begin{aligned} \dot{q}^i &= \frac{\partial H'}{\partial p_i} + v^a \frac{\partial \varphi_a}{\partial p_i} \\ -\dot{p}_i &= \frac{\partial H'}{\partial q^i} + v^a \frac{\partial \varphi_a}{\partial q^i} \\ & \varphi_a(q^i, p_i) = 0\,. \end{aligned} \tag{3.64}$$

in this and the following equations the index $a$ is summed in the range $1\,, \ldots\,, A$.

Take a variation of this trajectory of the form $(q^i(t) + \delta_0 q^i(t), p_i(t)) + \delta_0 p_i(t)$ that starts form the same point and satisfies the equations of motion with new functions $v^a(t) + \delta_0 v^a(t)$. Expanding equations (3.64) for small variations of the canonical variables and arbitrary function we have,

$$\delta_0 \dot{q}^i = \left(\delta_0 q^j \frac{\partial}{\partial q^j} + \delta_0 p_j \frac{\partial}{\partial p_j}\right) \frac{\partial H_T}{\partial p_i} + \delta_0 v^a \frac{\partial \varphi_a}{\partial p_i} \tag{3.65}$$

$$-\delta_0 \dot{p}_i = \left(\delta_0 q^j \frac{\partial}{\partial q^j} + \delta_0 p_j \frac{\partial}{\partial p_j}\right) \frac{\partial H_T}{\partial q^i} + \delta_0 v^a \frac{\partial \varphi_a}{\partial q^i} \tag{3.66}$$

$$\left(\delta_0 q^j \frac{\partial}{\partial q^j} + \delta_0 p_j \frac{\partial}{\partial p_j}\right) \varphi_a(q^i, p_i) = 0\,, \tag{3.67}$$

these are the conditions that the varied trajectories must satisfy to be dynamical.

The generators of the gauge transformations are functions of the form,

$$G = \sum_{n=0}^{k} \varepsilon^{(n)} G_n\,, \tag{3.68}$$

where $\varepsilon^{(n)}$ is the $n^{\text{th}}$-time derivative of $\varepsilon$.

And the variations in the canonical variables are given (refer to [74] for details) by,

$$\delta_0 q^i = \sum_{n=0}^{k} \varepsilon^{(n)} \{ q^i \, , \, G_n \} \, , \tag{3.69}$$

$$\delta_0 p_i = \sum_{n=0}^{k} \varepsilon^{(n)} \{ p_i \, , \, G_n \} \, . \tag{3.70}$$

With the $G_n$ determined recursively by the following conditions,

$$\begin{aligned} G_k &= \text{primary} \\ G_{k-1} + \{ G_k \, , \, H \} &= \text{primary} \\ &\vdots \\ G_0 + \{ G_1 \, , \, H \} &= \text{primary} \\ \{ G_0 \, , \, H \} &= \text{primary} \end{aligned} \tag{3.71}$$

We have therefore the following situation, all $G_n$ are FC, $G_{n-1}$ calculated from $G_n$ using (3.71). In practice we start with every primary FC and use (3.71) until $G_0$ is reached, that is until we find a quantity such that its Poissonn bracket with $H$ vanishes. Finally we note that $k$ is equal to the number of generations of secondary constraints. So if there are only primary and secondary constraints we know that $k = 1$.

## 3.11 A shortcut

This process just described is the *Dirac Procedure*. In special cases one can almost by inspection find out which are the variables and the Lagrange multipliers.

Namely, given an action for variables $Q$

$$S = \int_I L(Q, \dot{Q}) \, dt \, , \tag{3.72}$$

where $\dot{Q} = dQ/dt$, then the end-result of the Dirac procedure will be described by the action

$$S_D = \int_I dt \left[ P \, \dot{Q} - H_0(P, Q) - \lambda^a \, G_a(P, Q) - \mu^\alpha \, \theta_\alpha(P, Q) \right] \, , \tag{3.73}$$

where $P$ are the canonically conjugate momenta for the coordinates $Q$, $G_a$ are the First Class (FC) constraints, $\theta_\alpha$ are the Second Class (SC) constraints and $\lambda$ and $\mu$ are the corresponding Lagrange multipliers[2].

The FC constraints will satisfy

$$\{G_a, G_b\}_D = f_{ab}^{\ c}(P,Q)\, G_c \,, \tag{3.74}$$

and

$$\{G_a, H_0\}_D = h_a^b(P,Q)\, G_b \,, \tag{3.75}$$

where

$$\{A,B\}_D = \{A,B\} - \{A,\theta_\alpha\}\Delta^{\alpha\beta}\{\theta_\beta, B\} \,, \tag{3.76}$$

is the Dirac bracket. $\Delta^{\alpha\beta}$ is the inverse matrix of $\{\theta\alpha, \theta_\beta\}$ and the Poisson Bracket (PB) is defined as

$$\{A,B\} = \frac{\partial A}{\partial Q}\frac{\partial B}{\partial P} - \frac{\partial A}{\partial P}\frac{\partial B}{\partial Q} \,. \tag{3.77}$$

In particular, if one can write the action (3.72) in the form

$$S = \int_I dt \left[p\,\dot{q} - H(p,q) - \lambda^k\, G_k(p,q)\right] \,, \tag{3.78}$$

where $p \cup q \cup \lambda = Q$ and

$$\{G_k, G_l\}^* = f_{kl}^{\ m}(p,q)\, G_m \,, \tag{3.79}$$

where $\{,\}^*$ is the $(p,q)$ Poisson bracket, then from (3.73) it follows that (3.78) is a gauge-fixed form of $S_D$ where the second-class constraints have been eliminated and some of the phase-space coordinates have been set to zero. Hence the remaining FC constraints are given by $G_k$ and and the Hamiltonian is $H(p,q)$.

## 3.12 The Canonical Analysis of Classical Electrodynamics

As a first example we carry out the canonical analysis of the Maxwell vacuum theory. This is a very well known calculation see [70, 75].

[2]Here $Q$ denotes both the set of the coordinates and the corresponding vector. Hence $P\dot{Q}$ denotes the scalar product of vectors $P$ and $\dot{Q}$.

Let $M_4$ be the four dimensional Minkowski spacetime with metric $\eta_{\mu\nu}$.[3] We write the action as,

$$S = \int d^4x\mathcal{L} \tag{3.80}$$

$$= -\frac{1}{4}\int d^4x F_{\mu\nu}F^{\mu\nu} . \tag{3.81}$$

The field strength tensor is,

$$F_{\mu\nu} = \partial_\mu A_{\ \nu} - \partial_\nu A_{\ \mu} , \tag{3.82}$$

The field equations plus the Bianchi identity (Maxwell equations without sources) are

$$\partial_\mu F^{\mu\nu} = 0 , \tag{3.83}$$

$$\varepsilon^{\sigma\mu\nu\rho}\partial_\mu F_{\nu\rho} = 0 . \tag{3.84}$$

We are considering a second order formalism so the variables are,

$$A_\mu(x) , \tag{3.85}$$

and for the canonical momenta we have,

$$\pi^i(x) = \frac{\partial\mathcal{L}}{\partial(\partial_0 A_i)}$$

$$= -F^{0i} . \tag{3.86}$$

Note that using (3.82) the definition of $F^{\mu\nu}$ we have,

$$\pi^i(x) = \partial^0 A^i - \partial^i A^0 , \tag{3.87}$$

so that the momenta $\pi^i$ are invertible with respect to the "velocities" $\dot{A}_i$.

Note that both $A(x)$ and $\pi(x)$ are functions so that the phase space is infinite-dimensional.

The fundamental Poisson bracket between the variables and momenta are,

$$\{A^\mu(x), \pi_\nu(x')\} = \delta^\mu_\nu\delta^{(3)}(x - x') , \tag{3.88}$$

$$\{A^\mu(x), A_\nu(x')\} = 0 , \tag{3.89}$$

$$\{\pi^\mu(x), \pi_\nu(x')\} = 0 , \tag{3.90}$$

(3.91)

[3]We use Greek indices for flat spacetime in this and in the next sectioon to avoid collision with the Lie algebra indices in the Yang-Mills theory.

where these brackets are calculated at the same time. Since the field strength is the antisymmetric part of the derivative of the potential, the momentum conjugate to $A^0$ is always null. We therefore have,

$$P \equiv \pi^0 \approx 0\,. \tag{3.92}$$

which represents one constraint *per space point* sometimes denoted $1 \times \infty^3$.

The canonical Hamiltonian is:

$$H_c = \int_{\mathbb{R}^3} d^3x \, [\pi^\mu \partial_0 A_\mu - \mathcal{L}] \tag{3.93}$$

this gives,

$$H_c = \int_{\mathbb{R}^3} d^3x \left[ \frac{1}{4} F_{ij} F^{ij} - \frac{1}{2} \pi_i \pi^i - A_0 \partial_i \pi^i \right] \,, \tag{3.94}$$

We are discarding a surface term since the integrals are calculated in all three dimensional space and, we take the fields to decrease sufficiently fast so that the integral of a total divergence may be neglected. And the total Hamiltonian is,

$$H_T = H_c + \int_{\mathbb{R}^3} d^3x \lambda P\,, \tag{3.95}$$

where $\lambda(x)$ is a Lagrange multiplier.

The consistency condition for the primary constraint is,

$$\dot{P} \equiv \{P, H_T\} \,, \tag{3.96}$$

and it gives rise to a secondary constraint, Gauss's law

$$S \equiv \partial_i \pi^i \approx 0\,. \tag{3.97}$$

The consistency condition for the secondary constraint does not give any new constraints,

$$\dot{S} \equiv \{S, H_T\} = 0\,, \tag{3.98}$$

so there are no more constraints. The Lagrange multiplier $\lambda$ has a meaning easily found by calculating,

$$\dot{A}^0 \equiv \{A^0, H_T\} = \lambda\,, \tag{3.99}$$

and we note that $A^0$ itself is another multiplier because it features in $H_c$ in the form $A^0 S$, that is as a multiple of a secondary constraint.

The constraints are first class since,

$$\{P, S\} = 0\,. \tag{3.100}$$

The generator of gauge symmetry is,

$$G = \int \left(\dot{\varepsilon}\pi^0 - \varepsilon\partial_i\pi^i\right)\,, \tag{3.101}$$

and the gauge transformations are,

$$\delta_0 A^\mu = \partial^\mu\varepsilon\,, \tag{3.102}$$

$$\delta_0 \pi_\mu = 0\,. \tag{3.103}$$

These are the infinitesimal gauge transformations for the connection we found from the Lagrangian formalism.

In this case, the general relation between phase space dimension $2N$, number of first class constraints $FC$, number of second class constraints $SC$ and gauge fixing conditions $FC$ (since we know them to be equal to the number of first class constraints) and the number of degrees of freedom $n$ (all *per point*) becomes,

$$2n = 2N - FC - FC\,, \tag{3.104}$$

since there are no second class constraints.

Therefore, using $N = 4$ for the variable $A^\mu$ and $FC = 2$ one for $P$ and one for $S$ we come to a total number of *local* degrees of freedom $n = 2$.

## 3.13 Hamiltonian form of the Yang-Mills theory

The Hamiltonian analysis of the Yang-Mills theory is well known it may be found for example in, [8, 55, 76, 77].

Let $M_4$ be the four dimensional spacetime with metric $\eta_{\mu\nu}$. Let $G$ be a Lie group and $\mathfrak{g}$ its Lie algebra with generators such that,

$$[T_a, T_b] = f^c{}_{bc}T_c\,. \tag{3.105}$$

The structure constants satisfy the Jacobi identities

$$f^d{}_{ac}f^c{}_{be} = f^c_{ab}f^d{}_{ce} - f^c{}_{ae}f^d{}_{cb}\,. \tag{3.106}$$

For the purpose of the canonical analysis, we write the Yang-Mills theory (2.114) in the following way:

$$S = -\frac{1}{4}\int d^4x F_a^{\ \mu\nu} F^a_{\ \mu\nu} \tag{3.107}$$

where $a = 1, 2, \ldots, dim(\mathfrak{g})$.

$$F^a_{\ \mu\nu} = \partial_\mu A^a_{\ \nu} - \partial_\nu A^a_{\ \mu} + f^a_{\ bc} A^b_{\ \mu} A^c_{\ \nu}\,. \tag{3.108}$$

The field equations and Bianchi identities are

$$\nabla_\mu F^{a\,\mu\nu} = 0\,, \tag{3.109}$$
$$\varepsilon^{\sigma\mu\nu\rho}\nabla_\mu F^a_{\ \nu\rho} = 0\,. \tag{3.110}$$

where $\nabla_\mu v^a = \partial_\mu v^a + f^a_{\ bc} A^b_{\ \mu} v^c$.

The gauge transformations for the connection are given by (2.117) which for an infinitesimal $\alpha$ becomes,

$$\delta A_a^{\ \mu} = \nabla^\mu \alpha_a\,. \tag{3.111}$$

The $4dim(\mathfrak{g})$ variables are,

$$A^a_{\ \mu}(x)\,, \tag{3.112}$$

and the canonical momenta are:

$$\pi_a^{\ i}(x) = \frac{\partial \mathcal{L}}{\partial(\partial_0 A^a_{\ i})} = -F_a^{\ 0i}\,. \tag{3.113}$$

As in the case of the Maxwell theory, the momenta $\pi^i$ are invertible with respect to the "velocities" $\dot{A}^a_{\ i}$.

Since both $A^a_{\ \mu}(x)$ and $\pi_a^{\ \mu}(x)$ are functions the phase space is infinite-dimensional and has $8dim(\mathfrak{g}) \times \infty$ coordinates.

The variables and momenta obey the following fundamental (equal time) Poisson brackets ,

$$\left\{A_a^{\ \mu}(x), \pi^b_{\ \nu}(x')\right\} = \delta_a^b \delta_\nu^\mu \delta^{(3)}(x - x')\,, \tag{3.114}$$
$$\left\{A_a^{\ \mu}(x), A^b_{\ \nu}(x')\right\} = 0\,, \tag{3.115}$$
$$\left\{\pi_a^{\ \mu}(x), \pi^b_{\ \nu}(x')\right\} = 0\,, \tag{3.116}$$
$$\tag{3.117}$$

Like in the $U(1)$ case, the field strength is the antisymmetric part of the derivative of the potential therefore, the momentum conjugate to $A_a{}^0$ is always null. We therefore have,

$$P_a \equiv \pi_a{}^0 \approx 0\,. \tag{3.118}$$

this represents $dim(\mathfrak{g})$ constraints *per space point* or $dim(\mathfrak{g}) \times \infty^3$.

The canonical Hamiltonian is:

$$H_c = \int_{\mathbb{R}^3} d^3x \left[\pi_a{}^\mu \partial_0 A^a{}_\mu - \mathcal{L}\right] \tag{3.119}$$

this gives,

$$H_c = \int_{\mathbb{R}^3} d^3x \left[\frac{1}{4} F^a{}_{ij} F_a{}^{ij} - \frac{1}{2}\pi^a{}_i \pi_a{}^i - A^a{}_0 \nabla_i \pi_a{}^i\right]\,, \tag{3.120}$$

up to a boundary term.

And the total Hamiltonian is,

$$H_T = H_c + \int_{\mathbb{R}^3} d^3x \lambda^a P_a\,, \tag{3.121}$$

where $\lambda^a(x)$ is a Lagrange multiplier.

The consistency condition for the primary constraint is,

$$\dot{P}_a \equiv \{P_a, H_T\}\,, \tag{3.122}$$

and it gives rise to a secondary constraint, the generalized Gauss law

$$S_a \equiv \nabla_i \pi_a{}^i \approx 0\,. \tag{3.123}$$

This is really $dim(\mathfrak{g}) \times \infty^3$ constraints, but we will continue to speak of ′a constraint’ for simplicity.

The consistency condition for the secondary constraint does not give any new constraints,

$$\dot{S}_a \equiv \{S_a, H_T\} \approx 0\,, \tag{3.124}$$

so there are no more constraints.

The constraints are first class since,

$$\{S_a, S_b\} = f^c{}_{ab} S_c\,, \tag{3.125}$$

where we used (3.106). And the rest of the brackets are zero,

$$\{P, S_a\} = 0, \quad (3.126)$$
$$\{P, P\} = 0, \quad (3.127)$$

so we have a first class constrained system with a Poisson algebra,

$$\{\varphi_a, \varphi_b\} = f^c{}_{ab}\varphi_c, \quad (3.128)$$

where $\varphi_a$ are generic constraints. This algebra is isomorphic to the Lie algebra $\mathfrak{g}$ of the gauge group eq. (3.105).

The generator of gauge symmetry is,

$$G = \int d^3x \left(\dot{\varepsilon}^a P_a - \varepsilon^a S_a\right), \quad (3.129)$$

and the gauge transformations are,

$$\delta_0 A_a{}^\mu = \nabla^\mu \varepsilon_a, \quad (3.130)$$
$$\delta_0 \pi^a{}_\mu = 0. \quad (3.131)$$

These are the infinitesimal gauge transformations for the connection eq. (3.111) we found from the Lagrangian formalism.

Like in the previous case, the general relation between phase space dimension $2N$, number of first class constraints $FC$, number of second class constraints $SC$ and gauge fixing conditions $FC$ (since we know them to be equal to the number of first class constraints) and the number of degrees of freedom $n$ (all *per point*) becomes,

$$2n = 2N - FC - FC, \quad (3.132)$$

since there are no second class constraints.

Therefore, using $N = 4dim(\mathfrak{g})$ for the variable $A_a{}^\mu$ and $FC = 2dim(\mathfrak{g})$, with $dim(\mathfrak{g})$ for $P_a$ and, $dim(\mathfrak{g})$ for $S_a$ we come to a total number of *local* degrees of freedom $n = 2dim(\mathfrak{g})$.

## 3.14 Hamiltonian structure of *BF* theory

In this section we use the Dirac procedure to find the constraints of the *BF* theory, the analysis can be found in [78].

We specialize the *BF* action (2.146) to the case of the $SO(1,3)$ group. The Greek indices $\mu\,,\nu\,,\ldots = 0\,,1\,,2\,,3$, are space-time indices whereas $a\,,b\,,\ldots = 0\,,1\,,2\,,3$, are internal $SO(1,3)$ indices raised and lowered with the flat metric $\eta$.

The Lorentz connection is,

$$\omega^{ab} \equiv \omega^{ab}{}_{\mu} dx^{\mu}\,, \tag{3.133}$$

an $SO(1,3)$-valued one-form. And $B$ is

$$B^{ab} \equiv \frac{1}{2} B^{ab}{}_{\mu\nu} dx^{\mu} \wedge dx^{\nu}\,. \tag{3.134}$$

The curvature is,

$$R^{ab} \equiv \frac{1}{2} R^{ab}{}_{\mu\nu} dx^{\mu} \wedge dx^{\nu} \tag{3.135}$$

with,

$$R^{ab}{}_{\mu\nu} = \partial_{\mu}\omega^{ab}{}_{\nu} - \partial_{\nu}\omega^{ab}{}_{\mu} + \omega^{a}{}_{c\mu}\omega^{cb}{}_{\nu} - \omega^{a}{}_{c\nu}\omega^{cb}{}_{\mu}\,. \tag{3.136}$$

The action can be written in the form[4],

$$S = \int_M d^4x\, \varepsilon^{\mu\nu\rho\sigma}\, \frac{1}{4} B_{ab\mu\nu} R^{ab}{}_{\rho\sigma}\,. \tag{3.137}$$

where the four dimensional spacetime manifold has the topology $M = \Sigma \times \mathbb{R}$, with $\Sigma$ a 3-dimensional spacelike hypersurface.

The Lagrangian is:

$$L = \int_{\Sigma} d^3x\, \varepsilon^{\mu\nu\rho\sigma} \frac{1}{4} B_{ab\mu\nu} R^{ab}{}_{\rho\sigma}\,. \tag{3.138}$$

The variables in this canonical analysis are,

$$B^{ab}{}_{\mu\nu}\,, \qquad \omega^{ab}{}_{\mu}\,. \tag{3.139}$$

Variation of the action (3.137) whith respect to these variables gives the following equations of motion,

$$R^{ab}{}_{\mu\nu} = 0\,, \qquad \varepsilon^{\lambda\mu\nu\rho} \nabla_{\rho} B^{ab}{}_{\mu\nu} = 0\,. \tag{3.140}$$

[4]As is customary in the literature of this field, we use the letter $R$ for an $SO(1,3)$ Lie algebra valued curvature. This is the case, for example , of (8.4). In the general case we use the symbol $F$, e.g. (2.114).

The momenta are given by,

$$\begin{array}{rcccl} \pi(B)_{ab}{}^{\mu\nu} & = & \dfrac{\delta L}{\delta\partial_0 B^{ab}{}_{\mu\nu}} & = & 0\,, \\ \pi(\omega)_{ab}{}^{i} & = & \dfrac{\delta L}{\delta\partial_0 \omega^{ab}{}_{i}} & = & \varepsilon^{0ijk}B_{abjk}\,, \\ \pi(\omega)_{ab}{}^{0} & = & \dfrac{\delta L}{\delta\partial_0 \omega^{ab}{}_{0}} & = & 0\,, \end{array} \tag{3.141}$$

Since these momenta are not invertible with respect to the time derivatives of their conjugate variables, we identify the following primary constraints,

$$\begin{array}{rcl} P(B)_{ab}{}^{\mu\nu} & \equiv & \pi(B)_{ab}{}^{\mu\nu} \approx 0\,, \\ P(\omega)_{ab}{}^{\mu} & \equiv & \pi(\omega)_{ab}{}^{\mu} - \varepsilon^{0\mu\nu\rho}B_{ab\nu\rho} \approx 0\,. \end{array} \tag{3.142}$$

The simultaneous Poisson brackets[5] between the fields and their conjugate momenta,

$$\begin{array}{rcl} \{\, B^{ab}{}_{\mu\nu}\,,\, \pi(B)_{cd}{}^{\rho\sigma}\,\} & = & 4\delta^a_{[c}\delta^b_{d]}\delta^\rho_{[\mu}\delta^\sigma_{\nu]}\delta^{(3)}\,, \\ \{\, \omega^{ab}{}_{\mu}\,,\, \pi(\omega)_{cd}{}^{\nu}\,\} & = & 2\delta^a_{[c}\delta^b_{d]}\delta^\nu_{\mu}\delta^{(3)}\,, \end{array} \tag{3.145}$$

where $[ab]$ is given by (2.20).

We use (3.145) to calculate the algebra of primary constraints,

$$\{\, P(B)^{abjk}\,,\, P(\omega)_{cd}{}^{i}\,\} = 4\varepsilon^{0ijk}\delta^a_{[c}\delta^b_{d]}\delta^{(3)}, \tag{3.146}$$

The canonical, *on-shell* Hamiltonian:

$$H_c \;=\; \int_\Sigma d^3x \left[\frac{1}{4}\pi(B)_{ab}{}^{\mu\nu}\partial_0 B^{ab}{}_{\mu\nu} + \frac{1}{2}\pi(\omega)_{ab}{}^{\mu}\partial_0\omega^{ab}{}_{\mu}\right] - L\,. \tag{3.147}$$

this can be written in a form where all time derivatives are multiplied by primary constraints, and therefore drop out of the Hamiltonian. The resulting expression can be written in the form,

$$H_c \;=\; -\int d^3x\,\varepsilon^{0ijk}\left[\frac{1}{2}B_{ab0i}R^{ab}{}_{jk} + \frac{1}{2}\omega_{ab0}\nabla_i B^{ab}{}_{jk}\right]\,, \tag{3.148}$$

[5] In order to simplify the notation involving Poisson brackets, we will adopt the following convention. The left quantity in every Poisson bracket is assumed to be evaluated at the point $x = (t, x)$, while the right quantity at the point $x' = (t, x')$. In addition, we use the shorthand notation for the 3-dimensional Dirac delta function $\delta^{(3)} \equiv \delta^{(3)}(x - x')$. This allows us to write an explicit but bulky expression like

$$\{\, A(t,x)\,,\, B(t,x')\,\} = C(t,x)\delta^{(3)}(x-x') \tag{3.143}$$

more compactly as

$$\{\, A\,,\, B\,\} = C\delta^{(3)}\,, \tag{3.144}$$

without ambiguity. This notation will be used systematically unless explicitly stated otherwise.

up to a boundary term, note that this Hamiltonian does not depend on the momenta only on the fields and their spacial derivatives.

Like in the previous examples, we introduce a Lagrange multiplier $\lambda$ for each of the primary constraints. In this way, we construct the *off-shell*, total Hamiltonian:

$$H_T \;=\; H_c + \int d^3x \left[\frac{1}{2}\lambda(\omega)^{ab}{}_{\mu}P(\omega)_{ab}{}^{\mu} + \frac{1}{4}\lambda(B)^{ab}{}_{\mu\nu}P(B)_{ab}{}^{\mu\nu}\right] . \tag{3.149}$$

The next step in the Dirac procedure is to impose the consistency requirements for the primary constraints, that is,

$$\dot{P} \equiv \{P\,,\,H_T\} \approx 0\,. \tag{3.150}$$

We find that from the conditions:

$$\dot{P}(B)_{ab}{}^{0i} \approx 0\,, \qquad \dot{P}(\omega)_{ab}{}^{0} \approx 0\,, \tag{3.151}$$

arise respectively the following secondary constraints:

$$S(R)^{ab}{}_{jk} \;\equiv\; R^{ab}{}_{jk} \approx 0\,, \qquad S(B)^{ab} \;\equiv\; \varepsilon^{0ijk}\,\nabla_i B^{ab}{}_{jk} \approx 0\,. \tag{3.152}$$

The remaining consistency requirements,

$$\dot{P}(B)_{ab}{}^{jk} \approx 0\,, \qquad \dot{P}(\omega)_{ab}{}^{k} \approx 0\,, \tag{3.153}$$

determine the following multipliers:

$$\begin{aligned} \lambda(\omega)^{ab}{}_{i} &\approx \nabla_i \omega^{ab}{}_0\,, \\ \lambda(B)^{ab}{}_{ij} &\approx 2\nabla_{[j}B^{ab}{}_{i]0} + 2\omega^{[a}{}_{c0}B^{b]c}{}_{ij}\,. \end{aligned} \tag{3.154}$$

We therefore find that the Lagrange multipliers:

$$\lambda(\omega)^{ab}{}_{0}\,, \quad \lambda(B)^{ab}{}_{0i}\,, \tag{3.155}$$

remain undetermined.

The procedure now continues with the consistency conditions for the secondary constraints (3.152),

$$\dot{S}(R)^{ab}{}_{jk} \approx 0\,, \qquad \dot{S}(B)^{ab} \approx 0\,. \tag{3.156}$$

are identically satisfied. They do not produce new constraints or determine any remaining Lagrange multipliers.

We have now found all the constraints in the theory. The next step is to classify them into first and second class. The algebra of primary constraints (3.146) may be used to read of some of the secondary constraints. However the complete classification is not easy since constraints are unique only up to linear combinations. In order to find all first class constraints, we substitute all determined multipliers into the total Hamiltonian (3.149) and rewrite it in the form

$$\begin{aligned} H_T &= \int d^3x \left[ \frac{1}{2}\lambda(B)^{ab}{}_{0i}\, \varphi(B)_{ab}{}^{i} + \frac{1}{2}\lambda(\omega)^{ab}\, \varphi(\omega)_{ab} \right. \\ &\quad \left. -\frac{1}{2}B_{ab0i}\, \varphi(R)^{abi} - \frac{1}{2}\omega_{ab0}\, \varphi(\nabla B)^{ab} \right] . \end{aligned} \tag{3.157}$$

The quantities $\varphi$ are linear combinations of constraints, but must all be first class, since the total Hamiltonian has a weakly vanishing Poisson bracket with all constraints. Written in terms of primary and secondary constraints, they are:

$$\begin{aligned} \varphi(B)_{ab}{}^{i} &= P(B)_{ab}{}^{0i} , \\ \varphi(\omega)_{ab} &= P(\omega)_{ab}{}^{0} , \\ \varphi(R)^{abi} &= \varepsilon^{0ijk} S(R)^{ab}{}_{jk} - \nabla_j P(B)^{abij} , \\ \varphi(\nabla B)^{ab} &= S(B)^{ab} + \nabla_i P(\omega)^{abi} - B^{[a}{}_{cij} P(B)^{b]cij} . \end{aligned} \tag{3.158}$$

The remaining constraints are second class:

$$\chi(B)_{ab}{}^{jk} = P(B)_{ab}{}^{jk} , \quad \chi(\omega)_{ab}{}^{i} = P(\omega)_{ab}{}^{i} , \tag{3.159}$$

the algebra between the first class constraints is

$$\begin{aligned} \{ \varphi(R)^{abi} , \varphi(\nabla B)_{cd} \} &= -4\delta^{[a}_{[c} \varphi(R)^{b]}{}_{d]}{}^{i} \delta^{(3)} , \\ \{ \varphi(\nabla B)^{ab} , \varphi(\nabla B)_{cd} \} &= -4\delta^{[a}_{[c} \varphi(\nabla B)^{b]}{}_{d]} \delta^{(3)} . \end{aligned} \tag{3.160}$$

And the algebra between the second class constraints is, according to (3.146),

$$\{ \chi(B)^{abjk} , \chi(\omega)_{cd}{}^{i} \} = 4\varepsilon^{0ijk} \delta^{a}_{[c} \delta^{b}_{d]} \delta^{(3)} . \tag{3.161}$$

While the algebra between the first and second class constraints is

$$\begin{aligned} \{ \varphi(R)^{abi} , \chi(\omega)_{cd}{}^{j} \} &= 4\delta^{[a}_{[c} \chi(B)^{b]}{}_{d]}{}^{ij} \delta^{(3)} , \\ \{ \varphi(\nabla B)^{ab} , \chi(\omega)_{cd}{}^{i} \} &= 4\delta^{[a}_{[c} \chi(\omega)_{d]}{}^{b]i} \delta^{(3)} , \\ \{ \varphi(\nabla B)^{ab} , \chi(B)_{cd}{}^{jk} \} &= 4\delta^{[a}_{[c} \chi(B)_{d]}{}^{b]jk} \delta^{(3)} . \end{aligned} \tag{3.162}$$

All other Poisson brackets among $\varphi$ and $\chi$ are zero.

We see that the algebra is closed, and all Poisson brackets involving $\varphi$ constraints weakly vanish, confirming that all $\varphi$ are indeed first class. We see also that the constraint $\varphi(\nabla B)^{ab}$ is analogous to the Gauss constraint and generates the $SO(1,3)$ transformations.

To count the physical degrees of freedom we proceed in the following way. Given $N$ initial independent fields in the theory, the dimension of the phase space *per point* is $2N$. From this one subtracts the total number $F$ of first class constraints *per point*, the total number $S$ of second class constraints *per point*, and the total number $F$ of gauge fixing conditions *per point*. The result is the *local* dimension of the physical phase space, $2n$, where $n$ is the number of physical degrees of freedom. Thus we use the above formula,

$$n = N - F - \frac{S}{2} \,. \tag{3.163}$$

For the fields in the $BF$ theory we have that the number of independent components is,

| $\omega^{ab}{}_{\mu}$ | $B^{ab}{}_{\mu\nu}$ |
|---|---|
| 24 | 36 |

which gives the total $N = 60$.

Similarly, the number of independent components for the second class constraints is

| $\chi(B)_{ab}{}^{jk}$ | $\chi(\omega)_{ab}{}^{i}$ |
|---|---|
| 18 | 18 |

which gives the total $S = 36$.

For the first class constraints, there are relations among the components of some of the constraints called Bianchi identities (see Appendix B). In particular, not all components of $\varphi(R)^{abi}$ are independent. To see this, take the derivative of $\varphi(R)^{abi}$ to obtain

$$\nabla_i \varphi(R)^{abi} = \varepsilon^{0ijk} \nabla_i R^{ab}{}_{jk} + R^{c[a}{}_{ij} \pi(B)_c{}^{b]ij} \,. \tag{3.164}$$

The first term on the right-hand side is zero *off-shell* as a consequence of the second Bianchi identity (B.8). The second term on the right-hand side is also zero *off-shell*, since it is a product of two constraints,

$$R^{c[a}{}_{ij} \pi(B)_c{}^{b]ij} \equiv S(R)^{c[a}{}_{ij} P(B)_c{}^{b]ij} = 0 \,. \tag{3.165}$$

Therefore, we have the off-shell identity

$$\nabla_i \varphi(R)^{abi} = 0 \,, \tag{3.166}$$

which means that 6 components of $\varphi(R)^{abi}$ are not independent of the others.

Taking (3.166) into account, the number of independent components of the first class constraints is

| $\varphi(B)_{ab}{}^{i}$ | $\varphi(\omega)_{ab}$ | $\varphi(R)^{abi}$ | $\varphi(\nabla B)^{ab}$ |
|---|---|---|---|
| 18 | 6 | 18 − 6 | 6 |

which gives the total of $F = 42$.

Substituting $N$, $F$ and $S$ into (3.163), we obtain:

$$n = 60 - 42 - \frac{36}{2} = 0\,. \tag{3.167}$$

We conclude that (if the constraints are independent) the theory has no *local* degrees of freedom.

The gauge generator for the $BF$ theory is given by,

$$G = \int d^3x \left( (\nabla_0 \varepsilon^{ab}{}_{i}) \varphi(B)_{ab}{}^{i} + (\nabla_0 \varepsilon^{ab}) \varphi(\omega)_{ab} + \varepsilon_{abi} \varphi(R)^{abi} + \varepsilon_{ab} \varphi(\nabla B)^{ab} \right) \,. \tag{3.168}$$

And the gauge transformations are,

$$\begin{array}{lcl}
\delta_0 \omega^{ab}{}_{0} & = & \nabla_0 \varepsilon^{ab} \\
\delta_0 \omega^{ab}{}_{i} & = & -\nabla_i \varepsilon^{ab} \\
\delta_0 B^{ab}{}_{0i} & = & \nabla_0 \varepsilon^{ab}{}_{i} \\
\delta_0 B^{ab}{}_{ij} & = & \nabla_{[i} \varepsilon^{ab}{}_{j]} + \varepsilon^{[a|c|} B_{c}{}^{b]}{}_{ij} \\
 & & \\
\delta_0 \pi(\omega)_{ab}{}^{0} & = & 0 \\
\delta_0 \pi(\omega)_{ab}{}^{i} & = & \varepsilon^{ijk} \nabla_j \varepsilon_{abk} + 4\pi(B)_{[a}{}^{cik} \varepsilon_{|c|b]k} + 2\pi(\omega)_{[a|c|}{}^{i} \varepsilon^{c}{}_{b]} \\
\delta_0 \pi(B)_{ab}{}^{0i} & = & 0 \\
\delta_0 \pi(B)_{ab}{}^{ij} & = & -2\pi(B)_{[a|c|}{}^{ij} \varepsilon^{c}{}_{b]}\,.
\end{array} \tag{3.169}$$

# 4

# Hamiltonian formulation of General Relativity

The cannonical quantization program begins with the Hamiltonian analysis of the theory we want to quantize. Here we sketch the Arnowitt Deser Misner formalism (ADM) following [47] but restricting the discussion to the four dimensional case. Other sources are [30, 34, 79, 80], and an example of the original work is [81].

## 4.1 The Arnowitt Deser Misner formalism

Let $\sigma$ and $M$ be manifolds, the differentiable map,

$$\mathcal{Y} \; : \; \sigma \longrightarrow M \tag{4.1}$$

is called an immersion if the differential of $\mathcal{Y}$ is injective at every point of $\sigma$ (a submersion if the differential of $\mathcal{Y}$ is surjective at every point of $\sigma$). Furthermore, if $\mathcal{Y}$ is an injective immersion and a homeomorphism[1] onto its image $\mathcal{Y}(\sigma)$, then $\mathcal{Y}$ is an embedding of $\sigma$ in $M$ (refer to [82] for the relevant definitions and consequences).

The image of $\sigma$ by this embedding is a submanifold of $M$. In these conditions, $\mathcal{Y}$ is a diffeomorphism between $\sigma$ and $\Sigma = \mathcal{Y}(\sigma)$.

[1]See footnote 2 in chapter 2.

Our goal in this section is to write the Einstein-Hilbet action (1.35) (for $\Lambda = 0$) in canonical form. We avoid complications arising from boundary terms, by assuming that $M$ is spatially compact *without boundary*. For a treatment of the general case the reader is referred to [83]. We write the action in the form,

$$S_{EH} = \frac{1}{\kappa} \int_M \sqrt{-g} R \,, \tag{4.2}$$

where $\frac{1}{\kappa} = \frac{c^4}{16\pi G}$.

To achieve the above mentioned objective, we must restrict the topology of spacetime by assuming that it is of the form $M = \mathbb{R} \times \sigma$, where $\sigma$ is a fixed three-dimensional, compact manifold without boundary. In this case $\Sigma = Y(\sigma)$ is a 3-dimensional manifold in the 4-dimensional spacetime and therefore a hypersurface. This topological restriction can be related to global hyperbolicity of spacetime (see [47]).

Under this assumption a diffeomorphism exists,

$$\begin{aligned} Y: \quad \mathbb{R} \times \sigma &\to M \\ (t, x) &\mapsto Y(t,x) = Y_t(x) \,, \end{aligned} \tag{4.3}$$

here $Y_t : \sigma \longrightarrow M$ is a one-parameter family of embeddings (see figure 4.1), which foliates $M$ into hypersurfaces $\Sigma_t = Y_t(\sigma)$. Let $x^i$, for $i = 1, 2, 3$, and $X^\mu$ for $\mu = 0, 1, 2, 3$, be local coordinates in $\sigma$ and $M$ respectively. This foliation is not unique.

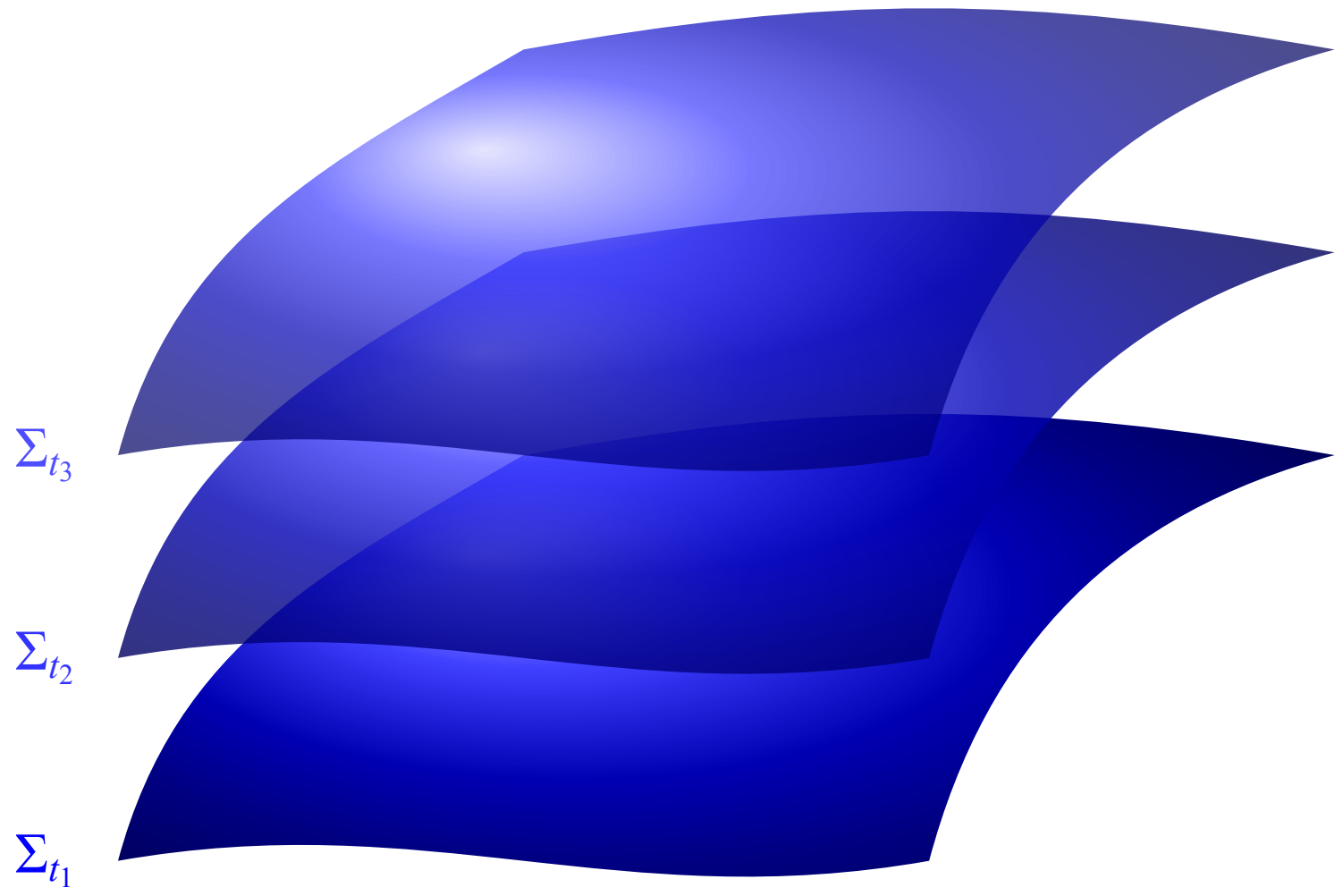

**Figure 4.1:** A one-parameter family of embeddings

The inverse of (4.3) is (obviously also a) a diffeomorphism

$$\begin{aligned} Y^{-1}: \quad M &\to \mathbb{R} \times \sigma \\ X &\mapsto (\tau(X), \tilde{\sigma}(X)) \ , \end{aligned} \tag{4.4}$$

which takes the form of a product of a time function $\tau : M \longrightarrow \mathbb{R}$, and a space function $\sigma : M \longrightarrow \sigma$.

The map (4.3) can be viewed as,

$$\begin{aligned} Y_x: \quad \mathbb{R} &\to M \\ t &\mapsto Y_x(t) = Y(t,x) \, , \end{aligned} \tag{4.5}$$

which for each $x \in \sigma$ defines a curve. We therefore have a one-parameter family of tangent vectors $\dot{Y}_x(t)$ with components $\left(\frac{\partial Y_x}{\partial t}\right)^\mu$ in local coordinates.

The deformation vector can be written in the form,

$$\frac{\partial Y^\mu(t,x)}{\partial t} =: N(X) n^\mu(X) + N^\mu(X) \tag{4.6}$$

where, $n^\mu$ is a unit normal vector to $\Sigma_t$. The coefficients of proportionality $N$ and $N^\mu$ are called *lapse* function and *shift* vector field respectively (the geometric situation is depicted in figure 4.2).

For a more precise characterization of the the foliations as dynamical variables and their relation to the diffeomorphism group refer to [84, 85].

We make the restriction to spacelike embeddings. The hypersurfaces $\Sigma_t$ are spacelike and the normal vector has the properties,

$$n(x)_\mu Y^\mu_{,i} = 0 \tag{4.7}$$

$$g_{\mu\nu} n^\mu n^\nu = -1 \, . \tag{4.8}$$

The first of these equations defines what it means to be spacelike, while the second is a normalization condition and says that the vector is timelike with respect to the metric $g$ on $M$.[2] Since we are dealing with spacelike hypersurfaces the deformation vector given by equation (4.6) is subject to the restriction $-N^2 + g_{\mu\nu} n^\mu n^\nu < 0$. Furthermore we have $N > 0$ everywhere for a future directed foliation.

The normal vector $n_\mu$ also has the following property,

$$n_\mu = F \tau_{,\mu} \, , \tag{4.9}$$

[2]Note we are using the $(- + ++)$ signature.

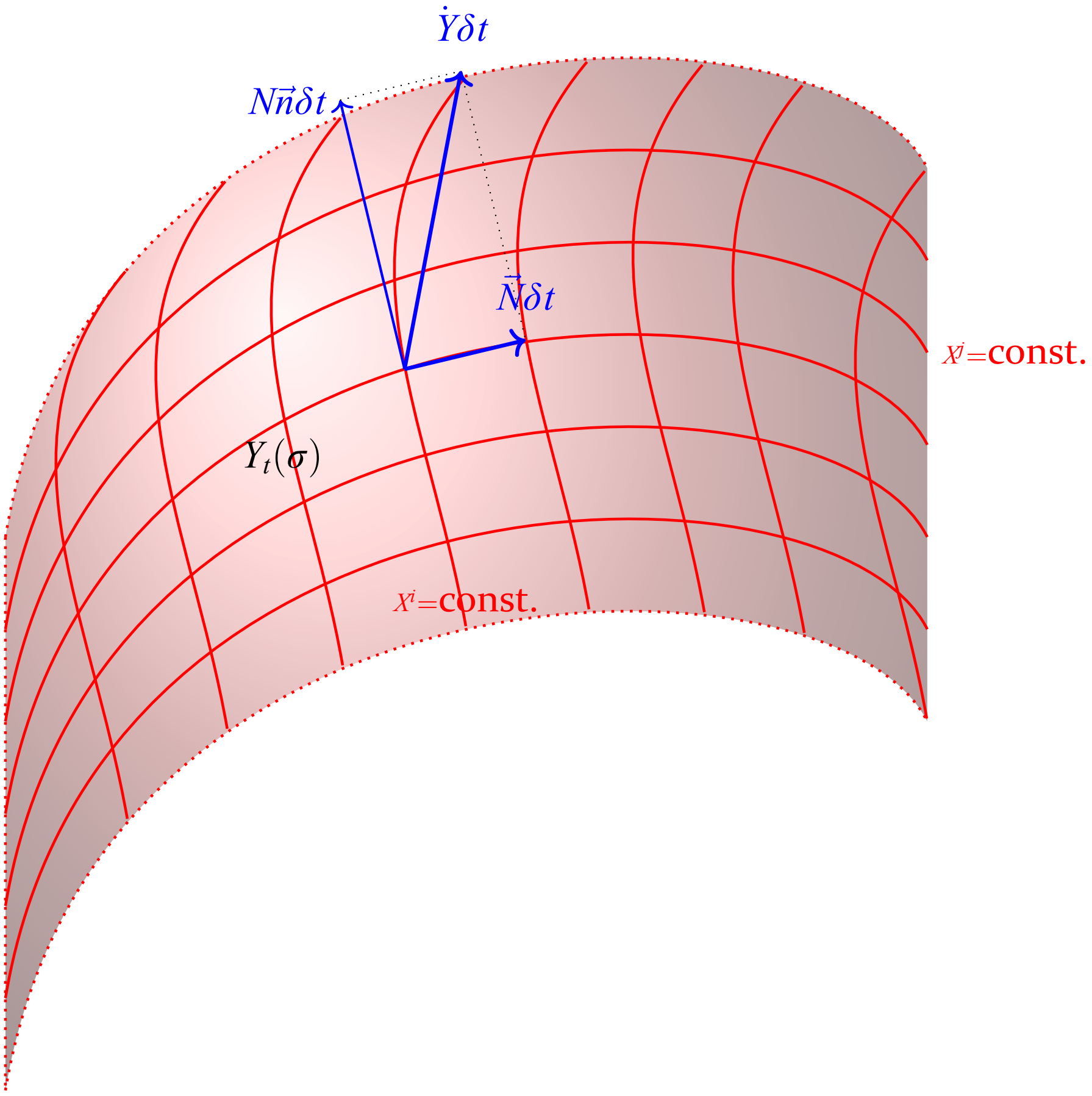

**Figure 4.2:** A spacelike hypersurface $\Sigma_t$ and the relation between lapse function $N$ and shift vector $\vec{N}$ given by equation (4.6).

where $F$ is a function.

For a spacelike embedding $Y$ and a 3-dimensional manifold $\sigma$, embedded in space-time $M$ through $\Sigma_t = Y_t(\sigma)$, we have the possibility to work either on $\sigma$ or on the hypersurface $\Sigma$, when developing the calculus of 'spacial tensor fields'. We choose to work on $\Sigma_t$ since this has the advantage that we can compare spacial tensor fields with arbitrary tensor fields *restricted* to $\Sigma$, because they are both tensor fields on a subset of space-time $M$.

We introduce the *first* and *second fundamental form* on $\Sigma$,

$$q_{\mu\nu} = g_{\mu\nu} + n_\mu n_\nu \quad K_{\mu\nu} = q^\rho_\mu q^\sigma_\nu \nabla_\rho n_\sigma \,, \tag{4.10}$$

where indices are raised and lowered with the metric $g_{\mu\nu}$. Note that $n_\mu n_\nu$ is a projector orthogonal to $\Sigma_t$ and so $g_{\mu\nu} + n_\mu n_\nu$ plays the role of a projector on this hyper-surface (see [70]). Since these tensors vanish whenever they are contracted with $n^\mu$, they are sometimes called 'spatial tensors'. Furthermore $K_{\mu\nu}$ called the *extrinsic curvature,* is a symmetric tensor satisfying the following property,

$$2K_{\mu\nu} = (\mathcal{L}_n q)_{\mu\nu} \,. \tag{4.11}$$

This can also be cast in the form,

$$\begin{aligned} K_{\mu\nu} &= \frac{1}{2N}(\mathcal{L}_{T-N}q)_{\mu\nu} \\ &= \frac{1}{2N}(\mathcal{L}_T q - \mathcal{L}_N q)_{\mu\nu} \end{aligned} \tag{4.12}$$

where we used, $n^\mu = (T^\mu - N^\mu)/N$. Note that, in equation (4.12) there is an extra term, proportional to the Lie derivative of the inverse of the lapse function, which vanishes because it is contracted with the spatial metric.

The derivative $D$ associated with the metric $q_{\mu\nu}$ is given by,

$$D_\mu f = q_\mu^\nu \nabla_\nu \tilde{f} \tag{4.13}$$

$$D_\mu u_\nu = q_\mu^\rho q_\nu^\sigma \nabla_\rho \tilde{u}_\sigma \,, \qquad \text{for } u_\mu n^\mu = 0 \,, \tag{4.14}$$

where, $\tilde{f}$ and $\tilde{u}$ are arbitrary smooth extensions of $f$ and $u$ respectively into a neighbourhood of $\Sigma_t$ in $M$. The derivative $D$ is compatible with the (Euclidean) metric $q$ i.e. $D_\mu q_{\nu\rho} = 0$ and, also has the property $D_{[\mu}D_{\nu]}f = 0$ for scalar functions $f$. The two properties are the three-dimensional counterparts of the metric compatibility and the no torsion conditions.

The Riemann curvature tensors $R^{(3)}$ and $R^{(4)}$ respectively for $D$ and $\nabla$ are related by,

$$R^{(3)}_{\mu\nu\rho\sigma} = q_\mu^{\mu'} q_\nu^{\nu'} q_\rho^{\rho'} q_\sigma^{\sigma'} R^{(4)}_{\mu'\nu'\rho'\sigma'} - 2K_{\rho[\mu}K_{\nu]\sigma} \,, \tag{4.15}$$

called the *Gauß equation*.

The corresponding Ricci scalars are related by the *Codacci*[3] *equation*

$$R^{(4)} = R^{(3)} + [K_{\mu\nu}K^{\mu\nu} - K^2] - 2\nabla_\mu (n^\nu \nabla_\nu n^\mu - n^\mu \nabla_\nu n^\nu) \,. \tag{4.16}$$

We can now use $Y_t$ to pull-back the metric and various other quantities. We define spatial vector fields on $\Sigma_t$ by,

$$Y_i^\mu(X) := Y^\mu_{,i}(x,t)_{|Y(x,t)=X} \,. \tag{4.17}$$

[3]Sometimes written Codazzi equation.

Using the first of (4.10) and the first of (4.7) we have,

$$q_{ij}(t,x) = (Y^\mu_{,i} Y^\nu_{,j} q_{\mu\nu})(Y(t,x)) = g_{\mu\nu}(Y(t,x)) Y^\mu_{,i}(t,x) Y^\nu_{,j}(t,x) \tag{4.18}$$

And for the extrinsic curvature,

$$K_{ij}(t,x) = (Y^\mu_{,i} Y^\nu_{,j} K_{\mu\nu})(Y(t,x)) = (Y^\mu_{,i} Y^\nu_{,j} \nabla_\mu n_\nu)(t,x) \,. \tag{4.19}$$

The inverse of $q_{ij}$ is given by,

$$q^{ij} = \frac{1}{2\det((q_{kl}))} \varepsilon^{imn} \varepsilon^{jpq} q_{mp} q_{nq} \,. \tag{4.20}$$

We can also define,

$$N(x,t) = N(Y(x,t)) \,, \quad N^i(x,t) = q^{ij}(x,t)(Y^\mu_{,j} g_{\mu\nu} N^\nu)(Y(x,t)) \,. \tag{4.21}$$

And we write the extrinsic curvature in the form,

$$K_{ij}(x,t) = \frac{1}{2N}(\dot{q}_{ij} - (\mathcal{L}_{\vec{N}} q)_{ij})(x,t) \,. \tag{4.22}$$

We can now express the line element in the new variables $q_{ij}, N, N^i$. It reads,

$$ds^2 = [-N^2 + q_{ij} N^i N^j] dt \otimes dt + q_{ij} N^j [dt \otimes dx^i + dx^i \otimes dt] + q_{ij} dx^i \otimes dx^j \,. \tag{4.23}$$

Or in matrix form,

$$Y^*(g) = \begin{pmatrix} N^i N^j q_{ij} - N^2 & N^j q_{ij} \\ N^j q_{ij} & q_{ij} \end{pmatrix} \,. \tag{4.24}$$

And finally, writing the volume element as $\sqrt{\det(q)} N dt d^3x$, the Einstein-Hilbert action in the ADM form reads,

$$S_{ADM} = \frac{1}{\kappa} \int_{\mathbb{R}} dt \int_\sigma d^3x \sqrt{\det(q)} N (R^{(3)} + K_{ij} K^{ij} - (K^i_i)^2]) \tag{4.25}$$

where we have dropped the total derivative coming from the Codacci equation (4.16) (for a treatment of these boundary terms refer to [47]).

## 4.2 Hamiltonian analysis of metric General Relativity

We now perform the canonical analysis of the Einstein-Hilbert action written in the form (4.25). We begin by noting the the action depends on the 'velocities' $\dot{q}_{ij}$ but not on $\dot{N}$ and $\dot{N}^i$. Using this observation, equation (4.22) and the fact that that $R$ does not contain time derivatives, we find the following momenta:

$$\begin{aligned} P^{ij}(t,x) &= \frac{\delta S_{ADM}}{\delta \dot{q}_{ij}(t,x)} = \frac{1}{\kappa}\sqrt{\det(q)}[K^{ij} - q^{ij}(K^l_l)] \\ \Pi(t,x) &= \frac{\delta S_{ADM}}{\delta \dot{N}(t,x)} = 0 \\ \Pi_i(t,x) &= \frac{\delta S_{ADM}}{\delta \dot{N}^i(t,x)} = 0\,. \end{aligned} \tag{4.26}$$

Note that we can invert $P^{ij}$ to solve for $\dot{q}_{ij}$ (with $N$ and $N^i$), but this is not possible for $\Pi$ and $\Pi_i$. We therefore have a singular system.

The *primary constraints* are:

$$C(t,x) = \Pi(t,x) = 0 \qquad C^i(t,x) = \Pi^i(t,x) = 0 \tag{4.27}$$

Using the following formulas,

$$\begin{aligned} \dot{q}_{ij} &= 2NK_{ij} + (\mathcal{L}_{\vec{N}}q)_{ij} \\ \dot{q}_{ij}P^{ij} &= (\mathcal{L}_{\vec{N}}q)_{ij}P^{ij} + 2N\frac{\sqrt{\det(q)}}{\kappa}[K_{ij}K^{ij} - K^2] \\ P_{ij}P^{ij} &= \frac{\det(q)}{\kappa^2}(K_{ij}K^{ij} + K^2) \\ P^2 &:= (P^i_i)^2 = \left(\frac{2\det(q)}{\kappa}\right)^2 K^2 \end{aligned} \tag{4.28}$$

the action can be written in the form,

$$S = \int_{\mathbb{R}} dt \int_\sigma d^3x\{\dot{q}_{ij}P^{ij} + \dot{N}\Pi + \dot{N}^i\Pi_i - [\lambda C + \lambda^i C_i + N^i H_i + NH\} \tag{4.29}$$

where

$$\begin{aligned} H_i &:= -2q_{ik}D_jP^{jk} \\ H &= \frac{\kappa}{\sqrt{\det(q)}}[q_{ik}q_{jl} - \frac{1}{2}q_{ij}q_{kl}]P^{ij}P^{kl} - \frac{\sqrt{\det(q)}}{\kappa}R^{(3)}\,. \end{aligned} \tag{4.30}$$

These are called the *spacial diffeomorphism constraint* and the *Hamiltonian constraint* respectively.

At this point, the phase space coordinates are,

$$q_{ij}\,,N\,,N^i\,,P^{ij}\,,\Pi\,,\Pi_i\,. \tag{4.31}$$

The fundamental Poisson brackets are,

$$\begin{aligned}\{\,N^i(t,x)\,,\,\Pi_j(t,x')\,\} &= \delta^i_j\delta^{(3)}(x,x')\,,\\ \{\,N(t,x)\,,\,\Pi(t,x')\,\} &= \delta^{(3)}(x,x')\,,\\ \{\,q_{ij}(t,x)\,,\,P^{kl}(t,x')\,\} &= \delta_i^{(k}\delta_j^{l)}\delta^{(3)}(x,x')\end{aligned} \tag{4.32}$$

with all other brackets vanishing[4].

The Hamiltonian can be identified as the term is square brackets in equation (4.29).

$$\begin{aligned}\mathbf{H} &= \int_\sigma d^3x\,\left[\lambda C + \lambda^i C_i + N^i H_i + NH\right] \\ &= C(\lambda) + \vec{C}(\vec{\lambda}) + \vec{H}(\vec{N}) + H(N)\,.\end{aligned} \tag{4.33}$$

Where we made the following identifications,

$$\begin{aligned} C(\lambda) &= \int_\sigma d^3x\lambda C\,, \qquad \vec{C}(\vec{\lambda}) = \int_\sigma d^3x\lambda^i C_i\,,\\ H(N) &= \int_\sigma d^3xNH\,, \qquad \vec{H}(\vec{N}) = \int_\sigma d^3xN^iH_i\,.\end{aligned} \tag{4.34}$$

For consistency, the primary constraints (4.27) are required to have a vanishing time derivative, that is:

$$\dot{C}(f) := \{\,\mathbf{H}\,,\,C(f)\,\} = 0 \qquad \dot{\vec{C}}(\vec{f}) := \{\,\mathbf{H}\,,\,\vec{C}(\vec{f})\,\} = 0 \tag{4.35}$$

for all ($t$-independent) smearing fields.

The conditions give,

$$\{\,C(f)\,,\,\mathbf{H}\,\} = H(f)\,. \qquad \{\,\vec{C}(\vec{f})\,,\,\mathbf{H}\,\} = \vec{H}(\vec{f})\,. \tag{4.36}$$

So we have the following *secondary constraints*

$$H(x,t) = 0 \qquad \text{and} \qquad H_i(x,t) = 0\,. \tag{4.37}$$

where $H$ and $H_i$ were defined in (4.30).

[4] In the last equation, we are exceptionally using $(i,j) = ij + ji$ with no preceding factor.

There are no tertiary constraint since,

$$\begin{aligned}\dot{\vec{H}}(\vec{f}) &\equiv \{\mathbf{H}, \vec{H}(\vec{f})\} = \vec{H}(\mathcal{L}_{\vec{N}}\vec{f}) - H(\mathcal{L}_{\vec{f}}N) = 0 \\ \dot{H}(f) &\equiv \{\mathbf{H}, H(f)\} = H(\mathcal{L}_{\vec{N}}f) + \vec{H}(\vec{N}(N,f,q)) = 0\,,\end{aligned} \tag{4.38}$$

where,

$$\vec{N}(f,f',q)^i = q^{ij}(ff'_{,j} - f'f_{,j})\,. \tag{4.39}$$

The conditions (4.38) weakly vanish if we use the constraints (4.37). The algebra of constraints is

$$\begin{aligned}\{\vec{H}(\vec{f}), \vec{H}(\vec{f}')\} &= \vec{H}(\mathcal{L}_{\vec{f}}\vec{f}') \\ \{\vec{H}(\vec{f}), H(f))\} &= H(\mathcal{L}_{\vec{f}}f) \\ \{H(f), H(f'))\} &= \vec{H}(\vec{N}(f,f',q))\end{aligned} \tag{4.40}$$

this is called the Dirac algebra (see [86]) or also the *hypersurface deformation algebra*. We immediately see that the constraints are first class since they form a closed Poisson algebra.

We note that because $\dot{N}^a = \lambda^a, \dot{N} = \lambda$ we can treat $N, N^i$ as *Lagrange multipliers* and drop all terms proportional to $C, C_i$ from the action (4.29), and get the following reduced action

$$S = \int_{\mathbb{R}} dt \int_{\sigma} d^3x\{\dot{q}_{ij}P^{ij} - [N^iH_i + NH]\} \tag{4.41}$$

this action, called the canonical (ADM) action (see ([81])), is equivalent to (4.29) in all that concerns $q_{ij}, P^{ij}$. From this action we find the following reduced Hamiltonian,

$$\mathbf{H} = \int_{\sigma} d^3x[N^iH_i + NH]\,. \tag{4.42}$$

We note also that the secondary constraints ((4.37)) are related to the Einstein equations (in the vacuum) $G_{\mu\nu} = 0$ by,

$$\begin{aligned}G_{\mu\nu}n^{\mu}n^{\nu} &= -\frac{H}{2\sqrt{\det(q)}} \\ G_{\mu\nu}n^{\mu}q^{\nu}_{\rho} &= \frac{H_{\rho}}{2\sqrt{\det(q)}}\,.\end{aligned} \tag{4.43}$$

The Hamilton equations (3.28) in this case take the specific form:

$$\dot{q}_{ij} = \{\, q_{ij}\,,\, \mathbf{H}\,\}\,, \tag{4.44}$$

$$\dot{P}^{ij} = \{\, P^{ij}\,,\, \mathbf{H}\,\} \tag{4.45}$$

The constraints (4.37) and the algebra (4.40) have the following geometrical interpretation:

- The constraint $\vec{H}(\vec{f})$ is the generator of spatial diffeomorphisms. The Lie algebra of the diffeomorphism group $Diff(\Sigma)$ is generated by the vector fields on $\Sigma$, with minus the commutator of a pair of vector fields

$$\left[\vec{N}_1, \vec{N}_2\right]^i = N_1^j N_{2\ ,j}^i - N_2^j N_{1\ ,j}^i \,, \tag{4.46}$$

  as the Lie bracket. The first of (4.40) shows that the map $N \longrightarrow -\vec{H}(\vec{N})$ is a homomorphism of the Lie algebra of $Diff(\Sigma)$ into the Poisson algegra of the theory.

  A direct calculation of the Poisson bracket of $\vec{H}(\vec{N})$ with $q_{ij}$ and $p^{ij}$ gives:

$$\{\, q_{ij}(x)\, ,\, \vec{H}(\vec{N})\, \} = (\mathcal{L}_{\vec{N}} q)_{ij}(x) \tag{4.47}$$
$$\{\, P^{ij}(x)\, ,\, \vec{H}(\vec{N})\, \} = (\mathcal{L}_{\vec{N}} P)^{ij}(x) \tag{4.48}$$

  which confirms $\vec{H}(\vec{N})$ as the generator of spacial diffeomorphisms.

- The constraint $H(f)$ generates normal deformations of the hypersurface $\Sigma$.

  This is confirmed by the following Poisson brackets,

$$\begin{aligned} \{\, q_{ij}(x)\, ,\, H(N)\, \} &= (\mathcal{L}_{N\vec{n}}\, q)_{ij}(x) \\ \{\, P^{\mu\nu}\, ,\, H(N)\, \} &= \frac{q^{\mu\nu} NH}{2} - N\sqrt{\det(q)}[q^{mu\rho} q^{\nu\sigma} - q^{mu\nu} q^{\rho\sigma}] R^{(4)}_{\rho\sigma} \\ &+\ \mathcal{L}_{N\vec{n}} P^{\mu\nu}\,. \end{aligned} \tag{4.49}$$

  Furthermore, in both the brackets the field equations must be used to obtain the geometric interpretation of $H(N)$ as generator of deformations normal to $\Sigma$. In the second equation of (4.49) we explicitly see that we must use the Einstein equations[5] $R^{(4)}_{\mu\nu} = 0$ (note that the constraints $H$ and $H_i$ are projections of the Einstein equations as given in equations (4.43) see also [47]) so this is an on-shell relation. In the first of these equations the field equations were used in the intermediate calculations to re-express $P$ in terms of $\dot{q}$ (see [47] for the details of the calculation), so this is also an on-shell relation.

[5]Note that in the vacuum, we may write $R_{\mu\nu} = 0$ or $G_{\mu\nu} = 0$ for the Einstein equations.

We add two things (refer to [79]) about the constraints (4.37) and the algebra (4.40). First, it is not the Lie algebra of the diffeomorphism group of spacetime *Diff*$(M)$ even though this was the the invariance group of the original theory. Second, it is not a Lie algebra at all! Given the presence of $q^{ij}$ on the right hand side of (4.39) and consequently in the last Poisson bracket in (4.40).

The counting of the degrees of freedom is the following. The canonical variables parametrizing the phase space are the fields $(q_{ij}, P^{ij})$ and they constitute 12 variables *per space point* (sometimes written $12 \times \infty^3$ variables). We have 4 constraints *per space point* (three for $H_i$ and one for $H$), furthermore since these constraints are first class, they generate a four parameter set of gauge transformations, *per space point*. Therefore we need four gauge fixing conditions *per point*. This way we are left with $(12 - 4 - 4) \times \infty^3$ variables[6] for the reduced phase space which correspond to two *local degrees of freedom*. This is consistent with the number of variables we found in section (1.4) for the linearised theory, corresponding to the the polarizations of planar gravitational waves.

[6]This is sometimes stated as: *first class constraints act twice.*

# 5
# Ashtekar variables

We now proceed with the canonical quantization of GR and examine its consequences. However, we take a slight change of stance and start using tetrads instead of the metric $g_{\mu\nu}$. This is intended to facilitate the transition to Ashtekar variables, considered a major breakthrough in this subject. Since tetrads and the metric are related, some parts of this chapter overlap with the the previous one, this is necessary to clarify the subsequent discussion. We follow [87, 88, 80].

To facilitate the use of tetrads (or triads in the Hamiltonian formalism) we introduce a tensor similar to (4.10) in the following way,

$$q^a{}_\mu = e^a{}_\mu + n^a n_\mu \,. \tag{5.1}$$

where $n^a = n^\mu e^a{}_\mu$. This has the properties $q^a{}_\mu n^\mu = 0$ and $q_a{}^\mu n^a = 0$.

Furthermore, we choose a decomposition compatible with the splitting of spacetime of the last chapter. That is, in terms of the pull-back of $e$:

$$Y^*(e)^a{}_\mu = \begin{pmatrix} N & 0 \\ N^\alpha & q^\alpha{}_i \end{pmatrix} , \tag{5.2}$$

where the group indices are split in $a \to (0, \alpha)$ where $\alpha = 1, 2, 3$, and $N^\alpha = N^i q^\alpha_i$. This is the *time gauge* which consists explicitly in demanding that the pull-back of $e^0{}_i$ vanishes (for a generalization see [89]).

The variables are the triads $q^\alpha{}_i$ and the momenta are functional derivatives of the Palatini action (2.139) with $\Lambda = 0$, written in these variables (refer to [90]). We have,

$$P_\alpha{}^i = \frac{\delta S}{\delta \dot{q}^\alpha{}_i} = \frac{1}{2}\tilde{q} q_\beta{}^i \left(K_\alpha{}^\beta - \delta_\alpha{}^\beta K\right) \tag{5.3}$$

Where, $\tilde{q} = \det(q^\alpha{}_i)$ and $K$ is the trace of $K_\alpha{}^\beta$, i.e. $K = K_\alpha{}^\alpha$. This is analogous to (4.26). where the tensor $K_{\alpha\beta}$ can be calculated from (4.10) in the following way,

$$\begin{aligned} K_{\mu\nu} &= q_\mu{}^\rho q_\nu{}^\sigma \nabla_\rho n_\sigma \\ &= q_\mu{}^\alpha q_\nu{}^\beta e_\alpha{}^\rho e_\beta{}^\sigma \nabla_\rho e^0{}_\sigma \\ &= q_\mu{}^\alpha q_\nu{}^\beta K_{\alpha\beta} \, . \end{aligned} \tag{5.4}$$

Where we used $e^0{}_\mu = n_\mu$ and the already mentioned property of the contraction of $q$ with the normal.

We note that, since in this section we have objects with 'spacial indices' $a\,, b\,, \ldots$, their derivatives are given by (2.134) and in therefore, in this section, the symbol $\nabla$ is an abbreviation of $\nabla^\omega$.

Furthermore, using (2.138) we have,

$$K_{\alpha\beta} = e_\alpha{}^\mu e_\beta{}^\nu \nabla_\mu e^0{}_\nu = e_\alpha{}^\mu \omega_{\beta 0\,\mu} \tag{5.5}$$

The extrinsic curvature can be written as a function of the momentum $P_{\alpha\beta}$,

$$K_{\alpha\beta} = \frac{2}{\tilde{q}} \left( P_{\alpha\beta} - \frac{1}{2} \delta_{\alpha\beta} P \right) \, . \tag{5.6}$$

In this equation, $P_{\alpha\beta}$ can be related to (5.3) by the use of the triad. Furthermore $P$ is the trace of $P_{\alpha\beta}$.

In the previous chapter we used a similar relation (4.28).

The symmetry $K_{\alpha\beta} = K_{\beta\alpha}$ gives the Lorentz constraint,

$$L_{\alpha\beta} = q_{[\alpha\,|i|} P_{\beta]}{}^i \approx 0 \, , \tag{5.7}$$

this is the generator of spatial rotations of the triads $q^\alpha{}_i$ [88].

The reduced Hamiltonian is given by equation (4.42)

$$\mathbf{H} = \int_\sigma d^3x [N^\alpha H_\alpha + NH] \, . \tag{5.8}$$

where the spacial diffeomorphism and Hamiltonian constraints are given respectively by the equivalent of equations (4.30),

$$\begin{aligned} H_\alpha &= -D_j P_\alpha^{\;j} \\ H &= \frac{1}{\tilde{q}} \left( P_{\alpha\beta} P^{\alpha\beta} - \frac{1}{2} P^2 \right) - \tilde{q} R^{(3)} \, . \end{aligned} \tag{5.9}$$

For the relation between the metric and triad canonical formulations see [47, 87].

The fundamental Poisson brackets (these are essentially the same as the (4.32) but written with triads) are,

$$\left\{q_{\alpha\,i}(x), P_{\beta}{}^{j}(x')\right\} \;=\; \delta_{\alpha\beta}\delta_i^j\delta^3(x-x')\,, \tag{5.10}$$

with all other brackets vanishing, for the position and momentum variables with the indices in the indicated positions.

Canonical Quantization, in the 'position space representation' is constructed by promoting the canonical position and momentum to operators. The triad acts diagonally as a multiplicative operator and the momentum acts as differentiation with respect to the triad. That is [1]

$$\begin{aligned} \hat{q}^{\alpha}{}_{i}\Psi\,[q] &\equiv q^{\alpha}{}_{i}\Psi\,[q] \\ \hat{P}_{\alpha}{}^{i}\Psi\,[q] &\equiv \frac{\hbar}{i}\frac{\delta}{\delta\, q^{\alpha}{}_{i}}\Psi\,[q]\;, \end{aligned} \tag{5.13}$$

in the following we will drop the ˆ from operators. In these formulas $\Psi\,[q]$ is the 'wave function' (functional to be more precise) of the system.

Through this quantization, the classical constraint equations (5.7) and the first of (5.9) become quantum constraint operators. These are called the 'kinematical constraints'. They read,

$$\begin{aligned} L_{\alpha\beta}\Psi\,[q] &= 0 \\ H_{\alpha}\Psi\,[q] &= 0\,. \end{aligned} \tag{5.14}$$

Dynamics is generated via the Hamiltonian constraint, by the *Wheeler-DeWitt (WDW) equation*

$$H\Psi\,[q] = 0\,. \tag{5.15}$$

This[2] is a highly singular and ill-defined equation. The two $P_{\alpha}{}^{i}$ in (5.9) become functional derivatives which in turn involve distributions. Their square is therefore singular [91] and needs to be regularized. The lack of an appropriate Hilbert space structure on the space of quantum states $\Psi(q)$ makes this theory virtually inviable.

[1]The same may, of course, be written in terms of the 3-metric $q_{ij}$ ant its conjugate momentum $P^{ij}$, the operators are

$$\hat{q}_{ij}\Psi\,[q] \;\equiv\; q_{ij}\Psi\,[q] \tag{5.11}$$

$$\hat{P}^{ij}\Psi\,[q] \;\equiv\; \frac{\hbar}{i}\frac{\delta}{\delta\, q_{ij}}\Psi\,[q]\;, \tag{5.12}$$

[2]The WdW equation, obtained after substituting (5.13) in the Hamiltonian constraint, the second equation of (5.9) see [88].

## 5.1 New variables for General Relativity

A breakthrough in this subject was achieved by Ashtekar who introduced a new set of variables [92]. These variables may be derived *ab initio* from the Holst action (more details may be found in Appendix C and [93, 94]). Here we consider the combination[3]:

$$\begin{aligned} A_{\alpha\, i} &\equiv -\frac{1}{2}\varepsilon_{\alpha\beta\gamma}\, \omega^{\beta\gamma}{}_{i} + \frac{\gamma}{\tilde{q}}\left(P_{\alpha\, i} - \frac{1}{2} q_{\alpha\, i} p\right) \\ &= -\frac{1}{2}\varepsilon_{\alpha\beta\gamma}\, \omega^{\beta\gamma}{}_{i} + \gamma K_{\alpha\, i}\,, \end{aligned} \tag{5.16}$$

where $\gamma \neq 0$ is the 'Barbero-Immirzi' parameter. It is believed to be important at the quantum level (by defining the fundamental units of areas and volumes, more details may be found in [12]) despite lacking physical significance classically.

There is a parallel here, to the $\theta$-parameter of the Yang-Mills theory. The $SU(2)$ Yang-Mills action can be extended — through the addition of the Pontryagin topological term (Cf. Appendix C) — to the Yang-Mills action,

$$S = \frac{1}{2g_{YM}^2} tr \int_M F \wedge *F + \theta tr \int_M F \wedge F\,. \tag{5.17}$$

It can be consistently argued (see [95]) that the parameters $\gamma$ and $\theta$ play analogous roles.

The variable conjugate to $A_{\alpha\, i}$ is the densitised inverse triad,

$$E_{\alpha}{}^{i} = \tilde{q}\, q_{\alpha}{}^{i}\,. \tag{5.18}$$

Note that,

$$\det(E_{\alpha}{}^{i}) = \tilde{q}^2\,. \tag{5.19}$$

The fundamental Poisson brackets are[4]:

$$\begin{aligned} \left\{A_{\alpha\, i}(x)\, A_{\beta\, j}(x')\right\} &= 0 \\ \left\{E_{\alpha}{}^{i}(x)\, E_{\beta}^{\,j}(x')\right\} &= 0 \\ \left\{A^{\alpha}{}_{i}(x)\, E_{\beta}^{\,j}(x')\right\} &= \gamma \delta^{\alpha}_{\beta}\delta^{j}_{i}\delta^{3}(x-x')\,. \end{aligned} \tag{5.20}$$

[3]Note that since we are usiing $\eta_{ab} = diag\{-1,1,1,1\}$ the spacial part of this metric is the Kronecker delta $\delta_{\alpha\beta}$. Consequently there is no need to distinguish between internal upper and lower indices e.g. $A_{\alpha\, i} \equiv A^{\alpha}{}_{i}$. However, we will try to keep all indices in their original positions, whenever it is possible.

[4]The parameter $\gamma$ is sometimes eliminated from these brackets through its placement in the definition of $E$, equation (5.18).

The Gauss constraint, in the variables (5.16) and (5.18), reads:

$$\nabla_i E_\alpha{}^i \equiv \partial_i E_\alpha{}^i + \varepsilon_{\alpha\beta\gamma} A^\beta{}_i A^{\gamma i} \approx 0\,. \tag{5.21}$$

This constraint implies the covariant constancy of the spacial triads and the Lorentz constraint (5.7).

Note that, although up to this point everything could be generalized to $n$ dimensions, equation (5.21) can only be written in three dimensions (more details can be found in [47] page 127).

The field strength is defined by

$$\begin{aligned} F^\alpha{}_{ij} &= \partial_i A^a{}_j - \partial_j A^a{}_i + \varepsilon^{abc} A_{bi} A_{cj} \\ &= -\frac{1}{2}\varepsilon^{abc} R_{bc\,ij} + \gamma\left(\nabla_i K^a{}_j - \nabla_j K^a{}_i\right) + \gamma^2 \varepsilon^{abc} K_{bi} K_{cj}\,, \end{aligned} \tag{5.22}$$

thus the diffeomorphism constraint (the first equation of (5.9))) becomes

$$E_\alpha{}^i F^\alpha{}_{ij} \approx 0\,. \tag{5.23}$$

The Hamiltonian constraint can be expressed in these variables using the relation,

$$\begin{aligned} \varepsilon^{\alpha\beta\gamma} E_\alpha{}^i E_\beta{}^i F_{\gamma\,ij} &= -\gamma^2\left(p^{\alpha\beta} p_{\alpha\beta} - \frac{1}{2}p^2\right) - \tilde{q}^2 R^{(3)} \\ &= \gamma^2 \tilde{q} H - \frac{1}{4}\left(1+\gamma^2\right)\tilde{q}^2\left(K^{\alpha\beta}K_{\alpha\beta} - K^2\right)\,. \end{aligned} \tag{5.24}$$

and choosing $\gamma = \pm i$. For this particular value of the Barbero-Immirzi parameter (and example of a similar construction for real $\gamma$ is [96]), the second term in (5.24) drops out, and the Hamiltonian constraint is completely written (except for an extra $\tilde{q}$) in terms the new variables. A very important feature of this form is that — unlike the second equation of (4.30) or the second equation of (5.9) — it is polynomial in these variables! Although the Ashtekar variables have other nice advantages, this is probably the most important: the polynomiality of the Hamiltonian constraint.

Sadly, this same choice, has brought into play a nearly insurmountable obstacle: the value $\gamma = \pm i$ implies a complexification of the phase space of GR, and obviously the real phase space must be later recovered using suitable reality conditions. While in the classical theory this is a fairly straightforward procedure, in the quantum formalism however, problems arose concerning the definition and imposition of appropriate hermiticity conditions on the states and operators [88].

If we take $\gamma$ to be a real, quantity, no reality conditions are necessary, but the the cancellation in the second term of eq (5.24) no longer takes place. As a consequence, we loose the polynomial character of the Hamiltonian constraint.

## 5.2 Connection representation and Kodama state

A quantization scheme based upon the the connection $A^{\alpha}{}_{i}$ may be set up and in this context the choice $\gamma = \pm i,$. The functional $\Psi[A]$ now depends on this connection, and the operators are,

$$\hat{A}^{\alpha}{}_{i}\Psi[A] = A^{\alpha}{}_{i}\Psi[A] \tag{5.25}$$

$$\hat{E}_{\alpha}^{\ j}\Psi[A] = \frac{\hbar}{i}\frac{\delta}{\delta A^{\alpha}{}_{j}}\Psi[A]\,. \tag{5.26}$$

In this connection representation the 3-metric (operator) becomes a lot less intuitive. It reads,

$$qq^{ij}(x)\Psi[A] = -\hbar^{2}\frac{\delta}{\delta A^{\alpha}{}_{i}(x)}\frac{\delta}{\delta A_{\alpha j}(x)}\Psi[A]\,, \tag{5.27}$$

where $q = \det(q_{ij})$. The volume operator is given by,

$$\hat{q}(x)\Psi[A] = \frac{i\hbar^{3}}{6}\varepsilon^{\alpha\beta\gamma}\varepsilon_{ijk}\frac{\delta}{\delta A^{\alpha}{}_{i}(x)}\frac{\delta}{\delta A^{\beta}{}_{j}(x)}\frac{\delta}{\delta A^{\gamma}{}_{k}(x)}\Psi[A]\,. \tag{5.28}$$

We must note that these operators are highly singular, mainly because they contain products of functional derivatives evaluated at the same point. However, ignoring these difficulties for now the WdW equation (5.15) becomes in this representation[5],

$$\varepsilon^{\alpha\beta\gamma}F_{\alpha jk}\frac{\delta}{\delta A^{\beta}{}_{j}}\frac{\delta}{\delta A^{\gamma}{}_{k}}\Psi[A] = 0\,. \tag{5.29}$$

this represents a somewhat simpler form of (5.15), then the one obtained substituting (5.13) in the Hamiltonian constraint, the second equation of (5.9). Since $F(A)$ and $\frac{\delta}{\delta A}$ are operators (analogous to $\hat{x}$ and $\hat{p}$ in the quantum mechanics of particles) they do not commute and their ordering is important.

Although there are no known solutions to eq (5.29), if we ad a Cosmological Constant term $\Lambda q$ and choose the ordering opposite to one in the last equation we get,

$$\varepsilon^{\alpha\beta\gamma}\frac{\delta}{\delta A^{\beta}{}_{j}}\frac{\delta}{\delta A^{\gamma}{}_{k}}\left(F_{\alpha jk} - \frac{i\hbar\Lambda}{6}\varepsilon_{ijk}\frac{\delta}{\delta A^{\alpha}{}_{i}}\right)\Psi_{\Lambda}[A] = 0\,, \tag{5.30}$$

[5]We are omitting the point dependency for simplicity. We are also allowing for an extra factor of $q$ and imposing $q \neq 0, \infty$ (see [88])

and after having complicated the problem, (at least formally) a solution exists! It is called the Kodama state [88, 97] and is given by,

$$\Psi_\Lambda[A] = \exp\left(\frac{i\hbar}{\Lambda}\int_\Sigma d^3x\mathcal{L}_{CS}\right), \tag{5.31}$$

where,

$$\mathcal{L}_{CS} = A\wedge dA + iA\wedge A\wedge A \ ,$$

is the Cheren-Simons Lagrangian density (see [98]).

The solution (5.31) has problems with its limits: both the flat limit — vanishing $\Lambda$ — and the 'semiclassical' limit — vanishing $\hbar$ — are hard to calculate and interpret.

## 5.3 Holonomies and Fluxes

As mentioned above, the WdW equation (5.29) is very complicated and hard to solve. This motivates the search for other variables. As it turns out, a better set of variables exists called *Holonomies and Fluxes*[6]. Given a curve $e$ in $\Sigma$, sometimes called a *link,* the holonomy is given by

$$h_e[A] = \mathcal{P}exp\int_e A = \mathcal{P}exp\int_e A^\alpha{}_i\tau_\alpha dx^i \tag{5.33}$$

where $\tau_\alpha = -\sigma_\alpha/2$, and $\sigma_\alpha$ are the Pauli matrices. The connection $A$ we are using takes values on the $\mathfrak{su}(2)$ Lie algebra and its path-ordered exponential, in turn, takes values in the $SU(2)$ Lie group. The connection transforms in the usual way — given by (2.86) — under the action of the $SU(2)$ group. The holonomy[7] $h_e[A]$, transforms under the action of this same group, in the following way,

$$h_e[A]\to h_e^g[A] = g(e(0))h_e[A]g^{-1}(e(1)) \quad \text{with} \quad g(e(0))\,, g^{-1}(e(1))\in SU(2)\,. \tag{5.34}$$

[6]The transition from the connection representation to the *Loop Representation* may be carried out using the *Loop Transform*. This a sort of infinite dimensional Fourier transform from the space of connections $A$ to the space of "loops" $e$. It is given by,

$$\Psi[e] = \int d\mu_0[A]\, tr\mathcal{P}e^{\int_e A}\,\Psi[A]\,. \tag{5.32}$$

We use a similar concept in Sec. 8.6 equation (8.118). More details about the Loop transform can be found in [99].

[7]Parallel transport to be precise.

This transformation under the action of $SU(2)$ is, in fact, a strong reason to choose the holonomy as a variable.

To parametrize the phase space of GR we need one more set of variables. For this we note that the (infinitesimal) flux vector may be written in the following way,

$$dF_\alpha = \frac{1}{2}\varepsilon_{\alpha\beta\gamma}\theta^\beta \wedge \theta^\gamma\,, \qquad \theta^\alpha = q^\alpha{}_i dx^i \tag{5.35}$$

or in terms of the variables $e$ $E$

$$dF_\alpha = \frac{1}{2}\varepsilon_{\alpha\beta\gamma}q^\beta{}_i q^\gamma{}_j dx^i \wedge dx^j = \frac{1}{2}\varepsilon_{ijk}E_\alpha{}^i dx^j \wedge dx^k\,. \tag{5.36}$$

Given a surface $S$ embedded in $\Sigma$, we have the following flux vector

$$F^\alpha{}_S[E] = \int_S dF^\alpha\,. \tag{5.37}$$

This vector may be smeared using a test function $f^\alpha$ giving,

$$F = \int_S \frac{1}{2}f^\alpha\varepsilon_{\alpha\beta\gamma}e^\beta{}_i e^\gamma{}_j dx^i \wedge dx^j = \int_S \frac{1}{2}f^\alpha\varepsilon_{ijk}E_\alpha{}^i dx^j \wedge dx^k\,. \tag{5.38}$$

The Flux variable (5.36) does not transform as a vector under gauge transformations. Since this is such an important feature, a modified definition of flux variable can be used (see [100, 101]).

Take an oriented path $e$ embedded in $\Sigma$ and a point $z \in e$. Take also a surface $S_e$ that is intersected by the path $e$ transversally such that $z = e \cap S_e$. We choose now, a set of paths $\pi_e\,:\,S_e \times [0,1] \to \Sigma$. These paths are in a one to one correspondence with the points $y$ of $e$ in such a way that they go from the source point $S(e)$ to $y \in S_e$. This means that $\pi(y,0) = S_e$ and $\pi(y,1) = y$. In these conditions we can define,

$$(\tilde{F}_\alpha)_{S,\pi_e}(E,A) \equiv \frac{1}{2}\int_S \varepsilon_{ijk}h_{\pi_e}E_\alpha{}^i h^{-1}{}_{\pi_e}dx^j \wedge dx^k \tag{5.39}$$

where, $h_{\pi_e}$ is the holonomy of the connection along the path $\pi_e$. The integral in $h_{\pi_e}$ goes from $S(\pi_e)$ to $y$. Note that this variable depends on $E$ and on the holonomies (of course though these also on the connection $A$).

The new form (5.39) of the flux, has a transformation under the $SU(2)$ group similar to (5.34). It reads,

$$(\tilde{F}_\alpha)_{S,\pi_e} \to (\tilde{F}_\alpha)^g_{S,\pi_e} = g(S(\pi_e))(\tilde{F}_\alpha)_{S,\pi_e}g^{-1}(S(\pi_e))\,. \tag{5.40}$$

The variables $(h_e[A])$ and $F$ (or $(\tilde{F}_\alpha)_{S,\pi_e}$) are used to describe the phase space.

To calculate the Poisson bracket between holonomies and fluxes we note that, if a curve $\phi$ intersects the surface $S$ tangentially at a point $p$ (this point splits the curve in two parts $\phi_1$ and $\phi_2$) then we have,

$$\left\{ (h_\phi[A])_{\alpha\beta}\ ,\ F[E,A] \right\} \ = \ \imath(e,S)\gamma f_a(P) \left( h_{\phi_1}[A]\tau^a h_{\phi_2}[A] \right)_{\alpha\beta} \tag{5.41}$$

and the intersection number

$$\imath(\phi,S) = \int_\phi dx^i \int_S dx^j dx^k \varepsilon_{ijk} \delta^{(3)}(x-y) \tag{5.42}$$

can take values $\pm 1$ or 0 and is a measure of how $\varphi$ intersects $S$. In particular, if the curve does not pierce the surface the Poisson bracket is null. All other Poisson brackets vanish.

If we choose to use the alternative form of the flux variable (5.39), the Poisson brackets of these fluxes no longer vanish (as in the case of the Poisson brackets between (5.37)) instead, we have a bracket of the form

$$\left\{ \tilde{F}^\alpha_{S_e\,\pi_e}, \tilde{F}^\beta_{S_{e'}\,\pi_{e'}} \right\} = \varepsilon^{\alpha\beta\gamma} \delta_{ee'} \tilde{F}_{\gamma S_e\,\pi_e}\ . \tag{5.43}$$

The brackets between $h$ and $\tilde{F}$ are

$$\left\{ \tilde{F}^\alpha_{S_e\,\pi_e}, h_{e'} \right\} = -\delta_{ee'}\tau^\alpha h_e + \delta_{ee'^{-1}}\tau^\alpha h_e \tag{5.44}$$

where $\tau^\alpha$ was defined in the text following (5.33) (details may be found in [100], and bout the absence of $\gamma$ in (5.43) and (5.44), see footnote 4 in this chapter).

## 5.4 The quantization process in LQG

Holonomies and fluxes, are the starting point of the LQG quantization formalism. In this context, appropriate Hilbert spaces and operators can be constructed see for example [12, 88]. In this section we describe some aspects of this quantization, however we do not mention (for the sake of brevity) the construction of operators like those associated to holonomies, fluxes, areas and volumes which are an integral part of LQG, details about these may be found e.g. in [12, 88, 102] and in references therein.

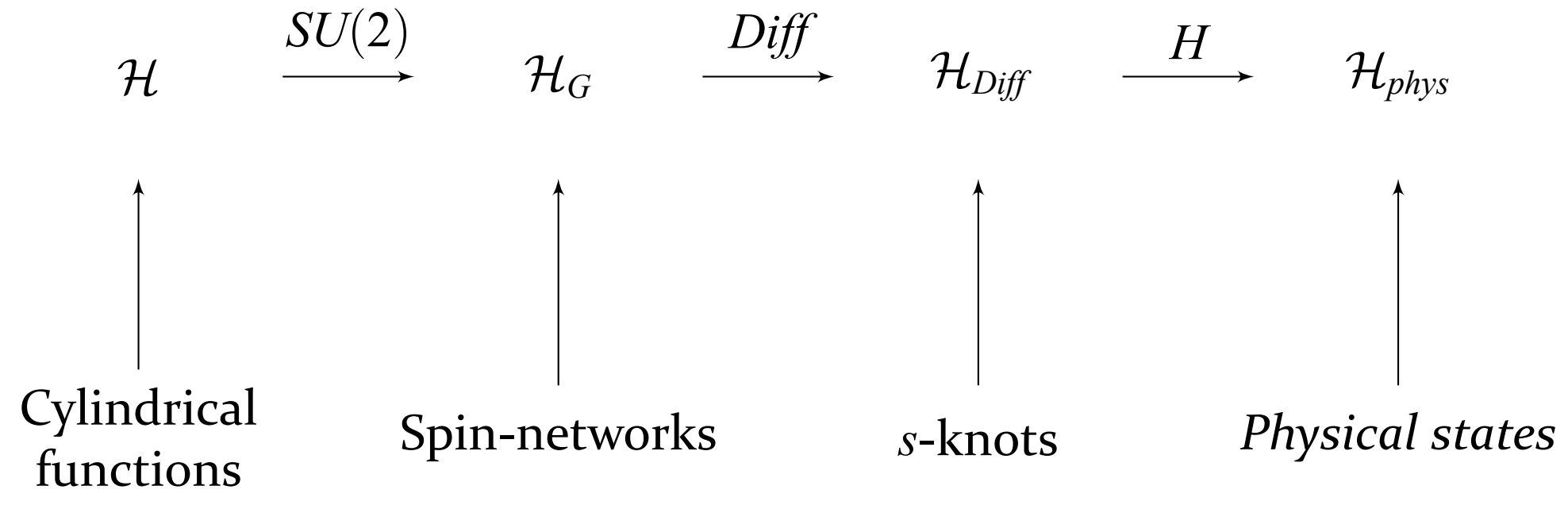

**Figure 5.1:** The general quantization scheme in Loop Quantum Gravity.

This quantization scheme starts from the phase space of GR expressed in terms of the real $SU(2)$ Barbero connection. Thus we have the Poisson brackets (5.20) and the constraints (5.21),(5.23) and (5.24) for real $\gamma$. It is important to stress that in the LQG quantization scheme (sometimes called Polymer Quantization), the variables to be promoted to operators are the holonomies (5.33) and fluxes (5.36) (or alternatively (5.39)), which are taken as variables in their own right. The Poisson brackets that go over to commutators are (5.41) (or alternatively (5.43) and (5.44)). Quantizing holonomies and fluxes has some analogies with QM where one may replace the Heisenberg operators $x$ and $p$ by Weyl operators $e^{i\alpha x}$ and $e^{i\beta p}$. In QM one may invoke the Stone–von Neumann theorem to guarantee the uniqueness of the quantization, that is, it makes no difference whether we quantize the Heisenberg or the Weyl algebra. However, the Stone–von Neumann theorem does not apply in the case of field theories (infinite degrees of freedom), and an alternative uniqueness theorem had to be found. In 2006 Lewandowski, Okolow, Salhmann and Thiemann proved such a theorem. The LOST[8] theorem guarantees such a uniqueness, provided that the quantization carries a unitary action of the diffeomophism group (see [103, 104] for details).

The quantization process continues with the construction of a *kinematical Hilbert space* $\mathcal{H}$ from functions of holonomies called *cylindrical functions*. We then impose the constraints (5.21),(5.23) and (5.24) successively. We then obtain from $\mathcal{H}$ the $SU(2)$-invariant Hilbert space $\mathcal{H}_G$ by imposing the Gauss constraint (5.21).Next, by imposing (5.23) the $SU(2)$ and diffeomorphism invariant Hilbert space $\mathcal{H}_{Diff}$ is obtained. And finally, the physical Hilbert space $\mathcal{H}_{phys}$ is constructed imposing the Hamiltonian constraint ((5.24)). This is depicted in figure 5.1.

An *ordered oriented graph* (we will use the simpler term *graph* to mean the same concept) embedded in $\Sigma$, is a collection $\Gamma$ of piecewise analytic oriented paths $e_l$ with $l = 1 \ldots L$. If we take a smooth function $f(U_1, \ldots, U_L)$ of $L$ $SU(2)$-group elements , the couple $(\Gamma, f)$ defines a function of $A$, with,

$$\Psi_{\Gamma,f}[A] = f(h_{e_1}[A], \ldots, h_{e_L}[A]) \ . \tag{5.45}$$

[8] The acronym formed from the initials of the names of the authors.

A function of this sort is called a *cylindrical function*.

For two cylindrical functions $\Psi_{\Gamma,f}[A]$ and $\Psi_{\Gamma,g}[A]$ with the same graph, a scalar product can be define by,

$$\langle \Psi_{\Gamma,f} | \Psi_{\Gamma,g} \rangle = \int_{SU(2)^L} dU_1 \ldots dU_L \overline{f(U_1 \ldots U_L)} g(U_1 \ldots U_L) \,, \tag{5.46}$$

where $dU$ denotes the Haar measure on the group $SU(2)$. This definition can be extended, in a consistent way, for cylindrical functions on different graphs.

The space of cylindrical functions is denoted *Cyl* and the kinematical Hilbert space $\mathcal{H}$ is the Cauchy conpletion of *Cyl* with respect to the norm of the inner product (5.46) (more details may be found in [12]).

The Gauss constraint (5.21) ensures the invariance under the internal $SU(2)$ rotations. It may be written in the following form,

$$C[\lambda] = \frac{1}{8\pi G\gamma} \int_\sigma d^3x \lambda^\alpha \nabla_i E_\alpha{}^i \,. \tag{5.47}$$

This constraint (5.47) can be promoted to an operator of the form,

$$(C[\lambda]\Psi)[A] = \frac{d}{d\varepsilon} \Psi[A - \varepsilon \nabla \lambda] \bigg|_{\varepsilon=0} \,. \tag{5.48}$$

The $SU(2)$ invariant states are those in the kernel of the operator (5.48) for arbitrary $\lambda$. The state $\mathcal{H}_G$ as the kernel of (5.48) can be obtained as the invariant subspace of $\mathcal{H}$ under finite transformations (see [102] and references therein for details).

A *spin network*, consists of a graph $\Gamma$ embedded in $\Sigma$, a finite number of edges (sometimes also called links) $e_l$ and vertices $v$ (the edges intersect at the vertices). Each edge caries a holonomy $h_e[A]$ (and consequently an irreducible representation $j$ of the $SU(2)$) and each vertex an intertwiner $i$, these may be regarded as generalized Clebsch–Gordan coefficients (see [12] for details). This process of assigning holonomies and intertwiner to links and vertices respectively is sometimes called *colouring*. In this way, a spin-network state $|\Gamma, j_l, i_n\rangle$ can be constructed (see [12, 102]. We will have more to say about the spin-network wave functions in section 8.6 see equation (8.115)).

To implement the spatial diffeomorphism constraint (5.23), a process a known as *group averaging* is employed. In general, group averaging involves finding an integration mesure $d\mu(\varphi)$ in the infinite dimensional diffeomorphism group $Diff(\Sigma)$. We would then have the following (formally) diffeomorphisms invariant state:

$$\Psi_\Gamma^{inv}[A] \stackrel{?}{=} \int_{Diff(\Sigma)} d\mu(\varphi)\Psi_\Gamma[A \circ \varphi]\,. \tag{5.49}$$

However, no measure of this sort is known in $Diff(\Sigma)$, and consequently the last expression is meaningless. In LQG, a modified version of the group averaging technique is used. The constraint can be formally solved in a much larger space, the algebraic dual $Cyl^*$ of the space of cylindrical functions. We then have,

$$Cyl \subset \mathcal{H} \subset Cyl^*\,. \tag{5.50}$$

We note that given a graph $\Gamma$, there is a subgroup $Diff_\Gamma$ of *Diff* that maps $\Gamma$ to itself, there is also a subgroup $TDiff_\Gamma$ of *Diff* that preserves every edge and its orientation. The quotient,

$$GS_\Gamma = Diff_\Gamma / TDiff_\Gamma\,, \tag{5.51}$$

is the *group of graph symmetries of* $\Gamma$. It permutes the ordering and/or flips the orientation of the edges of $\Gamma$ (see [102]).

In this way, following [102] we construct the general solution to the diffeomorphism constraint in two steps:

1. For a given graph $\Gamma$ the cylindrical functions with support on $\Gamma$ form a finite-dimensional subspace $\tilde{\mathcal{H}}_\Gamma$ of $\mathcal{H}$. The proper subspace $\mathcal{H}_\Gamma$ of $\tilde{\mathcal{H}}_\Gamma$ is the subspace spanned by the spin network states associated with $\Gamma$.

   Take a state $|\Psi_\Gamma\rangle \in \mathcal{H}_\Gamma$, a projector $\hat{P}_{Diff,\Gamma}$ form $\mathcal{H}_\Gamma$ to it's subspace invariant under $\hat{GS}_\Gamma$ is given by

$$\hat{P}_{Diff,\Gamma}\,|\Psi_\Gamma\rangle = \frac{1}{N_\Gamma}\sum_{\varphi\in GS_\Gamma}\hat{U}_\varphi\,|\Psi_\Gamma\rangle \tag{5.52}$$

   where $N_\Gamma$ is the number of $GS_\Gamma$ which is a finite group. The operator $\hat{U}_\varphi$ is given by,

$$\hat{U}_\varphi\,|\Psi_\Gamma\rangle = |\Psi_{\varphi\Gamma}\rangle\ . \tag{5.53}$$

2. Now, for any given state $|\Psi_\Gamma\rangle \in \mathcal{H}_\Gamma$ we can define the group averaged state $(\eta(\Psi_\Gamma)| \in Cyl^*$ by its action on arbitrary cylindrical functions $|\Phi_{\Gamma'}\rangle$ in the following way:

$$(\eta(\Psi_\Gamma)|\Phi_{\Gamma'}\rangle = \sum_{\varphi \in Diff/Diff_\Gamma} \langle \hat{U}_\varphi \hat{P}_{Diff,\Gamma} \Psi_\Gamma | \Phi_{\Gamma'} \rangle \tag{5.54}$$

   Where $\langle \cdot | \cdot \rangle$ is the inner product of $\mathcal{H}$.

The states of $\mathcal{H}_{Diff}$, are called *s-knots*. A knot is an equivalence class of unoriented graphs under diffeomorphisms. A coloured knot (a diagonalization process is necessary see [12]), i.e. one in which we associate a spin to each link and an intertwiner to each node is called an *s*-knot.

The implementation of the Hamiltonian constraint (5.24) which is supposed to reveal the quantum dynamics, is at the heart of the quantization procedure for LQG. One may think of following a procedure for the scalar constraint (5.24) similar to the one used for the spatial diffeomorphisms constraint. This however, runs into difficulties because the finite transformations generated by the Hamiltonian constraint are poorly understood even at the classical level. We then resort to a process of regularization of the classical expression, and then we promote this regularized classical constraint to a quantum operator. There are many difficulties related to this process, the complexity of (5.24) is the obvious one. Another one, pertains to the fact that if we are to satisfy the spatial diffeomorphism constraint (5.23), we must — as we saw above — transfer the action of the operator to the dual space $Cyl^*$. This in turn, implies that we must use a weaker notion of limit [88]. In any case, this is still an open question in LQG, and at the same time one of its limitations.

# Part II

# The *BFCG* theory

# 6 Categories and the *BFCG* theory

In this chapter, we present a brief introduction to the theory of categories. The standard reference about categories is [105], although in this Chapter we closely follow [106], from which we adopt the notation and conventions. Categories theory can be thought of as an attempt to treat processes (called *morphisms* in this context) on an equal footing to things (*objects* as they are called in category theory) [107]. The main interest of categories (and their higher order generalization) is that they serve as a language for formulating *Topological Quantum Field Theories* which, in turn are related to topological theories mentioned in section 2.7. After an exposition of the theory of categories, we present the *BFCG* theory which is the main concern of this thesis.

## 6.1 Categories

A *category* consists of:

- a collection of objects, which we denote by dots,

$$\underset{x}{\bullet} \tag{6.1}$$

- for any pair of objects $x$ and $y$ a collection of *morphisms* $f$ between them,

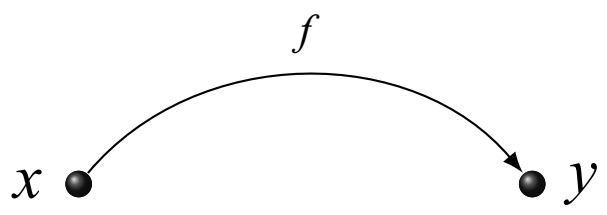

(6.2)

$x$ is called the source and $y$ the target of $f$.

- A *composition* exists for morphisms: given the morphisms $f$ and $g$, such that the target of $f$ is equal to the source of $g$

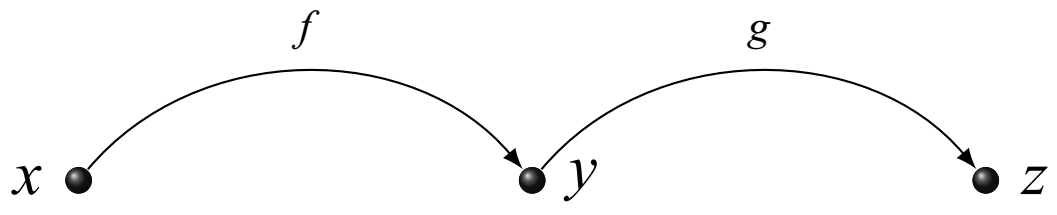

(6.3)

their composition is,

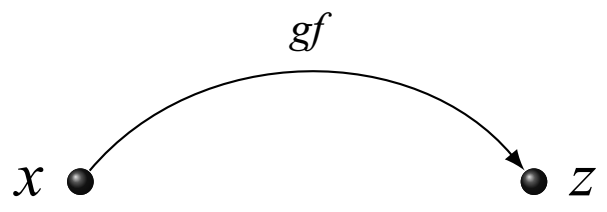

(6.4)

because $gf$ looks awkward, we write the last composition with reversed arrows which make composition look more natural,

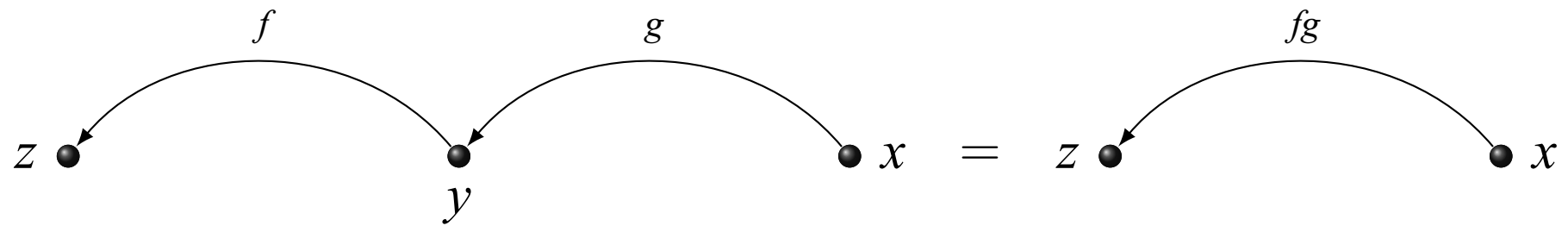

(6.5)

Composition of morphisms obeys an associative law,

$$(hg)f = h(gf)\,. \tag{6.6}$$

- For any object $x$ there is an *identity morphism* $1_x : x \longrightarrow x$, that satisfies the right and left unit laws:

$$1_y f = f = f\,1_x\,, \tag{6.7}$$

for any morphism $f : x \longrightarrow y$.

An example of a category is Set. The category Set has sets for objects and functions for morphisms. A second example is a group. In category theory, a group is a category with one object and for which all morphisms are invertible. The elements of the group are the morphisms of the category, and composition of morphisms is the group composition law.

Given categories $C$ and $D$ a *functor* $F : C \longrightarrow D$ consists of:

- a map $F$ sending objects in $C$ to objects in $D$;
- another map also called $F$ sending morphisms in $C$ to morphisms in $D$, that fulfils the following conditions:
  - for a morphism $f : x \longrightarrow y$ in $C$ we have, $F(f) : F(x) \longrightarrow F(y)$;
  - the map $F$ preservers composition, i.e. $F(fg) = F(f)F(g)$ when both sides are well defined;
  - the map $F$ preserves identities $F(1_x) = 1_{F(x)}$, for every object $x$ in $C$ (this property is a consequence of the others).

If the categories $C$ and $D$ are groups (one object categories with invertible morphisms) then, a functor $F : C \longrightarrow D$ is only a (group) homomorphism.

A connection, in this language, can be viewed as a functor from a category called the *path groupoid* $\mathcal{P}_1(M)$ to a Lie group G. For a smooth manifold $M$ the path groupoid is a category $\mathcal{P}_1(M)$ where,

- Objects are points of $M$.
- Morphisms are equivalence classes of a special kind of paths of $M$. To be precise morphisms are *thin homotopy classes of lazy paths* of $M$ (refer to [106] for the definitions).
- There is a composition for these classes of paths. If a thin homotopy class for the path $\gamma$ is denoted by $[\gamma]$ composition is denoted $[\gamma\delta] = [\gamma][\delta]$, and is related to the usual composition of paths (see [106]).
- For a point $x \in M$, the identity $1_x$ is the thin homotopy class of the constant path at $x$.

This functor from $\mathcal{P}_1(M)$ to $G$ (viewed as a category), will send every object of the path groupoid to the same object of the group category (after all there is only one object in the category of a group ). This functor, the connection, will also send every morphism of $\mathcal{P}_1(M)$ to a morphism of the category $G$ (a group element, since the morphisms of a group category are precisely the elements of the group). So the connection assigns a group element to each path on the space $M$. This is just what a connection does, it gives the parallel transport of of the wave function of a particle as it moves along a path.

## 6.2 2-Categories

The generalization of categories is given by the notion of 2-categories. These contain objects, morphisms and morphisms between morphisms called 2-*morphisms*.

To be precise a 2-category consists of:

- a collection of objects, which we denote by dots,

(6.8)

- for any pair of objects $x$ and $y$ a set of *morphisms* $f$ (the same as before but with inverted arrows) between them,

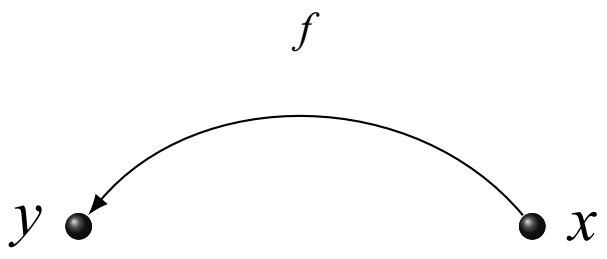

(6.9)

- for any pair of morphisms $f, g : x \longrightarrow y$ a set of 2-morphisms $\alpha : f \Longrightarrow g$,

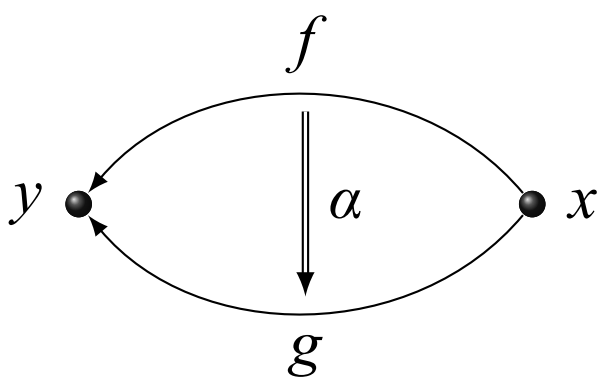

(6.10)

where $f$ is called the source of $\alpha$ and $g$ its target.

- morphisms can be compose like in the case of a category:

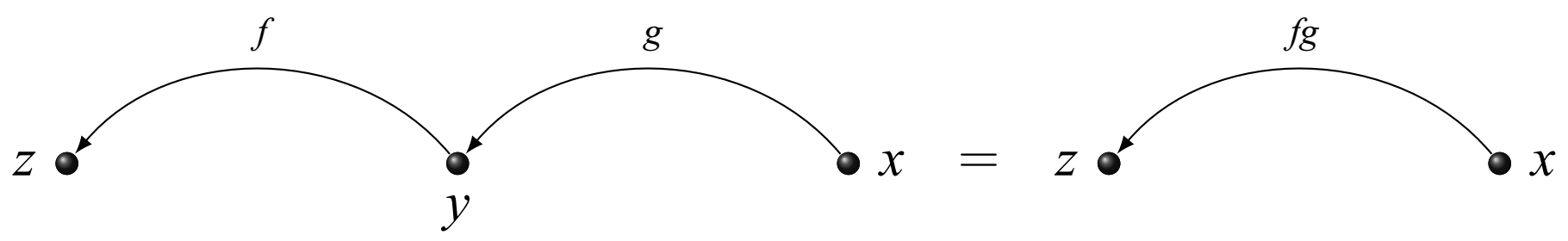

(6.11)

- while 2-morphism can be composed in two different ways:
  - vertically

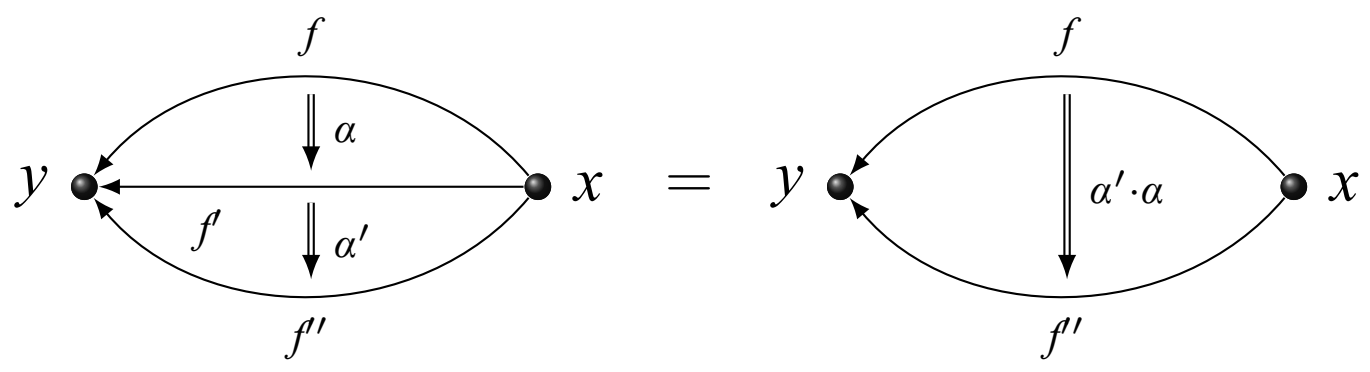

(6.12)

  - and horizontally

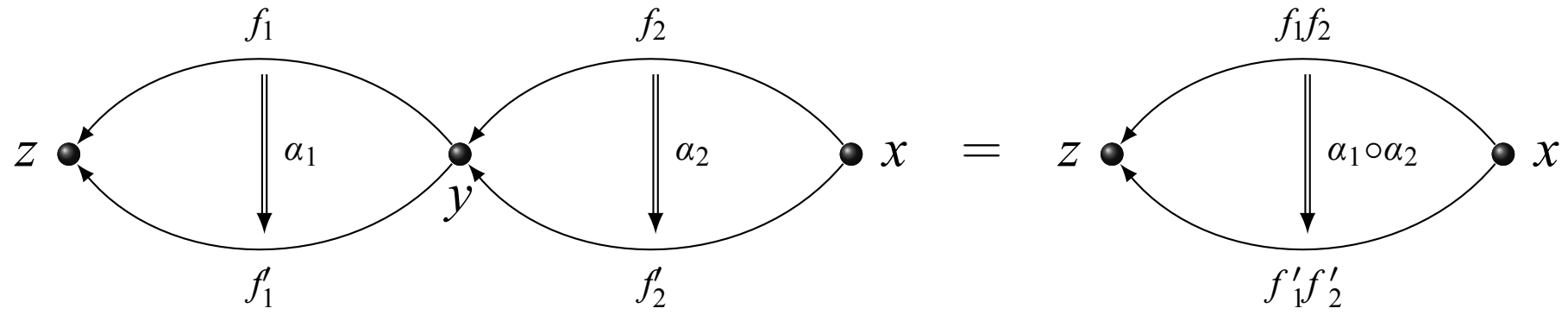

(6.13)

- composition of morphism is associative, and every object $x$ has a morphism,

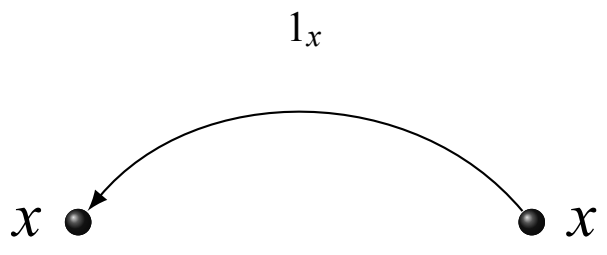

(6.14)

fulfilling the role of identity, as in the case of categories.

- Vertical composition is associative and every morphism $f$ has a 2-morphism,

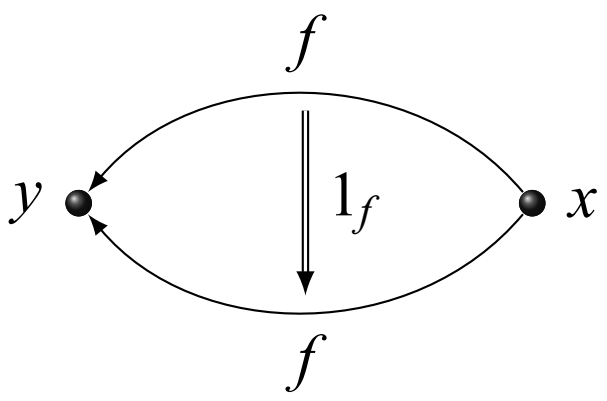

(6.15)

which is the identity for vertical composition.

- Horizontal composition is associative and the 2-morphism,

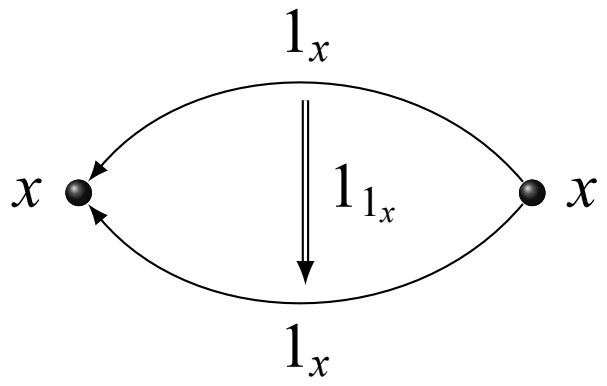

(6.16)

plays the role of identity for this composition.

- The vertical and horizontal compositions obey the *interchange law*:

$$(\alpha_1' \cdot \alpha_1) \circ (\alpha_2' \cdot \alpha_2) = (\alpha_1' \circ \alpha_2') \cdot (\alpha_1 \circ \alpha_2) \ , \tag{6.17}$$

that is a diagram of the form,

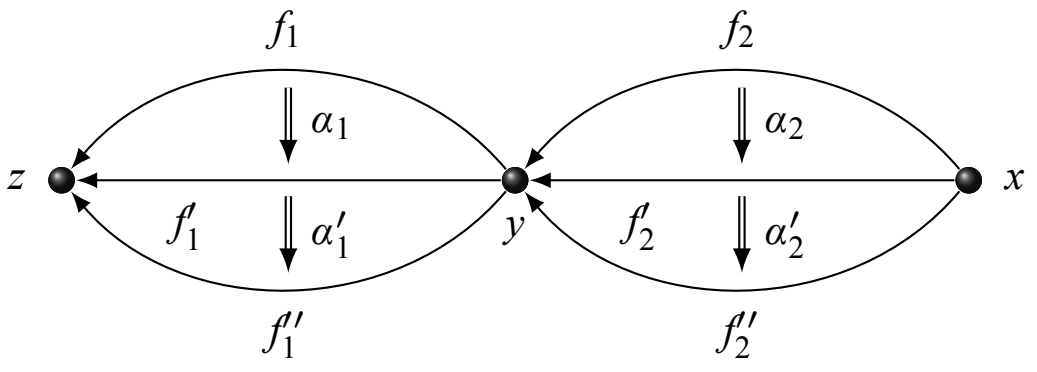

(6.18)

defines unambiguous 2-morphisms.

An example of a 2-category is a 2-group $\mathcal{G}$, which consists in:

- one object,

(6.19)

there is only one, so no label will be used.

- morphisms $g$,

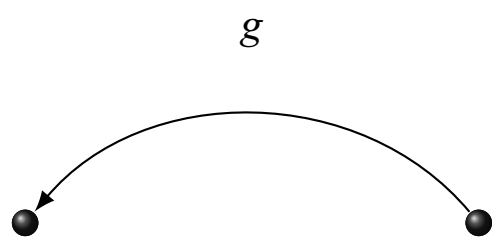

(6.20)

- and 2-morphisms

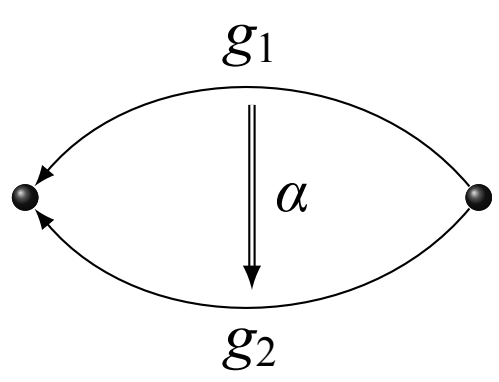

(6.21)

- the morphisms form a group under composition

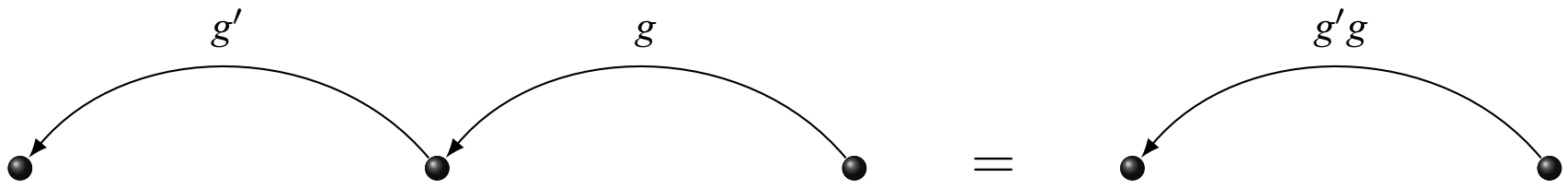

(6.22)

- the 2-morphisms form a group under horizontal composition,

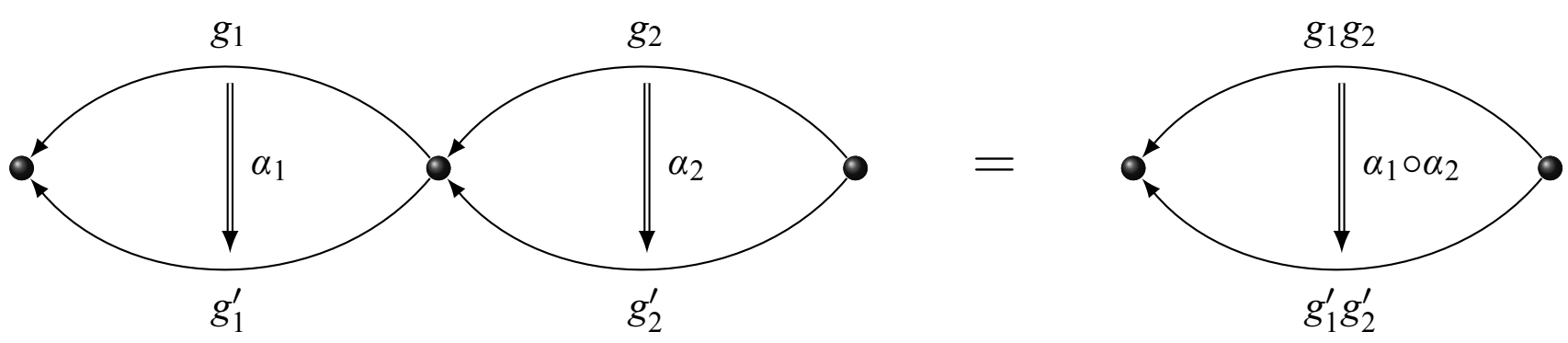

(6.23)

- 2-morphisms can also be composed vertically,

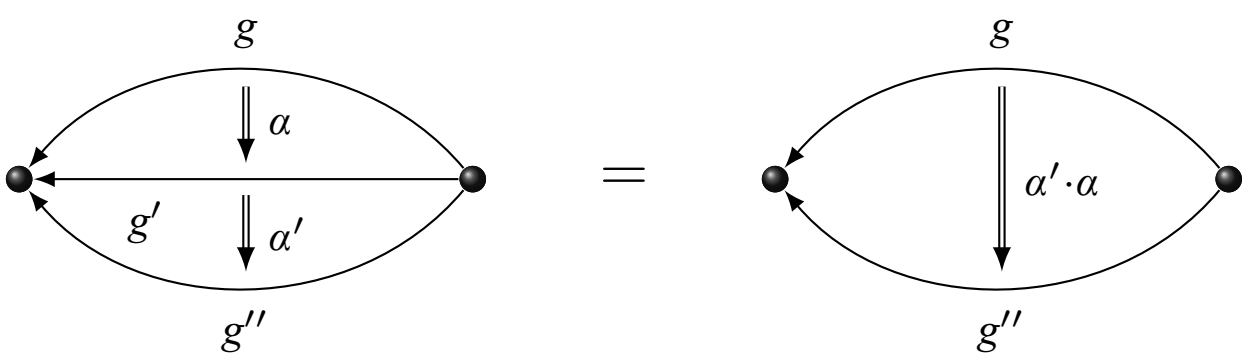

(6.24)

vertical composition is associative with identity and inverses. However the 2-morphism do not form a group under vertical composition.

Two-groups were originally introduced using the concept of Lie crossed module $(G, H, \partial, \triangleright)$. This structure is composed of,

- the morphisms of $\mathcal{G}$ form a group $G$ with group operation given by the composition law (6.22);
- the group $H$ is the set of 2-morphisms whose source is the identity $1_\bullet$,

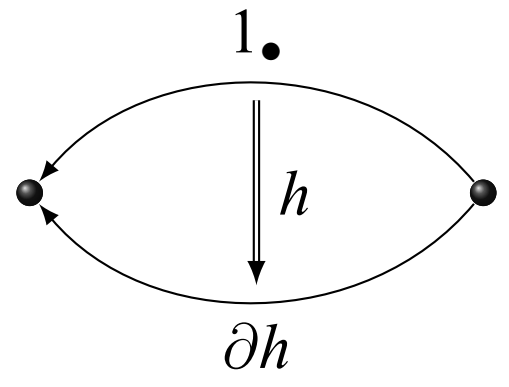

(6.25)

the group operation for $H$ is given by the following horizontal composition

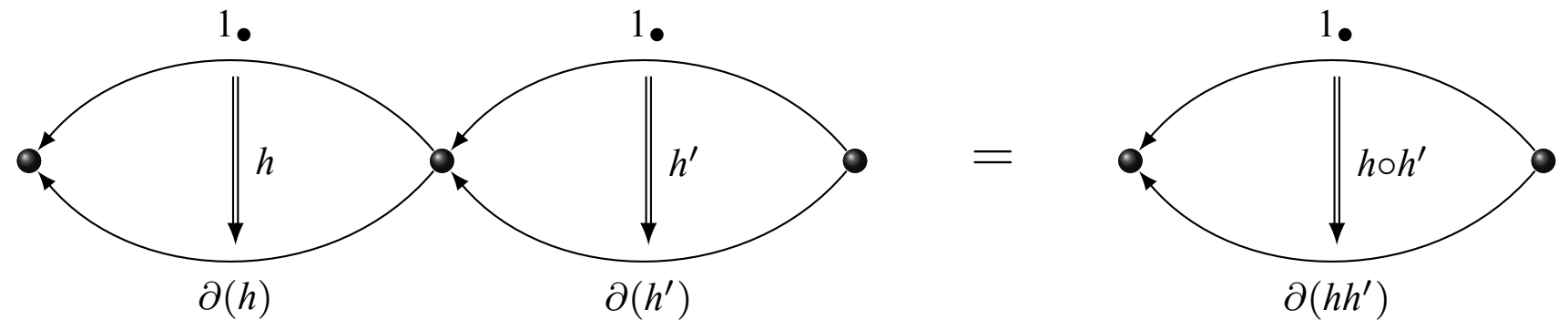

(6.26)

where we are using $hh'$ for the horizontal composition $h \circ h'$ of two elements of $H$. We will also denote the horizontal inverse of an element of $H$ by $h^{-1}$. From the definition of 2-category it follows that the map $\partial : H \longrightarrow G$ is a group homomorphism

$$\partial(hh') = \partial(h)\partial(h') \,. \tag{6.27}$$

That is this is a group homomorphism $\partial$ sending each 2-morphism in $H$ to its target.

- An action $\triangleright$ of $G$ on $H$ given by

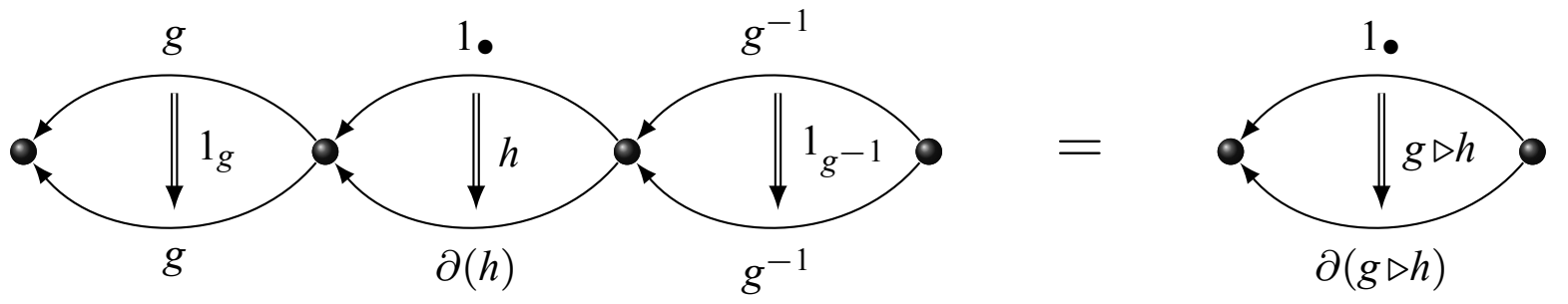

(6.28)

  this defines an 2-morphism in $H$ which we call $g \triangleright h$. For a given $g \in G$ the map $\triangleright$ is an automorphism of $H$, i.e. a one-to-one and onto function with,

$$g \triangleright (hh') = (g \triangleright h)(g \triangleright h') \,, \tag{6.29}$$

  note that we are omitting $\circ$ like we said before.

  The automorphisms of $H$ form a group called $Aut(H)$, sot that $\triangleright$ is a group homomorphism $\triangleright : G \longrightarrow Aut(H)$, which means,

$$(gg') \triangleright (h) = (g\triangleright)(g' \triangleright h) \,. \tag{6.30}$$

- We have also that $\partial$ is $G$-equivariant

$$\partial(g \triangleright h) = g\partial(h)g^{-1} \,, \quad \text{for all } g \in G \,, \quad h \in H \,. \tag{6.31}$$

- The Peiffer identity holds,

$$\partial(h) \triangleright h' = hh'h^{-1} \,, \quad \text{for all } h \,, h' \in H \,. \tag{6.32}$$

For the diagrams corresponding to these last two properties we refer the reader to [106]. We also note that the inverse path we took is also possible. We can recover a 2-group $\mathcal{G}$ form a Lie crossed module $(G, H, \partial, \triangleright)$, this way crossed modules are just a way of looking at 2-groups (see [106]).

A 2-functor is a map between categories that preserves all properties and composition laws. Given a pair of 2-categories $C$ and $D$, a 2-functor $F : C \longrightarrow D$ consists of:

- A map $F$ sending objects in $C$ to objects in $D$;

- another map $F$ sending morphisms in $C$ to morphisms in $D$;
- yet another map $F$ sending 2-morphisms in $C$ to 2-morphisms in $D$.
- Given a morphism $f : x \longrightarrow y$ in $C$ we have $F(f) : F(x) \longrightarrow F(y)$;
- the function $F$ preserves composition for morphisms and identity morphism, that is,

$$F(fg) = F(f)F(g)\,, \tag{6.33}$$
$$F(1_x) = 1_{F(x)}\,. \tag{6.34}$$

- Given a 2-morphism $\alpha : f \Longrightarrow g$ we have $F(\alpha) : F(f) \Longrightarrow F(g)$,
- $F$ preserves vertical and horizontal composition for 2-morphisms and identity 2-morphisms:

$$F(\alpha \cdot \beta) = F(\alpha) \cdot F(\beta)\,, \tag{6.35}$$
$$F(\alpha \circ \beta) = F(\alpha) \circ F(\beta)\,, \tag{6.36}$$
$$F(1_f) = 1_{F(f)}\,. \tag{6.37}$$

A 2-connection can (like in the case of a connection) be viewed as a 2-functor from a 2-category called the path 2-goupoid to a 2-group $\mathcal{G}$ (refer to [106] for details).

We will, however, take a more operative point of view and define a connection in the following way. Let $(G, H, \partial, \triangleright)$ be a Lie crossed module with $\partial : H \to G$ and $\triangleright : G \to Aut(H)$ group homomorphisms given in (6.27) and (6.30) and denoting with the same letter the corresponding homomorphism of the algebras. Given a smooth manifold $M$, a 2-connection can be defined as a pair of forms $(A, \beta)$ such that $A$ is a $\mathfrak{g}$-valued 1-form and $\beta$ is a $\mathfrak{h}$-valued 2-form (i.e. $A \in \Omega^1 \otimes \mathfrak{g}$ and $\beta \in \Omega^2 \otimes \mathfrak{h}$ ). One can associate to $(A, \beta)$ a curvature 2-form $\mathcal{F}$ and a curvature 3-form $\mathcal{G}$ as

$$\mathcal{F}_{(A,\beta)} = dA + A \wedge A - \partial\beta\,, \qquad \mathcal{G}_{(A,\beta)} = d\beta + A \wedge^{\triangleright} \beta\,. \tag{6.38}$$

Note that $\mathrm{F}_A = dA + A \wedge A$ is the curvature on the principal bundle $P_G(M)$ and $(A, \beta)$ is a 2-connection on a 2-bundle associated to the 2-group $(G, H)$.

There are two types of gauge transformations for a 2-connection $(A, \beta)$. Given a smooth map $\varphi\colon M \to G$, we can define maps

$$A \mapsto \varphi^{-1} A \varphi + \varphi^{-1} d\varphi \,, \qquad \beta \mapsto \varphi^{-1} \triangleright \beta \,, \tag{6.39}$$

which will be called a thin gauge transformation.

Similarly, given a 1-form $\eta$ on $M$ with values in $\mathfrak{h}$, we can define maps

$$A \mapsto A + \partial\eta \,, \qquad \beta \mapsto \beta + d\eta + A \wedge^{\triangleright} \eta + \eta \wedge \eta \,, \tag{6.40}$$

which will be called a fat gauge transformation.

The curvature F, the fake curvature $\mathcal{F}$ and the 3-form curvature $\mathcal{G}$ transform under a thin gauge transformation as

$$\mathrm{F}_A \mapsto \varphi^{-1} \mathrm{F}_A \varphi \ , \qquad \mathcal{F}_{(A,\beta)} \mapsto \varphi^{-1} \mathcal{F}_{(A,\beta)} \varphi \,, \qquad \mathcal{G}_{(A,\beta)} \mapsto \varphi^{-1} \triangleright \mathcal{G}_{(A,\beta)} \,, \tag{6.41}$$

while under a fat gauge transformation, they transform as

$$\mathrm{F}_A \mapsto \mathrm{F}_A + \partial \left( d\eta + A \wedge^{\triangleright} \eta + \eta \wedge \eta \right) \ , \tag{6.42}$$

$$\mathcal{F}_{(A,\beta)} \mapsto \mathcal{F}_{(A,\beta)} \,, \qquad \mathcal{G}_{(A,\beta)} \mapsto \mathcal{G}_{(A,\beta)} + \mathcal{F}_{(A,\beta)} \wedge^{\triangleright} \eta \,. \tag{6.43}$$

## 6.3 The *BFCG* theory

In the same way we can generalize the concepts of category and group, to those of respectively 2-category and 2-group, the *BF* theory can be *categorically generalized* to a 2-*BF* theory (see e.g. [108, 109]). This is part of a generalization of gauge theories called *Higher Gauge Theories* (for details refer to [110], see also [111] for a generalization of 2-form electrodynamics).

Specifically one can construct a topological theory of flat 2-connections by generalizing the *BF* action (2.146) to the Lie 2-group case. In the case when the homomorphism $\partial$ is trivial, the corresponding action was constructed in [108], while the action for the general case was constructed in [109]. The action for the *BFCG* theory is given by

$$S = \int_M \langle B \wedge \mathcal{F}_{(A,\beta)} \rangle_{\mathfrak{g}} + \int_M \langle C \wedge \mathcal{G}_{(A,\beta)} \rangle_{\mathfrak{h}} \,, \tag{6.44}$$

where $B$ is a 2-form taking values in $\mathfrak{g}$ (the Lie algebra of $G$) and $C$ is a 1-form taking values in $\mathfrak{h}$ (the Lie algebra of $H$). Also in (6.44) $\langle , \rangle_{\mathfrak{g}}$ and $\langle , \rangle_{\mathfrak{h}}$ are $G$-invariant, bilinear, non-degenerate and symmetric forms in the corresponding Lie algebras.

The *BFCG* action (6.44) will be invariant under a thin gauge transformation if

$$C \to \varphi^{-1} \triangleright C\,, \qquad B \to \varphi^{-1} B \varphi\,, \tag{6.45}$$

while the invariance under a fat gauge transformation is achieved if the fields $B$ and $C$ transform as

$$B \mapsto B + C \wedge^{\mathcal{T}} \eta\,, \qquad C \mapsto C\,. \tag{6.46}$$

The antisymmetric map $\mathcal{T}\colon \mathfrak{h} \times \mathfrak{h} \to \mathfrak{g}$ is defined as

$$\langle \mathcal{T}(u,v), Z\rangle_{\mathfrak{g}} = -\langle u, Z \triangleright v\rangle_{\mathfrak{h}}\,, \qquad u,v \in \mathfrak{h}\,,\ Z \in \mathfrak{g}\,. \tag{6.47}$$

Also note that $C\wedge^{\mathcal{T}}\eta$ is the antisymmetrization of $\mathcal{T}(C,\eta)$, see [109] for details.

The *BFCG* theory is the main concern of this thesis, the case where we use the Poincaré 2-group has applications to quantum gravity.

# 7

# Hamiltonian analysis of the *BFCG* theory for a generic Lie 2-group

In this chapter, following [112] we perform the canonical analysis of the *BFCG* action (6.44) for a generic Lie 2-group. The structure of this chapter is the following: in section 7.1 we write the BFCG action (6.44) as well as various other identities in component form and give a brief overview of the Lagrange equations of motion. In section 7.2 we preform a gauge fixed canonical analysis of this theory using a theorem proved in section 3.11. In section 7.3 we generalize the analysis of the previous section to a full Hamiltonian analysis using the Dirac procedure of chapter 3. Finally in section 7.4 we count the *local* degrees of freedom of the theory, and find it to be — under the assumptions made here, namely the independence of the constraints — a theory without degrees of freedom *per point*, i.e. a *topological field theory*.

## 7.1 Spacetime components of the *BFCG* action

To carry out the canonical analysis of (6.44), we need to write it in terms of the spacetime components of the relevant fields.

Let $T_a$ be a basis in $\mathfrak{g}$ and $\tau_\alpha$ a basis in $\mathfrak{h}$. The structure constants are defined by

$$[T_a, T_b] = f^c{}_{ab}\, T_c\,, \qquad [\tau_\alpha, \tau_\beta] = \varphi^\gamma{}_{\alpha\beta}\, \tau_\gamma\,. \tag{7.1}$$

The homomorphisms $\partial$ and $\triangleright$ (defined in equations (6.27) and (6.29) respectively) then act as

$$\partial\tau_\alpha = \partial_\alpha{}^a\, T_a\,, \qquad T_a \triangleright \tau_\alpha = \triangleright^\beta{}_{a\alpha}\,\tau_\beta\,, \tag{7.2}$$

and satisfy the following relations,

$$\triangleright^\beta{}_{a\alpha}\,\partial_\beta{}^b = \partial_\alpha{}^c f^b{}_{ac}\,, \qquad \partial_\alpha{}^a\ \triangleright^\gamma{}_{a\beta} = \varphi^\gamma{}_{\alpha\beta}\,. \tag{7.3}$$

Also, the following relation

$$f^a{}_{bc} \triangleright_{\alpha a\beta} = \triangleright_{\alpha[b|\gamma}\triangleright^\gamma{}_{|c]\beta}\,, \tag{7.4}$$

will be useful, where $X_{[bc]} = X_{bc} - X_{cb}$.

The structure constants satisfy the Jacobi identities

$$f^d{}_{ac} f^c{}_{be} = f^c{}_{a[b|} f^d{}_{c|e]}\,, \qquad \varphi^\gamma{}_{\alpha\delta}\,\varphi^\delta{}_{\beta\varepsilon} = \varphi^\delta{}_{\alpha[\beta|}\,\varphi^\gamma{}_{\delta|\varepsilon]}\,. \tag{7.5}$$

We have also

$$\langle X \wedge Y\rangle_{\mathfrak{g}} = X^a \wedge Y^b \langle T_a, T_b\rangle_{\mathfrak{g}} = X^a \wedge Y^b Q_{ab}\,, \tag{7.6}$$

and

$$\langle U \wedge V\rangle_{\mathfrak{h}} = U^\alpha \wedge V^\beta \langle \tau_\alpha, \tau_\beta\rangle_{\mathfrak{h}} = U^\alpha \wedge V^\beta q_{\alpha\beta}\,. \tag{7.7}$$

recall that $\langle,\rangle_{\mathfrak{g}}$ and $\langle,\rangle_{\mathfrak{h}}$ are $G$-invariant, bilinear, non-degenerate and symmetric forms in the corresponding Lie algebras. Therefore $Q$ and $q$ are matrices associated to these forms.

The fake curvature, defined in equation (6.38) can be written as

$$\mathcal{F}_{(A,\beta)} = \frac{1}{2}\mathcal{F}^b{}_{\mu\nu} T_b\, dx^\mu \wedge dx^\nu \tag{7.8}$$

where

$$\mathcal{F}^b{}_{\mu\nu} = \partial_\mu A^b{}_\nu - \partial_\nu A^b{}_\mu + f^b{}_{cd} A^c{}_\mu A^d{}_\nu - \partial_\alpha{}^b \beta^\alpha{}_{\mu\nu}\ . \tag{7.9}$$

The curvature 3-form (equation (6.38) on the right) can be written as

$$\mathcal{G}_{(A,\beta)} = \frac{1}{6}\mathcal{G}^\alpha{}_{\mu\nu\rho}\tau_\alpha dx^\mu \wedge dx^\nu \wedge dx^\rho \tag{7.10}$$

where,

$$\mathcal{G}^\alpha{}_{\mu\nu\rho} = \partial_{[\mu}\beta^\alpha{}_{\nu\rho]} + A^a{}_{[\mu}\beta^\gamma{}_{\nu\rho]} \triangleright^\alpha{}_{a\gamma}\,. \tag{7.11}$$

$X_{[\mu\nu\rho]}$ denotes a total antisymmetrization of indices, given by

$$\sum_{p\in S_3} (-1)^p X_{p(\mu\nu\rho)}\,. \tag{7.12}$$

where $p$ is a permutation and $(-1)^p$ is the parity of $p$ (cf. (2.20) with no factor in the denominator).

The *BFCG* action then becomes

$$S = \int_M d^4x\, \varepsilon^{\mu\nu\rho\sigma} \left( \frac{1}{4} B^a{}_{\mu\nu}\, \mathcal{F}^b{}_{\rho\sigma}\, Q_{ab} + \frac{1}{6}\, C^\alpha{}_\mu\, \mathcal{G}^\beta{}_{\nu\rho\sigma}\, q_{\alpha\beta} \right) . \tag{7.13}$$

To simplify notation we use $Q$ and $q$ to lower $\mathfrak{g}$ and $\mathfrak{h}$ Lie algebra indices respectively, for example,

$$B_b = B^a Q_{ab} \qquad \beta_\beta = \beta^\alpha q_{\alpha\beta} . \tag{7.14}$$

We also use the same symbol for $\partial_\alpha{}^a$ and $\partial^\alpha{}_a = q^{\alpha\beta}\partial_\beta{}^b Q_{ba}$ and $\triangleright_{a\alpha\gamma} = \triangleright^\beta{}_{a\gamma} q_{\beta\alpha}$. This last quantity (that was in fact already defined in (6.47)) is antisymmetric in $\alpha\gamma$, that is $\triangleright_{a\alpha\gamma} = -\triangleright_{\gamma a\alpha}$, and we have as a consequence

$$\begin{aligned} C^\alpha \nabla^\triangleright_\mu \beta_\alpha &= C^\alpha \left( \partial_\mu \beta_\alpha + A^a{}_\mu \triangleright_{a\alpha\gamma} \beta^\gamma \right) \\ &= -\left( \partial_\mu C_\alpha + A^a{}_\mu \triangleright_{a\alpha\gamma} C^\gamma \right) \beta^\alpha + \partial_\mu \left( C^\alpha \beta_\alpha \right) \\ &= -\nabla^\triangleright_\mu \left( C^\alpha \right) \beta_\alpha + \partial_\mu \left( C^\alpha \beta_\alpha \right) , \end{aligned} \tag{7.15}$$

where $\nabla^\triangleright_\mu$ is defined as the quantity in parenthesis.

The equations of motion are obtained by equating to zero the variational derivatives of the action with respect to all fields. The variational derivatives with respect to $B$ and $C$ give respectively,

$$\mathcal{F}^b{}_{\mu\nu} = 0 , \qquad \mathcal{G}^\alpha{}_{\mu\nu\rho} = 0 . \tag{7.16}$$

While the variational derivatives with respect to $A$ and $\beta$ give respectively,

$$\varepsilon^{\mu\nu\rho\sigma} \left( \nabla_\mu B_{a\nu\rho} + \beta^\alpha{}_{\mu\nu} \triangleright_{\alpha a\beta} C^\beta{}_\rho \right) = 0 , \quad \varepsilon^{\mu\nu\rho\sigma} \left( \nabla^\triangleright_\mu C^\alpha{}_\nu - \frac{1}{2} \partial^\alpha{}_a B^a{}_{\mu\nu} \right) = 0 . \tag{7.17}$$

We will also use the Bianchi identities (BI) associated to the 1-form fields $A$ and $C$. Namely, the corresponding 2-form curvatures

$$F^a = dA^a + f^a{}_{bc} A^b \wedge A^c , \qquad T^\alpha = dC^\alpha + \triangleright^\alpha{}_{a\beta} A^a \wedge C^\beta , \tag{7.18}$$

satisfy the following Bianchi identities

$$\varepsilon^{\lambda\mu\nu\rho}\, \nabla_\mu F^a{}_{\nu\rho} = 0 , \tag{7.19}$$

and

$$\varepsilon^{\lambda\mu\nu\rho} \left( \nabla^\triangleright_\mu T^\alpha{}_{\nu\rho} - \triangleright^\alpha{}_{a\beta} F^a{}_{\mu\nu} C^\beta{}_\rho \right) = 0 . \tag{7.20}$$

There are also the BI associated with the 2-form fields $B$ and $\beta$. The corresponding 3-form curvatures are given by

$$G^a = dB^a + f^a{}_{bc} A^b \wedge B^c \,, \qquad \mathcal{G}^\alpha = d\beta^\alpha + \triangleright^\alpha{}_{a\gamma} A^a \wedge \beta^\gamma \,, \tag{7.21}$$

so that

$$\varepsilon^{\lambda\mu\nu\rho} \left( \frac{2}{3} \nabla_\lambda G^a{}_{\mu\nu\rho} - f^a{}_{bc} F^b{}_{\lambda\mu} B^c{}_{\nu\rho} \right) = 0 \tag{7.22}$$

and

$$\varepsilon^{\lambda\mu\nu\rho} \left( \frac{2}{3} \nabla^{\triangleright}_\lambda \mathcal{G}^\alpha{}_{\mu\nu\rho} - \triangleright^\alpha{}_{a\gamma} F^a{}_{\lambda\mu} \beta^\gamma{}_{\nu\rho} \right) = 0 \,. \tag{7.23}$$

## 7.2 A gauge-fixed canonical analysis

We will assume (as was done in chapter 4) that $M = \Sigma \times \mathbb{R}$ and that $t$ is a coordinate on $\mathbb{R}$ while $\{x^i | i = 1, 2, 3\}$ is a local coordinate chart on $\Sigma$. We can split the *BFCG* fields into temporal and spatial components by using

$$x^\mu = (x^0, x^i) = (t, \vec{x}) \tag{7.24}$$

and $U_\mu = (U_0, U_i)$. For example

$$\partial_\mu U_\nu = (\partial_0 U_0, \partial_0 U_i, \partial_i U_0, \partial_i U_j) \tag{7.25}$$

and

$$\begin{aligned} &\varepsilon^{\mu\nu\rho\sigma} \partial_\mu U_\nu \partial_\rho V_\sigma = \\ &= \varepsilon^{0ijk} \partial_0 U_i \partial_j V_k + \varepsilon^{i0jk} \partial_i U_0 \partial_j V_k + \varepsilon^{ij0k} \partial_i U_j \partial_0 V_k + \varepsilon^{ijk0} \partial_i U_j \partial_k V_0 \\ &= \varepsilon^{ijk} \left( \dot{U}_i \partial_j V_k - \partial_i U_0 \partial_j V_k + \partial_i U_j \dot{V}_k - \partial_i U_j \partial_k V_0 \right) \,, \end{aligned} \tag{7.26}$$

where $\dot{X} = \partial_0 X$ and throughout the rest of this chapter, $\varepsilon^{ijk} \equiv \varepsilon^{0ijk}$.

The *BFCG* action can be then written as

$$S = \int_{t_1}^{t_2} dt \, L(t) \,, \tag{7.27}$$

where the Lagrangian $L$ is given by

$$L = \int_\Sigma d^3x \left[ \pi(A)_a{}^i \dot{A}^a{}_i + \frac{1}{2} \pi(\beta)_\alpha{}^{ij} \dot{\beta}^\alpha{}_{ij} \right] - H \tag{7.28}$$

and

$$
\begin{aligned}
H &= -\int_\Sigma d^3x \left[\frac{1}{2}\varepsilon^{ijk}B_{a\,0i}\mathcal{S}(\mathcal{F})^a{}_{jk} + C_{\alpha 0}\mathcal{S}(\mathcal{G})^\alpha\right.\\
&+ A^a{}_0\mathcal{S}(CB\beta)_a + \beta^\alpha{}_{k0}\mathcal{S}(CA)_\alpha{}^k\\
&+ \left.\partial_i\left(\pi(A)_a{}^iA^a{}_0 - \pi(\beta)_\alpha{}^{ij}\beta^\alpha{}_{j0}\right)\right] . \qquad (7.29)
\end{aligned}
$$

The fields $\pi(A)$ and $\pi(\beta)$ (which are the conjugate momenta to $A$ and $\beta$ respectively) are given by

$$
\pi(A)_a{}^i = \frac{1}{2}\,\varepsilon^{ijk}B_{a\,jk}\,, \qquad \pi(\beta)_\alpha{}^{ij} = -\,\varepsilon^{ijk}C_{\alpha\,k}\,, \qquad (7.30)
$$

while

$$
\begin{aligned}
\mathcal{S}(\mathcal{F})^a{}_{ij} &\equiv \mathcal{F}^a{}_{ij}\,,\\
\mathcal{S}(\mathcal{G})^\alpha &\equiv \frac{1}{6}\varepsilon^{ijk}\mathcal{G}^\alpha{}_{ijk}\,,\\
\mathcal{S}(BC\beta)_a &\equiv \nabla_k\pi(A)_a{}^k - \frac{1}{2}\pi(\beta)_\alpha{}^{jk}\rhd^\alpha{}_{a\beta}\beta^\beta{}_{jk}\,,\\
\mathcal{S}(CB)_\alpha{}^k &\equiv \frac{1}{2}\nabla^{\rhd}_j\pi(\beta)_\alpha{}^{jk} + \partial_\alpha{}^a\pi(A)_a{}^k\,.
\end{aligned} \qquad (7.31)
$$

From these equations we see that the *BFCG* Lagrangian has the form

$$
L = \sum_m P_m\dot{Q}_m - H = \sum_m P_m\dot{Q}_m - \sum_n \lambda_n G_n(P,Q)\,. \qquad (7.32)
$$

According to the theorem proved in section 3.11, (see also [113]) such a Lagrangian is a result of the Dirac procedure in the gauge $P(\lambda_n) = 0$ if the constraints $G_n(P,Q)$ are of the first class with respect to the $(P,Q)$ Poisson bracket, i.e. form a closed algebra under the Poison bracket (3.27), which can be equivalently defined by

$$
\{A,B\} = \sum_n\left(\frac{\partial A}{\partial Q_n}\frac{\partial B}{\partial P_n} - \frac{\partial A}{\partial P_n}\frac{\partial B}{\partial Q_n}\right)\,. \qquad (7.33)
$$

It is straightforward to verify that the constraints from (7.31) are of the first class, by using the Poisson bracket (7.33). The non-zero Poisson bracket are then given by

$$
\begin{aligned}
\{\,A^a{}_i(x)\,,\,\pi(A)_b{}^j(y)\,\} &= \delta^a_b\delta^j_i\delta^{(3)}(x-y)\,,\\
\{\,\beta^\alpha{}_{ij}(x)\,,\,\pi(\beta)_\beta{}^{kl}(y)\,\} &= \delta^\alpha_\beta\delta^k_{[i}\delta^l_{j]}\delta^{(3)}(x-y)\,,
\end{aligned} \qquad (7.34)
$$

where $x = \vec{x}$, $y = \vec{y}$ and $\delta^{(3)}(x-y)$ is the three-dimensional Dirac delta function.

The Poisson-bracket algebra for the constraints from (7.31) is then given by

$$\begin{array}{rcl} \{\, \mathcal{S}(\mathcal{F})^a{}_{ij}(x)\,,\, \mathcal{S}(BC\beta)_b(y)\,\} &=& 2f^a{}_{bc}\mathcal{S}(\mathcal{F})^c{}_{ij}(x)\,\delta^{(3)}(x-y)\,, \\ \{\, \mathcal{S}(\mathcal{G})^\alpha(x)\,,\, \mathcal{S}(CB)_\beta{}^k(y)\,\} &=& \varepsilon^{ijk} \triangleright^\alpha{}_{c\beta}\mathcal{S}(\mathcal{F})^c{}_{ij}(x)\,\delta^{(3)}(x-y)\,, \\ \{\, \mathcal{S}(BC\beta)_a(x)\,,\, \mathcal{S}(BC\beta)_b(y)\,\} &=& 2f^c{}_{ab}\mathcal{S}(BC\beta)_c(x)\,\delta^{(3)}(x-y)\,, \\ \{\, \mathcal{S}(\mathcal{G})^\alpha(x)\,,\, \mathcal{S}(BC\beta)_a(y)\,\} &=& 2\triangleright^\alpha{}_{a\beta}\mathcal{S}(\mathcal{G})^\beta(x)\,\delta^{(3)}(x-y)\,, \\ \{\, \mathcal{S}(CB)_\alpha{}^k(x)\,,\, \mathcal{S}(BC\beta)_a(y)\,\} &=& \triangleright^\beta{}_{a\alpha}\mathcal{S}(BC)_\beta{}^k(x)\,\delta^{(3)}(x-y)\,, \end{array} \tag{7.35}$$

which confirms that they are of the first class. Hence the constraints of the action (7.13) correspond to the constraints of the Dirac analysis in the gauge

$$\pi(B^a{}_{0i}) = \pi(C^\alpha{}_0) = \pi(A^a{}_0) = \pi(\beta^\alpha{}_{0i}) = 0\,. \tag{7.36}$$

## 7.3 The complete canonical analysis

The analysis in the previous section has an implicit gauge fixing. To see this, we can perform the complete canonical analysis by using the Dirac procedure, see [70]. For this we consider the Lagrangian

$$L = \int_\Sigma d^3x\, \varepsilon^{\mu\nu\rho\sigma} \left( \frac{1}{4}\, B^a{}_{\mu\nu}\, \mathcal{F}^b{}_{\rho\sigma}\, Q_{ab} + \frac{1}{6}\, C^\alpha{}_\mu\, \mathcal{G}^\beta{}_{\nu\rho\sigma}\, q_{\alpha\beta} \right)\,, \tag{7.37}$$

and calculate the momenta (the functional derivatives of the Lagrangian with respect to the time derivatives of the variables) for all variables $B^a{}_{\mu\nu}$, $A^a{}_\mu$, $C^\alpha{}_\mu$ and $\beta^\alpha{}_{\mu\nu}$,

$$\begin{array}{rclcl} \pi(B)_a{}^{\mu\nu} &=& \dfrac{\delta L}{\delta \partial_0 B^a{}_{\mu\nu}} &=& 0\,, \\ \pi(C)_\alpha{}^\mu &=& \dfrac{\delta L}{\delta \partial_0 C^\alpha{}_\mu} &=& 0\,, \\ \pi(A)_a{}^\mu &=& \dfrac{\delta L}{\delta \partial_0 A^a{}_\mu} &=& \dfrac{1}{2}\varepsilon^{0\mu\nu\rho}B_{a\nu\rho}\,, \\ \pi(\beta)_\alpha{}^{\mu\nu} &=& \dfrac{\delta L}{\delta \partial_0 \beta^\alpha{}_{\mu\nu}} &=& -\varepsilon^{0\mu\nu\rho}C_{\alpha\rho}\,. \end{array} \tag{7.38}$$

All of these momenta give rise to primary constraints since none of them can be inverted for the time derivatives of the variables,

$$\begin{array}{rcl} P(B)_a{}^{\mu\nu} &\equiv& \pi(B)_a{}^{\mu\nu} \approx 0\,, \\ P(C)_\alpha{}^\mu &\equiv& \pi(C)_\alpha{}^\mu \approx 0\,, \\ P(A)_a{}^\mu &\equiv& \pi(A)_a{}^\mu - \dfrac{1}{2}\varepsilon^{0\mu\nu\rho}B_{a\nu\rho} \approx 0\,, \\ P(\beta)_\alpha{}^{\mu\nu} &\equiv& \pi(\beta)_\alpha{}^{\mu\nu} + \varepsilon^{0\mu\nu\rho}C_{\alpha\rho} \approx 0\,. \end{array} \tag{7.39}$$

We use the weak equality "≈" for the equality that holds on a subspace of the phase space, while the equality that holds for any point of the phase space will be called "strong" and it is denoted by the usual symbol "=" (see section 3.5). We will also use the expressions "on-shell" and "off-shell" for strong and weak equalities, respectively.

We will use the following fundamental Poisson brackets

$$\begin{array}{lcl} \{ B^a{}_{\mu\nu}(x) \,,\, \pi(B)_b{}^{\rho\sigma}(y) \} &=& \delta^a_b \delta^\rho_{[\mu} \delta^\sigma_{\nu]} \, \delta^{(3)}(x-y) \,, \\ \{ C^\alpha{}_\mu(x) \,,\, \pi(C)_\beta{}^\nu(y) \} &=& \delta^\alpha_\beta \delta^\nu_\mu \, \delta^{(3)}(x-y) \,, \\ \{ A^a{}_\mu(x) \,,\, \pi(A)_b{}^\nu(y) \} &=& \delta^a_b \delta^\nu_\mu \, \delta^{(3)}(x-y) \,, \\ \{ \beta^\alpha{}_{\mu\nu}(x) \,,\, \pi(\beta)_\beta{}^{\rho\sigma}(y) \} &=& \delta^\alpha_\beta \, \delta^\rho_{[\mu} \delta^\sigma_{\nu]} \, \delta^{(3)}(x-y) \,, \end{array} \tag{7.40}$$

to calculate the algebra between the primary constraints. We obtain

$$\begin{array}{lcl} \{ P(B)_a{}^{jk}(x) \,,\, P(A)_b{}^i(y) \} &=& \varepsilon^{0ijk} \, Q_{ab}(x) \, \delta^{(3)}(x-y) \,, \\ \{ P(C)_\alpha{}^k(x) \,,\, P(\beta)_b{}^{ij}(y) \} &=& -\varepsilon^{0ijk} \, q_{ab}(x) \, \delta^{(3)}(x-y) \,, \end{array} \tag{7.41}$$

while all other Poisson brackets vanish.

The canonical on-shell Hamiltonian is defined by

$$\begin{array}{rcl} H_c &=& \displaystyle\int_\Sigma d^3x \left[ \frac{1}{2} \pi(B)_a{}^{\mu\nu} \partial_0 B^a{}_{\mu\nu} + \pi(C)_\alpha{}^\mu \partial_0 C^\alpha{}_\mu + \right. \\ && \left. + \pi(A)_a{}^\mu \partial_0 A^a{}_\mu + \displaystyle\frac{1}{2} \pi(\beta)_\alpha{}^{\mu\nu} \partial_0 \beta^\alpha{}_{\mu\nu} \right] - L \,. \end{array} \tag{7.42}$$

By using (7.37), (7.9) and (7.11) we can rewrite the Hamiltonian (7.42) in such a way that all the velocities are multiplied by the first class constraints. Therefore in an on-shell quantity they drop out, so that

$$\begin{array}{rcl} H_c &=& - \displaystyle\int d^3x \, \varepsilon^{0ijk} \left[ \frac{1}{2} B_{a0i} \mathcal{F}^a{}_{jk} + \frac{1}{6} C_{\alpha 0} \mathcal{G}^\alpha{}_{ijk} + \right. \\ && + \beta^\alpha{}_{0k} \left( \nabla^{\triangleright}_i C_{\alpha j} - \displaystyle\frac{1}{2} \partial_\alpha{}^a B_{a\,ij} \right) + \\ && \left. + \displaystyle\frac{1}{2} A^a{}_0 \left( \nabla_i B_{a\,jk} - C^\alpha{}_i \triangleright_{\alpha a \beta} \beta^\beta{}_{jk} \right) \right] \,. \end{array} \tag{7.43}$$

This expression does not depend on any of the canonical momenta and it contains only the fields and their spatial derivatives. By adding a Lagrange multiplier $\lambda$ for each of the primary constraints we can build the off-shell Hamiltonian, which is given by

$$
\begin{aligned}
H_T \;=\; H_c + \int d^3x \Big[ & \lambda(C)^\alpha{}_\mu P(C)_\alpha{}^\mu + \lambda(A)^a{}_\mu P(A)_a{}^\mu \\
& + \frac{1}{2}\lambda(B)^a{}_{\mu\nu} P(B)_a{}^{\mu\nu} + \frac{1}{2}\lambda(\beta)^\alpha{}_{\mu\nu} P(\beta)_\alpha{}^{\mu\nu} \Big] \,.
\end{aligned} \tag{7.44}
$$

Since the primary constraints must be preserved in time, we must impose the following requirement

$$
\dot{P} \equiv \{ P \,,\, H_T \} \approx 0 \,, \tag{7.45}
$$

for each primary constraint $P$. By using the consistency condition (7.45) for the primary constraints $P(B)_a{}^{0i}$, $P(C)_\alpha{}^0$, $P(\beta)_\alpha{}^{0i}$ and $P(A)_a{}^0$ we obtain the secondary constraints $\mathcal{S}$

$$
\begin{aligned}
\mathcal{S}(\mathcal{F})^a{}_{jk} &\equiv \mathcal{F}^a{}_{jk} \approx 0, \\
\mathcal{S}(\mathcal{G})^\alpha &\equiv \frac{1}{6}\varepsilon^{0ijk}\mathcal{G}^\alpha{}_{ijk} \approx 0, \\
\mathcal{S}(CB)_{\alpha ij} &\equiv \nabla^{\triangleright}_{[i|} C_{\alpha|j]} - \partial_\alpha{}^a B_{a\,ij} \approx 0, \\
\mathcal{S}(BC\beta)_a &\equiv \frac{1}{2}\varepsilon^{0ijk}\left( \nabla_i B_{ajk} - C^\alpha{}_i \triangleright_{\alpha a \beta} \beta^\beta{}_{jk} \right) \approx 0 \,.
\end{aligned} \tag{7.46}
$$

In the case of $P(B)_a{}^{jk}$, $P(C)_\alpha{}^k$, $P(\beta)_\alpha{}^{jk}$ and $P(A)_a{}^k$ the consistency condition determines the following Lagrange multipliers

$$
\begin{aligned}
\lambda(A)^a{}_i &\approx \nabla_i A^a{}_0 - \partial_\alpha{}^a \beta^\alpha{}_{i0} \,, \\
\lambda(\beta)^\alpha{}_{ij} &\approx \nabla^{\triangleright}_{[i|}\beta^\alpha{}_{0|j]} - \beta^\beta{}_{ij} \triangleright^\alpha{}_{a\beta} A^a{}_0 \,, \\
\lambda(C)^\alpha{}_i &\approx \nabla^{\triangleright}_i C^\alpha{}_0 + C^\beta{}_i \triangleright_{\beta a}{}^\alpha A^a{}_0 \,, \\
\lambda(B)^a{}_{ij} &\approx \nabla_{[i|} B^a{}_{0|j]} - C^\alpha{}_0 \triangleright_\alpha{}^a{}_\gamma \beta^\gamma{}_{ij} + f^a{}_{bc} A^b{}_0 B^c{}_{ij} - C^\alpha{}_{[i|} \triangleright_\alpha{}^a{}_\gamma \beta^\gamma{}_{0|j]} \,.
\end{aligned} \tag{7.47}
$$

The consistency conditions of the secondary constraints (7.46) turn out to be identically satisfied, and produce no new constraints. Note that the consistency conditions leave the Lagrange multipliers

$$
\lambda(A)^a{}_0 \,, \qquad \lambda(\beta)^\alpha{}_{0i} \,, \qquad \lambda(C)^\alpha{}_0 \,, \qquad \lambda(B)^a{}_{0i} \,, \tag{7.48}
$$

undetermined.

By using (7.47), the total Hamiltonian can be written as

$$
\begin{aligned}
H_T \;=\; \int_\Sigma d^3x \, \big[ & \lambda(B)^a{}_{0i}\, \varphi(B)_a{}^i + \lambda(C)^\alpha{}_0\, \varphi(C)_\alpha + \lambda(\beta)^\alpha{}_i\, \varphi(\beta)_\alpha{}^i \\
& + \lambda(A)^a\, \varphi(A)_a - B_{a0i}\, \varphi(\mathcal{F})^{ai} - C_{\alpha 0}\, \varphi(\mathcal{G})^\alpha \\
& - \beta_{\alpha 0i}\, \varphi(CB)^{\alpha i} - A_{a0}\, \varphi(BC\beta)^a \big] \,,
\end{aligned}
\tag{7.49}
$$

where

$$
\begin{aligned}
\varphi(B)_a{}^i &= P(B)_a{}^{0i}\,, \\
\varphi(C)_\alpha &= P(C)_\alpha{}^0\,, \\
\varphi(\beta)_\alpha{}^i &= P(\beta)_\alpha{}^{0i}\,, \\
\varphi(A)_{ab} &= P(A)_a{}^0\,, \\
\varphi(\mathcal{F})^{ai} &= \frac{1}{2}\varepsilon^{0ijk}\mathcal{S}(\mathcal{F})^a{}_{jk} - \nabla_j P(B)^{aij}\,, \\
\varphi(\mathcal{G})^\alpha &= \mathcal{S}(\mathcal{G})^\alpha + \nabla^{\triangleright}_i P(C)^{\alpha i} - \frac{1}{2}\beta_{\beta ij} \triangleright^\beta{}_a{}^\alpha P(B)^{aij}\,, \\
\varphi(CB)^{\alpha i} &= \frac{1}{2}\varepsilon^{0ijk}\mathcal{S}(CB)^\alpha{}_{jk} - \nabla^{\triangleright}_j P(\beta)^{\alpha ij} - C_{\beta j} \triangleright^\beta{}_a{}^\alpha P(B)^{aij} + \partial^\alpha{}_a P(A)^{ai}\,, \\
\varphi(BC\beta)^a &= \mathcal{S}(BC\beta)^a + \nabla_i P(A)^{ai} - \frac{1}{2} f^a{}_{bc} B^b{}_{ij} P(B)^{cij} \\
&\quad - C^\alpha{}_i \triangleright_\alpha{}^a{}_\beta P(e)^{\beta i} - \frac{1}{2}\beta^\alpha{}_{ij} \triangleright_\alpha{}^a{}_\beta P(\beta)^{\beta ij}\,,
\end{aligned}
\tag{7.50}
$$

are the first-class constraints, while

$$
\begin{aligned}
\chi(B)_a{}^{jk} = P(B)_a{}^{jk}\,, \qquad & \chi(C)_\alpha{}^i = P(C)_\alpha{}^i\,, \\
\chi(A)_a{}^i = P(A)_a{}^i\,, \qquad & \chi(\beta)_\alpha{}^{ij} = P(\beta)_\alpha{}^{ij}\,.
\end{aligned}
\tag{7.51}
$$

are the second-class constraints.

The Poisson bracket algebra of the first-class constraints is given by

$$
\begin{aligned}
\{\, \varphi(\mathcal{G})^\alpha(x)\,,\, \varphi(CB)^{\beta i}(y)\,\} &= \triangleright^\alpha{}_{a\beta}\, \varphi(\mathcal{F})^{ai}(x)\, \delta^{(3)}(x-y)\,, \\
\{\, \varphi(\mathcal{G})^\alpha(x)\,,\, \varphi(BC\beta)_a(y)\,\} &= 2 \triangleright^\alpha{}_{a\beta}\, \varphi(\mathcal{G})^\beta(x)\, \delta^{(3)}(x-y)\,, \\
\{\, \varphi(CB)_\alpha{}^k(x)\,,\, \varphi(BC\beta)_a(y)\,\} &= \triangleright^\beta{}_{a\alpha}\, \varphi(BC)_\beta{}^k(x)\, \delta^{(3)}(x-y)\,, \\
\{\, \varphi(\mathcal{F})^a{}_{ij}(x)\,,\, \varphi(BC\beta)_b(y)\,\} &= 2 f^a{}_{bc}\, \varphi(\mathcal{F})^c{}_{ij}(x)\, \delta^{(3)}(x-y)\,, \\
\{\, \varphi(BC\beta)_a(x)\,,\, \varphi(BC\beta)_b(y)\,\} &= 2 f^c{}_{ab}\, \varphi(BC\beta)_c(x)\, \delta^{(3)}(x-y)\,.
\end{aligned}
\tag{7.52}
$$

Whereas the Poisson bracket algebra between the first and the second-class constraints is given by

$$
\begin{aligned}
\{\,\varphi(\mathcal{F})^{ai}(x)\,,\,\chi(A)_b{}^j(y)\,\} &= -f^a{}_{bc}\,\chi(B)^{cij}(x)\,\delta^{(3)}(x-y)\,,\\
\{\,\varphi(\mathcal{G})^{\alpha}(x)\,,\,\chi(A)_a{}^i(y)\,\} &= -\rhd_{a\gamma}{}^{\alpha}\,\chi(C)^{\gamma i}(x)\,\delta^{(3)}(x-y)\,,\\
\{\,\varphi(\mathcal{G})^{\alpha}(x)\,,\,\chi(\beta)_\beta{}^{ij}(y)\,\} &= \rhd^{\alpha}{}_{a\beta}\,\chi(B)^{\gamma ij}(x)\,\delta^{(3)}(x-y)\,,\\
\{\,\varphi(CB)^{\alpha i}(x)\,,\,\chi(A)_a{}^j(y)\,\} &= -\rhd^{\alpha}{}_a{}^{\gamma}\,\chi(\beta)_\gamma{}^{ij}(x)\,\delta^{(3)}(x-y)\,,\\
\{\,\varphi(CB)^{\alpha i}(x)\,,\,\chi(C)_\beta{}^j(y)\,\} &= \rhd^{\alpha}{}_{a\beta}\,\chi(B)^{aij}(x)\delta^{(3)}(x-y)\,,\\
\{\,\varphi(BC\beta)^{a}(x)\,,\,\chi(A)_b{}^i(y)\,\} &= f^a{}_{bc}\,\chi(A)^{ci}(x)\,\delta^{(3)}(x-y)\,,\\
\{\,\varphi(BC\beta)^{a}(x)\,,\,\chi(\beta)_\alpha{}^{jk}(y)\,\} &= \rhd^{\gamma a}{}_{\alpha}\,\chi(\beta)_\gamma{}^{jk}(x)\,\delta^{(3)}(x-y)\,,\\
\{\,\varphi(BC\beta)^{a}(x)\,,\,\chi(C)_\alpha{}^i(y)\,\} &= -\rhd_{\alpha}{}^{a}{}_{\beta}\,\chi(C)^{\beta i}(x)\,\delta^{(3)}(x-y)\,,\\
\{\,\varphi(BC\beta)^{a}(x)\,,\,\chi(B)_b{}^{jk}(y)\,\} &= -f^a{}_{bc}\,\chi(B)^{cjk}(x)\,\delta^{(3)}(x-y)\,.
\end{aligned}
\tag{7.53}
$$

The elimination of the second class constraints can be achieved by using the Dirac brackets (DB). It can be shown that the DB algebra of the FC constraints is the same as the PB algebra (7.52).

Note that the constraints (7.50) and the algebra (7.52) reduce respectively to (7.31) and (7.35), if we consider the second-class constraints (7.51) as gauge-fixing conditions.

## 7.4 The physical degrees of freedom

In this section we will show[1] that the structure of the constraints implies that there are no local degrees of freedom in the *BFCG* theory for a generic Lie 2-group. In general case, if there are $N$ initial fields in the theory and there are $F$ independent first-class constraints per space point and $S$ independent second-class constraints per space point, then the number of local degrees of freedom (DOF), i.e. the number of independent field components, is given by

$$
n = N - F - \frac{S}{2}\,. \tag{7.54}
$$

[1]We are, however, assuming without proof that the constraints are independent.

The formula (7.54) is a consequence of the fact that $S$ second-class constraints are equivalent to vanishing of $S/2$ canonical coordinates and $S/2$ of their momenta. The $F$ first-class constraints are equivalent to vanishing of $F$ canonical coordinates, and since the first-class constraints generate the gauge symmetries, we can impose $F$ gauge-fixing conditions for the corresponding $F$ canonical momenta. Consequently there are $2N - 2F - S$ independent canonical coordinates and momenta and therefore $2n = 2N - 2F - S$.

In our case, $N$ can be determined from the table

| $A^a{}_\mu$ | $\beta^\alpha{}_{\mu\nu}$ | $C^\alpha{}_\mu$ | $B^a{}_{\mu\nu}$ |
|---|---|---|---|
| $4p$ | $6q$ | $4q$ | $6p$ |

where $p$ is the dimensionality of the Lie group $G$ and $q$ is the dimensionality of the Lie group $H$. Consequently $N = 10(p+q)$.

Similarly, the number of independent components for the second class constraints is determined by the table

| $\chi(B)_a{}^{jk}$ | $\chi(C)_\alpha{}^i$ | $\chi(A)_a{}^i$ | $\chi(\beta)_\alpha{}^{ij}$ |
|---|---|---|---|
| $3p$ | $3q$ | $3p$ | $3q$ |

so that $S = 6(p+q)$.

The first-class constraints are not all independent, since they satisfy the following relations

$$\nabla_i \varphi(\mathcal{F})_a{}^i + \frac{1}{2}\,\partial_{\alpha a}\,\varphi(\mathcal{G})^\alpha - \frac{1}{2}\,\partial^\alpha{}_a\,\nabla^\triangleright_i \chi(C)_\alpha{}^i - \frac{1}{2} f_{abc}\,\partial_\alpha{}^b\,\beta^\alpha{}_{ij}\,\chi(B)^{c\,ij} = 0\,, \tag{7.55}$$

$$\begin{aligned} &\nabla^\triangleright_i \varphi(CB)_\alpha{}^i - \frac{1}{2}\,C_{\beta i} \triangleright^\beta{}_{a\alpha}\varphi(\mathcal{F})^{ai} + \partial_{\alpha a} S(BC\beta)^a + \frac{1}{2} F^b{}_{ij} \triangleright_{\alpha b\gamma}\,\chi(\beta)^{\gamma ij} + \\ &\qquad + T^\beta{}_{jk} \triangleright_{\beta a\alpha}\,\chi(B)^{ajk} - \partial_{\alpha a}\nabla_i \chi(A)^{a\,i} = 0\,. \end{aligned} \tag{7.56}$$

One can show that

$$\begin{aligned} \nabla_i \varphi(\mathcal{F})_a{}^i + \frac{1}{2}\partial_{\alpha a}\varphi(\mathcal{G})^\alpha - \frac{1}{2}\partial^\alpha{}_a \nabla^\triangleright_i \chi(C)_\alpha{}^i - \frac{1}{2} f_{abc}\partial_\alpha{}^b \beta^\alpha{}_{ij}\chi(B)^{c\,ij} &= \\ = \varepsilon^{ijk}\nabla_i F_{ajk}\,, \end{aligned} \tag{7.57}$$

which gives (7.55) because $\varepsilon^{ijk}\,\nabla_i F^a_{jk} = 0$ is the $\lambda = 0$ component of the BI (7.19). In the same way

$$\begin{aligned} & \nabla^\triangleright_i \varphi(CB)_\alpha{}^i - \frac{1}{2}\,C_{\beta i} \triangleright^\beta{}_{a\alpha}\varphi(\mathcal{F})^{ai} + \partial_{\alpha a} S(BC\beta)^a + \\ & + \frac{1}{2} F^b{}_{ij} \triangleright_{\alpha b\gamma}\,\chi(\beta)^{\gamma ij} + T^\beta{}_{jk} \triangleright_{\beta a\alpha}\,\chi(B)^{ajk} - \partial_{\alpha a}\nabla_i \chi(A)^{a\,i} \\ = \; & \varepsilon^{ijk}\left(\nabla^\triangleright_i T_{\alpha jk} - \triangleright_{\alpha a\beta} F^a_{jk} C^\beta_i\right)\,. \end{aligned} \tag{7.58}$$

The right-hand side of the equation (7.58) is the $\lambda = 0$ component of the Bianchi identity (7.20), so that (7.58) gives the relation (7.56).

As discussed in [114] for the case of the Poincaré 2-group, only the $\lambda = 0$ components of the BI give new restrictions on the canonical variables, because those BI do not contain the time derivatives of the fields. The BI components with $\lambda \neq 0$ will contain the time derivatives of the fields and hence must be consequences of the equations of motion (EOM). Related to this is the fact that the Bianchi identities associated to the 2-forms $\beta$ and $B$ do not induce any new relations among the constraints, see [114]. Namely, the corresponding BI (7.22) and (7.23) contain the time derivatives of the fields, so that the equations (7.22) and (7.23) are necessarily consequences of the EOM, and hence do not represent additional restrictions on the canonical variables.

The number of components of the first-class constraints can be obtained from the table

| $\varphi(B)_a{}^i$ | $\varphi(C)_\alpha$ | $\varphi(\beta)_\alpha{}^i$ | $\varphi(A)_a$ | $\varphi(\mathcal{F})^{ai}$ | $\varphi(\mathcal{G})^\alpha$ | $\varphi(CB)^{\alpha i}$ | $\varphi(BC\beta)^a$ |
|---|---|---|---|---|---|---|---|
| $3p$ | $q$ | $3q$ | $p$ | $3p$ | $q$ | $3q$ | $p$ |

The number of independent components for the first-class constraints is given by

$$F = 8(p+q) - p - q = 7(p+q)\,,$$

where we have subtracted the $p$ relations (7.55) and the $q$ relations (7.56). Therefore,

$$n = 10(p+q) - 7(p+q) - \frac{6(p+q)}{2} = 0\,, \tag{7.59}$$

and consequently, under the present assumptions (see footnote 1) there are no local DOF in the *BFCG* theory. Hence the physical DOF are global, and can be identified with the coordinates on the moduli space of the flat 2-connections on the 3-manifold $\Sigma$, see [113] for the case of the Poincaré 2-group.

# 8

# Hamiltonian analysis of the *BFCG* theory for the Poincaré 2-group

In this chapter we present the the *BFCG* theory for the Poincaré 2-group and a canonical analysis of this theory. This is original work published in [113, 114].

The structure of this chapter is as follows.. In section 8.1 we give the motivation for the *BFCG* theory for the Poincaré 2-group. In section 8.2 we introduce in detail the actions for the *BFCG* theory and the topological Poincaré gauge theory — as well as the relations between them —, we give a short overview of the Lagrange equations of motion, and prepare for the Hamiltonian analysis. The bulk of the Hamiltonian analysis is done in section 8.3. We evaluate the conjugate momenta for the fields, obtain the primary constraints and construct the Hamiltonian of the theory. Then we impose consistency conditions on the primary constraints, which leads to secondary constraints and some determined Lagrange multipliers. The consistency conditions of the secondary constraints turn out to be satisfied identically, and do not introduce any new constraints. The constraints are then separated into first and second class, and their algebra is computed. Finally, the number of physical degrees of freedom is calculated, and ends up being zero, confirming that the theory is indeed topological. Building on these results, in section 8.4 we construct the Dirac brackets, which facilitate the elimination of the second class constraints from the theory, leading to the reduction of the phase space. Section 8.5 is devoted to the study of the properties of the reduced phase space, with the emphasis on the differences between the *BFCG* model and the topological Poincaré gauge theory. . Appendix B provides some technical details about the derivation and the discussion of the Bianchi identities, used in the main text.

Regarding notation in this Chapter the capital Latin indices $A, B, C, \ldots$ represent multi-index notation, and are used to count the second class constraints. Antisymmetrization is denoted with the square brackets around the indices (Cf. equation (2.20)),

$$A_{[ab]} = \frac{1}{2}\left(A_{ab} - A_{ba}\right) . \tag{8.1}$$

About the notation and conventions for Poisson brackets see footnote 5 in chapter 3.

## 8.1 The *BFCG* theory

The canonical formulation of General Relativity is generally viewed as a prerequisite for a non-perturbative and background-metric independent quantization of this theory (see chapter 5). When using the spatial metric and its canonically conjugate momentum as the degrees of freedom for the gravitational field, one obtains a non-polynomial Hamiltonian Constraint (4.30) and (5.9). Consequently the corresponding operator in the canonical quantization yields the WdW equation (5.15) which is a complex mathematical object because it is difficult to define in a rigorous way, and generally nearly impossible to solve.

The situation improves if the Ashtekar variables are used, see section 5.1. These are given by, equation (5.16), an $SU(2)$ complex connection on the spatial manifold and its canonically conjugate momentum (5.18). One then obtains a polynomial Hamiltonian Constraint (equation (5.24) with the choice $\gamma = \pm i$), but since the connection is complex, this introduces an additional non-polynomial constraint, the reality condition, which makes the quantization complicated. One can also use the real Ashtekar connection (see section 5.4 and [96]) but then the constraint (5.24) becomes again non-polynomial.

In any case, the fact that the basic canonical variables are analogous to the ones used in the $SU(2)$ Yang-Mills gauge theory , makes it possible (as we saw in chapter 5) to use the holonomy (5.33) and the flux (5.35) (or its gauge invariant modification (5.39)) variables, which leads to spin-network variables and LQG.

The difficulties stymying progress in the Hamiltonian approach led to the development of a path-integral quantization method known as *spin-foam models,* (see [115]).

Just as a spin-network (or $s$-knot) represents a discretized 3-dimensional space, a spin-foam may be viewed as a discretization of space-time. In a related way, spin-foams may be regarded as the time evolution of spin-networks.

A *spin-foam* (further details can be found in [116, 117]) $\mathcal{F} = (\Delta^*, j_f, i_e)$, consists of:

- A two-dimensional cellular complex $\Delta^*$[1] that is, a combinatorial structure consisting of faces $f$ intersecting at edges $e$ which in turn meet at vertices $v$;
- the faces $f$ are coloured by irreducible group representations $j_f$;
- and the edges are coloured by intertwiners $i_e$.

We can look at the colouring mentioned above as some kinematical information inherited from the boudary states. These are one dimensional cellular complexes (graphs) obtained by intersecting the spin-foam with codimension-one surfaces. The graphs are coloured in the way described in section 5.4. So we have a spin-foam which encodes information about space-time with the boundaries being spin-networks (which represent a discretization of space). In this way, we can think of the sum over different spin-foams as a discrete path integral over possible geometries, interpolating between given boundary information.

To calculate transition amplitudes we must assign complex amplitudes $A_f(j_f)$ to faces, $A_e(j_{f\supset e}, i_e)$ to edges and $A_v(j_{f\supset v}, i_{e\supset v})$ to vertices of $\mathcal{F}$. Where the notation $f \supset e$ means that the faces $f$ meet at edge $e$, and likewise for the others.

The transition amplitude, for a spin-foam with a 2-complex $\Delta^*$ interpolating between two graphs $S_1$ and $S_2$, is given by,

$$W(\Delta^*) \equiv \int d\mu_{\{j_f\}} \int d\mu_{\{i_e\}} \prod_f A_f(j_f) \prod_e A_e(j_{f\supset e}, i_e) \prod_v A_v(j_{f\supset v}, i_{e\supset v}) \,. \quad (8.2)$$

In this expression, $d\mu_{\{j_f\}}$ and $d\mu_{\{i_e\}}$ are the integration measures depending on the group theoretical data used to colour the 2-complex. However, if the spin foam model is constructed with a compact Lie Group, then the integrals become discrete sums over all the representations colouring the faces and intertwiners colouring the the edges of the two-dimensional complex.

Spin foams replace the problems besetting the Hamiltonian formulation of LQG, with difficulties of equal — if not greater — magnitude. First of all, there is the problem of the divergence and regularization of (8.2). There is also the issue of the classical limit of a spin foam model [118] and that of the coupling of fermionic matter [118, 119]. These questions are related to the fact that the edge-lengths, or the tetrads, are not always defined in a spin-foam model of quantum gravity. We therefore see a need for a model of QG were the tetrads (the edge-lengths) are present.

[1]Sometimes the dual two-dimensional complex is used (see [117] for the definition).

The *BFCG* theory (6.44), a categorical generalization of the *BF* model, has exactly the property we need, if we choose to build the theory upon the Pincaré 2-group.

The importance of the *BFCG* theory for the Poincaré 2-group, defined by the choice $G = SO(3,1)$ and $H = \mathbb{R}^4$ (see section 6.2 equations (6.22) and (6.25) for the definitions), lies in the fact that one can construct the action for GR by simply adding an additional term to the *BFCG* action, called the simplicity constraint (2.152), thus we have the following action:

$$S = \int_{\mathcal{M}} \langle B \wedge R \rangle_{\mathfrak{g}} + \langle e \wedge G \rangle_{\mathfrak{h}} - \langle \varphi \wedge (B - \star(e \wedge e)) \rangle_{\mathfrak{g}} \,. \tag{8.3}$$

Here we have made the identifications $C \equiv e$ and $F \equiv R$, since in the case of the Poincaré 2-group these fields have the interpretation of the tetrad field[2] and the curvature two-form for the spin connection $A \equiv \omega$. The $\mathfrak{g}$-valued two-form $\varphi$ is an additional Lagrange multiplier, featuring in the simplicity constraint term. The $\star$ is the Hodge dual operator for the Minkowski space. See [119] for details.

The constrained *BFCG* theory (8.3) see also appendix D is in full analogy to the Plebanski model (2.153), where GR is constructed by enforcing a suitable simplicity constraint upon the *BF* theory based on the Lorentz group. However, in contrast to the Plebanski model, the constrained *BFCG* model has one big advantage. Namely, the Lagrange multiplier $C$ has the interpretation of the tetrad field, which is therefore explicitly present in the topological sector of the action. This is not the case for the Plebanski model, where the simplicity constraint merely infers the implicit existence of the tetrad fields. Upon the covariant quantization, the Plebanski action gives rise to spin-foam models, while the constrained *BFCG* action gives rise to the spin-cube model. These are categorical generalizations of spin-foam models, they consist of a coloured 3-complex. The colouring in turn is given by objects, morphisms and 2-morphisms of a 2-category representation of the relevant 2-group (for details the reader is referred to [118]) . Then, the explicit presence of the tetrad in (8.3) enables us to easily couple matter fields to gravity in the spin-cube model, in contrast to spin-foam models where this is a notoriously hard problem.

As a classical theory, the constrained *BFCG* action lends itself also to the canonical quantization programme. In the canonical approach (see Chap. 3), the first and crucial step is to perform the Hamiltonian analysis of the theory, study the algebra of constraints, and eliminate second class constraints from the theory. However, due to the technical complexity of the Hamiltonian analysis, it is wise to discuss the pure *BFCG* theory first, leaving the constrained theory for later. That is the aim of this Chapter largely based in [114]. We will perform the full Hamiltonian analysis of the unconstrained *BFCG* theory based on the Poincaré 2-group, as a preparation for the more complicated case of the constrained *BFCG* theory (8.3). The similar analysis has been done for the *BF* theory in [78], and our analysis represents the generalization of that work to the *BFCG* case. The analysis here can also be viewed as a special case of the one we presented in chapter 7. However, in our work we took the opposite route, beginning with the Poincaré 2-group case and later, generalizing some parts of this to the case of a generic Lie 2-group. Furthermore, since our main interest is quantum gravity it is instructive to see the specific details of the analysis unfolding when we apply the Dirac procedure to the topological *BFCG* theory for the Poincaré 2-group.

[2] This identification is base on the fact that the transformation properties of the one-forms $C^a$ are the same as those of the tetrad one forms $e^a$ (see [119] for details) under local Lorentz and diffeomorphism transformations.

We should note that the Hamiltonian analysis of the *BFCG* model has been done in a gauge-fixed form in [113]. Here we improve those results by providing a gauge-invariant canonical analysis.

There is an interesting relationship between the *BFCG* theory for the Poincaré 2-group on one hand, and the topological Poincaré gauge theory on the other. Perhaps surprisingly, the two theories are equivalent, while their Hamiltonian structure is vastly different. This was discussed to an extent in [113], but the full Hamiltonian analysis presented here is naturally suited to a more complete comparison of the Hamiltonian formulations for the *BFCG* theory and the topological Poincaré gauge theory. The results of this comparison are especially intriguing, and provide additional insight into the structure of the theory.

## 8.2 BFCG action

The *BFCG* theory for the Poincaré 2-group is defined by the action

$$S_{BFCG} = \int_{\mathcal{M}} B_{ab} \wedge R^{ab} + e^a \wedge G_a \,. \tag{8.4}$$

The variables of this action are the one-forms $e^a$, $\omega^{ab}$ and the two-forms $B^{ab}$, $\beta^a$. The curvatures $R^{ab}$ and $G^a$ are the field strengths of the 2-connection $(\omega^{ab}, \beta^a)$,

$$R^{ab} = d\omega^{ab} + \omega^a{}_c \wedge \omega^{cb} \,, \tag{8.5}$$

$$G^a = \nabla\beta^a \equiv d\beta^a + \omega^a{}_b \wedge \beta^b \,. \tag{8.6}$$

The fields $B^{ab}$ and $e^a$ play the role of the Lagrange multipliers.

It is also convenient to introduce torsion as the field strength for the tetrad $e^a$,

$$T^a = \nabla e^a \equiv de^a + \omega^a{}_b \wedge e^b \,. \tag{8.7}$$

Then, performing a partial integration in the second term in (8.4) and using the Stokes theorem one can rewrite the action as

$$S_{TPGT} = \int_{\mathcal{M}} B_{ab} \wedge R^{ab} + \beta^a \wedge T_a - \int_{\partial\mathcal{M}} e^a \wedge \beta_a \,, \tag{8.8}$$

where now $B^{ab}$ and $\beta^a$ play the role of Lagrange multipliers. Aside from the immaterial boundary term, this action represents the topological Poincaré gauge theory (TPGT). In order to fully appreciate the relationship between the two theories in the sense of the Hamiltonian analysis, let us introduce a parameter $\xi \in \mathbb{R}$ and rewrite the action as

$$S = \int_{\mathcal{M}} B_{ab} \wedge R^{ab} + \xi e^a \wedge G_a + (1-\xi)\beta^a \wedge T_a \,, \tag{8.9}$$

where we have dropped the boundary term. It is obvious that the action (8.9) is a convenient interpolation between (8.4) and (8.8), to which it reduces for the choices $\xi = 1$ and $\xi = 0$, respectively. The action (8.9) will therefore be the starting point for the Hamiltonian analysis.

It is also clear that all three actions (8.4), (8.8) and (8.9) give rise to the same set of equations of motion, since these do not depend on the boundary. Taking the variation of (8.9) with respect to all the variables, one obtains

$$\begin{array}{ll} \delta B: & R^{ab} = 0\,, \\ \delta\beta: & T^a = 0\,, \\ \delta e: & G^a = 0\,, \\ \delta\omega: & \nabla B^{ab} - e^{[a} \wedge \beta^{b]} = 0\,, \end{array} \tag{8.10}$$

where

$$\nabla B^{ab} \equiv dB^{ab} + \omega^a{}_c \wedge B^{cb} + \omega^b{}_c \wedge B^{ac} \,. \tag{8.11}$$

The first two equations of motion are equivalent to the Einstein vacuum field equations, while the third and fourth determine $\beta^a$ and $B^{ab}$. As we shall see later, there are no local propagating degrees of freedom in the theory.

Finally, for the convenience of the Hamiltonian analysis, we need to rewrite both the action and the equations of motion in a local coordinate frame. Choosing $dx^\mu$ as basis one-forms, we can expand the fields in the standard fashion:

$$e^a = e^a{}_\mu dx^\mu \,, \qquad \omega^{ab} = \omega^{ab}{}_\mu dx^\mu \,, \tag{8.12}$$

$$B^{ab} = \frac{1}{2} B^{ab}{}_{\mu\nu} dx^\mu \wedge dx^\nu \,, \qquad \beta^a = \frac{1}{2}\beta^a{}_{\mu\nu} dx^\mu \wedge dx^\nu \,. \tag{8.13}$$

Similarly, the field strengths for $\omega$, $e$ and $\beta$ are

$$\begin{aligned} R^{ab} &= \frac{1}{2} R^{ab}{}_{\mu\nu} dx^\mu \wedge dx^\nu \,, \\ T^a &= \frac{1}{2} T^a{}_{\mu\nu} dx^\mu \wedge dx^\nu \,, \\ G^a &= \frac{1}{6} G^a{}_{\mu\nu\rho} dx^\mu \wedge dx^\nu \wedge dx^\rho \,. \end{aligned} \tag{8.14}$$

Using the relations (8.5), (8.6) and (8.7), we can write the component equations

$$\begin{aligned} R^{ab}{}_{\mu\nu} &= \partial_\mu \omega^{ab}{}_\nu - \partial_\nu \omega^{ab}{}_\mu + \omega^a{}_{c\mu}\omega^{cb}{}_\nu - \omega^a{}_{c\nu}\omega^{cb}{}_\mu \,, \\ T^a{}_{\mu\nu} &= \partial_\mu e^a{}_\nu - \partial_\nu e^a{}_\mu + \omega^a{}_{b\mu} e^b{}_\nu - \omega^a{}_{b\nu} e^b{}_\mu \,, \\ G^a{}_{\mu\nu\rho} &= \partial_\mu \beta^a{}_{\nu\rho} + \partial_\nu \beta^a{}_{\rho\mu} + \partial_\rho \beta^a{}_{\mu\nu} + \omega^a{}_{b\mu}\beta^b{}_{\nu\rho} + \omega^a{}_{b\nu}\beta^b{}_{\rho\mu} + \omega^a{}_{b\rho}\beta^b{}_{\mu\nu} \,. \end{aligned} \tag{8.15}$$

Substituting expansions (8.12), (8.13) and (8.14) into the action, we obtain

$$S = \int_{\mathcal{M}} d^4x \, \varepsilon^{\mu\nu\rho\sigma} \left[ \frac{1}{4} B_{ab\mu\nu} R^{ab}{}_{\rho\sigma} + \frac{\xi}{6} e_{a\mu} G^a{}_{\nu\rho\sigma} + \frac{1-\xi}{4} \beta_{a\mu\nu} T^a{}_{\rho\sigma} \right] . \tag{8.16}$$

Assuming that the spacetime manifold has the topology $\mathcal{M} = \Sigma \times \mathbb{R}$, where $\Sigma$ is a 3-dimensional spacelike hypersurface, from the above action we can read off the Lagrangian, which is the integral of the Lagrangian density over the hypersurface $\Sigma$:

$$L = \int_{\Sigma} d^3x \, \varepsilon^{\mu\nu\rho\sigma} \left[ \frac{1}{4} B_{ab\mu\nu} R^{ab}{}_{\rho\sigma} + \frac{\xi}{6} e_{a\mu} G^a{}_{\nu\rho\sigma} + \frac{1-\xi}{4} \beta_{a\mu\nu} T^a{}_{\rho\sigma} \right] . \tag{8.17}$$

Finally, the component form of equations of motion is:

$$\begin{aligned} & R^{ab}{}_{\mu\nu} = 0 \,, \qquad T^a{}_{\mu\nu} = 0 \,, \qquad G^a{}_{\mu\nu\rho} = 0 \,, \\ & \varepsilon^{\lambda\mu\nu\rho} \left[ \nabla_\rho B^{ab}{}_{\mu\nu} - e^{[a}{}_\rho \beta^{b]}{}_{\mu\nu} \right] = 0 \,. \end{aligned} \tag{8.18}$$

## 8.3 Hamiltonian analysis

Now we turn to the Hamiltonian analysis of the *BFCG* theory. A review of the general formalism can be found in chapter 3. In addition, the equivalent procedure for the ordinary *BF* theory has been done in [78] see also section 3.14.

As a first step, we calculate the momenta $\pi$ corresponding to the field variables $B^{ab}{}_{\mu\nu}$, $e^a{}_\mu$, $\omega^{ab}{}_\mu$ and $\beta^a{}_{\mu\nu}$. Differentiating the Lagrangian with respect to the time derivative of the appropriate fields, we obtain the momenta as follows:

$$\begin{aligned} \pi(B)_{ab}{}^{\mu\nu} &= \frac{\delta L}{\delta \partial_0 B^{ab}{}_{\mu\nu}} = 0\,, \\ \pi(e)_a{}^\mu &= \frac{\delta L}{\delta \partial_0 e^a{}_\mu} = \frac{1-\xi}{2}\varepsilon^{0\mu\nu\rho}\beta_{a\nu\rho}\,, \\ \pi(\omega)_{ab}{}^\mu &= \frac{\delta L}{\delta \partial_0 \omega^{ab}{}_\mu} = \varepsilon^{0\mu\nu\rho}B_{ab\nu\rho}\,, \\ \pi(\beta)_a{}^{\mu\nu} &= \frac{\delta L}{\delta \partial_0 \beta^a{}_{\mu\nu}} = -\xi\varepsilon^{0\mu\nu\rho}e_{a\rho}\,. \end{aligned} \tag{8.19}$$

None of the momenta can be solved for the corresponding "velocities", so they all give rise to primary constraints:

$$\begin{aligned} P(B)_{ab}{}^{\mu\nu} &\equiv \pi(B)_{ab}{}^{\mu\nu} \approx 0\,, \\ P(e)_a{}^\mu &\equiv \pi(e)_a{}^\mu - \frac{1-\xi}{2}\varepsilon^{0\mu\nu\rho}\beta_{a\nu\rho} \approx 0\,, \\ P(\omega)_{ab}{}^\mu &\equiv \pi(\omega)_{ab}{}^\mu - \varepsilon^{0\mu\nu\rho}B_{ab\nu\rho} \approx 0\,, \\ P(\beta)_a{}^{\mu\nu} &\equiv \pi(\beta)_a{}^{\mu\nu} + \xi\varepsilon^{0\mu\nu\rho}e_{a\rho} \approx 0\,. \end{aligned} \tag{8.20}$$

The weak, on-shell equality is denoted "$\approx$", as opposed to the strong, off-shell equality which is denoted by the usual symbol "$=$".

Next we introduce the fundamental simultaneous Poisson brackets between the fields and their conjugate momenta,

$$\begin{aligned} \{\, B^{ab}{}_{\mu\nu}\,,\, \pi(B)_{cd}{}^{\rho\sigma} \,\} &= 4\delta^a_{[c}\delta^b_{d]}\delta^\rho_{[\mu}\delta^\sigma_{\nu]}\delta^{(3)}\,, \\ \{\, e^a{}_\mu\,,\, \pi(e)_b{}^\nu \,\} &= \delta^a_b\delta^\nu_\mu\delta^{(3)}\,, \\ \{\, \omega^{ab}{}_\mu\,,\, \pi(\omega)_{cd}{}^\nu \,\} &= 2\delta^a_{[c}\delta^b_{d]}\delta^\nu_\mu\delta^{(3)}\,, \\ \{\, \beta^a{}_{\mu\nu}\,,\, \pi(\beta)_b{}^{\rho\sigma} \,\} &= 2\delta^a_b\delta^\rho_{[\mu}\delta^\sigma_{\nu]}\delta^{(3)}\,, \end{aligned} \tag{8.21}$$

and we employ them to calculate the algebra of primary constraints,

$$\begin{array}{rcl} \{\, P(B)^{abjk} \,,\, P(\omega)_{cd}{}^{i} \,\} &=& 4\varepsilon^{0ijk}\delta^a_{[c}\delta^b_{d]}\delta^{(3)} , \\ \{\, P(e)^{ak} \,,\, P(\beta)_b{}^{ij} \,\} &=& -\varepsilon^{0ijk}\delta^a_b\delta^{(3)} , \end{array} \tag{8.22}$$

while all other Poisson brackets vanish. Note that the algebra of primary constraints is independent of $\xi$.

Next we construct the canonical, on-shell Hamiltonian:

$$\begin{array}{rcl} H_c &=& \displaystyle\int_\Sigma d^3x \left[ \frac{1}{4}\pi(B)_{ab}{}^{\mu\nu}\partial_0 B^{ab}{}_{\mu\nu} + \pi(e)_a{}^\mu \partial_0 e^a{}_\mu + \right. \\ && \left. + \displaystyle\frac{1}{2}\pi(\omega)_{ab}{}^\mu \partial_0 \omega^{ab}{}_\mu + \frac{1}{2}\pi(\beta)_a{}^{\mu\nu}\partial_0\beta^a{}_{\mu\nu} \right] - L \,. \end{array} \tag{8.23}$$

The factors $1/4$ and $1/2$ are introduced to prevent overcounting of variables. Using (8.15) and (8.17), one can rearrange the expressions such that all velocities are multiplied by primary constraints, and therefore vanish from the Hamiltonian. After some algebra, the resulting expression can be written as

$$\begin{array}{rcl} H_c &=& -\displaystyle\int d^3x\, \varepsilon^{0ijk} \left[ \frac{1}{2} B_{ab0i} R^{ab}{}_{jk} + \frac{1}{6} e_{a0} G^a{}_{ijk} + \right. \\ && \left. + \displaystyle\frac{1}{2}\beta_{a0k} T^a{}_{ij} + \frac{1}{2}\omega_{ab0}\left( \nabla_i B^{ab}{}_{jk} - e^a{}_i \beta^b{}_{jk} \right) \right] \,, \end{array} \tag{8.24}$$

up to a boundary term. The canonical Hamiltonian does not depend on any momenta, but only on fields and their spatial derivatives. Also, note that it does not depend on $\xi$ either. Finally, introducing Lagrange multipliers $\lambda$ for each of the primary constraints, we construct the total, off-shell Hamiltonian:

$$\begin{array}{rcl} H_T &=& H_c + \displaystyle\int d^3x \left[ \lambda(e)^a{}_\mu P(e)_a{}^\mu + \frac{1}{2}\lambda(\omega)^{ab}{}_\mu P(\omega)_{ab}{}^\mu \right. \\ && \left. + \displaystyle\frac{1}{4}\lambda(B)^{ab}{}_{\mu\nu} P(B)_{ab}{}^{\mu\nu} + \frac{1}{2}\lambda(\beta)^a{}_{\mu\nu} P(\beta)_a{}^{\mu\nu} \right] \,. \end{array} \tag{8.25}$$

We proceed with the calculation of the consistency requirements for the primary constraints,

$$\dot{P} \equiv \{\, P \,,\, H_T \,\} \approx 0 \,. \tag{8.26}$$

Half of the consistency requirements will give the secondary constraints $S$, while the other half will determine some of the multipliers $\lambda$. In particular, requiring that

$$\begin{array}{ll} \dot{P}(B)_{ab}{}^{0i} \approx 0 \,, & \dot{P}(e)_a{}^0 \approx 0 \,, \\ \dot{P}(\beta)_a{}^{0i} \approx 0 \,, & \dot{P}(\omega)_{ab}{}^0 \approx 0 \,, \end{array} \tag{8.27}$$

we obtain the following secondary constraints:

$$
\begin{array}{lcl}
S(R)^{ab}{}_{jk} & \equiv & R^{ab}{}_{jk} \approx 0, \\
S(G)^{a} & \equiv & \varepsilon^{0ijk} G^{a}{}_{ijk} \approx 0, \\
S(T)^{a}{}_{ij} & \equiv & T^{a}{}_{ij} \approx 0, \\
S(B)^{ab} & \equiv & \varepsilon^{0ijk} \left( \nabla_i B^{ab}{}_{jk} - e^{[a}{}_i \beta^{b]}{}_{jk} \right) \approx 0.
\end{array}
\tag{8.28}
$$

The remaining consistency conditions for the primary constraints,

$$
\begin{array}{ll}
\dot{P}(B)_{ab}{}^{jk} \approx 0\,, & \dot{P}(e)_{a}{}^{k} \approx 0\,, \\
\dot{P}(\beta)_{a}{}^{jk} \approx 0\,, & \dot{P}(\omega)_{ab}{}^{k} \approx 0\,,
\end{array}
\tag{8.29}
$$

determine the following multipliers:

$$
\begin{array}{lcl}
\lambda(\omega)^{ab}{}_{i} & \approx & \nabla_i \omega^{ab}{}_0\,, \\
\lambda(\beta)^{a}{}_{ij} & \approx & 2\nabla_{[j} \beta^{a}{}_{i]0} - \omega^{a}{}_{b0} \beta^{b}{}_{ij}\,, \\
\lambda(e)^{a}{}_{i} & \approx & \nabla_i e^{a}{}_0 - \omega^{a}{}_{b0} e^{b}{}_i\,, \\
\lambda(B)^{ab}{}_{ij} & \approx & 2\nabla_{[j} B^{ab}{}_{i]0} + 2\omega^{[a}{}_{c0} B^{b]c}{}_{ij} + e^{[a}{}_0 \beta^{b]}{}_{ij} + e^{[a}{}_j \beta^{b]}{}_{0i} - e^{[a}{}_i \beta^{b]}{}_{0j}\,.
\end{array}
\tag{8.30}
$$

This leaves the multipliers

$$
\lambda(\omega)^{ab}{}_0\,, \quad \lambda(\beta)^{a}{}_{0i}\,, \quad \lambda(e)^{a}{}_0\,, \quad \lambda(B)^{ab}{}_{0i}\,,
\tag{8.31}
$$

undetermined.

As the next step we impose the consistency conditions for the secondary constraints (8.28),

$$
\begin{array}{ll}
\dot{S}(R)^{ab}{}_{jk} \approx 0\,, & \dot{S}(G)^{a} \approx 0\,, \\
\dot{S}(T)^{a}{}_{ij} \approx 0\,, & \dot{S}(B)^{ab} \approx 0\,.
\end{array}
\tag{8.32}
$$

After a straightforward but lengthy calculation, it turns out that all these conditions are identically satisfied, producing no new constraints and determining no additional multipliers. Therefore, at this point all the consistency conditions have been exhausted.

Once we have found all the constraints in the theory, we need to classify them into first and second class. While some of the second class constraints can already be read from (8.22), the classification is not easy since constraints are unique only up to linear combinations. The most efficient way to tabulate all first class constraints is to substitute all determined multipliers into the total Hamiltonian (8.25) and rewrite it in the form

$$\begin{aligned} H_T &= \int d^3x \left[ \frac{1}{2}\lambda(B)^{ab}{}_{0i}\,\varphi(B)_{ab}{}^{i} + \lambda(e)^a{}_0\,\varphi(e)_a + \lambda(\beta)^a{}_i\,\varphi(\beta)_a{}^i \right. \\ & \quad + \frac{1}{2}\lambda(\omega)^{ab}\,\varphi(\omega)_{ab} - \frac{1}{2}B_{ab0i}\,\varphi(R)^{abi} - e_{a0}\,\varphi(G)^a \\ & \quad \left. - \beta_{a0i}\,\varphi(T)^{ai} - \frac{1}{2}\omega_{ab0}\,\varphi(\nabla B)^{ab} \right] . \end{aligned} \tag{8.33}$$

The quantities $\varphi$ are linear combinations of constraints, but must all be first class, since the total Hamiltonian weakly commutes with all constraints. Written in terms of primary and secondary constraints, they are:

$$\begin{aligned} \varphi(B)_{ab}{}^i &= P(B)_{ab}{}^{0i}, \\ \varphi(e)_a &= P(e)_a{}^0, \\ \varphi(\beta)_a{}^i &= P(\beta)_a{}^{0i}, \\ \varphi(\omega)_{ab} &= P(\omega)_{ab}{}^0, \\ \varphi(R)^{abi} &= \varepsilon^{0ijk}S(R)^{ab}{}_{jk} - \nabla_j P(B)^{abij}, \\ \varphi(G)^a &= \frac{1}{6}S(G)^a + \nabla_i P(e)^{ai} - \frac{1}{4}\beta_{bij}P(B)^{abij}, \\ \varphi(T)^{ai} &= \frac{1}{2}\varepsilon^{0ijk}S(T)^a{}_{jk} - \nabla_j P(\beta)^{aij} + \frac{1}{2}e_{bj}P(B)^{abij}, \\ \varphi(\nabla B)^{ab} &= S(B)^{ab} + \nabla_i P(\omega)^{abi} - B^{[a}{}_{cij}P(B)^{b]cij} \\ & \quad - 2e^{[a}{}_i P(e)^{b]i} - \beta^{[a}{}_{ij}P(\beta)^{b]ij} . \end{aligned} \tag{8.34}$$

These are the first class constraints in the theory. The remaining constraints are second class:

$$\begin{aligned} \chi(B)_{ab}{}^{jk} = P(B)_{ab}{}^{jk}, & \qquad \chi(e)_a{}^i = P(e)_a{}^i, \\ \chi(\omega)_{ab}{}^i = P(\omega)_{ab}{}^i, & \qquad \chi(\beta)_a{}^{ij} = P(\beta)_a{}^{ij} . \end{aligned} \tag{8.35}$$

In order to calculate the full algebra of constraints, it is convenient to express them as functions of fundamental variables, as follows:

$$
\begin{aligned}
\varphi(B)_{ab}{}^{i} &= \pi(B)_{ab}{}^{0i}\,, \\
\varphi(e)_{a} &= \pi(e)_{a}{}^{0}\,, \\
\varphi(\beta)_{a}{}^{i} &= \pi(\beta)_{a}{}^{0i}\,, \\
\varphi(\omega)_{ab} &= \pi(\omega)_{ab}{}^{0}\,, \\
\varphi(R)^{abi} &= \varepsilon^{0ijk}R^{ab}{}_{jk} - \nabla_j \pi(B)^{abij}\,, \\
\varphi(G)^{a} &= \frac{\xi}{6}\varepsilon^{0ijk}G^{a}{}_{ijk} + \nabla_i \pi(e)^{ai} - \frac{1}{4}\beta_{bij}\pi(B)^{abij}\,, \\
\varphi(T)^{ai} &= \frac{1-\xi}{2}\varepsilon^{0ijk}T^{a}{}_{jk} - \nabla_j \pi(\beta)^{aij} + \frac{1}{2}e_{bj}\pi(B)^{abij}, \\
\varphi(\nabla B)^{ab} &= \nabla_i \pi(\omega)^{abi} - B^{[a}{}_{cij}\pi(B)^{b]cij} - 2e^{[a}{}_{i}\pi(e)^{b]i} - \beta^{[a}{}_{ij}\pi(\beta)^{b]ij}\,,
\end{aligned}
\tag{8.36}
$$

and

$$
\begin{aligned}
\chi(B)_{ab}{}^{jk} &= \pi(B)_{ab}{}^{jk}\,, \\
\chi(e)_{a}{}^{i} &= \pi(e)_{a}{}^{i} - \frac{1-\xi}{2}\varepsilon^{0ijk}\beta_{ajk}\,, \\
\chi(\omega)_{ab}{}^{i} &= \pi(\omega)_{ab}{}^{i} - \varepsilon^{0ijk}B_{abjk}\,, \\
\chi(\beta)_{a}{}^{ij} &= \pi(\beta)_{a}{}^{ij} + \xi\varepsilon^{0ijk}e_{ak}\,.
\end{aligned}
\tag{8.37}
$$

The algebra between the first class constraints is then

$$
\begin{aligned}
\{\,\varphi(G)^{a}\,,\,\varphi(T)^{bi}\,\} &= -\varphi(R)^{abi}\delta^{(3)}\,, \\
\{\,\varphi(G)^{a}\,,\,\varphi(\nabla B)_{cd}\,\} &= 2\delta^{a}_{[c}\varphi(G)_{d]}\delta^{(3)}\,, \\
\{\,\varphi(T)^{ai}\,,\,\varphi(\nabla B)_{cd}\,\} &= 2\delta^{a}_{[c}\varphi(T)_{d]}{}^{i}\delta^{(3)}\,, \\
\{\,\varphi(R)^{abi}\,,\,\varphi(\nabla B)_{cd}\,\} &= -4\delta^{[a}_{[c}\varphi(R)^{b]}{}_{d]}{}^{i}\delta^{(3)}\,, \\
\{\,\varphi(\nabla B)^{ab}\,,\,\varphi(\nabla B)_{cd}\,\} &= -4\delta^{[a}_{[c}\varphi(\nabla B)^{b]}{}_{d]}\delta^{(3)}\,,
\end{aligned}
\tag{8.38}
$$

the algebra between the second class constraints is, according to (8.22),

$$
\begin{aligned}
\{\,\chi(B)^{abjk}\,,\,\chi(\omega)_{cd}{}^{i}\,\} &= 4\varepsilon^{0ijk}\delta^{a}_{[c}\delta^{b}_{d]}\delta^{(3)}\,, \\
\{\,\chi(e)^{ak}\,,\,\chi(\beta)_{b}{}^{ij}\,\} &= -\varepsilon^{0ijk}\delta^{a}_{b}\delta^{(3)}\,,
\end{aligned}
\tag{8.39}
$$

while the algebra between the first and second class constraints is

$$
\begin{aligned}
\{\, \varphi(R)^{abi} \,,\, \chi(\omega)_{cd}{}^{j} \,\} &= 4\delta^{[a}_{[c}\chi(B)^{b]}{}_{d]}{}^{ij}\delta^{(3)} \,, \\
\{\, \varphi(G)^{a} \,,\, \chi(\omega)_{cd}{}^{i} \,\} &= 2\delta^{a}_{[c}\chi(e)_{d]}{}^{i}\delta^{(3)} \,, \\
\{\, \varphi(G)^{a} \,,\, \chi(\beta)_{c}{}^{jk} \,\} &= -\tfrac{1}{2}\chi(B)^{a}{}_{c}{}^{jk}\delta^{(3)} \,, \\
\{\, \varphi(T)^{ai} \,,\, \chi(\omega)_{cd}{}^{j} \,\} &= -2\delta^{a}_{[c}\chi(\beta)_{d]}{}^{ij}\delta^{(3)} \,, \\
\{\, \varphi(T)^{ai} \,,\, \chi(e)_{b}{}^{j} \,\} &= \tfrac{1}{2}\chi(B)^{a}{}_{b}{}^{ij}\delta^{(3)} \,, \\
\{\, \varphi(\nabla B)^{ab} \,,\, \chi(\omega)_{cd}{}^{i} \,\} &= 4\delta^{[a}_{[c}\chi(\omega)_{d]}{}^{b]i}\delta^{(3)} \,, \\
\{\, \varphi(\nabla B)^{ab} \,,\, \chi(\beta)_{c}{}^{jk} \,\} &= -2\delta^{[a}_{c}\chi(\beta)^{b]jk}\delta^{(3)} \,, \\
\{\, \varphi(\nabla B)^{ab} \,,\, \chi(e)_{c}{}^{i} \,\} &= -2\delta^{[a}_{c}\chi(e)^{b]i}\delta^{(3)} \,, \\
\{\, \varphi(\nabla B)^{ab} \,,\, \chi(B)_{cd}{}^{jk} \,\} &= 4\delta^{[a}_{[c}\chi(B)_{d]}{}^{b]jk}\delta^{(3)} \,.
\end{aligned}
\tag{8.40}
$$

All other Poisson brackets among $\varphi$ and $\chi$ are zero.

We see that the algebra is closed, and all Poisson brackets involving $\varphi$ constraints weakly vanish, confirming that all $\varphi$ are indeed first class. Also, the Poisson brackets between $\chi$ constraints do not weakly vanish, confirming that $\chi$ are indeed second class. Finally, note that the structure constants do not depend on $\xi$, despite the fact that the constraints $\varphi$ and $\chi$ do.

The last main step in the Hamiltonian analysis is the counting of the physical degrees of freedom. Given $N$ initial independent fields in the theory, the dimension of the phase space *per point* is $2N$. From this one subtracts the total number $F$ of first class constraints *per point* , the total number $S$ of second class constraints *per point*, and the total number $F$ of gauge fixing conditions *per point*. The result is the dimension of the phase space *per point*, $2n$, where $n$ is the number of *local* degrees of freedom. Thus we have the general formula (see section 3.9),

$$n = N - F - \frac{S}{2} \,. \tag{8.41}$$

The number of independent field components for each of the fundamental fields is

| $\omega^{ab}{}_{\mu}$ | $\beta^{a}{}_{\mu\nu}$ | $e^{a}{}_{\mu}$ | $B^{ab}{}_{\mu\nu}$ |
|---|---|---|---|
| 24 | 24 | 16 | 36 |

which gives the total $N = 100$. Similarly, the number of independent components for the second class constraints is

| $\chi(B)_{ab}{}^{jk}$ | $\chi(e)_a{}^i$ | $\chi(\omega)_{ab}{}^i$ | $\chi(\beta)_a{}^{ij}$ |
|---|---|---|---|
| 18 | 12 | 18 | 12 |

which gives the total $S = 60$. Regarding the first class constraints, the situation is a little more complicated, due to the presence of Bianchi identities (see the Appendix). In particular, not all components of $\varphi(R)^{abi}$ and $\varphi(T)^{ai}$ are independent. To see this, take the derivative of $\varphi(R)^{abi}$ to obtain

$$\nabla_i \varphi(R)^{abi} = \varepsilon^{0ijk} \nabla_i R^{ab}{}_{jk} + R^{c[a}{}_{ij} \pi(B)_c{}^{b]ij} \,. \tag{8.42}$$

The first term on the right-hand side is zero off-shell as a consequence of the second Bianchi identity (B.8). The second term on the right-hand side is also zero off-shell, since it is a product of two constraints,

$$R^{c[a}{}_{ij} \pi(B)_c{}^{b]ij} \equiv S(R)^{c[a}{}_{ij} P(B)_c{}^{b]ij} = 0 \,. \tag{8.43}$$

Therefore, we have the off-shell identity

$$\nabla_i \varphi(R)^{abi} = 0 \,, \tag{8.44}$$

which means that 6 components of $\varphi(R)^{abi}$ are not independent of the others. In a similar fashion, we can calculate the following linear combination:

$$\begin{aligned} \nabla_i \varphi(T)^{ai} - \frac{1}{2} e_{bi} \varphi(R)^{abi} = & \\ = & \frac{1-\xi}{2} \varepsilon^{0ijk} \left[ \nabla_i T^a{}_{jk} - R^{ab}{}_{ij} e_{bk} \right] \\ & - \frac{1}{2} S(R)^{ac}{}_{ij} \chi(B)_c{}^{ij} + \frac{1}{4} S(T)_{bij} P(B)^{abij} \,. \end{aligned} \tag{8.45}$$

The term in the square brackets is zero off-shell as a consequence of the first Bianchi identity (B.7). Additionally, the remaining two terms are products of constraints, and therefore also zero off-shell. Thus we have another off-shell identity,

$$\nabla_i \varphi(T)^{ai} - \frac{1}{2} e_{bi} \varphi(R)^{abi} = 0 \,, \tag{8.46}$$

which means that 4 components of $\varphi(T)^{ai}$ are not independent of the others.

Taking (8.44) and (8.46) into account, the number of independent components of the first class constraints is

| $\varphi(B)_{ab}{}^i$ | $\varphi(e)_a$ | $\varphi(\beta)_a{}^i$ | $\varphi(\omega)_{ab}$ | $\varphi(R)^{abi}$ | $\varphi(G)^a$ | $\varphi(T)^{ai}$ | $\varphi(\nabla B)^{ab}$ |
|---|---|---|---|---|---|---|---|
| 18 | 4 | 12 | 6 | $18-6$ | 4 | $12-4$ | 6 |

which gives the total of $F = 70$. Finally, substituting $N$, $F$ and $S$ into (8.41), we obtain:

$$n = 100 - 70 - \frac{60}{2} = 0\,. \tag{8.47}$$

We conclude that the theory has no physical degrees of freedom[3].

As the final point of the analysis, we note that one can introduce the following canonical transformation on the phase space of the theory:

$$\begin{aligned} \pi(\beta)_a{}^{ij} &\;\to\; \tilde{\pi}(\beta)_a{}^{ij} = \pi(\beta)_a{}^{ij} + (1-2\xi)\,\varepsilon^{0ijk} e_{ak}\,, \\ \pi(e)_a{}^{i} &\;\to\; \tilde{\pi}(e)_a{}^{i} = \pi(e)_a{}^{i} + \left(\frac{1}{2}-\xi\right)\varepsilon^{0ijk}\beta_{ajk}\,, \end{aligned} \tag{8.48}$$

while all other fields and momenta map identically onto themselves. It is easy to check that this change of variables is indeed canonical, since it does not change the Poisson structure. Moreover, the Hamiltonian (8.33) and the primary and secondary constraints (8.36) and (8.37) all transform such that

$$\xi \to \tilde{\xi} = 1 - \xi\,. \tag{8.49}$$

This is a symmetry of the action (8.9) up to the boundary term, since

$$S[1-\xi] = S[\xi] - \int_{\partial\mathcal{M}} e^a \wedge \beta_a\,. \tag{8.50}$$

At the level of the full phase space, the canonical transformation (8.48) therefore maps between $\xi$ and $1-\xi$, in particular between the *BFCG* theory ($\xi = 1$) and the TPGT theory ($\xi = 0$). Nevertheless, after the elimination of the second class constraints and the phase space reduction, the situation will be more complicated, as we shall see in section 8.5. The canonical transformation between the *BFCG* and TPGT will still exist, but it will be singular in a certain sense, and not expressible in the generic form (8.48). This will be discussed in detail in section 8.5.

The gauge generator for the *BFCG* theory specialized to the Poincaré 2-group is given by,

$$\begin{aligned} G \;=\; & \int d^3x\, \Big( (\nabla_0 \varepsilon^{ab}{}_i)\varphi(B)_{ab}{}^i + (\nabla_0\varepsilon^{ab})\varphi(\omega)_{ab} + (\nabla_0\varepsilon^a{}_i)\varphi(\beta)_a{}^{0i} + \\ +\; & (\nabla_0\varepsilon^a)\varphi(e)_a + \varepsilon_{abi}\varphi(R)^{abi} + \varepsilon_{ab}\varphi(\nabla B)^{ab} + \varepsilon_a\varphi(G)^a + \\ +\; & \varepsilon_a{}^i\varphi(T)^a{}_i + \varepsilon^a e^b{}_0\varphi(B)_{ab}{}^{0i} + \varepsilon^{ab}B_{[a|c|\,0i}\varphi(B)^c{}_{b]}{}^{0i} + \\ *\; & \varepsilon^{ab}\beta_{[a\,0i}\varphi(\beta)_{b]}{}^{0i} \Big) \end{aligned} \tag{8.51}$$

[3]Like in the last chapter, we stress that we have not provided an explicit proof of the independence of the constraints we found in this analysis.

And the gauge transformations can be found from (8.51) using (8.36) and (3.69). They are,

$$
\begin{array}{lcl}
\delta_0 \omega^{ab}{}_0 & = & \nabla_0 \varepsilon^{ab} \\
\delta_0 \omega^{ab}{}_i & = & -\nabla_i \varepsilon^{ab} \\
\delta_0 B^{ab}{}_{0i} & = & \nabla_0 \varepsilon^{ab}{}_i \\
\delta_0 B^{ab}{}_{ij} & = & \nabla_{[i} \varepsilon^{ab}{}_{j]} + \varepsilon^{[a|c|} B_c{}^{b]}{}_{ij} + \varepsilon^{[a} \beta^{b]}{}_{ij} + \varepsilon^{[a}{}_{[i} e^{b]}{}_{j]} \\
\delta_0 e^a{}_0 & = & \nabla_0 \varepsilon^a \\
\delta_0 \omega^a{}_i & = & -\nabla_i \varepsilon^a + \varepsilon^{ab} e_{b\,i} \\
\delta_0 \beta^a{}_{0i} & = & \nabla_0 \varepsilon^a{}_i \\
\delta_0 \beta^a{}_{ij} & = & \nabla_{[i} \varepsilon^a{}_{j]} + \varepsilon^{ab} \beta_{b\,ij} \\
 & & \\
\delta_0 \pi(\omega)_{ab}{}^0 & = & 0 \\
\delta_0 \pi(\omega)_{ab}{}^i & = & \varepsilon^{ijk} \nabla_j \varepsilon_{abk} + 4\pi(B)_{[a}{}^{cik} \varepsilon_{|c|b]k} + 2\pi(\omega)_{[a|c|}{}^i \varepsilon^c{}_{b]} \\
 & + & 2\pi(e)_{[a}{}^i \varepsilon_{b]} + 2\pi(\beta)_{[a}{}^{ij} \varepsilon_{b]j} \\
\delta_0 \pi(B)_{ab}{}^{0i} & = & 0 \\
\delta_0 \pi(B)_{ab}{}^{ij} & = & -2\pi(B)_{[a|c|}{}^{ij} \varepsilon^c{}_{b]} \\
\delta_0 \pi(e)_a{}^0 & = & 0 \\
\delta_0 \pi(e)_a{}^i & = & \dfrac{1-\xi}{2} \varepsilon^{0ijk} \nabla_i \varepsilon_{ak} + \dfrac{1}{2} \pi(B)_a{}^{b\,ij} \varepsilon_{b\,j} + \varepsilon_{ab} \pi(e)^{b\,i} \\
\delta_0 \pi(\beta)_a{}^{0i} & = & 0 \\
\delta_0 \pi(\beta)_a{}^{ij} & = & \dfrac{\xi}{6} \varepsilon^{0ijk} \nabla_i \varepsilon_a - \varepsilon_b \pi(B)_a{}^{bij} + \varepsilon_{ab} \pi(\beta)^{b\,ij} \\
, . & &
\end{array}
\tag{8.52}
$$

These transformation are an extension of (6.39) (6.40) (6.45) (6.46) for the case of the Poincaré 2-group. Thus we have a very close parallel to the *BF* case (see section 3.14 and [78]).

## 8.4 Dirac brackets

After the Hamiltonian analysis has been completed, we proceed to eliminate the second class constraints from the theory. This is done by introducing the Dirac brackets, defined as:

$$
\begin{aligned}
\{\, F(t,x)\,,\, G(t,x')\,\}_D &= \{\, F(t,x)\,,\, G(t,x')\,\} - \\
&- \int_\Sigma d^3y \int_\Sigma d^3y' \{\, F(t,x)\,,\, \chi^A(t,y)\,\} \Delta^{-1}_{AB}(t,y,y') \{\, \chi^B(t,y')\,,\, G(t,x')\,\}\,,
\end{aligned}
\tag{8.53}
$$

where $F$ and $G$ are some functions of the phase space variables, while the kernel $\Delta^{-1}_{AB}(t,y,y')$ is the inverse of

$$\Delta^{AB}(t,y,y') \equiv \{\, \chi^A(t,y)\,,\, \chi^B(t,y')\,\}\,. \tag{8.54}$$

The multi-indices $A$ and $B$ count all 60 independent second class constraints.

In order to evaluate the kernel $\Delta^{-1}$ and make the general definition (8.53) more manageable, we proceed in several steps. First, from the Poisson brackets (8.39) we see that $\Delta^{AB}(t,x,y)$ is diagonal in the space variables $x$ and $y$, i.e. it can be written as

$$\Delta^{AB}(t,x,y) = \Delta^{AB}(t)\delta^{(3)}(x-y)\,. \tag{8.55}$$

That means that its inverse will also be diagonal in those variables,

$$\Delta^{-1}_{AB}(t,y,y') = \Delta^{-1}_{AB}(t)\delta^{(3)}(y-y')\,, \tag{8.56}$$

so that

$$\int_\Sigma d^3y\; \Delta^{AB}(t,x,y)\Delta^{-1}_{BC}(t,y,y') = \delta^A_C\delta^{(3)}(x-y')\,, \tag{8.57}$$

provided that

$$\Delta^{AB}(t)\Delta^{-1}_{BC}(t) = \delta^A_C\,. \tag{8.58}$$

From now on we will drop the explicit dependence of time from the notation of these matrices, for convenience. Substituting (8.56) into (8.53) and integrating over $y'$, the Dirac brackets can be written in a simpler form

$$\{\,F\,,\,G\}_D = \{\,F\,,\,G\,\} - \int_\Sigma d^3y\{\,F\,,\,\chi^A(y)\,\}\Delta^{-1}_{AB}\{\,\chi^B(y)\,,\,G\,\}\,, \tag{8.59}$$

where we have again simplified the notation by implicitly assuming appropriate spacetime dependence of variables $F$ and $G$.

As a second step, if we rewrite $\chi^A$ and $\chi^B$ as quadruples

$$\begin{array}{rcl} \chi^A &=& \left(\chi(B)^{abij}, \chi(\omega)^{abi}, \chi(e)^{ai}, \chi(\beta)^{aij}\right)\,, \\ \chi^B &=& \left(\chi(B)^{cdmn}, \chi(\omega)^{cdm}, \chi(e)^{cm}, \chi(\beta)^{cmn}\right)\,, \end{array} \tag{8.60}$$

we can write the matrix $\Delta^{AB}$ in the block-diagonal form,

$$\Delta^{AB} = \left(\begin{array}{cc|cc} & \Delta^{abij|cdm} & & \\ \Delta^{abi|cdmn} & & & \\ \hline & & & \Delta^{ai|cmn} \\ & & \Delta^{aij|cm} & \end{array}\right)\,, \tag{8.61}$$

where we have used vertical bars to separate row from column indices, and the blank entries in the matrix are assumed to be zero by convention. According to (8.39) we have

$$\begin{array}{rcl} \Delta^{abij|cdm} & = & 4\varepsilon^{0mij}\eta^{a[c}\eta^{d]b} \,, \\ \Delta^{abi|cdmn} & = & -4\varepsilon^{0imn}\eta^{a[c}\eta^{d]b} \,, \\ \Delta^{ai|cmn} & = & -\varepsilon^{0mni}\eta^{ac} \,, \\ \Delta^{aij|cm} & = & \varepsilon^{0ijm}\eta^{ac} \,. \end{array} \tag{8.62}$$

The inverse matrix $\Delta^{-1}_{AB}$ then has a similar form,

$$\Delta^{-1}_{AB} = \left( \begin{array}{cc|cc} & \Delta^{-1}_{abij|cdm} & & \\ \Delta^{-1}_{abi|cdmn} & & & \\ \hline & & & \Delta^{-1}_{ai|cmn} \\ & & \Delta^{-1}_{aij|cm} & \end{array} \right) . \tag{8.63}$$

Using this, from (8.58) one can obtain the equations

$$\begin{array}{rcl} \frac{1}{2}\Delta^{abij|cdm}\Delta^{-1}_{cdm|a'b'i'j'} & = & 4\delta^a_{[a'}\delta^b_{b']}\delta^i_{[i'}\delta^j_{j']} \,, \\ \frac{1}{4}\Delta^{abi|cdmn}\Delta^{-1}_{cdmn|a'b'i'} & = & 2\delta^i_{i'}\delta^a_{[a'}\delta^b_{b']} \,, \\ \frac{1}{2}\Delta^{ai|cmn}\Delta^{-1}_{cmn|a'i'} & = & \delta^a_{a'}\delta^i_{i'} \,, \\ \Delta^{aij|cm}\Delta^{-1}_{cm|a'i'j'} & = & 2\delta^a_{a'}\delta^i_{[i'}\delta^j_{j']} \,, \end{array} \tag{8.64}$$

and then using (8.62) one can solve them to obtain the components of the inverse matrix,

$$\begin{array}{rcl} \Delta^{-1}_{abij|cdm} & = & \varepsilon_{0ijm}\eta_{a[c}\eta_{d]b} \,, \\ \Delta^{-1}_{abi|cdmn} & = & -\varepsilon_{0imn}\eta_{a[c}\eta_{d]b} \,, \\ \Delta^{-1}_{ai|cmn} & = & -\varepsilon_{0imn}\eta_{ac} \,, \\ \Delta^{-1}_{aij|cm} & = & \varepsilon_{0ijm}\eta_{ac} \,. \end{array} \tag{8.65}$$

Here we have defined $\varepsilon_{0123} \equiv -\varepsilon^{0123}$.

Finally, the third step is to substitute the matrix $\Delta^{-1}_{AB}$ into (8.59) in order to obtain an explicit expression for the Dirac brackets:

$$\{F, G\}_D = \{F, G\} - \frac{1}{2}\varepsilon_{0ijk}\int_\Sigma d^3y\, K^{ijk}(F,G,y) \,, \tag{8.66}$$

where the kernel $K^{ijk}(F,G,y)$ is

$$
\begin{aligned}
K^{ijk}(F,G,y) \;=\;& \frac{1}{4}\{\, F\,,\, \chi(B)^{abij}(y)\,\}\{\, \chi(\omega)_{ab}{}^{k}(y)\,,\, G\,\} \\
& -\frac{1}{4}\{\, F\,,\, \chi(\omega)^{abi}(y)\,\}\{\, \chi(B)_{ab}{}^{jk}(y)\,,\, G\,\} \\
& -\{\, F\,,\, \chi(e)^{ai}(y)\,\}\{\, \chi(\beta)_{a}{}^{jk}(y)\,,\, G\,\} \\
& +\{\, F\,,\, \chi(\beta)^{aij}(y)\,\}\{\, \chi(e)_{a}{}^{k}(y)\,,\, G\,\}\,.
\end{aligned}
\tag{8.67}
$$

Having constructed the Dirac brackets, the next task is to express the constraint algebra in terms of them. This has two main consequences. The first is that the Dirac bracket between any quantity and any second class constraint is automatically zero, by construction. This is obvious from the definition (8.53). The second is that after passing from Poisson brackets to Dirac brackets, the second class constraints can be set equal to zero off-shell, giving rise to the reduction of the phase space to one of its hypersurfaces. We will now concentrate on the constraint algebra, while the reduction of the phase space will be discussed in detail in the next section.

Looking at the algebra of constraints, (8.38), (8.39) and (8.40), we see that it has the following rough structure:

$$
\{\, \varphi\,,\, \varphi\,\} = \varphi\,, \qquad \{\, \chi\,,\, \chi\,\} = \Delta\,, \qquad \{\, \varphi\,,\, \chi\,\} = \chi\,, \tag{8.68}
$$

for non-zero brackets. With all other brackets are zero,

$$
\{\, \varphi\,,\, \varphi\,\} = 0\,, \qquad \{\, \chi\,,\, \chi\,\} = 0\,, \qquad \{\, \varphi\,,\, \chi\,\} = 0\,. \tag{8.69}
$$

Knowing that the Dirac brackets between an arbitrary quantity and a second class constraint is zero by construction, we can immediately conclude that

$$
\{\, \chi\,,\, \chi\,\}_D = 0\,, \qquad \{\, \varphi\,,\, \chi\,\}_D = 0\,, \tag{8.70}
$$

which leaves only $\{\, \varphi\,,\, \varphi\,\}_D$ to be discussed. For this, a schematic calculation gives:

$$
\begin{aligned}
\{\, \varphi\,,\, \varphi\,\}_D \;&=\; \{\, \varphi\,,\, \varphi\,\} - \int \{\, \varphi\,,\, \chi\,\}\Delta^{-1}\{\, \chi\,,\, \varphi\,\} \\
&=\; \{\, \varphi\,,\, \varphi\,\} + \int \chi\Delta^{-1}\chi \\
&=\; \{\, \varphi\,,\, \varphi\,\}\,,
\end{aligned}
\tag{8.71}
$$

due to the fact that the product of two constraints is zero off-shell. This is actually a proof that the algebra of primary constraints does not change when we pass from Poisson brackets to Dirac brackets. Therefore, we have:

$$
\begin{aligned}
\{\,\varphi(G)^a\,,\,\varphi(T)^{bi}\,\}_D &= -\varphi(R)^{abi}\delta^{(3)}\,,\\
\{\,\varphi(G)^a\,,\,\varphi(\nabla B)_{cd}\,\}_D &= 2\delta^a_{[c}\varphi(G)_{d]}\delta^{(3)}\,,\\
\{\,\varphi(T)^{ai}\,,\,\varphi(\nabla B)_{cd}\,\}_D &= 2\delta^a_{[c}\varphi(T)_{d]}{}^i\delta^{(3)}\,,\\
\{\,\varphi(R)^{abi}\,,\,\varphi(\nabla B)_{cd}\,\}_D &= -4\delta^{[a}_{[c}\varphi(R)^{b]}{}_{d]}{}^i\delta^{(3)}\,,\\
\{\,\varphi(\nabla B)^{ab}\,,\,\varphi(\nabla B)_{cd}\,\}_D &= -4\delta^{[a}_{[c}\varphi(\nabla B)^{b]}{}_{d]}\delta^{(3)}\,,
\end{aligned}
\tag{8.72}
$$

while all other $\{\,\varphi\,,\,\varphi\,\}_D$ vanish.

As a final point, note also that for an arbitrary quantity $A$ we have:

$$
\begin{aligned}
\{A\,,\,H_T\}_D &= \{A\,,\,H_T\} - \int\{A\,,\,\chi\}\Delta^{-1}\{\chi\,,\,H_T\}\\
&= \{A\,,\,H_T\} - \int\{A\,,\,\chi\}\Delta^{-1}\{\chi\,,\,\varphi\}\\
&= \{A\,,\,H_T\} - \int\{A\,,\,\chi\}\Delta^{-1}\chi\,,
\end{aligned}
\tag{8.73}
$$

where we have used the fact that the total Hamiltonian (8.33) is a linear combination of first class constraints. The result can be rewritten as

$$
\dot{A} = \{A\,,\,H_T\}_D + \int\{A\,,\,\chi\}\Delta^{-1}\chi\,,
\tag{8.74}
$$

which becomes the standard-looking equation of motion

$$
\dot{A} = \{A\,,\,H_T\}_D
\tag{8.75}
$$

when one reduces the phase space by promoting $\chi \approx 0$ to off-shell equalities $\chi = 0$. This reduction is the subject of the next section.

## 8.5 Phase space reduction

The purpose of introducing Dirac brackets is to remove the second class constraints from the theory. When we use exclusively Dirac brackets, no result depends on second class constraints, and we can project all phase space points to the hypersurface defined by strong equalities $\chi = 0$, reducing its dimension[4] from $2N$ to $2N - S$, without changing any physical property of the theory, in particular without breaking its gauge symmetry. In our case, from (8.37) we have the following off-shell equations:

$$\begin{aligned} \pi(B)_{ab}{}^{jk} &= 0\,, \\ \pi(\omega)_{ab}{}^{i} - \varepsilon^{0ijk} B_{abjk} &= 0\,, \\ \pi(e)_{a}{}^{i} - \frac{1-\xi}{2}\varepsilon^{0ijk}\beta_{ajk} &= 0\,, \\ \pi(\beta)_{a}{}^{ij} + \xi\varepsilon^{0ijk} e_{ak} &= 0\,. \end{aligned} \tag{8.76}$$

We will discuss these equations in two steps, first by analyzing the two equations independent of $\xi$, and then discussing the $\xi$-dependent equations, which are more complicated. The first two equations can be rewritten as:

$$\pi(B)_{ab}{}^{ij} = 0\,, \qquad B^{ab}{}_{ij} = -\frac{1}{2}\varepsilon_{0ijk}\pi(\omega)^{abk}\,. \tag{8.77}$$

Note that we have expressed two conjugate variables in terms of other variables in the phase space. This reduces the full phase space to a hypersurface defined by these equations, namely orthogonal to the directions of $\pi(B)_{ab}{}^{ij}$ and diagonal in the directions of corresponding $(B, \pi(\omega))$ planes. Given that we have eliminated 36 phase space variables *per point* (see footnote 4), the dimension of the hypersurface is $200 - 36 = 164$, since the dimension of the full phase space was $2N = 200$. On this hypersurface the expressions for some of the first class constraints (8.36) simplify. In particular, the first four constraints remain unaffected, while the final four constraints become:

$$\begin{aligned} \varphi(R)^{abi} &= \varepsilon^{0ijk} R^{ab}{}_{jk}\,, \\ \varphi(G)^{a} &= \frac{\xi}{6}\varepsilon^{0ijk} G^{a}{}_{ijk} + \nabla_i \pi(e)^{ai}\,, \\ \varphi(T)^{ai} &= \frac{1-\xi}{2}\varepsilon^{0ijk} T^{a}{}_{jk} - \nabla_j \pi(\beta)^{aij}\,, \\ \varphi(\nabla B)^{ab} &= \nabla_i \pi(\omega)^{abi} - 2e^{[a}{}_{i}\pi(e)^{b]i} - \beta^{[a}{}_{ij}\pi(\beta)^{b]ij}\,. \end{aligned} \tag{8.78}$$

Let us now turn to $\xi$-dependent equations in (8.76). We immediately see that there are two mutually incompatible cases, namely $\xi \neq 0$ and $\xi \neq 1$. In the $\xi \neq 0$

[4]Strictly speaking, the phase space if of infinite dimension. In this section, we mean dimension *per space point*.

case, we see that we can solve the equations for $e$ and $\pi(e)$,

$$e^a{}_i = \frac{1}{2\xi}\varepsilon_{0ijk}\pi(\beta)^{ajk}\,, \qquad \pi(e)_a{}^i = \frac{1-\xi}{2}\varepsilon^{0ijk}\beta_{ajk}\,, \tag{8.79}$$

again expressing two conjugate variables in terms of remaining ones. This reduces the dimension of the (*local*) hypersurface even further, down to $164 - 24 = 140$. If one does not impose any gauge fixing in the theory, this is the minimal dimension of the hypersurface, since we have in total $S = 60$ second class constraints. The final three first class constraints simplify even further, as follows:

$$\begin{aligned} \varphi(G)^a &= \frac{1}{6}\varepsilon^{0ijk}G^a{}_{ijk}\,, \\ \varphi(T)^{ai} &= -\frac{1}{\xi}\nabla_j\pi(\beta)^{aij}\,, \\ \varphi(\nabla B)^{ab} &= \nabla_i\pi(\omega)^{abi} - \frac{1}{\xi}\beta^{[a}{}_{ij}\pi(\beta)^{b]ij}\,. \end{aligned} \tag{8.80}$$

We should stress that none of these equations make sense in the case $\xi = 0$, i.e. for the topological Poincaré gauge theory (8.8), but are completely valid for the case $\xi = 1$, which represents the *BFCG* theory (8.4).

Alternatively, in the $\xi \neq 1$ case, it is not a good idea to solve (8.76) for $e, \pi(e)$, since this cannot be done if $\xi = 0$. Instead, we can solve for $\beta$ and $\pi(\beta)$,

$$\beta^a{}_{ij} = \frac{1}{\xi - 1}\varepsilon_{0ijk}\pi(e)^{ak}\,, \qquad \pi(\beta)_a{}^{ij} = -\xi\varepsilon^{0ijk}e_{ak}\,. \tag{8.81}$$

This time the phase space reduces to another *local* hypersurface, different from the previous one, but again of the same dimension $164 - 24 = 140$. The final three first class constraints simplify again, however not to (8.80), but to:

$$\begin{aligned} \varphi(G)^a &= \frac{1}{1-\xi}\nabla_i\pi(e)^{ai}\,, \\ \varphi(T)^{ai} &= \frac{1}{2}\varepsilon^{0ijk}T^a{}_{jk}\,, \\ \varphi(\nabla B)^{ab} &= \nabla_i\pi(\omega)^{abi} - \frac{2}{1-\xi}e^{[a}{}_i\pi(e)^{b]i}\,. \end{aligned} \tag{8.82}$$

In this case the choice $\xi = 1$ does not make sense, which means that the *BFCG* theory case is excluded. Nevertheless, the case $\xi = 0$ is included, describing topological Poincaré gauge theory.

It is interesting to ask what happens in the case of generic $\xi$, when it is neither zero nor one. In that case one can solve (8.76) either for $(e, \pi(e))$ or for $(\beta, \pi(\beta))$.

The constraints can be expressed in either form (8.80) or (8.82). It is important to note, though, that the resulting hypersurface depends on the choice of $\xi$. In this case one can calculate the Dirac brackets

$$\{ e^a{}_i \,,\, \pi(e)_b{}^j \}_D = (1-\xi)\delta^a_b \delta^j_i \delta^{(3)} \tag{8.83}$$

and

$$\{ \beta^a{}_{ij} \,,\, \pi(\beta)_b{}^{mn} \}_D = 2\xi \delta^a_b \delta^m_{[i} \delta^n_{j]} \delta^{(3)} \,. \tag{8.84}$$

It is then easy to verify that (8.81) is a canonical transformation from $(e, \pi(e))$ to $(\beta, \pi(\beta))$, with (8.79) being the inverse transformation. In particular, as long as $\xi \neq 0, 1$, this transformation maps (8.82) to (8.80), and in addition maps (8.83) to (8.84), justifying its canonical nature.

However, the cases $\xi = 0$ and $\xi = 1$ are singular, and the canonical transformation (8.81), (8.79) does not make sense for either of those. Nevertheless, there exists a singular canonical transformation which maps between those two cases (see [113]), given as:

$$\beta^a{}_{ij} = -\varepsilon_{0ijk} \pi(e)^{ak} \,, \qquad \pi(\beta)_a{}^{ij} = -\varepsilon^{0ijk} e_{ak} \,. \tag{8.85}$$

In particular, for the case $\xi = 0$ the Dirac bracket (8.83) evaluates to

$$\{ e^a{}_i \,,\, \pi(e)_b{}^j \}_D = \delta^a_b \delta^j_i \delta^{(3)} \,, \tag{8.86}$$

while the constraints (8.82) become

$$\begin{aligned} \varphi(G)^a &= \nabla_i \pi(e)^{ai} \,, \\ \varphi(T)^{ai} &= \frac{1}{2} \varepsilon^{0ijk} T^a{}_{jk} \,, \\ \varphi(\nabla B)^{ab} &= \nabla_i \pi(\omega)^{abi} - 2e^{[a}{}_i \pi(e)^{b]i} \,. \end{aligned} \tag{8.87}$$

On the other hand, when $\xi = 1$ we have from (8.84)

$$\{ \beta^a{}_{ij} \,,\, \pi(\beta)_b{}^{mn} \}_D = 2 \delta^a_b \delta^m_{[i} \delta^n_{j]} \delta^{(3)} \,, \tag{8.88}$$

and from (8.80) we obtain

$$\begin{aligned} \varphi(G)^a &= \frac{1}{6} \varepsilon^{0ijk} G^a{}_{ijk} \,, \\ \varphi(T)^{ai} &= -\nabla_j \pi(\beta)^{aij} \,, \\ \varphi(\nabla B)^{ab} &= \nabla_i \pi(\omega)^{abi} - \beta^{[a}{}_{ij} \pi(\beta)^{b]ij} \,. \end{aligned} \tag{8.89}$$

The transformation (8.85) then maps (8.86) to (8.88) and (8.87) to (8.89), with its inverse mapping everything the other way around.

To sum up, we have the following general situation. For a generic $\xi$ one can write the theory using either $(e, \pi(e))$ variables or $(\beta, \pi(\beta))$ variables, and there is a canonical transformation (8.81) connecting these two sets of variables, for the same value of $\xi$. For the singular cases $\xi = 0$ and $\xi = 1$ one does not have a choice which variables to use, but there exists the canonical transformation (8.85) which maps the $\xi = 0$ theory into the $\xi = 1$ theory, and vice versa. This canonical transformation is called singular because it cannot be obtained as a solution of the second class constraints (8.76). In contrast to (8.81), which maps between various variables on the same *local* hypersurface determined by the choice of $\xi$, the singular canonical transformation maps between two different hypersurfaces determined by choices $\xi = 0$ and $\xi = 1$. This establishes the relationship between the canonical structure of the *BFCG* model and the canonical structure of the topological Poincaré gauge theory.

We should also discuss the status of the canonical transformation (8.48), which maps the full phase space onto itself, while inducing the transformation $\xi \to 1-\xi$. Note that (8.48) assumes that both momenta $\pi(e)_a{}^i$ and $\pi(\beta)_a{}^{ij}$ are variables in the phase space. However, after the reduction of the phase space using either (8.79) or (8.81), one of the two momenta is eliminated, and is not a variable in the reduced phase space. Therefore, one cannot formulate the canonical transformation (8.48) as it stands, on the reduced phase space. Geometrically, every choice of $\xi$ specifies one particular reduced phase space, and the symmetry $\xi \to 1 - \xi$ now maps between different spaces. While it is possible to construct a set of canonical transformations analogous to (8.48), in the sense of the map $\xi \to 1 - \xi$, these are not derivable from (8.48). In particular, in the case of reduction (8.79), the change of variables

$$\pi(\beta)_a{}^{ij} \to \tilde{\pi}(\beta)_a{}^{ij} = \frac{\xi}{1-\xi}\pi(\beta)_a{}^{ij} \tag{8.90}$$

implements the map $\xi \to 1 - \xi$ in the constraints (8.80). Similarly, in the case of reduction (8.81), the change of variables

$$\pi(e)_a{}^i \to \tilde{\pi}(e)_a{}^i = \frac{1-\xi}{\xi}\pi(e)_a{}^i\,, \tag{8.91}$$

implements the same map in the constraints (8.82). Note that neither (8.90) nor (8.91) can be derived from (8.48), and also that both transformations are defined only for $\xi \neq 0, 1$. Finally, for the case $\xi = 0 \to \xi = 1$ and vice versa, we have the canonical transformation (8.85), which also cannot be inferred from (8.48). We can of course infer the existence of all these canonical transformations from the fact that the map $\xi \to 1-\xi$ is a symmetry of the theory, see (8.50). But the actual forms of the transformations cannot be obtained from each other, due to the different sets of variables in respective phase spaces.

## 8.6 Canonical quantization

Since the equivalence (8.8) holds, we may quantize the *BF* formulation for the Poincaé group (see [113]) of the *BFCG* theory.

Taking a Poincaé connection

$$A(x) = A^I(x)X_I = \omega^{ab}(x)\,J_{ab} + e^a(x)\,P_a\,, \tag{8.92}$$

where $J$ and $P$ satisfy the Poincaré Lie algebra

$$[J_{ab}, J_{cd}] = \eta_{[a|[c}J_{d]|b]}\,, \quad [P_a, J_{bc}] = \eta_{a[b}P_{c]}\,, \quad [P_a, P_b] = 0\,. \tag{8.93}$$

The curvature can be written,

$$F = F^I X_I = \left(dA^I + f^I{}_{JK}A^J \wedge A^K\right)X_I \tag{8.94}$$

$$= R^{ab}J_{ab} + T^a P_a\,. \tag{8.95}$$

where the $R$ and $T$ are given by (8.5) and (8.7) respectively.

The *BFCG* theory may thus be written tin the form,

$$S \cong \int_{\mathcal{M}} \left(B^{ab}\wedge R_{ab} + T^a \wedge \beta_a\right) \tag{8.96}$$

$$= \int_{\mathcal{M}} B^I \wedge F_I\,, \tag{8.97}$$

where,

$$B^I = \left(B^{ab}, \beta^a\right)\,, \quad F_I = \left(R_{ab}, T_a\right)\,. \tag{8.98}$$

The canonical analysis can be performed using the method in Sec. (3.11) (see also [113]). This is also the case $\xi \neq 1$ discussed in the previous section (see equations (8.81) and (8.82) for $\xi = 0$). We repeat the calculation for definiteness.

The Lagrangian density can be written as

$$\mathcal{L} = \pi_{ab}{}^i\,\dot{\omega}^{ab}{}_i + p^i_a\,\dot{e}^a_i - \tilde{\mathcal{H}}\,, \tag{8.99}$$

the momenta are given by,

$$\pi_{ab}{}^{i} = \frac{1}{2}\varepsilon^{ijk} B_{abjk}\,, \quad p_{a}{}^{i} = \frac{1}{2}\varepsilon^{ijk}\beta_{ajk}\,, \tag{8.100}$$

and $\tilde{\mathcal{H}}$ by

$$\begin{aligned} \tilde{\mathcal{H}} &= -\Bigg[\frac{1}{2}\varepsilon^{ijk} B^{ab}{}_{0i} R_{abjk} + e^{a}{}_{0}\nabla_{i} p_{a}{}^{i} + \\ &+ \omega^{ab}{}_{0}\left(\nabla_{i}\pi_{ab}{}^{i} - e_{[a|i}p_{b]}{}^{i}\right) + \frac{1}{2}\varepsilon^{ijk}\beta_{a0i}T^{a}{}_{jk}\Bigg]\,. \end{aligned} \tag{8.101}$$

we therefore find the following constraints

$$\tilde{\mathcal{C}}_{1}{}^{abi} \equiv \frac{1}{2}\varepsilon^{ijk} R^{ab}{}_{jk} = 0, \tag{8.102}$$

$$\tilde{\mathcal{C}}_{2}{}^{ai} \equiv \frac{1}{2}\varepsilon^{ijk} T^{a}{}_{jk} = 0 \tag{8.103}$$

$$\tilde{\mathcal{G}}_{1\,ab} \equiv \nabla_{i}\pi_{ab}{}^{i} - e_{[a|i}p_{b]}{}^{i} = 0\,, \tag{8.104}$$

$$\tilde{\mathcal{G}}_{2\,a} \equiv \nabla_{i} p_{a}{}^{i} = 0\,. \tag{8.105}$$

and the constraint algebra is

$$\begin{aligned} \left\{\tilde{\mathcal{C}}_{2}{}^{a}(\vec{x})\,,\tilde{\mathcal{G}}_{2\,b}{}^{i}(\vec{y})\right\} &= -4\tilde{\mathcal{C}}_{1}{}^{a}{}_{b}{}^{i}\delta^{3}(x-y) \\ \left\{\tilde{\mathcal{C}}_{2}{}^{a}(\vec{x})\,,\mathcal{G}_{1\,cd}(\vec{y})\right\} &= \delta^{a}{}_{[c}\tilde{\mathcal{C}}_{2d]}\delta^{3}(x-y) \\ \left\{\tilde{\mathcal{C}}_{1}{}^{abi}(\vec{x})\,,\tilde{\mathcal{G}}_{1\,cd}(\vec{y})\right\} &= -4\delta^{[a}{}_{[c}\tilde{\mathcal{C}}_{1}{}^{b]}{}_{d]}{}^{i}\delta^{3}(x-y) \\ \left\{\tilde{\mathcal{G}}_{1\,ab}(\vec{x})\,,\tilde{\mathcal{G}}_{1}{}^{cd}(\vec{y})\right\} &= 4\delta^{[c}{}_{[a}\tilde{\mathcal{G}}_{1}{}^{d]}{}_{b]}\delta^{3}(x-y)\,, \\ \left\{\tilde{\mathcal{G}}_{1\,ab}(\vec{x})\,,\tilde{\mathcal{G}}_{2}{}^{c}(\vec{y})\right\} &= -\delta^{c}{}_{[a}\tilde{\mathcal{G}}_{2\ \,b]}\delta^{3}(x-y)\,. \end{aligned} \tag{8.106}$$

Given a set of canonical variables $\{(p_k, q_k)|\, k \in K\}$, one can define a quantization based on a representation of the corresponding Heisenberg algebra in the Hilbert space $\mathcal{H}_0 = L_2\left(\mathbb{R}^{|K|}\right)$ such that

$$\hat{p}_k\,\Psi(q) = i\,\frac{\partial\Psi(q)}{\partial q^k}\,, \quad \hat{q}_k\,\Psi(q) = q_k\,\Psi(q)\,. \tag{8.107}$$

We will refer to the representation (8.107) as the quantization in the $q$ basis. The results of the previous sections imply that the canonical quantization of the Poincaré *BFCG* theory in the 2-connection basis $(\omega, \beta)$, can be related to the canonical quantization of the Poncarè *BF* theory in the $(\omega, e)$ basis. Since $\beta$ is canonically conjugate to $e$, by performing a functional Fourirer transform, we obtain

$$\Psi(\omega,\beta) = \int \mathcal{D}e\,\Phi(\omega, e)\exp\left(i\int_{\Sigma}\beta^{a}\wedge e_{a}\right)\,. \tag{8.108}$$

Furthermore, $\Phi(\omega, e) \equiv \Phi(A)$ is a solution of a quantum version of the Poincarè BF constraints. For any *BF* theory, the canonical pair $(A^I{}_i, E_I{}^i)$ can be represented by the operators

$$\hat{E}_I{}^i(x)\,\Phi(A) = i\frac{\delta\Phi}{\delta A^I{}_i(x)}\,, \quad \hat{A}^I{}_i(x)\,\Phi(A) = A^I{}_i(x)\,\Phi(A)\,, \tag{8.109}$$

so that the Gauss constraint

$$\hat{G}_I\,\Phi(A) = \partial_i\left(\frac{\delta\Phi}{\delta A^I{}_i(x)}\right) + f_{IJ}{}^K A^J{}_i(x)\,\frac{\delta\Phi}{\delta A^K{}_i(x)} = 0\,, \tag{8.110}$$

is equivalent to

$$\Phi(A) = \Phi(\tilde{A}) \tag{8.111}$$

where $\tilde{A} = A + d\lambda + [A, \lambda]$ is the infinitesimal gauge-transform of $A$. This implies that $\Phi(A)$ must be a gauge-invariant functional, while the vanishing curvature constraint

$$F(A(x))\Phi(A) = 0 \tag{8.112}$$

implies

$$\Phi(A) = \prod_x \delta(F_x)\varphi(A)\,, \tag{8.113}$$

i.e. $\Phi(A)$ has a non-zero support on flat connections.

Consequently any gauge-invariant functional of flat Poincaré connections on $\Sigma$, $\Phi(\omega_0, e_0)$, is a solution. The space of $\Phi(\omega_0, e_0)$, which we denote as $\mathcal{H}_0$, is the space of functions on the moduli space of flat connections on $\Sigma$ for the Poincarè group $ISO(1,3)$, which we denote as $MS(ISO(3,1))$. It is easy to see that

$$MS(ISO(3,1)) = VB[MS(SO(3,1))]\,, \tag{8.114}$$

where VB is the vector bundle such that the fibre at a point $\omega_0$ of $MS(SO(3,1))$ is the solution space of the vanishing torsion $de_0 + \omega_0 \wedge e_0 = 0$.

In $\mathcal{H}_0$ we can introduce a basis of spin-network wavefunctions. Let $A$ be a connection for a Lie group $G$ on $\Sigma$, and let $\gamma$ be a graph in $\Sigma$. Given the irreducible representations *irreps* $\Lambda_l$ of $G$ associated to the edges of $\gamma$ and the corresponding intertwiners $\iota_v$ associated to the vertices of $\gamma$, one can construct the spin-network wavefunctions

$$W_{\hat{\gamma}}(A) = Tr\left(\prod_{v\in\gamma} C^{(\iota_v)} \prod_{l\in\gamma} D^{(\Lambda_l)}(A)\right) \equiv \langle A|\hat{\gamma}\rangle\,, \tag{8.115}$$

where $D^{(\Lambda_l)}(A)$ is the holonomy for the line-segment $l$, $C^{(\iota)}$ are the intertwiner coefficients and $\hat{\gamma} = (\gamma, \Lambda, \iota)$ denotes a spin network associated to a graph $\gamma$.

Note that when $A$ is a flat connection, than (8.115) is invariant under a homotopy of the graph $\gamma$, so that we can label the spin-network wavefunctions by combinatorial (abstract) graphs $\gamma$.

In the case of a non-compact group there is a technical difficulty when constructing the spin-network wavefunctions. Namely, if one uses the unitary irreps (UIR), these are infinite-dimensional, and one has to insure that the trace in (8.115) is convergent. In the Poincaré group case, we will consider the massive UIRs, which are labelled by a pair $(M,j)$, where $M > 0$ is the mass and $j \in \mathbf{Z}_+/2$ is an $SU(2)$ spin. In this case

$$D^{(M,j)}_{q,m';p,m}(\omega, a) = e^{i(\Lambda_\omega p)\cdot a}\, D^{(j)}_{m'm}(W(\omega,p))\, \delta^3(\vec{q} - \vec{\Lambda_\omega p})\,, \tag{8.116}$$

where $p = (p_0, \vec{p}) = (\sqrt{(\vec{p})^2 + M^2}, \vec{p})$, $D^{(j)}$ is a spin-$j$ rotation matrix and $W(\omega,p)$ is the Wigner rotation, see [120].

By requiring that $W_{\hat{\gamma}}(A)$ form a basis in $\mathcal{H}_0$, we obtain

$$|\Psi\rangle = \int DA\, |A\rangle\langle A|\Psi\rangle = \sum_{\hat{\gamma}} |\hat{\gamma}\rangle\langle\hat{\gamma}|\Psi\rangle \tag{8.117}$$

and

$$\langle\hat{\gamma}|\Psi\rangle = \int DA\, \langle\hat{\gamma}|A\rangle\langle A|\Psi\rangle = \int DA\, W^*_{\hat{\gamma}}(A)\, \Psi(A)\,. \tag{8.118}$$

The last formula is known as the loop transform.

Since we are dealing with a Lie 2-group, one would like to generalize the spin-network wavefunctions for the case of a 2-connection $(\omega, \beta)$. The categorical nature of a 2-group implies that one can associate 2-group representations to a 2-complex. Namely, if $(\omega, \beta)$ is a 2-connection for a Lie 2-group $(G, H)$ on $\Sigma$, then given a 2-complex $\Gamma$ in $\Sigma$, one can associate the 2-group representations $L_f$ to the faces $f$ of $\Gamma$. The corresponding 1-intertwiners $\Lambda_l$ can be associated to the edges of $\Gamma$, while the corresponding 2-intertwiners $\iota_v$ can be associated to the vertices of $\Gamma$. Hence we obtain a spin foam $\hat{\Gamma} = (\Gamma, L, \Lambda, \iota)$.

For example, in the 2-Poincaré group case, there is a class of representations labelled by a positive number $L$, see [121]. The intertwiners for three such representations, $L_1, L_2, L_3$, are labelled by integers $m$ if $L_k$ satisfy the triangle inequalities strongly. The $m$'s label the *irreps* of an $SO(2)$ group, which leaves the triangle $(L_1, L_2, L_3)$, embedded in $\mathbb{R}^4$, invariant. The 2-intertwiners for the $m$'s are trivial and $L_k$ in this case can be identified with the edge-lengths of a triangle, see [119]. If $L_k$ are collinear, i.e. $L_1 = L_2 + L_3$, the invariance group is $SO(3)$ and the corresponding intertwiners are the $SU(2)$ spins $j$ while the 2-intertwiners are the $SU(2)$ intertwiners. In this case the $L_k$ look like particle masses, but then it is not clear what would be the geometrical interpretation of these masses.

A spin-foam wavefunction should be an appropriate generalization of the spin-network wavefunction (8.115) such that the spin-foam wavefunction includes the surface holonomies associated with the spin-foam faces $f$. Let us embed $\Gamma$ into a triangulation of the spatial manifold and let $\omega$ and $\beta$ be piece-wise constant in the appropriate cells of the triangulation. If $g_l = \exp(\omega_l J)$ and $h_f = \exp(\beta_f P)$, then the formula for the surface holonomy $h_p$ for the surface of a polyhedron $p$, is given by

$$h_p = \prod_{f\in\partial p} g_{l(f)} \triangleright h_f \,, \tag{8.119}$$

where $g_{l(f)}$ can be calculated by representing the $\partial p$ surface as a composition of 2-morphisms $(g_l, h_f)$ from some 1-morphism $g_{l'}$ $(l' \in p)$ to itself, see [108] for the case of a tetrahedron.

Hence we expect that

$$W_{\hat\Gamma}(\omega,\beta) = Tr\left(\prod_{v\in\Gamma} C^{(\iota_v)} \prod_{l\in\Gamma} D^{(\Lambda_l)}(\omega) \prod_{f\in\Gamma} D^{(L_f)}(\omega,\beta)\right) \,, \tag{8.120}$$

where

$$D^{(L_f)}(\omega,\beta) = D^{(L_f)}(g_{l(f)} \triangleright h_f) \,. \tag{8.121}$$

In the Poincare 2-group case, the representation matrix (8.121) is of the type $1\times 1$, because $H$ is an abelian group. The analysis in [119] suggests that

$$D^{(L_f)}(\omega,\beta) = \exp\left(i\vec{L}_f \cdot g(\omega_l)\vec{\beta}_f\right) \,, \tag{8.122}$$

where $\vec{L}_f$ is a 4-vector satisfying $L_f^2 = \vec{L}_f \cdot \vec{L}_f = \eta_{ab} L_f^a L_f^b$ and $\eta$ is a flat Minkowski metric.

A related problem consists in the fact that no analogue of the Peter-Weyl (PW) theorem is known for the case of a 2-group. The PW theorem states that a basis on the Hilbert space of square integrable functions on a Lie group $G$ is given by the matrix elements of the irreducible representations of the group (see [12]), that is

$$\varphi(g) = \sum_{\Lambda} \sum_{\alpha_\Lambda,\beta_\lambda} \tilde\varphi_\Lambda^{\alpha_\Lambda\beta_\lambda} D^{(\Lambda)}_{\alpha_\Lambda\beta_\lambda}(g) = \sum_{\Lambda} \langle \tilde\Phi_\Lambda \,, D^{(\Lambda)}(g)\rangle \,, \tag{8.123}$$

where $\varphi$ is a function on a Lie group $G$ and

$$\tilde{\varphi}_{\Lambda}^{\alpha_\Lambda \beta_\lambda} = \int_G dg\, \bar{D}^{(\Lambda)}_{\alpha_\Lambda \beta_\lambda}(g)\, \varphi(g)\,. \tag{8.124}$$

Note that in the case of the Poincaré 2-group, the relation (8.108) can give some clues. Let us consider again piece-wise constant fields on a triangulated manifold. The Poincaré group holonomy for an edge $\varepsilon$ is given by $g'_\varepsilon = \exp(\omega_\varepsilon J + e_\varepsilon P)$, so that a function $\varphi(g'_\varepsilon) = \Phi(\omega_\varepsilon, e_\varepsilon)$ can be expanded by using the generalization of the PW theorem for the Poincaré group

$$\Phi(\omega_\varepsilon, e_\varepsilon) = \int_0^\infty dM \sum_j \langle \tilde{\Phi}_{M,j}\,,\, D^{(M,j)}(\omega_\varepsilon, e_\varepsilon)\rangle\,. \tag{8.125}$$

Consequently

$$\begin{aligned}
\Psi(\omega_\varepsilon, \beta_f) &= \int_{\mathbf{R}^4} d^4 e_l\, \mu(e_\varepsilon)\, e^{i\vec{\beta}_f \cdot \vec{e}_\varepsilon}\, \Phi(\omega_\varepsilon, e_\varepsilon) \\
&= \int_0^\infty dM \sum_j \langle \tilde{\Phi}_{M,j}\,,\, \int_{\mathbf{R}^4} d^4 e_\varepsilon\, \mu(e_\varepsilon)\, e^{i\vec{\beta}_f \cdot \vec{e}_\varepsilon}\, D^{(M,j)}(\omega_\varepsilon, e_\varepsilon)\rangle \\
&= \int_0^\infty dM \sum_j \langle \tilde{\Phi}_{M,j}\,,\, \tilde{D}^{(M,j)}(\omega_\varepsilon, \beta_f)\rangle\,,
\end{aligned} \tag{8.126}$$

where $\mu$ is some appropriately chosen measure and $f$ is the face dual to an edge $\varepsilon$.

If $\Gamma$ is a tetrahedron, then by comparing (8.126) to (8.120) one concludes that $\Lambda_\varepsilon = j_\varepsilon$ and that there should be a relationship between an $L_\Delta$ and the three $M_f$ for the dual faces for the edges of a triangle $\Delta$. Furthermore, there should be a relationship between the functions $D^{(j_\varepsilon)}(g_\varepsilon) D^{(L_\Delta)}(g_{\varepsilon'}, h_\Delta)$ on $\Gamma$ and the functions $\tilde{D}^{(M_f, j_\varepsilon)}(g'_\varepsilon)$ on $\Gamma$.

# 9

# Summary of results and conclusions

In this chapter we summarize the results and present our conclusions.

In Part I of this thesis, we discussed the motivation to carry out research in the area of quantum gravity. This motivation comes either form conceptual issues — three of the four known interaction have been quantized, but gravity has not — as well as from black hole singularities and the primordial singularity. One expects that quantum gravitational effects will be important for the description of singularities. We also explore some key methods and tools that form the mathematical language of gauge theories. The canonical analysis of such theories, using the Dirac procedure, was also reviewed.

Canonical quantization, uses the Hamiltonian structure of General Relativity, which was also the object of a chapter in this thesis. The quantization of Einstein's theory, was originally studied using the metric and its canonically conjugate momentum (or alternatively the tetrad and its conjugate momentum). This approach, does not lead to a complete quantum theory due, in particular, to the complicated non-polynomial character of the Hamiltonian constraint.

This motivated the search for alternative variables that would make the Hamiltonian constraint polynomial. The complex (self-dual) Ashtekar variables do, in fact, have this property but they introduce a new constraint — the reality constraint — that is equally problematic. The real version of these variables obviously avoids the reality constraint but fails to make the Hamiltonian constraint polynomial.

In Loop Quantum Gravity (LQG),a pair of variables — namely the holonomies and the (appropriately modified) fluxes based on Ashtekar's (real) variables — are quantized. Although LQG is a consistent well established theory, whose build-ing blocks are *spin networks* (or *s-knots*), some problems remain open. The most notable of these unsolved issues is, (again) the question of the Hamiltonian constraint which encodes the dynamical aspects of Quantum Gravity. The physical states are obtained by solving the quantum constraint. Very little progress has been achieved in this direction.

The spin foam formalism was introduced as a way of trying to circumvent this issue. It can be understood as a *sum over histories* type of formalism. Or alternatively, a way to make sense of Feynman's path integral formalism in the context of Quantum Gravity, and calculate transition amplitudes between boundary states. However, these spin foam models have unsolved problems of their own. Among these questions are the semiclassical limit, and the coupling of fermionic matter. Both of these can, in fact, be related to the absence of the tetrads in the topological sector of the action.

In an effort to tackle these problems the *BFCG* theory (6.44) was introduced. It is a categorical generalization (see chapter 6) of the *BF* model (2.146). Just as the *BF* theory is the gauge theory of some Lie group $G$, the *BFCG* (or 2-*BF*) theory is the *higher gauge theory* for some Lie 2-group $\mathcal{G}$.

We preformed the canonical analysis of the *BFCG* theory for a generic Lie 2-group in chapter 7. This analysis (see also [112]) implies that the *BFCG* action (7.13) is a topological field theory, i.e. an action which is diffeomorphism invariant and has no local degrees of freedom. The propagating degrees of freedom are global and the corresponding configuration space is the moduli space of flat 2-connections for the *BFCG* Lie 2-group on the spatial manifold $\Sigma$.

In chapter 8 we studied the Hamiltonian structure of the *BFCG* model (8.4), and its relationship to the topological Poincaré gauge theory (8.8)(see [113, 114]). In section 8.2 we defined both theories, proved the equivalence of their respective actions and Lagrange equations of motion, and introduced a generalized action (8.9) which depends on a real parameter $\xi$ and which reduces to the *BFCG* theory for $\xi = 1$, while it reduces to the topological Poincaré gauge theory for $\xi = 0$. In this way, we could perform the Hamiltonian analysis of both theories simultaneously. Because of this action (8.9) we can say that, although chapter 7 is more general then chapter 8 in the sense that we use a generic Lie 2-group in the former and the Poincaré 2-group in the latter, it is also possible to say that chapter 8 generalizes chapter 7 in a different direction, because the action we used (8.9) is an interpolation between the *BFCG* theory and a Poincaré gauge theory.

The canonical analysis of (8.9) was done in section 8.3, where it was established that (under the assumptions we made) there are no physical (local) degrees of freedom, equation (8.47), the algebra of constraints (8.38), (8.39) and (8.40) has been computed, and the Hamiltonian of the theory written as a linear combination of first class constraints, equation (8.33). In section 8.4 we have constructed the Dirac brackets (8.66), which facilitate the elimination of the second class constraints from the theory without breaking its gauge symmetry. The geometrical consequences of this were explored in section 8.5. The elimination of second class constraints projects the phase space onto one of its hypersurfaces, depending on the choice of the parameter $\xi$. Only at this point the difference between the *BFCG* theory and the topological Poincaré gauge theory becomes observable, since they live on distinct hypersurfaces. For the generic hypersurface, when $\xi \neq 0, 1$, the second class constraints can be solved in terms of different sets of variables, and for every choice of $\xi$ there is a canonical transformation which maps between these, mapping the hypersurface $\xi$ onto itself. However, this was not possible for the singular cases $\xi = 0, 1$. Instead, in these two cases there exists a singular canonical transformation which maps the $\xi = 0$ hypersurface to $\xi = 1$ hypersurface and vice versa, establishing the equivalence between the *BFCG* theory and the topological Poincaré gauge theory, and clarifying the relationship between their respective variables.

The results obtained in Chapter 8 (published in [114]) represent the straightforward generalization of those obtained in [78]. The analysis of *BF* theory based on the Lorentz group is generalized by the *BFCG* theory based on the Poincaré 2-group. Some of the material presented overlaps with [113], but also improves on those previous results, in the following important ways. First, in this Chapter we have performed the full Hamiltonian analysis, as opposed to the shorthand procedure used in [113] and described in section 3.11 . This facilitates a better basis for the ongoing analysis of the constrained *BFCG* theory (8.3) (see appendix D), whose relevance is very high since it is equivalent to GR. Second, the procedure used in [113] was performed by employing a partial gauge fixing. This makes the calculations much simpler, but prevents us from computing the full algebra of constraints, in particular (8.39) and (8.40). In this chapter the gauge symmetry was kept intact, the full algebra of constraints has been computed, and proved to be closed. And third, in this chapter we have given a more detailed analysis of the relationship between the *BFCG* model and the topological Poincaré gauge theory, providing a better insight into the geometry of the reduced (local) phase spaces for both theories.

The quantization of the *BFCG* topological theory ((8.9) with $\xi = 1$) may be carried out in the *BF* formulation (see section 8.6). In this case the physical Hilbert space is given by the space of square-integrable functions on the moduli space of flat connections. To proceed further on must introduce the spin-network basis, by constructing the spin-network wave functions for the Poincaré group. An inter-esting problem will be to investigate the relation between the spin-network basis and the spin-foam basis.

As far as the canonical quantization of GR in the spin-foam basis is concerned, this requires a canonical formulation of the constrained *BFCG* theory based on the 2-connection variables $(\omega, \beta)$ and their momenta. This in work in progress, which we report in appendix D. However, the structure of the GR constraints is such that the short-cut procedure based on the space-time decomposition of the fields in the action does not work, and one has to perform the full Dirac procedure.

We present here the Hamiltonian written in terms of the first class constraint of the constrained theory. It reads,

$$H_T = \int d^3\vec{x} \left[ \frac{1}{2}\lambda(\omega)^{ab}{}_0\, \Phi(\omega)_{ab} + \lambda(e)^a{}_0\, \Phi(e)_a + \frac{1}{2}\omega^{ab}{}_0\, \Phi(T)_{ab} + e^a{}_0\, \Phi(R)_a \right] . \quad (9.1)$$

Where the first class constraint are given by,

$$\begin{aligned}
\Phi(\omega)^{ab} &= P(\omega)^{ab0} , \\
\Phi(e)_a &= P(e)_a{}^0 + \frac{1}{2} R^{cd}{}_{ij} F^{fbij}{}_{acdk} P(\varphi)_{fb}{}^{0k} + \varepsilon_{abcd} e^b{}_k P(B)^{cd0k} , \\
\Phi(T)^{ab} &= 4\varepsilon^{abcd} e_{ci} S(T)_d{}^i - \nabla_i S(Bee)^{abi} + \varepsilon^{0ijk} e^{[a}{}_i T(\beta)^{b]}{}_{jk} + 2\varepsilon^{abcd} e^f{}_i e_{cj} P(B)_{fd}{}^{ij} \\
&\quad - \nabla_i P(\omega)^{abi} + 2e^{[a}{}_i P(e)^{b]i} - R^{[ac}{}_{ij} P(\varphi)_c{}^{b]ij} , \\
\Phi(R)_a &= -S(eR)_a + R^c{}_{hij} \omega^{hd}{}_0 F^{fbij}{}_{acdk} P(\varphi)_{fb}{}^{0k} + R^{cd}{}_{ij} \frac{\partial F^{fbij}{}_{acdk}}{\partial e^h{}_m} \left( \nabla_m e^h{}_0 - \omega^h{}_{g0} e^g{}_m \right) P(\varphi)_{fb}{}^{0k} \\
&\quad + \frac{1}{2} R^{cd}{}_{ij} F^{fbij}{}_{acdk} \left[ S(Bee)_{fb}{}^k + P(\omega)_{fb}{}^k + \nabla_m P(\varphi)_{fb}{}^{km} - 2\nabla_m \left( e^e{}_0 F^{ghmk}{}_{efbn} P(\varphi)_{gh}{}^{0n} \right) \right] \\
&\quad - \varepsilon^{0ijk} \nabla_i T(\beta)_{ajk} + \varepsilon_{abcd} e^b{}_i \nabla_j P(B)^{cdij} - \nabla_i P(e)_a{}^i + \varepsilon_{abcd} \left( \nabla_k e^b{}_0 - \omega^b{}_{f0} e^f{}_k \right) P(B)^{cd0k} .
\end{aligned} \quad (9.2)$$

In this way one would generalize the LQG spin-network basis to a spin-foam basis, and we expect that the corresponding Hamiltonian constraint may be simpler to solve. The definite advantage over the LQG formalism is that one can construct a wavefunction which is a function of the triads $\tilde{e}$ and the connection $\tilde{\omega}$, so that it will be easier to perform the semi-classical analysis.

Concerning the subject of future work, several avenues may be followed. The first we wish to mention is the equivalence of the *BFCG* model with cosmological constant, to the MacDowell-Mansouri (MM) theory of Gravity [122]. This equivalence may bring important insights into the problem of the quantization of the *BFCG* theory.

Also under way, is a modified gravity theory based on the *BFCG* model. The MM theory — and consequently the *BFCG* theory, by virtue of the above mentioned equivalence — may be deformed by the addition of a 'potential' term to the action. A study of the cosmological models contained in this classical theory is in progress.

A different route for the generalization of the *BFCG* model is to study it for 2-groups other then the Poincaré 2-group. This has been done for example for the Euclidean 2-group (see e.g. [123]). Furthermore, we know from section 4.2 of [106] that, we can construct a 2-group (or equivalently a Lie crossed module) from: i) any Lie group $G$; ii) any vector space $H$; iii) the representation $\triangleright$ of $G$ on $H$ and, iv) a trivial map $\partial$. The Poicaré 2-group corresponds to the choice $G = SO(3,1)$ and $H = \mathbb{R}^4$. It would be interesting to investigate the choice of the *(anti) de Sitter group* $(SO(3,2))SO(4,1)$ for the Lie group $G$ and $H = \mathbb{R}^5$.

In conclusion, we may say that the *BFCG* theory — especially in the constrained action (8.3) — constitute an important contribute to the study of the problem of Quantum Gravity. The theory is based on a 2-group, a categorical generalization of the notion of group. It is therefore in the forefront of research in this area.

Progress in the Hamiltonian quantization of the theory is hindered by the fact that higher category theory, is still under construction especially in some aspects of representation theory e.g. the generalization to 2-categories of the Peter–Weyl theorem. However the covariant approach [124] to quantization base on the generalization of spin foams to spin cubes is certainly a promising line of research.

# A

# Lie algebra valued differential forms

Lie algebra valued differential forms have some interesting properties [125, 126]. Take $\mathfrak{g}$ to be an $n$-dimensional Lie algebra over the field $\mathbb{R}$. If $\{g_a,\ i = 1, \ldots, n\}$ is a basis for $\mathfrak{g}$, then a Lie algebra valued differential $r$-form is an element of the tensor product $\Omega^1(M) \otimes \mathfrak{g}$, on a manifold $M$. It may be written:

$$\alpha \equiv \alpha^a \otimes g_a = \frac{1}{r!}\alpha^a_{\mu_1 \ldots \mu_r} dx^{\mu_1} \wedge \ldots \wedge dx^{\mu_r} \otimes g_a \,, \tag{A.1}$$

where summation over $a$ from 1 to $n$, is implied.

The exterior derivative and star operator act in the same way as above, only in the $\alpha^a$ element of $\Omega^1(M)$ that is

$$d\alpha = d\alpha^a \otimes g_a \tag{A.2}$$

$$*\alpha = *\alpha^a \otimes g_a \,. \tag{A.3}$$

The commutator of Lie algebra valued $r$ and $s$ forms $\alpha$ and $\beta$ respectively is

$$[\alpha, \beta] \equiv (\alpha^a \wedge \beta^b) \otimes [g_a, g_b] \,, \tag{A.4}$$

where $a$ and $b$ are summed in the range $1, \ldots, n$.

This commutator has the following properties

$$\begin{aligned}
[\alpha, \beta + \gamma] &= [\alpha, \beta] + [\alpha, \gamma] \\
[\alpha, \beta] &= (-1)^{rs+1} [\beta, \alpha] \\
(-1)^{sr} [\alpha, [\beta, \gamma]] &+ (-1)^{st} [\beta, [\gamma, \alpha]] + (-1)^{rt} [\gamma, [\alpha, \beta]] = 0 \,.
\end{aligned} \tag{A.5}$$

And we have the following results

$$\begin{aligned}
[\alpha, *\alpha] &= 0 \\
[\alpha, [\alpha, \alpha]] &= 0 \\
d\,[\alpha, \beta] &= [d\alpha, \beta] + (-1)^r\,[\alpha, d\beta]\ ,
\end{aligned} \tag{A.6}$$

where $r$ is the degree of the form $\alpha$.

If we regard these forms *as matrices,* ([46, 54]) aenother product may be defined in the following way:

$$\alpha \wedge \beta = \alpha^a \wedge \beta^b (g_a g_b)\,. \tag{A.7}$$

This means that like in the case of of (A.4) we take the wedge product of the forms i.e. $\alpha^a \wedge \beta^b$ but unlike the commutator where we take the Lie bracket of the elements of the basis of he Lie algebra, we merely multiply these together using the group multiplication that is $g_a g_b$.

This is related to the commutator by

$$[\alpha, \beta] = \alpha \wedge \beta - (-1)^{rs} \beta \wedge \alpha\,, \tag{A.8}$$

where $\alpha$ is an $r$-form and, $\beta$ is an $s$-form.

The product has the properties,

$$(\alpha + \beta) \wedge (\gamma + \delta) = \alpha \wedge \gamma + \alpha \wedge \delta + \beta \wedge \gamma + \beta \wedge \delta \tag{A.9}$$

$$d(\alpha \wedge \beta) = d\alpha \wedge \beta + (-1)^r \alpha \wedge d\beta\,. \tag{A.10}$$

Given the one form $A$

$$A = A^a{}_\mu\, g_a \otimes dx^\mu\,, \tag{A.11}$$

we have,

$$\begin{aligned}
A \wedge A &= A^a \wedge A^b \otimes (g_a g_b) \\
&= \frac{1}{2} A^a \wedge A^b \otimes [g_a, g_b] \\
&= \frac{1}{2}\,[A, A]\ .
\end{aligned} \tag{A.12}$$

We can associate to $A$ a curvature two-form $F$,

$$F = dA + \frac{1}{2}\,[A, A] \tag{A.13}$$

$$= dA + A \wedge A\,, \tag{A.14}$$

and a covariant derivative,

$$\nabla_A \alpha = d\alpha + \frac{1}{2}[A, \alpha] \quad \text{(A.15)}$$

$$\nabla_A \beta = d\beta + [A, \beta] \ , \quad \text{(A.16)}$$

for a one form $\alpha$ (or any odd form in general) and for any two form $\beta$ (or any even form).

In this language we may write,

$$\nabla_A(\nabla\alpha) = \frac{1}{2}[F, \alpha] \ , \quad \text{(A.17)}$$

we also have the Bianchi identities:

$$\nabla_A F = 0 \quad \text{(A.18)}$$

$$\nabla_A T = \frac{1}{2}[F, e] \quad \text{(A.19)}$$

where the torsion is,

$$T = \nabla_A \alpha = de + \frac{1}{2}[A, e] \quad \text{(A.20)}$$

# B
# Bianchi identities in component form

Recalling the definitions of the torsion and curvature 2-forms,

$$T^a = de^a + \omega^a{}_b \wedge e^b \,, \qquad R^{ab} = d\omega^{ab} + \omega^a{}_c \wedge \omega^{cb} \,, \tag{B.1}$$

one can take the exterior derivative of $T^a$ and $R^a$, and use the property $dd \equiv 0$ to obtain the following two identities:

$$\begin{array}{rcl} \nabla T^a & \equiv & dT^a + \omega^a{}_b \wedge T^b = R^a{}_b \wedge e^b \,, \\ \nabla R^{ab} & \equiv & dR^{ab} + \omega^a{}_c \wedge R^{cb} + \omega^b{}_c \wedge R^{ac} = 0 \,. \end{array} \tag{B.2}$$

These two identities are universally valid for torsion and curvature, and are called Bianchi identities. By expanding all quantities into components as

$$T^a = \frac{1}{2} T^a{}_{\mu\nu} dx^\mu \wedge dx^\nu \,, \qquad R^{ab} = \frac{1}{2} R^{ab}{}_{\mu\nu} dx^\mu \wedge dx^\nu \,, \tag{B.3}$$

$$e^a = e^a{}_\mu dx^\mu \,, \qquad \omega^{ab} = \omega^{ab}{}_\mu dx^\mu \,, \tag{B.4}$$

and using the formula $dx^\mu \wedge dx^\nu \wedge dx^\rho \wedge dx^\sigma = \varepsilon^{\mu\nu\rho\sigma} d^4x$, one can rewrite the Bianchi identities in component form as

$$\varepsilon^{\lambda\mu\nu\rho} \left( \nabla_\mu T^a{}_{\nu\rho} - R^a{}_{b\mu\nu} e^b{}_\rho \right) = 0 \,, \tag{B.5}$$

and

$$\varepsilon^{\lambda\mu\nu\rho} \nabla_\mu R^{ab}{}_{\nu\rho} = 0 \,. \tag{B.6}$$

For the purpose of Hamiltonian analysis, one can split the Bianchi identities into those which do not feature a time derivative and those that do. The time-independent pieces are obtained by taking $\lambda = 0$ components:

$$\varepsilon^{0ijk}\left(\nabla_i T^a{}_{jk} - R^a{}_{bij}e^b{}_k\right) = 0\,, \tag{B.7}$$

$$\varepsilon^{0ijk}\nabla_i R^{ab}{}_{jk} = 0\,. \tag{B.8}$$

These identities are valid as off-shell, strong equalities for every spacelike slice in spacetime, and can be enforced in all calculations involving the Hamiltonian analysis. The time-dependent pieces are obtained by taking $\lambda = i$ components:

$$\varepsilon^{0ijk}\left(\nabla_0 T^a{}_{jk} - 2\nabla_j T^a{}_{0k} - 2R^a{}_{b0j}e^b{}_k - R^a{}_{bjk}e^b{}_0\right) = 0\,, \tag{B.9}$$

and

$$\varepsilon^{0ijk}\left(\nabla_0 R^{ab}{}_{jk} - 2\nabla_j R^{ab}{}_{0k}\right) = 0\,. \tag{B.10}$$

Due to the fact that they connect geometries of different spacelike slices in spacetime, they cannot be enforced off-shell. Instead, they can be derived from the Hamiltonian equations of motion of the theory.

In light of the Bianchi identities, we should note that the action (8.4) features two more fields, $\beta^a$ and $B^{ab}$, which also have field strengths $G^a$ and $\nabla B^{ab}$, and for which one can similarly derive Bianchi-like identities,

$$\nabla G^a = R^a{}_b \wedge \beta^b\,, \qquad \nabla^2 B^{ab} = R^a{}_c \wedge B^{cb} + R^b{}_c \wedge B^{ac}\,. \tag{B.11}$$

However, due to the fact that both $\beta^a$ and $B^{ab}$ are two-forms, in 4-dimensional spacetime these identities will be single-component equations, with no free spacetime indices,

$$\varepsilon^{\lambda\mu\nu\rho}\left(\frac{2}{3}\nabla_\lambda G^a{}_{\mu\nu\rho} - R^a{}_{b\mu\nu}\beta^b{}_{\nu\rho}\right) = 0\,, \tag{B.12}$$

and similarly for $\nabla^2 B^{ab}$. Therefore, these equations necessarily feature time derivatives of the fields, and do not have a purely spatial counterpart to (B.7) and (B.8). In this sense, like the time-dependent pieces of the Bianchi identities, they do not enforce any restrictions in the sense of the Hamiltonian analysis, but can instead be derived from the equations of motion and expressions for the Lagrange multipliers.

# C

# Palatini action in differential form notation and the Holst modification

Tn this appendix we prove the equivalence of the Palatini form of the Einstein-Hilbert action (1.39) and the differential form version of this action (2.139) which we called the Einstein-Cartan action. We take (2.139) without cosmological term, but we will comment on the $\Lambda$ later.

We begin with the well known expression for the determinant $e$ of $e^a{}_\mu$ that is,

$$\varepsilon_{abcd} e^a{}_\mu e^b{}_\nu e^c{}_\rho e^d{}_\sigma = e \varepsilon_{\mu\nu\rho\sigma} \,, \tag{C.1}$$

and contract it with $e_f{}^\lambda$ twice, we get

$$\begin{aligned} \varepsilon_{abcd} e^a{}_\mu e^b{}_\nu e^c{}_\rho e^d{}_\sigma &= e \varepsilon_{\mu\nu\rho\sigma} \\ \varepsilon_{abcd} e^a{}_\mu e^b{}_\nu &= e \varepsilon_{\mu\nu\rho\sigma} e_c{}^\rho e_d{}^\sigma \\ \varepsilon^{\mu\nu\rho\sigma} \varepsilon_{abcd} e^a{}_\mu e^b{}_\nu &= 2 e e_c{}^{[\rho} e_d{}^{\sigma]} \,. \end{aligned} \tag{C.2}$$

Where in the second line we used $e^{f'}{}_\lambda e_f{}^\lambda = \delta^{f'}_f$, and in the last line we contracted both sides of the equation with $\varepsilon^{\mu\nu\rho\sigma}$.

Note that, to find $e_f{}^\lambda$ the tetrads must be invertible, i.e.

$$\det(e^a{}_\mu) \neq 0 \,. \tag{C.3}$$

We also have from,

$$g_{\mu\nu} = e^a{}_\mu e_{a\,\nu} \,, \tag{C.4}$$

the following relation,

$$|\det(g_{\mu\nu})| = |g| = e^2 = \det(e^a{}_\mu)^2 \,. \tag{C.5}$$

Starting form the Einstein-Hilbert action, we can use (C.2) and get,

$$\begin{aligned} S &= \frac{1}{2k}\int_M d^4x\sqrt{-g}R \\ &= \frac{1}{2k}\int_M d^4x e e_a{}^\mu e_b{}^\nu R^{ab}{}_{\mu\nu} \\ &= \frac{1}{4k}\int_M d^4x \varepsilon^{\mu\nu\rho\sigma}\varepsilon_{abcd} e^a{}_\mu e^b{}_\nu R^{cd}{}_{\rho\sigma} \\ &= \frac{1}{4k}\int_M \varepsilon_{abcd} e^a \wedge e^b \wedge R^{cd} \,. \end{aligned} \tag{C.6}$$

Note that in the last line we no longer write $d^4x$ since we are integrating a 4-form.

To add a cosmological term to the action (C.6) we can use (C.1) in the form,

$$e = \frac{1}{4!}\varepsilon^{abcd}\varepsilon_{\mu\nu\rho\sigma} e^a{}_\mu e^b{}_\nu e^c{}_\rho e^d{}_\sigma \,, \tag{C.7}$$

we then have

$$2\Lambda\sqrt{|g|} = \frac{\Lambda}{6}\varepsilon_{abcd} e^a \wedge e^b \wedge e^c \wedge e^d \,. \tag{C.8}$$

which together with (C.6) proves the equivalence of (1.39) and (2.139).

To the action (C.6) topological terms may be added (refer to [127]) these terms do not change the field equations. They are,

$$\begin{array}{ll} \text{Pontryagin term} & F^{ab} \wedge F_{ab} \\ \text{Euler term} & \varepsilon_{abcd} F^{ab} \wedge F^{cd} \\ \text{Nieh Yan term} & T^a \wedge T_a - e^a \wedge e^b \wedge F_{ab} \\ \text{Holst term} & e^a \wedge e^b \wedge F_{ab} \end{array} \,. \tag{C.9}$$

## C.1 The Holst action

Using the Holst term, the last in (C.9), we can build the following action,

$$\begin{aligned} S_{Holst} &= \frac{1}{4k}\int_M \varepsilon_{abcd} e^a \wedge e^b \wedge R^{cd} - \frac{1}{2k\gamma}\int_M e^a \wedge e^b \wedge R_{ab} \\ &= \frac{1}{4k}\int_M \left(\varepsilon_{abcd} - \frac{1}{\gamma}\eta_{[a|c|}\eta_{b]d}\right) e^a \wedge e^b \wedge R^{cd} \\ &= \frac{1}{4k}\int_M P_{abcd}\, e^a \wedge e^b \wedge R^{cd} \end{aligned} \tag{C.10}$$

this is called the Holst action.

Like in the case of (2.140) (for vanishing $\Lambda$) variation of (C.10) yields,

$$P_{abcd}\nabla\left(e^a \wedge e^b\right) = 0 \tag{C.11}$$

$$P_{abcd} e^a \wedge R^{bc} = 0\,. \tag{C.12}$$

The first equation above determines the Levi-Civita connection (note that $P_{abcd}$ is invertible) whereas the second equation is Einstein's equation $\varepsilon_{abcd} e^a \wedge R^{bc} = 0$ plus a term that vanishes due to symmetries of the Riemann tensor (for further details refer to [80]).

# D

# Hamiltonian analysis of the constrained *BFCG* theory

We give in this appendix, some results concerning the Hamiltonian analysis of the constrained *BFCG* theory. This work is still in progress, in collaboration with A. Miković and M. Vojinović, therefore some parts of the analysis are lacking at this point.

## D.1 Constrained *BFCG* action

The constrained *BFCG* action (8.3) for the Poincaré 2-group can be written more explicitly as

$$S_{GR} = \int_{\mathcal{M}} B_{ab} \wedge R^{ab} + e^a \wedge G_a - \varphi^{ab} \wedge \left(B_{ab} - \varepsilon_{abcd}\, e^c \wedge e^d\right) \,. \tag{D.1}$$

The variables of this action are the one-forms $e^a$, $\omega^{ab}$ and the two-forms $B^{ab}$, $\beta^a$ and $\varphi^{ab}$. The curvatures $R^{ab}$ and $G^a$ are the field strengths of the 2-connection $(\omega^{ab}, \beta^a)$,

$$R^{ab} = d\omega^{ab} + \omega^a{}_c \wedge \omega^{cb} \,, \tag{D.2}$$

$$G^a = \nabla \beta^a \equiv d\beta^a + \omega^a{}_b \wedge \beta^b \,. \tag{D.3}$$

The fields $B^{ab}$ and $e^a$ play the role of the Lagrange multipliers. Usually one would denote the latter multiplier as $C^a$, but we shall instead label it as $e^a$ since it will be interpreted as the tetrad field. Similarly, the usual notation for the connection one-form and its field strength is $A$ and $F$ respectively, but in our case they are denoted $\omega^{ab}$ and $R^{ab}$, since they are interpreted as the spin connection and the Riemann curvature two-form.

It is also convenient to introduce torsion as the field strength for the tetrad $e^a$,

$$T^a = \nabla e^a \equiv de^a + \omega^a{}_b \wedge e^b \,. \tag{D.4}$$

Then, performing a partial integration in the second term in (D.1) and using the Stokes theorem one can rewrite the action as

$$S_{PGT} = \int_{\mathcal{M}} B_{ab} \wedge R^{ab} + \beta^a \wedge T_a - \varphi^{ab} \wedge \left( B_{ab} - \varepsilon_{abcd}\, e^c \wedge e^d \right) - \int_{\partial\mathcal{M}} e^a \wedge \beta_a \,, \tag{D.5}$$

where now $B^{ab}$ and $\beta^a$ play the role of Lagrange multipliers. Aside from the immaterial boundary term, this action belongs to the class of Poincaré gauge theories (PGT). In order to fully appreciate the relationship between the two theories in the sense of the Hamiltonian analysis, let us introduce a parameter $\xi \in \mathbb{R}$ and rewrite the action as

$$S = \int_{\mathcal{M}} B_{ab} \wedge R^{ab} + \xi e^a \wedge G_a + (1-\xi)\beta^a \wedge T_a - \varphi^{ab} \wedge \left( B_{ab} - \varepsilon_{abcd}\, e^c \wedge e^d \right) \,, \tag{D.6}$$

where we have dropped the boundary term. It is obvious that the action (D.6) is a convenient interpolation between (D.1) and (D.5), to which it reduces for the choices $\xi = 1$ and $\xi = 0$, respectively. The action (D.6) will therefore be the starting point for the Hamiltonian analysis.

It is also clear that all three actions (D.1), (D.5) and (D.6) give rise to the same set of equations of motion, since these do not depend on the boundary. Taking the variation of (D.6) with respect to all the variables, one obtains

$$\delta B : \quad R^{ab} - \varphi^{ab} = 0 \,, \tag{D.7}$$

$$\delta\beta : \quad T^a = 0 \,, \tag{D.8}$$

$$\delta e : \quad G_a + 2\varepsilon_{abcd}\, \varphi^{bc} \wedge e^d = 0 \,, \tag{D.9}$$

$$\delta\omega : \quad \nabla B^{ab} - e^{[a} \wedge \beta^{b]} = 0 \,, \tag{D.10}$$

$$\delta\varphi : \quad B_{ab} - \varepsilon_{abcd}\, e^c \wedge e^d = 0 \,, \tag{D.11}$$

where the covariant exterior derivative of $B^{ab}$ is defined as

$$\nabla B^{ab} \equiv dB^{ab} + \omega^a{}_c \wedge B^{cb} + \omega^b{}_c \wedge B^{ac} \,. \tag{D.12}$$

One can simplify the equations of motion in the following way. Taking the covariant exterior derivative of (D.11) and using (D.8) one obtains $\nabla B^{ab} = 0$. Substituting this into (D.10) one further obtains $e^{[a} \wedge \beta^{b]} = 0$. Under the assumption that $\det(e^a{}_\mu) \neq 0$, it follows that $\beta^a = 0$ (see Appendix in [119] for a proof), and therefore also $G^a = 0$. As a consequence, we see that the equations of motion (D.7) – (D.11) are equivalent to the following system:

- the equation that determines the multiplier $\varphi^{ab}$ in terms of curvature,

$$\varphi^{ab} = R^{ab} , \tag{D.13}$$

- the equation that determines the multiplier $B^{ab}$ in terms of tetrads,

$$B_{ab} = \varepsilon_{abcd}\, e^c \wedge e^d , \tag{D.14}$$

- the equation that determines $\beta^a$,

$$\beta^a = 0 , \tag{D.15}$$

- the equation for the torsion,

$$T^a = 0 , \tag{D.16}$$

- and the Einstein field equation,

$$\varepsilon_{abcd} R^{bc} \wedge e^d = 0 . \tag{D.17}$$

Finally, for the convenience of the Hamiltonian analysis, we need to rewrite both the action and the equations of motion in a local coordinate frame. Choosing $dx^\mu$ as basis one-forms, we can expand the fields in the standard fashion:

$$e^a = e^a{}_\mu dx^\mu , \qquad \omega^{ab} = \omega^{ab}{}_\mu dx^\mu , \tag{D.18}$$

$$B^{ab} = \frac{1}{2} B^{ab}{}_{\mu\nu} dx^\mu \wedge dx^\nu , \qquad \beta^a = \frac{1}{2}\beta^a{}_{\mu\nu} dx^\mu \wedge dx^\nu , \qquad \varphi^{ab} = \frac{1}{2}\varphi^{ab}{}_{\mu\nu} dx^\mu \wedge dx^\nu . \tag{D.19}$$

Similarly, the field strengths for $\omega$, $e$ and $\beta$ are

$$\begin{aligned} R^{ab} &= \frac{1}{2}R^{ab}{}_{\mu\nu}dx^\mu \wedge dx^\nu \,, \\ T^a &= \frac{1}{2}T^a{}_{\mu\nu}dx^\mu \wedge dx^\nu \,, \\ G^a &= \frac{1}{6}G^a{}_{\mu\nu\rho}dx^\mu \wedge dx^\nu \wedge dx^\rho \,. \end{aligned} \tag{D.20}$$

Using the relations (D.2), (D.3) and (D.4), we can write the component equations

$$\begin{aligned} R^{ab}{}_{\mu\nu} &= \partial_\mu \omega^{ab}{}_\nu - \partial_\nu \omega^{ab}{}_\mu + \omega^a{}_{c\mu}\omega^{cb}{}_\nu - \omega^a{}_{c\nu}\omega^{cb}{}_\mu \,, \\ T^a{}_{\mu\nu} &= \partial_\mu e^a{}_\nu - \partial_\nu e^a{}_\mu + \omega^a{}_{b\mu}e^b{}_\nu - \omega^a{}_{b\nu}e^b{}_\mu \,, \\ G^a{}_{\mu\nu\rho} &= \partial_\mu \beta^a{}_{\nu\rho} + \partial_\nu \beta^a{}_{\rho\mu} + \partial_\rho \beta^a{}_{\mu\nu} + \omega^a{}_{b\mu}\beta^b{}_{\nu\rho} + \omega^a{}_{b\nu}\beta^b{}_{\rho\mu} + \omega^a{}_{b\rho}\beta^b{}_{\mu\nu} \,. \end{aligned} \tag{D.21}$$

Substituting expansions (D.18), (D.19) and (D.20) into the action, we obtain

$$S = \int_{\mathcal{M}} d^4x\, \varepsilon^{\mu\nu\rho\sigma} \left[ \frac{1}{4} B_{ab\mu\nu}R^{ab}{}_{\rho\sigma} + \frac{\xi}{6} e_{a\mu}G^a{}_{\nu\rho\sigma} + \frac{1-\xi}{4}\beta_{a\mu\nu}T^a{}_{\rho\sigma} - \frac{1}{4}\varphi^{ab}{}_{\mu\nu}\left( B_{ab\rho\sigma} - 2\varepsilon_{abcd}\, e^c{}_\rho e^d{}_\sigma \right) \right] \,. \tag{D.22}$$

Assuming that the spacetime manifold has the topology $\mathcal{M} = \Sigma \times \mathbb{R}$, where $\Sigma$ is a 3-dimensional spacelike hypersurface, from the above action we can read off the Lagrangian, which is the integral of the Lagrangian density over the hypersurface $\Sigma$:

$$L = \int_{\Sigma} d^3x\, \varepsilon^{\mu\nu\rho\sigma} \left[ \frac{1}{4} B_{ab\mu\nu}R^{ab}{}_{\rho\sigma} + \frac{\xi}{6} e_{a\mu}G^a{}_{\nu\rho\sigma} + \frac{1-\xi}{4}\beta_{a\mu\nu}T^a{}_{\rho\sigma} - \frac{1}{4}\varphi^{ab}{}_{\mu\nu}\left( B_{ab\rho\sigma} - 2\varepsilon_{abcd}\, e^c{}_\rho e^d{}_\sigma \right) \right] \,. \tag{D.23}$$

Finally, the component form of equations of motion (D.13) – (D.17) is:

$$\begin{gathered} \varphi^{ab}{}_{\mu\nu} = R^{ab}{}_{\mu\nu} \,, \qquad B_{ab\mu\nu} = 2\varepsilon_{abcd}\, e^c{}_\mu e^d{}_\nu \,, \qquad \beta^a{}_{\mu\nu} = 0 \,, \\ T^a{}_{\mu\nu} = 0 \,, \qquad \varepsilon^{\lambda\mu\nu\rho}\varepsilon_{abcd} R^{bc}{}_{\mu\nu}e^d{}_\rho = 0 \,. \end{gathered} \tag{D.24}$$

The coupling of the constrained *BFCG* theory to fermionic matter is given by the following action,

$$S_m = S + S_D + S_{\beta\psi} \,, \tag{D.25}$$

where $S$ is equation (D.6) (with $\xi = 1$ for simplicity), $S_{\beta\psi}$ reads,

$$S_{\beta\psi} = i\kappa_2 \int \varepsilon_{abcd}e^a \wedge e^b \wedge \beta^c\, \bar{\psi}\gamma_5\gamma^d\psi \,, \tag{D.26}$$

and $S_D$ was given in equation (2.143). By varying $S_m$ with respect to $B$, $e$, $\omega$, $\beta$, $\varphi$ and $\bar{\psi}$, respectively, we obtain

$$R_{ab} - \varphi_{ab} = 0\,, \tag{D.27}$$

$$\nabla\beta_a + \varepsilon_{abcd}e^b \wedge \left[2\varphi^{cd} - \frac{3i\kappa_1}{2}\beta^c\bar{\psi}\gamma_5\gamma^d\psi + 3i\kappa_1 e^c \wedge \bar{\psi}\left(\gamma^d \overset{\rightarrow}{\nabla} - \overset{\leftarrow}{\nabla}\gamma^d + \frac{im}{6}e^d\right)\psi\right] = 0\,, \tag{D.28}$$

$$\nabla B_{ab} - e_{[a} \wedge \beta_{b]} - 2\kappa_2\varepsilon_{abcd}e^c \wedge s^d = 0\,, \tag{D.29}$$

$$\nabla e_a + \kappa_2 s_a = 0\,, \tag{D.30}$$

$$B_{ab} - \varepsilon_{abcd}e^c \wedge e^d = 0\,, \tag{D.31}$$

$$i\kappa_1\varepsilon_{abcd}e^a \wedge e^b \wedge \left(2e^c \wedge \gamma^d\nabla + \frac{im}{2}e^c \wedge e^d - 3(\nabla e^c)\gamma^d - \frac{3}{4}\beta^c\gamma_5\gamma^d\right)\psi = 0\,. \tag{D.32}$$

These equations can be shown to be equivalent to the Einstein equation and the Dirac equation (see [119]).

## D.2 Hamiltonian analysis

Now we turn to the Hamiltonian analysis. A review of the general formalism can be found in Chapter 3. In addition, the equivalent procedure for the topological *BFCG* theory has been done in chapters 7 and 8.

The notation and conventions in this appendix are the same as those in chapter 8.

As a first step, we calculate the momenta $\pi$ corresponding to the field variables $B^{ab}{}_{\mu\nu}$, $\varphi^{ab}{}_{\mu\nu}$, $e^a{}_\mu$, $\omega^{ab}{}_\mu$ and $\beta^a{}_{\mu\nu}$. Differentiating the Lagrangian with respect to the time derivative of the appropriate fields, we obtain the momenta as follows:

$$
\begin{aligned}
\pi(B)_{ab}{}^{\mu\nu} &= \frac{\delta L}{\delta \partial_0 B^{ab}{}_{\mu\nu}} = 0\,, \\
\pi(\varphi)_{ab}{}^{\mu\nu} &= \frac{\delta L}{\delta \partial_0 \varphi^{ab}{}_{\mu\nu}} = 0\,, \\
\pi(e)_a{}^{\mu} &= \frac{\delta L}{\delta \partial_0 e^{a}{}_{\mu}} = \frac{1-\xi}{2}\varepsilon^{0\mu\nu\rho}\beta_{a\nu\rho}\,, \\
\pi(\omega)_{ab}{}^{\mu} &= \frac{\delta L}{\delta \partial_0 \omega^{ab}{}_{\mu}} = \varepsilon^{0\mu\nu\rho}B_{ab\nu\rho}\,, \\
\pi(\beta)_a{}^{\mu\nu} &= \frac{\delta L}{\delta \partial_0 \beta^{a}{}_{\mu\nu}} = -\xi\varepsilon^{0\mu\nu\rho}e_{a\rho}\,.
\end{aligned}
\tag{D.33}
$$

None of the momenta can be solved for the corresponding "velocities", so they all give rise to primary constraints:

$$
\begin{aligned}
P(B)_{ab}{}^{\mu\nu} &\equiv \pi(B)_{ab}{}^{\mu\nu} \approx 0\,, \\
P(\varphi)_{ab}{}^{\mu\nu} &\equiv \pi(\varphi)_{ab}{}^{\mu\nu} \approx 0\,, \\
P(e)_a{}^{\mu} &\equiv \pi(e)_a{}^{\mu} - \frac{1-\xi}{2}\varepsilon^{0\mu\nu\rho}\beta_{a\nu\rho} \approx 0\,, \\
P(\omega)_{ab}{}^{\mu} &\equiv \pi(\omega)_{ab}{}^{\mu} - \varepsilon^{0\mu\nu\rho}B_{ab\nu\rho} \approx 0\,, \\
P(\beta)_a{}^{\mu\nu} &\equiv \pi(\beta)_a{}^{\mu\nu} + \xi\varepsilon^{0\mu\nu\rho}e_{a\rho} \approx 0\,.
\end{aligned}
\tag{D.34}
$$

As usual, the weak, on-shell equality is denoted "$\approx$", as opposed to the strong, off-shell equality which is denoted by the usual symbol "$=$".

Next we introduce the fundamental simultaneous Poisson brackets between the fields and their conjugate momenta,

$$
\begin{aligned}
\{\, B^{ab}{}_{\mu\nu}\,,\, \pi(B)_{cd}{}^{\rho\sigma} \,\} &= 4\delta^{a}_{[c}\delta^{b}_{d]}\delta^{\rho}_{[\mu}\delta^{\sigma}_{\nu]}\delta^{(3)}\,, \\
\{\, \varphi^{ab}{}_{\mu\nu}\,,\, \pi(\varphi)_{cd}{}^{\rho\sigma} \,\} &= 4\delta^{a}_{[c}\delta^{b}_{d]}\delta^{\rho}_{[\mu}\delta^{\sigma}_{\nu]}\delta^{(3)}\,, \\
\{\, e^{a}{}_{\mu}\,,\, \pi(e)_{b}{}^{\nu} \,\} &= \delta^{a}_{b}\delta^{\nu}_{\mu}\delta^{(3)}\,, \\
\{\, \omega^{ab}{}_{\mu}\,,\, \pi(\omega)_{cd}{}^{\nu} \,\} &= 2\delta^{a}_{[c}\delta^{b}_{d]}\delta^{\nu}_{\mu}\delta^{(3)}\,, \\
\{\, \beta^{a}{}_{\mu\nu}\,,\, \pi(\beta)_{b}{}^{\rho\sigma} \,\} &= 2\delta^{a}_{b}\delta^{\rho}_{[\mu}\delta^{\sigma}_{\nu]}\delta^{(3)}\,,
\end{aligned}
\tag{D.35}
$$

and we employ them to calculate the algebra of primary constraints,

$$
\begin{aligned}
\{\, P(B)^{abjk}\,,\, P(\omega)_{cd}{}^{i} \,\} &= 4\varepsilon^{0ijk}\delta^{a}_{[c}\delta^{b}_{d]}\delta^{(3)}, \\
\{\, P(e)^{ak}\,,\, P(\beta)_{b}{}^{ij} \,\} &= -\varepsilon^{0ijk}\delta^{a}_{b}\delta^{(3)},
\end{aligned}
\tag{D.36}
$$

while all other Poisson brackets vanish. Note that the algebra of primary constraints is independent of $\xi$.

Next we construct the canonical, on-shell Hamiltonian:

$$\begin{aligned} H_c \;=\;& \int_\Sigma d^3\vec{x}\left[\frac{1}{4}\pi(B)_{ab}{}^{\mu\nu}\partial_0 B^{ab}{}_{\mu\nu} + \frac{1}{4}\pi(\varphi)_{ab}{}^{\mu\nu}\partial_0\varphi^{ab}{}_{\mu\nu} + \right. \\ &+ \left. \pi(e)_a{}^\mu\partial_0 e^a{}_\mu + \frac{1}{2}\pi(\omega)_{ab}{}^\mu\partial_0\omega^{ab}{}_\mu + \frac{1}{2}\pi(\beta)_a{}^{\mu\nu}\partial_0\beta^a{}_{\mu\nu}\right] - L\,. \end{aligned} \tag{D.37}$$

The factors 1/4 and 1/2 are introduced to prevent overcounting of variables. Using (D.21) and (D.23), one can rearrange the expressions such that all velocities are multiplied by primary constraints, and therefore vanish from the Hamiltonian. After some algebra, the resulting expression can be written as

$$\begin{aligned} H_c \;=\;& -\int_\Sigma d^3\vec{x}\,\varepsilon^{0ijk}\left[\frac{1}{2}B_{ab0i}\left(R^{ab}{}_{jk} - \varphi^{ab}{}_{jk}\right) + e^a{}_0\left(\frac{1}{6}G_{aijk} + \varepsilon_{abcd}\,\varphi^{bc}{}_{ij}e^d{}_k\right) + \right. \\ &+ \left. \frac{1}{2}\beta_{a0k}T^a{}_{ij} + \frac{1}{2}\omega_{ab0}\left(\nabla_i B^{ab}{}_{jk} - e^a{}_i\beta^b{}_{jk}\right) - \frac{1}{2}\varphi^{ab}{}_{0i}\left(B_{abjk} - 2\varepsilon_{abcd}\,e^c{}_j e^d{}_k\right)\right]\,, \end{aligned} \tag{D.38}$$

up to a boundary term. The canonical Hamiltonian does not depend on any momenta, but only on fields and their spatial derivatives. Also, note that it does not depend on $\xi$ either. Finally, introducing Lagrange multipliers $\lambda$ for each of the primary constraints, we construct the total, off-shell Hamiltonian:

$$\begin{aligned} H_T \;=\;& H_c + \int_\Sigma d^3\vec{x}\left[\frac{1}{4}\lambda(B)^{ab}{}_{\mu\nu}P(B)_{ab}{}^{\mu\nu} + \frac{1}{4}\lambda(\varphi)^{ab}{}_{\mu\nu}P(\varphi)_{ab}{}^{\mu\nu} + \right. \\ & \left. \lambda(e)^a{}_\mu P(e)_a{}^\mu + \frac{1}{2}\lambda(\omega)^{ab}{}_\mu P(\omega)_{ab}{}^\mu + \frac{1}{2}\lambda(\beta)^a{}_{\mu\nu}P(\beta)_a{}^{\mu\nu}\right]\,. \end{aligned} \tag{D.39}$$

We proceed with the calculation of the consistency requirements for the constraints. The consistency requirement is that the time derivative of each constraint (or equivalently its Poisson bracket with the total Hamiltonian (D.39)) must vanish on-shell. This requirement can either give rise to a new constraint, or determine some multiplier, or be satisfied identically. In our case, the consistency requiremets give rise to a complicated chain structure, depicted in the following diagram:

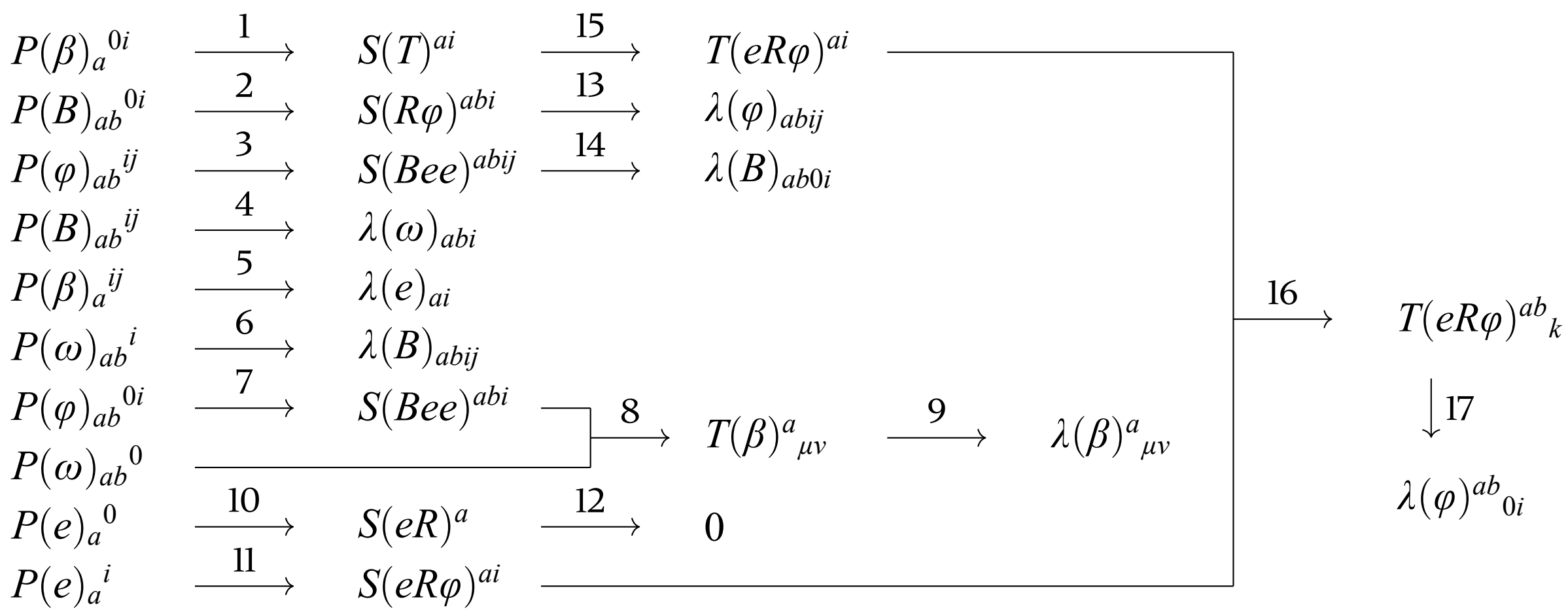

Here every arrow represents one consistency requirement, and numbers on the arrows denote the order in which we will discuss them. Steps 8 and 16 involve multiple constraints simultaneously, and will require special consideration. Primary, secondary and tertiary constraints are denoted as $P$, $S$ and $T$, respectively.

We begin by discussing consistency conditions 1–7,

$$\begin{gathered}\dot{P}(\beta)_a{}^{0i} \approx 0\,, \qquad \dot{P}(B)_{ab}{}^{0i} \approx 0\,, \qquad \dot{P}(\varphi)_{ab}{}^{ij} \approx 0\,, \qquad \dot{P}(\varphi)_{ab}{}^{0i} \approx 0\,, \\ \dot{P}(B)_{ab}{}^{ij} \approx 0\,, \qquad \dot{P}(\beta)_a{}^{ij} \approx 0\,, \qquad \dot{P}(\omega)_{ab}{}^{i} \approx 0\,. \end{gathered} \tag{D.40}$$

Calculating the corresponding Poisson brackets with the total Hamiltonian, these give rise to the following secondary constraints,

$$\begin{array}{lcl} S(T)^{ai} & \equiv & \varepsilon^{0ijk} T^a{}_{jk} \approx 0\,, \\ S(R\varphi)^{abi} & \equiv & \varepsilon^{0ijk}\left(R^{ab}{}_{jk} - \varphi^{ab}{}_{jk}\right) \approx 0\,, \\ S(Bee)^{abij} & \equiv & \varepsilon^{0ijk}\left(B^{ab}{}_{0k} - 2\varepsilon^{abcd}\, e_{c0} e_{dk}\right) \approx 0\,, \\ S(Bee)^{abi} & \equiv & \varepsilon^{0ijk}\left(B^{ab}{}_{jk} - 2\varepsilon^{abcd}\, e_{cj} e_{dk}\right) \approx 0\,, \end{array} \tag{D.41}$$

and determine the following multipliers,

$$\begin{array}{lcl} \lambda(\omega)^{ab}{}_i & \approx & \nabla_i \omega^{ab}{}_0 + \varphi^{ab}{}_{0i}\,, \\ \lambda(e)^a{}_i & \approx & \nabla_i e^a{}_0 - \omega^a{}_{b0} e^b{}_i\,, \\ \lambda(B)^{ab}{}_{ij} & \approx & 4\varepsilon^{abcd}\left(\nabla_{[i} e_{c0} - \omega_{cf0} e^f{}_{[i}\right) e_{dj]} + e^{[a}{}_0 \beta^{b]}{}_{ij} - 2 e^{[a}{}_{[i} \beta^{b]}{}_{0j]}\,. \end{array} \tag{D.42}$$

In step 8 we discuss the consistency conditions

$$\dot{S}(Bee)^{abi} \approx 0\,, \qquad \dot{P}(\omega)_{ab}{}^0 \approx 0\,, \tag{D.43}$$

simultaneously. Calculating the time derivatives, we obtain

$$\varepsilon^{0ijk}\left(e^{[a}{}_0 \beta^{b]}{}_{jk} - 2 e^{[a}{}_j \beta^{b]}{}_{0k}\right) \approx 0\,, \qquad \varepsilon^{0ijk}\, e^{[a}{}_i \beta^{b]}{}_{jk} \approx 0\,, \tag{D.44}$$

which can be jointly written as a covariant equation

$$\varepsilon^{\mu\nu\rho\sigma}\, e^{[a}{}_{\nu}\beta^{b]}{}_{\rho\sigma} \approx 0\,. \tag{D.45}$$

With the assumption that $\det(e^a{}_\mu) \neq 0$, this can be solved for $\beta^a$, giving a set of very simple tertiary constraints:

$$T(\beta)^a{}_{\mu\nu} \equiv \beta^a{}_{\mu\nu} \approx 0\,. \tag{D.46}$$

At this point we can immediately analyze the consistency step 9 as well. Taking the time derivative of (D.46), one easily determines the corresponding multipliers,

$$\lambda(\beta)^a{}_{\mu\nu} \approx 0\,. \tag{D.47}$$

Next, in steps 10 and 11, from the consistency conditions for the remaining two primary constraints,

$$\dot{P}(e)_a{}^0 \approx 0\,, \qquad \dot{P}(e)_a{}^i \approx 0\,, \tag{D.48}$$

we obtain two new secondary constraints,

$$\begin{array}{rcl} S(eR)_a & \equiv & \varepsilon^{0ijk}\varepsilon_{abcd}e^b{}_iR^{cd}{}_{jk} \approx 0\,, \\ S(eR\varphi)_a{}^i & \equiv & \varepsilon^{0ijk}\varepsilon_{abcd}\left(e^b{}_0R^{cd}{}_{jk} - 2e^b{}_j\varphi^{cd}{}_{0k}\right) \approx 0\,. \end{array} \tag{D.49}$$

In step 12 we need to discuss the consistency condition for the constraint $S(eR)_a$. After a straightforward but tedious calculation, one eventually ends up with the following expression:

$$\dot{S}(eR)_a = \nabla_i S(eR\varphi)_a{}^i + \omega^b{}_{a0}S(eR)_b + 2\varepsilon_{abcd}\varphi^{cd}{}_{0k}S(T)^{bk}\,, \tag{D.50}$$

up to terms proportional to primary constraints. Since the time derivative is already expressed as a linear combination of constraints, the consistency condition is trivially satisfied, which is denoted with a zero in the diagram above.

Moving on to steps 13, 14 and 15, the consistency conditions

$$\dot{S}(R\varphi)^{abi} \approx 0\,, \qquad \dot{S}(Bee)^{abij} \approx 0\,, \qquad \dot{S}(T)^{ai} \approx 0\,, \tag{D.51}$$

determine the multipliers

$$\begin{array}{rcl} \lambda(\varphi)^{ab}{}_{jk} & \approx & 2\omega^{[a}{}_{c0}R^{b]c}{}_{jk} + 2\nabla_{[j}\varphi^{ab}{}_{0k]}\,, \\ \lambda(B)_{ab0k} & \approx & 2\varepsilon_{abcd}\left[e^d{}_k\lambda(e)^c{}_0 - e^d{}_0\nabla_k e^c{}_0 + \omega^c{}_{f0}e^d{}_0e^f{}_k\right]\,, \end{array} \tag{D.52}$$

and another tertiary constraint

$$T(eR\varphi)^{ai} \equiv \varepsilon^{0ijk}\left(R^{ab}{}_{jk}e_{b0} + 2\varphi^{ab}{}_{0j}e_{bk}\right) \approx 0\,. \tag{D.53}$$

At this point there are only two constraints, $T(eR\varphi)^{ai}$ and $S(eR\varphi)^{ai}$, whose consistency conditions have not been discussed yet. To this end, note that these two constraints can be rewritten into a very similar form,

$$\begin{aligned} S(eR\varphi)_a{}^i &= \varepsilon_{abcd}\varepsilon^{0ijk}\left(e^b{}_0 R^{cd}{}_{jk} - 2e^b{}_j \varphi^{cd}{}_{0k}\right)\,, \\ T(eR\varphi)_a{}^i &= \eta_{ac}\eta_{bd}\varepsilon^{0ijk}\left(e^b{}_0 R^{cd}{}_{jk} - 2e^b{}_j \varphi^{cd}{}_{0k}\right)\,, \end{aligned} \tag{D.54}$$

where the identical expression in parentheses is contracted with $\varepsilon_{abcd}$ in the first constraint and with $\eta_{ac}\eta_{bd}$ in the second. This suggests that we should discuss their consistency conditions simultaneously. As suggested with the step 16 in the diagram above, we will first rewrite these 24 constaints (D.54) into a system of $18+6$ constraints (to be denoted $T(eR\varphi)_{abk}$ and $T(eR\varphi)_{jk}$ respectively) as follows. Given that the tetrad $e^a{}_\mu$ is nondegenerate, we can freely multiply the constraints with it and split the index $\mu$ into space and time components. The $\mu = 0$ part is

$$\begin{aligned} e^a{}_0 S(eR\varphi)_a{}^i &= -2\varepsilon_{abcd}\varepsilon^{0ijk} e^a{}_0 e^b{}_j \varphi^{cd}{}_{0k}\,, \\ e^a{}_0 T(eR\varphi)_a{}^i &= -2\eta_{ac}\eta_{bd}\varepsilon^{0ijk} e^a{}_0 e^b{}_j \varphi^{cd}{}_{0k}\,, \end{aligned} \tag{D.55}$$

where the curvature terms have automatically vanished, while the $\mu = m$ part is

$$\begin{aligned} e^a{}_m S(eR\varphi)_a{}^i &= e^a{}_m \varepsilon_{abcd}\varepsilon^{0ijk}\left(e^b{}_0 R^{cd}{}_{jk} - 2e^b{}_j \varphi^{cd}{}_{0k}\right)\,, \\ e^a{}_m T(eR\varphi)_a{}^i &= e^a{}_m \eta_{ac}\eta_{bd}\varepsilon^{0ijk}\left(e^b{}_0 R^{cd}{}_{jk} - 2e^b{}_j \varphi^{cd}{}_{0k}\right)\,. \end{aligned} \tag{D.56}$$

The system of 18 constraints (D.56) can be shown to be equivalent to the following constraint:

$$T(eR\varphi)^{ab}{}_k \equiv \varphi^{ab}{}_{0k} - e^f{}_0 R^{cd}{}_{ij} F^{abij}{}_{fcdk}\,, \tag{D.57}$$

where $F^{abij}{}_{fcdk}$ is a complicated function of $e^a{}_i$ only. The proof that the system (D.56) is equivalent to (D.57) is given in section D.5, and the explicit expression for $F^{abij}{}_{fcdk}$ is given in equation (D.118). Second, introducing the shorthand notation $K_{abcd} \in \{\varepsilon_{abcd}, \eta_{ac}\eta_{bd}\}$ and using (D.57), we define

$$T(eR\varphi)^i \equiv -2K_{abcd}\varepsilon^{0ijk} e^a{}_0 e^b{}_j e^f{}_0 R^{gh}{}_{mn} F^{cdmn}{}_{fghk}\,, \tag{D.58}$$

which represents a set of $3+3=6$ constraints equivalent to (D.55). However, a straightforward and meticulous (albeit very long) calculation shows that the expression (D.58) is already a linear combination of known constraints and Bianchi identities, and is thus already weakly equal to zero. Therefore, $T(eR\varphi)^i$ is not a new independent constraint, and its consistency condition is automatically satisfied.

Summing up the step 16, we have replaced the set of constraints (D.54) by an equivalent set (D.57). It thus follows that the consistency conditions for $S(eR\varphi)_a{}^i$ and $T(eR\varphi)_a{}^i$ are equivalent to the consistency condition for $T(eR\varphi)^{ab}{}_k$. Consequently, in step 17, we find that the consistency condition

$$\dot{T}(eR\varphi)^{ab}{}_k \approx 0 \tag{D.59}$$

determines the multiplier $\lambda(\varphi)^{ab}{}_{0k}$ as

$$\begin{aligned}\lambda(\varphi)^{ab}{}_{0k} &\approx \lambda(e)^f{}_0 R^{cd}{}_{ij} F^{abij}{}_{fcdk} + 2e^f{}_0 \left[ R^c{}_{hij}\omega^{hd}{}_0 + \nabla_i \varphi^{cd}{}_{0j} \right] F^{abij}{}_{fcdk} \\ &+ e^f{}_0 R^{cd}{}_{ij} \frac{\partial F^{abij}{}_{fcdk}}{\partial e^h{}_m} \left( \nabla_m e^h{}_0 - \omega^h{}_{g0} e^g{}_m \right) .\end{aligned} \tag{D.60}$$

This concludes the consistency procedure for all constraints.

Let us sum up the results of the consistency procedure. We have determined the full set of constraints and multipliers as follows: the primary constraints are

$$P(B)_{ab}{}^{\mu\nu}\,, \qquad P(\varphi)_{ab}{}^{\mu\nu}\,, \qquad P(\beta)_a{}^{\mu\nu}\,, \qquad P(\omega)_{ab}{}^{\mu}\,, \qquad P(e)_a{}^{\mu}\,, \tag{D.61}$$

the secondary constraints are

$$S(T)^{ai}\,, \qquad S(R\varphi)^{abi}\,, \qquad S(Bee)^{abij}\,, \qquad S(Bee)^{abi}\,, \qquad S(eR)^{a}\,, \tag{D.62}$$

while the tertiary constraints are

$$T(\beta)^a{}_{\mu\nu}\,, \qquad T(eR\varphi)^{ab}{}_i\,. \tag{D.63}$$

In addition, the determined multipliers are

$$\lambda(B)^{ab}{}_{\mu\nu}\,, \qquad \lambda(\varphi)^{ab}{}_{\mu\nu}\,, \qquad \lambda(\beta)^a{}_{\mu\nu}\,, \qquad \lambda(\omega)^{ab}{}_i\,, \qquad \lambda(e)^a{}_i\,, \tag{D.64}$$

while the remaining undetermined multipliers are

$$\lambda(\omega)^{ab}{}_0\,, \qquad \lambda(e)^a{}_0\,. \tag{D.65}$$

In total, there are — this counting is done *per point* — $C = 248$ constraints, and 10 undetermined multipliers, corresponding to the 10 parameters of the local Poincaré symmetry of the action.

## D.3 Local degrees of freedom

Once we have found all the constraints in the theory, we need to classify them into first and second class. While some of the second class constraints can already be read from (D.36), the classification is not easy since constraints are unique only up to linear combinations. The most efficient way to tabulate all first class constraints is to substitute all determined multipliers into the total Hamiltonian (D.39) and rewrite it in the form

$$H_T = \int d^3\vec{x}\left[\frac{1}{2}\lambda(\omega)^{ab}{}_0\,\Phi(\omega)_{ab} + \lambda(e)^a{}_0\,\Phi(e)_a + \frac{1}{2}\omega^{ab}{}_0\,\Phi(T)_{ab} + e^a{}_0\,\Phi(R)_a\right]. \tag{D.66}$$

The quantities $\Phi$ are linear combinations of constraints, but must all be first class, since the total Hamiltonian weakly commutes with all constraints. Written in terms of primary and secondary constraints, they are:

$$\begin{aligned}
\Phi(\omega)^{ab} &= P(\omega)^{ab0}\,,\\
\Phi(e)_a &= P(e)_a{}^0 + \frac{1}{2}R^{cd}{}_{ij}F^{fbij}{}_{acdk}P(\varphi)_{fb}{}^{0k} + \varepsilon_{abcd}e^b{}_kP(B)^{cd0k}\,,\\
\Phi(T)^{ab} &= 4\varepsilon^{abcd}e_{ci}S(T)_d{}^i - \nabla_iS(Bee)^{abi} + \varepsilon^{0ijk}e^{[a}{}_iT(\beta)^{b]}{}_{jk} + 2\varepsilon^{abcd}e^f{}_ie_{cj}P(B)_{fd}{}^{ij}\\
&\quad -\nabla_iP(\omega)^{abi} + 2e^{[a}{}_iP(e)^{b]i} - R^{[ac}{}_{ij}P(\varphi)_c{}^{b]ij}\,,\\
\Phi(R)_a &= -S(eR)_a + R^c{}_{hij}\omega^{hd}{}_0F^{fbij}{}_{acdk}P(\varphi)_{fb}{}^{0k} + R^{cd}{}_{ij}\frac{\partial F^{fbij}{}_{acdk}}{\partial e^h{}_m}\left(\nabla_me^h{}_0 - \omega^h{}_{g0}e^g{}_m\right)P(\varphi)_{fb}{}^{0k}\\
&\quad +\frac{1}{2}R^{cd}{}_{ij}F^{fbij}{}_{acdk}\left[S(Bee)_{fb}{}^k + P(\omega)_{fb}{}^k + \nabla_mP(\varphi)_{fb}{}^{km} - 2\nabla_m\left(e^e{}_0F^{ghmk}{}_{efbn}P(\varphi)_{gh}{}^{0n}\right)\right]\\
&\quad -\varepsilon^{0ijk}\nabla_iT(\beta)_{ajk} + \varepsilon_{abcd}e^b{}_i\nabla_jP(B)^{cdij} - \nabla_iP(e)_a{}^i + \varepsilon_{abcd}\left(\nabla_ke^b{}_0 - \omega^b{}_{f0}e^f{}_k\right)P(B)^{cd0k}\,.
\end{aligned} \tag{D.67}$$

These are the first class constraints in the theory. The remaining constraints are second class:

$$\begin{aligned}
\chi(B)_{ab}{}^{\mu\nu} &= P(B)_{ab}{}^{\mu\nu}\,, & \chi(T)^{ai} &= S(T)^{ai}\,, & &\\
\chi(\varphi)_{ab}{}^{\mu\nu} &= P(\varphi)_{ab}{}^{\mu\nu}\,, & \chi(R\varphi)^{abi} &= S(R\varphi)^{abi}\,, & \chi(\beta)^a{}_{\mu\nu} &= T(\beta)^a{}_{\mu\nu}\,,\\
\chi(\beta)_a{}^{\mu\nu} &= P(\beta)_a{}^{\mu\nu}\,, & \chi(Bee)^{abij} &= S(Bee)^{abij}\,, & \chi(eR\varphi)^{ab}{}_i &= T(eR\varphi)^{ab}{}_i\,.\\
\chi(\omega)_{ab}{}^i &= P(\omega)_{ab}{}^i\,, & \chi(Bee)^{abi} &= S(Bee)^{abi}\,, & &\\
\chi(e)_a{}^i &= P(e)_a{}^i\,, & & & &
\end{aligned} \tag{D.68}$$

Note that $\chi(\beta)_a{}^{\mu\nu}$ and $\chi(\beta)^a{}_{\mu\nu}$ are different constraints, despite similar notation. Of course, there is no possibility of confusion since we will never raise or lower spacetime indices of these constraints in the rest of this analysis. Also, note that despite the fact that there are 12 components of $\chi(T)^{ai}$, only 6 of them can be considered second class, since the other 6 are part of the first class constraint $\Phi(T)^{ab}$.

At this point we can count the physical degrees of freedom, the general formula is

(3.63). We repeat it here for definiteness:

$$n = N - F - \frac{S}{2} \,. \tag{D.69}$$

The number of independent field components[1] for each of the fundamental fields is

| $\omega^{ab}{}_{\mu}$ | $\beta^{a}{}_{\mu\nu}$ | $e^{a}{}_{\mu}$ | $B^{ab}{}_{\mu\nu}$ | $\varphi^{ab}{}_{\mu\nu}$ |
|---|---|---|---|---|
| 24 | 24 | 16 | 36 | 36 |

which gives the total $N = 136$. The number of components of the first class constraints is

| $\Phi(e)_a$ | $\Phi(\omega)_{ab}$ | $\Phi(R)^a$ | $\Phi(T)^{ab}$ |
|---|---|---|---|
| 4 | 6 | 4 | 6 |

which gives the total of $F = 20$. Similarly, the number of components for the second class constraints is

| $\chi(B)_{ab}{}^{\mu\nu}$ | $\chi(\varphi)_{ab}{}^{\mu\nu}$ | $\chi(\beta)_{a}{}^{\mu\nu}$ | $\chi(\omega)_{ab}{}^{i}$ | $\chi(e)_{a}{}^{i}$ | $\chi(T)^{ai}$ | $\chi(R\varphi)^{abi}$ | $\chi(Bee)^{abij}$ | $\chi(Bee)^{abi}$ | $\chi(\beta)^{a}{}_{\mu\nu}$ | $\chi(eR\varphi)^{a}$ |
|---|---|---|---|---|---|---|---|---|---|---|
| 36 | 36 | 24 | 18 | 12 | $12-6$ | 18 | 18 | 18 | 24 | 18 |

where we have denoted that only 6 of the total 12 components of $\chi(T)^{ai}$ are independent. Thus the total number of independent second class constraints is $S = 228$. This number can also be deduced as the difference between the previously counted total number of constraints $C = 248$ and the number of first class constraints $F = 20$.

Finally, substituting $N$, $F$ and $S$ into (D.69), we obtain:

$$n = 136 - 20 - \frac{228}{2} = 2 \,. \tag{D.70}$$

We conclude that the theory has two physical (local) degrees of freedom, as expected for general relativity. We further stress that we assumed that there is no extra dependence of the constraints. That is with the relation between $\chi(T)^{ai}$ and $\Phi(T)^{ab}$ we mentioned above the constraint are assume to be irreducible.

[1]Once again, we stress that we are counting components *per point* and consequently finding *local* degrees of freedom. In this appendix the terms *per point* and *local* wil be take to be implied.

## D.4 Inverse tetrad and metric

We perform the split of the group indices into space and time components as $a = (0, \underline{a})$, and write the tetrad $e^a{}_\mu$ as a $1+3$ matrix

$$e^a{}_\mu = \left[\begin{array}{c|c} e^0{}_0 & e^0{}_m \\ \hline e^{\underline{a}}{}_0 & e^{\underline{a}}{}_m \end{array}\right]. \tag{D.71}$$

Then the inverse tetrad $e^\mu{}_b$ can be expressed in terms of the $3D$ inverse tetrad ${}^{(3)}e^m{}_{\underline{b}}$ as

$$e^\mu{}_b = \left[\begin{array}{c|c} \dfrac{1}{\alpha} & -\dfrac{1}{\alpha}\,{}^{(3)}e^m{}_{\underline{b}}\, e^0{}_m \\ \hline -\dfrac{1}{\alpha}\,{}^{(3)}e^m{}_{\underline{a}}\, e^{\underline{a}}{}_0 & {}^{(3)}e^m{}_{\underline{b}} + \dfrac{1}{\alpha}\left({}^{(3)}e^m{}_{\underline{a}}\, e^{\underline{a}}{}_0\right)\left({}^{(3)}e^k{}_{\underline{b}}\, e^0{}_k\right) \end{array}\right], \tag{D.72}$$

where

$$\alpha \equiv e^0{}_0 - e^0{}_k\, {}^{(3)}e^k{}_{\underline{a}}\, e^{\underline{a}}{}_0 \tag{D.73}$$

is the $1 \times 1$ Schur complement [128] of the $4 \times 4$ matrix $e^a{}_\mu$. By definition, the $3D$ tetrad satisfies the identities

$$e^{\underline{a}}{}_m\, {}^{(3)}e^m{}_{\underline{b}} = \delta^{\underline{a}}_{\underline{b}}, \qquad e^{\underline{a}}{}_m\, {}^{(3)}e^n{}_{\underline{a}} = \delta^n_m. \tag{D.74}$$

In addition, if we denote $e \equiv \det e^a{}_\mu$ and ${}^{(3)}e \equiv \det e^{\underline{a}}{}_m$, the Schur complement $\alpha$ satisfies the Schur determinant formula

$$e = \alpha\, {}^{(3)}e\,, \tag{D.75}$$

which can be proved as follows.

Given any square matrix divided into blocks as

$$\Delta = \left[\begin{array}{c|c} A & B \\ \hline C & M \end{array}\right] \tag{D.76}$$

such that $A$ and $M$ are square matrices and $M$ has an inverse, we can use the Aitken block diagonalization formula [128]

$$\left[\begin{array}{c|c} I & -BM^{-1} \\ \hline & I \end{array}\right] \left[\begin{array}{c|c} A & B \\ \hline C & M \end{array}\right] \left[\begin{array}{c|c} I & \\ \hline -M^{-1}C & I \end{array}\right] = \left[\begin{array}{c|c} S & \\ \hline & M \end{array}\right], \tag{D.77}$$

where

$$S = A - BM^{-1}C \tag{D.78}$$

is called the Schur complement of the matrix $\Delta$. The Aitken formula can be written in the compact form

$$P\Delta Q = S \oplus M, \tag{D.79}$$

where $P$ and $Q$ are the above triangular matrices. Taking the determinant, we obtain

$$\det P \det \Delta \det Q = \det S \det M. \tag{D.80}$$

Since the determinant of a triangular matrix is the product of its diagonal elements, we have $\det P = \det Q = 1$, which then gives the famous Schur determinant formula:

$$\det \Delta = \det S \det M. \tag{D.81}$$

Now, performing the $1+3$ block splitting of the tetrad matrix $\Delta = [e^a{}_\mu]_{4\times 4}$, we obtain the Schur complement $S = [\alpha]_{1\times 1}$, while $M = [e^{\underline{a}}{}_m]_{3\times 3}$. The Schur determinant formula then gives

$$e = \alpha \, {}^{(3)}e \,, \tag{D.82}$$

which completes the proof.

Similarly to the tetrad, one can perform a $1+3$ split of the metric $g_{\mu\nu}$,

$$g_{\mu\nu} = \left[\begin{array}{c|c} g_{00} & g_{0j} \\ \hline g_{i0} & g_{ij} \end{array}\right]. \tag{D.83}$$

The inverse metric $g^{\mu\nu}$ can be expressed in terms of the $3D$ inverse metric ${}^{(3)}g^{ij}$ as

$$g^{\mu\nu} = \left[\begin{array}{c|c} \dfrac{1}{\beta} & -\dfrac{1}{\beta}\, {}^{(3)}g^{in} g_{0i} \\ \hline -\dfrac{1}{\beta}\, {}^{(3)}g^{mj} g_{0j} & {}^{(3)}g^{mn} + \dfrac{1}{\beta} \left({}^{(3)}g^{mj} g_{0j}\right)\left({}^{(3)}g^{in} g_{0i}\right) \end{array}\right], \tag{D.84}$$

where

$$\beta \equiv g_{00} - g_{0i} \, {}^{(3)}g^{ij} g_{0j} \tag{D.85}$$

is the $1 \times 1$ Schur complement of $g_{\mu\nu}$. By definition, the $3D$ metric satisfies the identity

$$g_{ij} \, {}^{(3)}g^{jk} = \delta_i^k. \tag{D.86}$$

In addition, if we denote $g \equiv \det g_{\mu\nu}$ and ${}^{(3)}g \equiv \det g_{ij}$, the Schur complement $\beta$ satisfies the Schur determinant formula

$$g = \beta\, {}^{(3)}g. \tag{D.87}$$

The components of the metric can of course be written in terms of the components of the tetrad,

$$g_{\mu\nu} = \eta_{ab} e^a{}_\mu e^b{}_\nu. \tag{D.88}$$

Regarding the inverse metric, the only nontrivial identity is between ${}^{(3)}g^{ij}$ and ${}^{(3)}e^i{}_{\underline{a}}$. Introducing the convenient notation $e_{\underline{a}} \equiv {}^{(3)}e^i{}_{\underline{a}}\, e^0{}_i$, it reads:

$${}^{(3)}g^{ij} = {}^{(3)}e^i{}_{\underline{a}}\, {}^{(3)}e^j{}_{\underline{b}} \left[ \eta^{\underline{ab}} + \frac{e^{\underline{a}} e^{\underline{b}}}{1 - e_{\underline{c}} e^{\underline{c}}} \right]. \tag{D.89}$$

The relationship between determinants and Schur complements is:

$$g = -e^2, \qquad {}^{(3)}g = \left({}^{(3)}e\right)^2 \left(1 - e_{\underline{a}} e^{\underline{a}}\right), \qquad \beta = \frac{\alpha^2}{e_{\underline{a}} e^{\underline{a}} - 1}. \tag{D.90}$$

Finally, there is one more useful identity,

$$g_{0j}\, {}^{(3)}g^{ij} = {}^{(3)}e^i{}_{\underline{a}}\, e^{\underline{a}}{}_0 - \frac{\alpha}{1 - e_{\underline{b}} e^{\underline{b}}}\, {}^{(3)}e^i{}_{\underline{a}}\, e^{\underline{a}}, \tag{D.91}$$

which can be easily proved with some patient calculation and the other identities above.

## D.5 Solving the system of equations

In order to show that the constraints (D.56) are equivalent to the constraint (D.57), we proceed as follows. Introducing the shorthand notation $K_{abcd} \in \{\varepsilon_{abcd}, \eta_{ac}\eta_{bd}\}$, we can rewrite (D.56) in a convenient form

$$e^a{}_m K_{abcd} \varepsilon^{0ijk} \left( e^b{}_0 R^{cd}{}_{jk} - 2 e^b{}_j \varphi^{cd}{}_{0k} \right) \approx 0\,. \tag{D.92}$$

Next we multiply it with the Levi-Civita symbol $\varepsilon_{0iln}$ in order to cancel the $\varepsilon^{0ijk}$, relabel the index $m \to i$ and obtain

$$K_{abcd} \left( e^a{}_i e^b{}_j \varphi^{cd}{}_{0k} - e^a{}_i e^b{}_k \varphi^{cd}{}_{0j} \right) \approx K_{abcd} e^a{}_i e^b{}_0 R^{cd}{}_{jk}\,. \tag{D.93}$$

The antisymmetrization in $jk$ indices can be eliminated by writing each equation three times with cyclic permutations of indices $ijk$, then adding the first two permutations and subtracting the third. This gives:

$$K_{abcd}e^a{}_i e^b{}_j \varphi^{cd}{}_{0k} \approx K_{abcd}e^a{}_0 \left[\frac{1}{2}e^b{}_k R^{cd}{}_{ij} - e^b{}_{[i}R^{cd}{}_{j]k}\right] . \tag{D.94}$$

Introducing the shorthand notation $P_{ijk}$ and $Q_{ijk}$ for the expression on the right-hand side as

$$P_{ijk} \equiv \eta_{ac}\eta_{bd}e^a{}_0 \left[\frac{1}{2}e^b{}_k R^{cd}{}_{ij} - e^b{}_{[i}R^{cd}{}_{j]k}\right] , \qquad Q_{ijk} \equiv \varepsilon_{abcd}e^a{}_0 \left[\frac{1}{2}e^b{}_k R^{cd}{}_{ij} - e^b{}_{[i}R^{cd}{}_{j]k}\right] , \tag{D.95}$$

our system can be rewritten as

$$\eta_{ac}\eta_{bd}e^a{}_i e^b{}_j \varphi^{cd}{}_{0k} \approx P_{ijk} , \qquad \varepsilon_{abcd}e^a{}_i e^b{}_j \varphi^{cd}{}_{0k} \approx Q_{ijk} . \tag{D.96}$$

This system consists of 18 equations for the 18 variables $\varphi^{ab}{}_{0k}$. We look for a solution in the form

$$\varphi^{cd}{}_{0k} = A^{cdmn}P_{mnk} + B^{cdmn}Q_{mnk} , \tag{D.97}$$

where the coefficients $A^{cdmn}$ and $B^{cdmn}$ are to be determined, for arbitrarily given values of $P_{ijk}$ and $Q_{ijk}$. Substituting (D.97) into (D.96) we obtain

$$\begin{aligned} \left[\eta_{ac}\eta_{bd}e^a{}_i e^b{}_j A^{cdmn} - \delta_i^{[m}\delta_j^{n]}\right] P_{mnk} + \left[\eta_{ac}\eta_{bd}e^a{}_i e^b{}_j B^{cdmn}\right] Q_{mnk} &\approx 0 , \\ \left[\varepsilon_{abcd}e^a{}_i e^b{}_j A^{cdmn}\right] P_{mnk} + \left[\varepsilon_{abcd}e^a{}_i e^b{}_j B^{cdmn} - \delta_i^{[m}\delta_j^{n]}\right] Q_{mnk} &\approx 0 . \end{aligned} \tag{D.98}$$

Since $P_{mnk}$ and $Q_{mnk}$ are considered arbitrary, the expressions in the brackets must vanish, giving the following equations for $A^{cdmn}$,

$$\eta_{ac}\eta_{bd}e^a{}_i e^b{}_j A^{cdmn} \approx \delta_i^{[m}\delta_j^{n]} , \qquad \varepsilon_{abcd}e^a{}_i e^b{}_j A^{cdmn} \approx 0 , \tag{D.99}$$

and for $B^{cdmn}$,

$$\eta_{ac}\eta_{bd}e^a{}_i e^b{}_j B^{cdmn} \approx 0 , \qquad \varepsilon_{abcd}e^a{}_i e^b{}_j B^{cdmn} \approx \delta_i^{[m}\delta_j^{n]} . \tag{D.100}$$

Focus first on (D.99). The first equation can be rewritten in the form

$$e_{ci}e_{dj}A^{cdmn} \approx \delta_i^{[m}\delta_j^{n]} , \tag{D.101}$$

and we want to rewrite the second equation in a similar form as well. In order to do that, we need to get rid of the Levi-Civita symbol on the left-hand side, by virtue of the identity

$$\det(e_{a\mu})\varepsilon_{abcd} = \varepsilon^{\mu\nu\rho\sigma}e_{a\mu}e_{b\nu}e_{c\rho}e_{d\sigma} . \tag{D.102}$$

Noting that $\det(e_{a\mu}) = \det(\eta_{ab}e^b{}_\mu) = -\det(e^a{}_\mu) = -e$ and introducing the metric $g_{\mu\nu} \equiv e^a{}_\mu e_{a\nu}$, we can multiply this identity with $e^a{}_i e^b{}_j$ to obtain:

$$\varepsilon_{abcd}e^a{}_i e^b{}_j = -\frac{1}{e}\varepsilon^{\mu\nu\rho\sigma}g_{\mu i}g_{\nu j}e_{c\rho}e_{d\sigma} \,. \tag{D.103}$$

Substituting this into the second equation in (D.99) gives

$$\varepsilon^{\mu\nu\rho\sigma}g_{\mu i}g_{\nu j}e_{c\rho}e_{d\sigma}A^{cdmn} \approx 0 \,. \tag{D.104}$$

Next we expand the $\rho$ and $\sigma$ indices into space and time components as $\rho = (0, k)$ and $\sigma = (0, l)$ to obtain

$$2\varepsilon^{\mu\nu 0l}g_{\mu i}g_{\nu j}e_{c0}e_{dl}A^{cdmn} + \varepsilon^{\mu\nu kl}g_{\mu i}g_{\nu j}e_{ck}e_{dl}A^{cdmn} \approx 0 \,. \tag{D.105}$$

The second term on the left can be evaluated using (D.101), which gives:

$$2\varepsilon^{\mu\nu 0l}g_{\mu i}g_{\nu j}e_{c0}e_{dl}A^{cdmn} + \varepsilon^{\mu\nu mn}g_{\mu i}g_{\nu j} \approx 0 \,. \tag{D.106}$$

The Levi-Civita symbol in the first term is nonzero only if $\mu\nu$ are spatial indices, so we can write

$$2\varepsilon^{rs0l}g_{ri}g_{sj}e_{c0}e_{dl}A^{cdmn} + \varepsilon^{\mu\nu mn}g_{\mu i}g_{\nu j} \approx 0 \,. \tag{D.107}$$

At this point we need to introduce $3D$ inverse metric, ${}^{(3)}g^{ij}$, and to split the group indices into $3+1$ form $a = (0, \underline{a})$, see section D.4. Multiplying (D.107) with two inverse spatial metrics and another Levi-Civita symbol, we can finally rewrite it as:

$$e_{c0}e_{di}A^{cdmn} \approx g_{0j}\,{}^{(3)}g^{j[m}\delta^{n]}_i \,. \tag{D.108}$$

The goal of all these transformations was to rewrite the system (D.99) into the form

$$e_{ci}e_{dj}A^{cdmn} \approx \delta^{[m}_i\delta^{n]}_j \,, \qquad e_{c0}e_{di}A^{cdmn} \approx g_{0j}\,{}^{(3)}g^{j[m}\delta^{n]}_i \,. \tag{D.109}$$

At this point we can expand the group indices on the left-hand side into $3+1$ form, to obtain:

$$e_{\underline{c}i}e_{\underline{d}j}A^{\underline{c}\underline{d}mn} + \left(e^0{}_j e_{\underline{d}i} - e^0{}_i e_{\underline{d}j}\right)A^{0\underline{d}mn} \approx \delta^{[m}_i\delta^{n]}_j \,, \tag{D.110}$$

$$e_{\underline{c}0}e_{\underline{d}j}A^{\underline{c}\underline{d}mn} + \left(e^0{}_j e_{\underline{d}0} - e^0{}_0 e_{\underline{d}j}\right)A^{0\underline{d}mn} \approx g_{0k}\,{}^{(3)}g^{k[m}\delta^{n]}_j \,. \tag{D.111}$$

Now we multiply (D.110) with ${}^{(3)}e^i{}_{\underline{a}}\, e^{\underline{a}}{}_0$ and subtract it from (D.111). The first terms on the left cancel, and (D.111) becomes

$$-\alpha e_{\underline{d}j}A^{0\underline{d}mn} \approx g_{0k}\,{}^{(3)}g^{k[m}\delta^{n]}_j - {}^{(3)}e^{[m}{}_{\underline{a}}\,\delta^{n]}_j e^{\underline{a}}{}_0 \,, \tag{D.112}$$

where $\alpha$ is the $1 \times 1$ Schur complement matrix of the tetrad $e^a{}_\mu$ (see section D.4). Multiplying with another inverse $3D$ tetrad and using the identity (D.91), we finally obtain the first half of the coefficients $A$:

$$A^{0\underline{a}mn} \approx \frac{1}{1 - e^{\underline{c}} e_{\underline{c}}} \, {}^{(3)}e^{[m}{}_{\underline{d}} \, {}^{(3)}e^{n]\underline{a}} \, e^{\underline{d}} \, . \tag{D.113}$$

Finally, substituting this back into (D.110) and multiplying with two more inverse $3D$ tetrads we obtain the second half of the coefficients $A$:

$$A^{\underline{a}\underline{b}mn} \approx {}^{(3)}e^{[m\underline{a}} \, {}^{(3)}e^{n]\underline{b}} + \frac{e^{\underline{d}}}{1 - e_{\underline{c}} e^{\underline{c}}} \left[ e^{\underline{a}} \, {}^{(3)}e^{[m}{}_{\underline{d}} \, {}^{(3)}e^{n]\underline{b}} - e^{\underline{b}} \, {}^{(3)}e^{[m}{}_{\underline{d}} \, {}^{(3)}e^{n]\underline{a}} \right] . \tag{D.114}$$

Next we turn to the system (D.100) for coefficients $B$. The method to solve it is completely analogous to the above method of solving (D.99), and we will not repeat all the steps, but rather only quote the final result:

$$B^{0\underline{b}mn} \approx \frac{1}{4} \varepsilon^{0\underline{b}\underline{c}\underline{d}} \left[ {}^{(3)}e^{m}{}_{\underline{c}} \, {}^{(3)}e^{n}{}_{\underline{d}} + 2 \, {}^{(3)}e^{[m}{}_{\underline{a}} \, {}^{(3)}e^{n]}{}_{\underline{d}} \frac{e^{\underline{a}} e_{\underline{c}}}{1 - e_{\underline{f}} e^{\underline{f}}} \right] , \tag{D.115}$$

and

$$B^{\underline{a}\underline{b}mn} \approx \frac{1}{2} \frac{1}{1 - e_{\underline{f}} e^{\underline{f}}} \varepsilon^{0\underline{a}\underline{b}\underline{c}} \, {}^{(3)}e^{[m}{}_{\underline{c}} \, {}^{(3)}e^{n]}{}_{\underline{d}} \, e^{\underline{d}} \, . \tag{D.116}$$

To conclude, by determining the $A$ and $B$ coefficients in (D.97) we have managed to solve the original system of equations (D.92) for $\varphi^{ab}{}_{0k}$. Substituting (D.95) into (D.97) the expression for $\varphi^{ab}{}_{0k}$ can be arranged into the form

$$\varphi^{ab}{}_{0k} \approx e^{f}{}_{0} R^{cd}{}_{mn} F^{abmn}{}_{fcdk} \, , \tag{D.117}$$

where

$$F^{abmn}{}_{fcdk} \equiv \frac{1}{2} \left[ A^{abmn} \eta_{fc} e_{dk} - 2 A^{abim} \eta_{fc} e_{di} \delta^{n}_{k} + B^{abmn} \varepsilon_{fhcd} e^{h}{}_{k} - 2 B^{abim} \varepsilon_{fhcd} e^{h}{}_{i} \delta^{n}_{k} \right] , \tag{D.118}$$

and coefficients $A$ and $B$ are specified by (D.113), (D.114), (D.115) and (D.116). Note that (D.118) depends only on $e^a{}_i$ components of the metric (in a very complicated way), while the dependence of $\varphi^{ab}{}_{0k}$ on $e^a{}_0$ and $\omega^{ab}{}_i$ is factored out in (D.117).

## D.6 Levi-Civita identity

The identity for the Levi-Civita symbol in 4 dimensions used in this appendix is:

$$A_{[a} \varepsilon_{b]cdf} C^c D^d F^f = -\frac{1}{2} \varepsilon_{abcd} A_f \left[ C^d D^f F^c + C^c D^d F^f + C^f D^c F^d \right] . \tag{D.119}$$

The proof goes as follows. Denote the left-hand side of the identity as

$$K_{ab} \equiv A_{[a}\varepsilon_{b]cdf}C^c D^d F^f \tag{D.120}$$

and take the dual to obtain:

$$\varepsilon^{aba'b'}K_{ab} = \varepsilon^{aba'b'}\varepsilon_{bcdf}A_a C^c D^d F^f. \tag{D.121}$$

Next expand the product of two Levi-Civita symbols into Kronecker deltas and use them to contract the vectors $A$, $C$, $D$ and $F$:

$$\varepsilon^{aba'b'}K_{ab} = 2\left[(A\cdot D)F^{[a'}C^{b']} + (A\cdot F)C^{[a'}D^{b']} + (A\cdot C)D^{[a'}F^{b']}\right]. \tag{D.122}$$

Now take the dual again, i.e. contract with $\varepsilon_{a'b'cd}$ to obtain

$$-4K_{cd} = \varepsilon_{a'b'cd}\varepsilon^{aba'b'}K_{ab} = 2\varepsilon_{a'b'cd}\left[(A\cdot D)F^{[a'}C^{b']} + (A\cdot F)C^{[a'}D^{b']} + (A\cdot C)D^{[a'}F^{b']}\right]. \tag{D.123}$$

Finally, multiply by $-1/4$ and relabel the indices to obtain

$$K_{ab} = -\frac{1}{2}\varepsilon_{abcd}A_f\left[C^d D^f F^c + C^c D^d F^f + C^f D^c F^d\right], \tag{D.124}$$

which proves the identity.